\documentclass{aa}

\usepackage{graphicx}
\usepackage{amsmath}
\usepackage[varg]{txfonts}
\usepackage{natbib}
\usepackage{booktabs}
\usepackage{hyperref}
\usepackage{orcidlink}
\usepackage{placeins}
\bibpunct{(}{)}{;}{a}{}{,}

\hypersetup{
    colorlinks=true,
    linkcolor=blue,
    citecolor=blue,
    urlcolor=blue
}

\newcommand{\kms}{\mbox{km\,s$^{-1}$}}
\newcommand{\sUVW}{\sigma_{\mathrm{UVW}}}

\newcommand{\LHa}{L_{\mathrm{H}\alpha}/L_{\mathrm{bol}}}
\newcommand{\EW}{\mathrm{EW}}
\newcommand{\chiH}{\chi_{\mathrm{H}\alpha}}
\newcommand{\um}{\mbox{$\mu$m}}
\usepackage{xcolor}

\AtBeginDocument{\nolinenumbers}
\let\linenumbers\relax
\begin{document}
\nolinenumbers

\newcommand{\Lbin}{${\geq}$L3}
\title{\emph{Euclid} DR1 Legacy Science on Ultracool Dwarfs: H$\alpha$ activity, kinematic age and rotation of M7--L1 dwarfs from DESI DR1, SPHEREx and \emph{Euclid}}
\titlerunning{H$\alpha$ activity, kinematic age and rotation of M7--L1 dwarfs}

\author{
    Ramarao Tata\orcidlink{0009-0005-1055-5848}\inst{1}
    \and
    Eduardo L. Mart\'in\inst{2}
    \and
    Herv\'e Bouy\orcidlink{0000-0002-7084-487X}\inst{3}
    \and
    Adam J. Burgasser\orcidlink{0000-0002-6523-9536}\inst{4}
    \and
    Nicolas Lodieu\orcidlink{0000-0002-3612-8968}\inst{2}
    \and
    Basmah Riaz\orcidlink{0000-0003-3863-4052}\inst{5}
    \and
    N.~Phan-Bao\orcidlink{0000-0002-4481-0094}\inst{6,7}
    \and
    Nikola Vitas\inst{2}
    \and
    Nafise Sedighi\orcidlink{0009-0001-9574-8585}\inst{2}
}
\authorrunning{R. Tata et al.}

\institute{
    Department of Physics and Astronomy, Ohio University, Athens, OH 45701, USA\\
    \email{tatar@ohio.edu}
    \and
    Instituto de Astrof\'isica de Canarias, E-38205 La Laguna, Tenerife, Spain
    \and
    Laboratoire d'Astrophysique de Bordeaux, Univ. Bordeaux, CNRS, B18N,
    all\'ee Geoffroy Saint-Hilaire, 33615 Pessac, France
    \and
    Department of Astronomy \& Astrophysics, University of California San Diego, La Jolla, CA 92093, USA
    \and
    Universit\"ats-Sternwarte M\"unchen, Ludwig-Maximilians-Universit\"at M\"unchen, Scheinerstra\ss e 1, 81679 M\"unchen, Germany
    \and
    Department of Physics, International University, VNU-HCM, Ho Chi Minh City, Vietnam
    \and
    Viet Nam National University Ho Chi Minh City, Vietnam
}

\date{Received --; accepted --}

\newcommand{\sxNparent}{10\,786}
\newcommand{\sxNsample}{8939}
\newcommand{\sxNmsix}{1815}
\newcommand{\sxNext}{10\,786}
\newcommand{\sxNdet}{9618}
\newcommand{\sxNextSample}{8939}
\newcommand{\sxNdetSample}{8101}
\newcommand{\sxDetPc}{89}
\newcommand{\sxDetBrightPc}{98}
\newcommand{\sxDetFaintPc}{77}
\newcommand{\sxSnrBright}{49}
\newcommand{\sxSnrFaint}{5}
\newcommand{\sxNfaintExt}{4483}
\newcommand{\sxNfaint}{4483}
\newcommand{\sxRJ}{0.85}
\newcommand{\sxRH}{0.86}
\newcommand{\sxRK}{0.86}
\newcommand{\sxRWone}{0.91}
\newcommand{\sxRWtwo}{0.87}
\newcommand{\sxRmad}{0.07}
\newcommand{\sxRn}{3166}
\newcommand{\sxRnHi}{3919}
\newcommand{\sxScaleTwoM}{0.865}
\newcommand{\sxScaleWone}{0.912}
\newcommand{\sxScaleN}{1735}
\newcommand{\sxNcompared}{10\,637}
\newcommand{\sxNmaterial}{6050}
\newcommand{\sxNadded}{4757}
\newcommand{\sxNown}{1584}
\newcommand{\sxNoverlay}{6374}
\newcommand{\sxNsat}{8}
\newcommand{\sxNnb}{168}
\newcommand{\sxNdup}{10}
\newcommand{\sxNunmod}{1032}
\newcommand{\sxNzgal}{7}
\newcommand{\sxNqa}{10\,786}
\newcommand{\sxNabsz}{530}
\newcommand{\sxAbszLpc}{100}
\newcommand{\sxNnotSearched}{102}
\newcommand{\sxNclean}{7705}
\newcommand{\sxOneFramePc}{52}
\newcommand{\sxScatFormal}{2.6}
\newcommand{\sxFbolN}{44}
\newcommand{\sxFbolMed}{\ensuremath{-0.041}}
\newcommand{\sxFbolMad}{0.09}
\newcommand{\sxFbolMedOld}{\ensuremath{-0.060}}
\newcommand{\sxFbolShift}{0.0002}
\newcommand{\sxFbolShiftHi}{0.001}
\newcommand{\sxObsMed}{100}
\newcommand{\sxObsLo}{92}
\newcommand{\sxObsHi}{116}
\newcommand{\sxNlbol}{3634}
\newcommand{\sxLoffCorrLo}{\ensuremath{+0.15}}
\newcommand{\sxLoffCorrHi}{\ensuremath{+0.24}}
\newcommand{\sxLoffTmplLo}{\ensuremath{+0.19}}
\newcommand{\sxLoffTmplHi}{\ensuremath{+0.44}}
\newcommand{\sxMJCorrLo}{\ensuremath{-0.65}}
\newcommand{\sxMJCorrHi}{\ensuremath{-0.51}}
\newcommand{\sxMJTmplLo}{\ensuremath{-1.09}}
\newcommand{\sxMJTmplHi}{\ensuremath{-0.60}}
\newcommand{\sxHtwoOseven}{0.85}
\newcommand{\sxHtwoOnine}{0.79}
\newcommand{\sxHtwoOLone}{0.64}
\newcommand{\sxHtwoOLsix}{0.50}
\newcommand{\sxHtwoOrho}{\ensuremath{-0.38}}
\newcommand{\sxHtwoOn}{4517}
\newcommand{\sxHtwoOninerho}{\ensuremath{-0.29}}
\newcommand{\sxHtwoOtwosevenrho}{\ensuremath{-0.35}}
\newcommand{\sxCOrho}{\ensuremath{-0.09}}
\newcommand{\sxCOfourrho}{\ensuremath{-0.30}}
\newcommand{\sxCOfourMseven}{0.62}
\newcommand{\sxCOfourLone}{0.55}
\newcommand{\sxCOfourBTMseven}{0.58}
\newcommand{\sxCOfourBTLone}{0.33}
\newcommand{\sxKJseven}{1.04}
\newcommand{\sxKJnine}{1.19}
\newcommand{\sxKJLone}{1.69}
\newcommand{\sxKJLsix}{2.27}
\newcommand{\sxKJrho}{0.49}
\newcommand{\sxCHlo}{0.98}
\newcommand{\sxCHhi}{1.01}
\newcommand{\sxCHLsix}{0.87}
\newcommand{\sxCHLone}{0.96}
\newcommand{\sxCHLoneBT}{0.92}
\newcommand{\sxCOfourLsix}{0.41}
\newcommand{\sxCOfourBTLsix}{0.35}
\newcommand{\sxHtwoOBTnine}{0.70}
\newcommand{\sxHtwoOBTLone}{0.53}
\newcommand{\sxNsearchedTypes}{M6--L}
\newcommand{\sxCHLsixN}{7}
\newcommand{\sxCHLsixBT}{0.96}
\newcommand{\sxCHrho}{\ensuremath{-0.10}}
\newcommand{\sxCHp}{0.00}
\newcommand{\sxRbtLlo}{1.23}
\newcommand{\sxRbtLhi}{1.94}
\newcommand{\sxRbtMseven}{1.14}
\newcommand{\sxRsoLlo}{0.82}
\newcommand{\sxRsoLhi}{0.92}
\newcommand{\sxRbtcLlo}{1.18}
\newcommand{\sxRbtcLhi}{2.10}
\newcommand{\sxMJpubLo}{\ensuremath{-0.63}}
\newcommand{\sxMJpubHi}{\ensuremath{-0.44}}
\newcommand{\sxMJpubN}{606}
\newcommand{\sxMJvCorrLo}{\ensuremath{-0.61}}
\newcommand{\sxMJvCorrHi}{\ensuremath{-0.43}}
\newcommand{\sxMJvCorrN}{108}
\newcommand{\sxMJvTmplLo}{\ensuremath{-0.85}}
\newcommand{\sxMJvTmplHi}{\ensuremath{-0.53}}
\newcommand{\sxMJvTmplN}{134}
\newcommand{\sxSigLo}{0.04}
\newcommand{\sxSigHi}{0.07}
\newcommand{\sxSpreadLo}{0.08}
\newcommand{\sxSpreadHi}{0.15}
\newcommand{\sxNullStdLo}{0.98}
\newcommand{\sxNullStdHi}{1.20}
\newcommand{\sxNullMadLo}{0.86}
\newcommand{\sxNullMadHi}{1.06}
\newcommand{\sxNullMeanLo}{\ensuremath{-0.15}}
\newcommand{\sxNullMeanHi}{0.06}
\newcommand{\sxNullN}{32\,418}
\newcommand{\sxInjMax}{0.36}
\newcommand{\sxXsearched}{7603}
\newcommand{\sxXref}{4737}
\newcommand{\sxXfaThree}{0.55}
\newcommand{\sxXmeasured}{7317}
\newcommand{\sxXexpThree}{40}
\newcommand{\sxXcand}{0}
\newcommand{\sxXtent}{22}
\newcommand{\sxXnullThree}{0.9}
\newcommand{\sxXsigma}{1.00}
\newcommand{\sxXflagged}{18}
\newcommand{\sxXgal}{7}
\newcommand{\sxXjointRef}{4259}
\newcommand{\sxXjointPass}{12}
\newcommand{\sxXjointExp}{21}
\newcommand{\sxXjointHalfN}{2124}
\newcommand{\sxXjointHalfPass}{7}
\newcommand{\sxXjointHalfExp}{24}
\newcommand{\sxAn}{9}
\newcommand{\sxAblend}{5}
\newcommand{\sxAnot}{2}
\newcommand{\sxAunclear}{2}
\newcommand{\sxAstill}{none}
\newcommand{\sxYn}{14}
\newcommand{\sxYmed}{1.1}
\newcommand{\sxYfield}{0.2}
\newcommand{\sxYmax}{6.3}
\newcommand{\sxNindex}{7229}
\newcommand{\sxNnorm}{3741}
\newcommand{\sxStatusN}{10\,786}

\newcommand{\rfrac}[1]{\csname rfrac#1\endcsname}
\newcommand{\rcountfrac}[1]{\csname rcountfrac#1\endcsname}
\expandafter\def\csname rfractype_within_one_old\endcsname{\ensuremath{97.7_{-0.7}^{+0.6}\%}}
\expandafter\def\csname rcountfractype_within_one_old\endcsname{\shortstack{516/528\\\ensuremath{97.7_{-0.7}^{+0.6}\%}}}
\expandafter\def\csname rfracM9_exact_old\endcsname{\ensuremath{88.8_{-3.8}^{+2.9}\%}}
\expandafter\def\csname rcountfracM9_exact_old\endcsname{\shortstack{79/89\\\ensuremath{88.8_{-3.8}^{+2.9}\%}}}
\expandafter\def\csname rfractype_within_one_corrected\endcsname{\ensuremath{97.3_{-0.8}^{+0.6}\%}}
\expandafter\def\csname rcountfractype_within_one_corrected\endcsname{\shortstack{514/528\\\ensuremath{97.3_{-0.8}^{+0.6}\%}}}
\expandafter\def\csname rfracM9_exact_corrected\endcsname{\ensuremath{52.8_{-5.3}^{+5.2}\%}}
\expandafter\def\csname rcountfracM9_exact_corrected\endcsname{\shortstack{47/89\\\ensuremath{52.8_{-5.3}^{+5.2}\%}}}
\expandafter\def\csname rfraclocus_old_M7\endcsname{\ensuremath{67.7_{-1.2}^{+1.2}\%}}
\expandafter\def\csname rcountfraclocus_old_M7\endcsname{\shortstack{993/1466\\\ensuremath{67.7_{-1.2}^{+1.2}\%}}}
\expandafter\def\csname rfraclocus_old_M8\endcsname{\ensuremath{41.9_{-1.2}^{+1.2}\%}}
\expandafter\def\csname rcountfraclocus_old_M8\endcsname{\shortstack{738/1760\\\ensuremath{41.9_{-1.2}^{+1.2}\%}}}
\expandafter\def\csname rfraclocus_old_M9\endcsname{\ensuremath{35.9_{-1.6}^{+1.7}\%}}
\expandafter\def\csname rcountfraclocus_old_M9\endcsname{\shortstack{294/820\\\ensuremath{35.9_{-1.6}^{+1.7}\%}}}
\expandafter\def\csname rfraclocus_old_L0\endcsname{\ensuremath{8.9_{-3.4}^{+5.1}\%}}
\expandafter\def\csname rcountfraclocus_old_L0\endcsname{\shortstack{4/45\\\ensuremath{8.9_{-3.4}^{+5.1}\%}}}
\expandafter\def\csname rfraclocus_old_L1\endcsname{\ensuremath{11.8_{-4.4}^{+6.6}\%}}
\expandafter\def\csname rcountfraclocus_old_L1\endcsname{\shortstack{4/34\\\ensuremath{11.8_{-4.4}^{+6.6}\%}}}
\expandafter\def\csname rfraclocus_corrected_M6\endcsname{\ensuremath{10.8_{-0.9}^{+1.0}\%}}
\expandafter\def\csname rcountfraclocus_corrected_M6\endcsname{\shortstack{119/1098\\\ensuremath{10.8_{-0.9}^{+1.0}\%}}}
\expandafter\def\csname rfraclocus_corrected_M7\endcsname{\ensuremath{8.2_{-0.7}^{+0.7}\%}}
\expandafter\def\csname rcountfraclocus_corrected_M7\endcsname{\shortstack{121/1470\\\ensuremath{8.2_{-0.7}^{+0.7}\%}}}
\expandafter\def\csname rfraclocus_corrected_M8\endcsname{\ensuremath{4.7_{-0.6}^{+0.7}\%}}
\expandafter\def\csname rcountfraclocus_corrected_M8\endcsname{\shortstack{47/992\\\ensuremath{4.7_{-0.6}^{+0.7}\%}}}
\expandafter\def\csname rfraclocus_corrected_M9\endcsname{\ensuremath{22.5_{-2.1}^{+2.2}\%}}
\expandafter\def\csname rcountfraclocus_corrected_M9\endcsname{\shortstack{82/365\\\ensuremath{22.5_{-2.1}^{+2.2}\%}}}
\expandafter\def\csname rfraclocus_corrected_L0\endcsname{\ensuremath{39.4_{-3.8}^{+3.9}\%}}
\expandafter\def\csname rcountfraclocus_corrected_L0\endcsname{\shortstack{63/160\\\ensuremath{39.4_{-3.8}^{+3.9}\%}}}
\expandafter\def\csname rfraclocus_corrected_L1\endcsname{\ensuremath{15.0_{-4.8}^{+6.5}\%}}
\expandafter\def\csname rcountfraclocus_corrected_L1\endcsname{\shortstack{6/40\\\ensuremath{15.0_{-4.8}^{+6.5}\%}}}
\expandafter\def\csname rfracmissing_fbol\endcsname{\ensuremath{17.1_{-0.6}^{+0.6}\%}}
\expandafter\def\csname rcountfracmissing_fbol\endcsname{\shortstack{762/4461\\\ensuremath{17.1_{-0.6}^{+0.6}\%}}}
\expandafter\def\csname rfracthick_M6_active\endcsname{\ensuremath{1.5_{-0.5}^{+0.8}\%}}
\expandafter\def\csname rcountfracthick_M6_active\endcsname{\shortstack{5/332\\\ensuremath{1.5_{-0.5}^{+0.8}\%}}}
\expandafter\def\csname rfracthick_M6_inactive\endcsname{\ensuremath{12.2_{-2.1}^{+2.5}\%}}
\expandafter\def\csname rcountfracthick_M6_inactive\endcsname{\shortstack{24/197\\\ensuremath{12.2_{-2.1}^{+2.5}\%}}}
\expandafter\def\csname rfracthick_M7_active\endcsname{\ensuremath{0.6_{-0.3}^{+0.6}\%}}
\expandafter\def\csname rcountfracthick_M7_active\endcsname{\shortstack{2/353\\\ensuremath{0.6_{-0.3}^{+0.6}\%}}}
\expandafter\def\csname rfracthick_M7_inactive\endcsname{\ensuremath{11.1_{-2.2}^{+2.6}\%}}
\expandafter\def\csname rcountfracthick_M7_inactive\endcsname{\shortstack{19/171\\\ensuremath{11.1_{-2.2}^{+2.6}\%}}}
\expandafter\def\csname rfracthick_M8_active\endcsname{\ensuremath{1.2_{-0.6}^{+1.2}\%}}
\expandafter\def\csname rcountfracthick_M8_active\endcsname{\shortstack{2/169\\\ensuremath{1.2_{-0.6}^{+1.2}\%}}}
\expandafter\def\csname rfracthick_M8_inactive\endcsname{\ensuremath{8.0_{-2.6}^{+3.7}\%}}
\expandafter\def\csname rcountfracthick_M8_inactive\endcsname{\shortstack{6/75\\\ensuremath{8.0_{-2.6}^{+3.7}\%}}}
\expandafter\def\csname rfracthick_M9_active\endcsname{\ensuremath{7.5_{-2.9}^{+4.4}\%}}
\expandafter\def\csname rcountfracthick_M9_active\endcsname{\shortstack{4/53\\\ensuremath{7.5_{-2.9}^{+4.4}\%}}}
\expandafter\def\csname rfracthick_M9_inactive\endcsname{\ensuremath{2.9_{-1.8}^{+4.5}\%}}
\expandafter\def\csname rcountfracthick_M9_inactive\endcsname{\shortstack{1/34\\\ensuremath{2.9_{-1.8}^{+4.5}\%}}}
\expandafter\def\csname rfracfalse_line_window\endcsname{\ensuremath{0.0_{-0.0}^{+0.1}\%}}
\expandafter\def\csname rcountfracfalse_line_window\endcsname{\shortstack{6/13906\\\ensuremath{0.0_{-0.0}^{+0.1}\%}}}
\expandafter\def\csname rfracpurity_original_L_audit\endcsname{\ensuremath{100.0_{-6.0}^{+0.0}\%}}
\expandafter\def\csname rcountfracpurity_original_L_audit\endcsname{\shortstack{60/60\\\ensuremath{100.0_{-6.0}^{+0.0}\%}}}
\expandafter\def\csname rfracblue_blends_original_L_audit\endcsname{\ensuremath{5.0_{-2.1}^{+3.6}\%}}
\expandafter\def\csname rcountfracblue_blends_original_L_audit\endcsname{\shortstack{3/60\\\ensuremath{5.0_{-2.1}^{+3.6}\%}}}
\expandafter\def\csname rfracgiant_G_test\endcsname{\ensuremath{0.0_{-0.0}^{+0.1}\%}}
\expandafter\def\csname rcountfracgiant_G_test\endcsname{\shortstack{0/4272\\\ensuremath{0.0_{-0.0}^{+0.1}\%}}}
\expandafter\def\csname rfracredrock_old_L\endcsname{\ensuremath{81.7_{-1.9}^{+1.8}\%}}
\expandafter\def\csname rcountfracredrock_old_L\endcsname{\shortstack{353/432\\\ensuremath{81.7_{-1.9}^{+1.8}\%}}}
\expandafter\def\csname rfracredrock_new_L\endcsname{\ensuremath{14.7_{-1.4}^{+1.5}\%}}
\expandafter\def\csname rcountfracredrock_new_L\endcsname{\shortstack{87/590\\\ensuremath{14.7_{-1.4}^{+1.5}\%}}}
\expandafter\def\csname rfracredrock_inspected\endcsname{\ensuremath{63.3_{-19.5}^{+16.7}\%}}
\expandafter\def\csname rcountfracredrock_inspected\endcsname{\shortstack{19/30\\\ensuremath{63.3_{-19.5}^{+16.7}\%}}}
\expandafter\def\csname rfrachot_tail_60pc\endcsname{\ensuremath{9.2_{-1.5}^{+1.7}\%}}
\expandafter\def\csname rcountfrachot_tail_60pc\endcsname{\shortstack{30/326\\\ensuremath{9.2_{-1.5}^{+1.7}\%}}}
\expandafter\def\csname rfracM9_weak_inactive\endcsname{\ensuremath{52.9_{-8.5}^{+8.3}\%}}
\expandafter\def\csname rcountfracM9_weak_inactive\endcsname{\shortstack{18/34\\\ensuremath{52.9_{-8.5}^{+8.3}\%}}}
\expandafter\def\csname rfracrotator_lt1d\endcsname{\ensuremath{69.7_{-3.8}^{+3.6}\%}}
\expandafter\def\csname rcountfracrotator_lt1d\endcsname{\shortstack{106/152\\\ensuremath{69.7_{-3.8}^{+3.6}\%}}}
\expandafter\def\csname rfracgagne_overlap\endcsname{\ensuremath{6.2_{-0.2}^{+0.3}\%}}
\expandafter\def\csname rcountfracgagne_overlap\endcsname{\shortstack{557/8939\\\ensuremath{6.2_{-0.2}^{+0.3}\%}}}
\expandafter\def\csname rfracgagne_M6\endcsname{\ensuremath{0.0_{-0.0}^{+0.1}\%}}
\expandafter\def\csname rcountfracgagne_M6\endcsname{\shortstack{0/1815\\\ensuremath{0.0_{-0.0}^{+0.1}\%}}}
\expandafter\def\csname rfracL_MG_old\endcsname{\ensuremath{100.0_{-6.6}^{+0.0}\%}}
\expandafter\def\csname rcountfracL_MG_old\endcsname{\shortstack{14/14\\\ensuremath{100.0_{-6.6}^{+0.0}\%}}}
\expandafter\def\csname rfracL_MG_all\endcsname{\ensuremath{62.2_{-5.7}^{+5.4}\%}}
\expandafter\def\csname rcountfracL_MG_all\endcsname{\shortstack{46/74\\\ensuremath{62.2_{-5.7}^{+5.4}\%}}}
\expandafter\def\csname rfracK0s_old_schmidt_EW_0.75_M7\endcsname{\ensuremath{71.2_{-4.3}^{+4.0}\%}}
\expandafter\def\csname rcountfracK0s_old_schmidt_EW_0.75_M7\endcsname{\shortstack{84/118\\\ensuremath{71.2_{-4.3}^{+4.0}\%}}}
\expandafter\def\csname rfracK0s_old_schmidt_EW_0.75_M8\endcsname{\ensuremath{91.7_{-11.4}^{+5.1}\%}}
\expandafter\def\csname rcountfracK0s_old_schmidt_EW_0.75_M8\endcsname{\shortstack{11/12\\\ensuremath{91.7_{-11.4}^{+5.1}\%}}}
\expandafter\def\csname rfracK0s_old_schmidt_EW_0.75_M9\endcsname{--}
\expandafter\def\csname rcountfracK0s_old_schmidt_EW_0.75_M9\endcsname{\shortstack{0/0\\--}}
\expandafter\def\csname rfracK0s_old_schmidt_EW_0.75_L0-L1\endcsname{--}
\expandafter\def\csname rcountfracK0s_old_schmidt_EW_0.75_L0-L1\endcsname{\shortstack{0/0\\--}}
\expandafter\def\csname rfracK0s_old_schmidt_EW_1_M7\endcsname{\ensuremath{66.6_{-2.3}^{+2.2}\%}}
\expandafter\def\csname rcountfracK0s_old_schmidt_EW_1_M7\endcsname{\shortstack{287/431\\\ensuremath{66.6_{-2.3}^{+2.2}\%}}}
\expandafter\def\csname rfracK0s_old_schmidt_EW_1_M8\endcsname{\ensuremath{78.7_{-5.7}^{+4.7}\%}}
\expandafter\def\csname rcountfracK0s_old_schmidt_EW_1_M8\endcsname{\shortstack{48/61\\\ensuremath{78.7_{-5.7}^{+4.7}\%}}}
\expandafter\def\csname rfracK0s_old_schmidt_EW_1_M9\endcsname{\ensuremath{100.0_{-49.7}^{+0.0}\%}}
\expandafter\def\csname rcountfracK0s_old_schmidt_EW_1_M9\endcsname{\shortstack{1/1\\\ensuremath{100.0_{-49.7}^{+0.0}\%}}}
\expandafter\def\csname rfracK0s_old_schmidt_EW_1_L0-L1\endcsname{--}
\expandafter\def\csname rcountfracK0s_old_schmidt_EW_1_L0-L1\endcsname{\shortstack{0/0\\--}}
\expandafter\def\csname rfracK0s_old_schmidt_EW_2_M7\endcsname{\ensuremath{58.5_{-1.4}^{+1.4}\%}}
\expandafter\def\csname rcountfracK0s_old_schmidt_EW_2_M7\endcsname{\shortstack{700/1196\\\ensuremath{58.5_{-1.4}^{+1.4}\%}}}
\expandafter\def\csname rfracK0s_old_schmidt_EW_2_M8\endcsname{\ensuremath{70.9_{-2.3}^{+2.2}\%}}
\expandafter\def\csname rcountfracK0s_old_schmidt_EW_2_M8\endcsname{\shortstack{285/402\\\ensuremath{70.9_{-2.3}^{+2.2}\%}}}
\expandafter\def\csname rfracK0s_old_schmidt_EW_2_M9\endcsname{\ensuremath{97.6_{-3.7}^{+1.5}\%}}
\expandafter\def\csname rcountfracK0s_old_schmidt_EW_2_M9\endcsname{\shortstack{40/41\\\ensuremath{97.6_{-3.7}^{+1.5}\%}}}
\expandafter\def\csname rfracK0s_old_schmidt_EW_2_L0-L1\endcsname{\ensuremath{100.0_{-33.1}^{+0.0}\%}}
\expandafter\def\csname rcountfracK0s_old_schmidt_EW_2_L0-L1\endcsname{\shortstack{2/2\\\ensuremath{100.0_{-33.1}^{+0.0}\%}}}
\expandafter\def\csname rfracK0s_old_schmidt_EW_5_M7\endcsname{\ensuremath{34.1_{-1.3}^{+1.3}\%}}
\expandafter\def\csname rcountfracK0s_old_schmidt_EW_5_M7\endcsname{\shortstack{469/1374\\\ensuremath{34.1_{-1.3}^{+1.3}\%}}}
\expandafter\def\csname rfracK0s_old_schmidt_EW_5_M8\endcsname{\ensuremath{49.4_{-1.5}^{+1.5}\%}}
\expandafter\def\csname rcountfracK0s_old_schmidt_EW_5_M8\endcsname{\shortstack{540/1094\\\ensuremath{49.4_{-1.5}^{+1.5}\%}}}
\expandafter\def\csname rfracK0s_old_schmidt_EW_5_M9\endcsname{\ensuremath{67.2_{-3.5}^{+3.3}\%}}
\expandafter\def\csname rcountfracK0s_old_schmidt_EW_5_M9\endcsname{\shortstack{129/192\\\ensuremath{67.2_{-3.5}^{+3.3}\%}}}
\expandafter\def\csname rfracK0s_old_schmidt_EW_5_L0-L1\endcsname{\ensuremath{18.2_{-8.8}^{+14.0}\%}}
\expandafter\def\csname rcountfracK0s_old_schmidt_EW_5_L0-L1\endcsname{\shortstack{2/11\\\ensuremath{18.2_{-8.8}^{+14.0}\%}}}
\expandafter\def\csname rfracK0s_old_schmidt_Lambda_-5_M7\endcsname{\ensuremath{67.3_{-2.5}^{+2.4}\%}}
\expandafter\def\csname rcountfracK0s_old_schmidt_Lambda_-5_M7\endcsname{\shortstack{241/358\\\ensuremath{67.3_{-2.5}^{+2.4}\%}}}
\expandafter\def\csname rfracK0s_old_schmidt_Lambda_-5_M8\endcsname{\ensuremath{69.8_{-2.1}^{+2.0}\%}}
\expandafter\def\csname rcountfracK0s_old_schmidt_Lambda_-5_M8\endcsname{\shortstack{353/506\\\ensuremath{69.8_{-2.1}^{+2.0}\%}}}
\expandafter\def\csname rfracK0s_old_schmidt_Lambda_-5_M9\endcsname{\ensuremath{80.3_{-3.7}^{+3.3}\%}}
\expandafter\def\csname rcountfracK0s_old_schmidt_Lambda_-5_M9\endcsname{\shortstack{102/127\\\ensuremath{80.3_{-3.7}^{+3.3}\%}}}
\expandafter\def\csname rfracK0s_old_schmidt_Lambda_-5_L0-L1\endcsname{\ensuremath{20.0_{-9.6}^{+15.0}\%}}
\expandafter\def\csname rcountfracK0s_old_schmidt_Lambda_-5_L0-L1\endcsname{\shortstack{2/10\\\ensuremath{20.0_{-9.6}^{+15.0}\%}}}
\expandafter\def\csname rfracK0s_old_schmidt_Lambda_-4.7_M7\endcsname{\ensuremath{58.9_{-1.4}^{+1.4}\%}}
\expandafter\def\csname rcountfracK0s_old_schmidt_Lambda_-4.7_M7\endcsname{\shortstack{684/1161\\\ensuremath{58.9_{-1.4}^{+1.4}\%}}}
\expandafter\def\csname rfracK0s_old_schmidt_Lambda_-4.7_M8\endcsname{\ensuremath{53.4_{-1.6}^{+1.5}\%}}
\expandafter\def\csname rcountfracK0s_old_schmidt_Lambda_-4.7_M8\endcsname{\shortstack{546/1023\\\ensuremath{53.4_{-1.6}^{+1.5}\%}}}
\expandafter\def\csname rfracK0s_old_schmidt_Lambda_-4.7_M9\endcsname{\ensuremath{42.1_{-2.7}^{+2.8}\%}}
\expandafter\def\csname rcountfracK0s_old_schmidt_Lambda_-4.7_M9\endcsname{\shortstack{133/316\\\ensuremath{42.1_{-2.7}^{+2.8}\%}}}
\expandafter\def\csname rfracK0s_old_schmidt_Lambda_-4.7_L0-L1\endcsname{\ensuremath{11.1_{-4.7}^{+7.4}\%}}
\expandafter\def\csname rcountfracK0s_old_schmidt_Lambda_-4.7_L0-L1\endcsname{\shortstack{3/27\\\ensuremath{11.1_{-4.7}^{+7.4}\%}}}
\expandafter\def\csname rfracK0s_old_schmidt_Lambda_-4.5_M7\endcsname{\ensuremath{54.7_{-1.4}^{+1.4}\%}}
\expandafter\def\csname rcountfracK0s_old_schmidt_Lambda_-4.5_M7\endcsname{\shortstack{722/1319\\\ensuremath{54.7_{-1.4}^{+1.4}\%}}}
\expandafter\def\csname rfracK0s_old_schmidt_Lambda_-4.5_M8\endcsname{\ensuremath{30.7_{-1.2}^{+1.3}\%}}
\expandafter\def\csname rcountfracK0s_old_schmidt_Lambda_-4.5_M8\endcsname{\shortstack{409/1333\\\ensuremath{30.7_{-1.2}^{+1.3}\%}}}
\expandafter\def\csname rfracK0s_old_schmidt_Lambda_-4.5_M9\endcsname{\ensuremath{21.2_{-1.7}^{+1.8}\%}}
\expandafter\def\csname rcountfracK0s_old_schmidt_Lambda_-4.5_M9\endcsname{\shortstack{109/514\\\ensuremath{21.2_{-1.7}^{+1.8}\%}}}
\expandafter\def\csname rfracK0s_old_schmidt_Lambda_-4.5_L0-L1\endcsname{\ensuremath{8.7_{-3.3}^{+5.0}\%}}
\expandafter\def\csname rcountfracK0s_old_schmidt_Lambda_-4.5_L0-L1\endcsname{\shortstack{4/46\\\ensuremath{8.7_{-3.3}^{+5.0}\%}}}
\expandafter\def\csname rfracK0s_old_schmidt_Lambda_-4.3_M7\endcsname{\ensuremath{37.9_{-1.3}^{+1.3}\%}}
\expandafter\def\csname rcountfracK0s_old_schmidt_Lambda_-4.3_M7\endcsname{\shortstack{519/1369\\\ensuremath{37.9_{-1.3}^{+1.3}\%}}}
\expandafter\def\csname rfracK0s_old_schmidt_Lambda_-4.3_M8\endcsname{\ensuremath{11.5_{-0.7}^{+0.8}\%}}
\expandafter\def\csname rcountfracK0s_old_schmidt_Lambda_-4.3_M8\endcsname{\shortstack{194/1693\\\ensuremath{11.5_{-0.7}^{+0.8}\%}}}
\expandafter\def\csname rfracK0s_old_schmidt_Lambda_-4.3_M9\endcsname{\ensuremath{7.3_{-0.9}^{+1.0}\%}}
\expandafter\def\csname rcountfracK0s_old_schmidt_Lambda_-4.3_M9\endcsname{\shortstack{53/727\\\ensuremath{7.3_{-0.9}^{+1.0}\%}}}
\expandafter\def\csname rfracK0s_old_schmidt_Lambda_-4.3_L0-L1\endcsname{\ensuremath{4.0_{-1.7}^{+2.9}\%}}
\expandafter\def\csname rcountfracK0s_old_schmidt_Lambda_-4.3_L0-L1\endcsname{\shortstack{3/75\\\ensuremath{4.0_{-1.7}^{+2.9}\%}}}
\expandafter\def\csname rfracK0e_old_chiobj_EW_0.75_M7\endcsname{\ensuremath{71.2_{-4.3}^{+4.0}\%}}
\expandafter\def\csname rcountfracK0e_old_chiobj_EW_0.75_M7\endcsname{\shortstack{84/118\\\ensuremath{71.2_{-4.3}^{+4.0}\%}}}
\expandafter\def\csname rfracK0e_old_chiobj_EW_0.75_M8\endcsname{\ensuremath{91.7_{-11.4}^{+5.1}\%}}
\expandafter\def\csname rcountfracK0e_old_chiobj_EW_0.75_M8\endcsname{\shortstack{11/12\\\ensuremath{91.7_{-11.4}^{+5.1}\%}}}
\expandafter\def\csname rfracK0e_old_chiobj_EW_0.75_M9\endcsname{--}
\expandafter\def\csname rcountfracK0e_old_chiobj_EW_0.75_M9\endcsname{\shortstack{0/0\\--}}
\expandafter\def\csname rfracK0e_old_chiobj_EW_0.75_L0-L1\endcsname{--}
\expandafter\def\csname rcountfracK0e_old_chiobj_EW_0.75_L0-L1\endcsname{\shortstack{0/0\\--}}
\expandafter\def\csname rfracK0e_old_chiobj_EW_1_M7\endcsname{\ensuremath{66.6_{-2.3}^{+2.2}\%}}
\expandafter\def\csname rcountfracK0e_old_chiobj_EW_1_M7\endcsname{\shortstack{287/431\\\ensuremath{66.6_{-2.3}^{+2.2}\%}}}
\expandafter\def\csname rfracK0e_old_chiobj_EW_1_M8\endcsname{\ensuremath{78.7_{-5.7}^{+4.7}\%}}
\expandafter\def\csname rcountfracK0e_old_chiobj_EW_1_M8\endcsname{\shortstack{48/61\\\ensuremath{78.7_{-5.7}^{+4.7}\%}}}
\expandafter\def\csname rfracK0e_old_chiobj_EW_1_M9\endcsname{\ensuremath{100.0_{-49.7}^{+0.0}\%}}
\expandafter\def\csname rcountfracK0e_old_chiobj_EW_1_M9\endcsname{\shortstack{1/1\\\ensuremath{100.0_{-49.7}^{+0.0}\%}}}
\expandafter\def\csname rfracK0e_old_chiobj_EW_1_L0-L1\endcsname{--}
\expandafter\def\csname rcountfracK0e_old_chiobj_EW_1_L0-L1\endcsname{\shortstack{0/0\\--}}
\expandafter\def\csname rfracK0e_old_chiobj_EW_2_M7\endcsname{\ensuremath{58.5_{-1.4}^{+1.4}\%}}
\expandafter\def\csname rcountfracK0e_old_chiobj_EW_2_M7\endcsname{\shortstack{700/1196\\\ensuremath{58.5_{-1.4}^{+1.4}\%}}}
\expandafter\def\csname rfracK0e_old_chiobj_EW_2_M8\endcsname{\ensuremath{70.9_{-2.3}^{+2.2}\%}}
\expandafter\def\csname rcountfracK0e_old_chiobj_EW_2_M8\endcsname{\shortstack{285/402\\\ensuremath{70.9_{-2.3}^{+2.2}\%}}}
\expandafter\def\csname rfracK0e_old_chiobj_EW_2_M9\endcsname{\ensuremath{97.6_{-3.7}^{+1.5}\%}}
\expandafter\def\csname rcountfracK0e_old_chiobj_EW_2_M9\endcsname{\shortstack{40/41\\\ensuremath{97.6_{-3.7}^{+1.5}\%}}}
\expandafter\def\csname rfracK0e_old_chiobj_EW_2_L0-L1\endcsname{\ensuremath{100.0_{-33.1}^{+0.0}\%}}
\expandafter\def\csname rcountfracK0e_old_chiobj_EW_2_L0-L1\endcsname{\shortstack{2/2\\\ensuremath{100.0_{-33.1}^{+0.0}\%}}}
\expandafter\def\csname rfracK0e_old_chiobj_EW_5_M7\endcsname{\ensuremath{34.1_{-1.3}^{+1.3}\%}}
\expandafter\def\csname rcountfracK0e_old_chiobj_EW_5_M7\endcsname{\shortstack{469/1374\\\ensuremath{34.1_{-1.3}^{+1.3}\%}}}
\expandafter\def\csname rfracK0e_old_chiobj_EW_5_M8\endcsname{\ensuremath{49.4_{-1.5}^{+1.5}\%}}
\expandafter\def\csname rcountfracK0e_old_chiobj_EW_5_M8\endcsname{\shortstack{540/1094\\\ensuremath{49.4_{-1.5}^{+1.5}\%}}}
\expandafter\def\csname rfracK0e_old_chiobj_EW_5_M9\endcsname{\ensuremath{67.2_{-3.5}^{+3.3}\%}}
\expandafter\def\csname rcountfracK0e_old_chiobj_EW_5_M9\endcsname{\shortstack{129/192\\\ensuremath{67.2_{-3.5}^{+3.3}\%}}}
\expandafter\def\csname rfracK0e_old_chiobj_EW_5_L0-L1\endcsname{\ensuremath{18.2_{-8.8}^{+14.0}\%}}
\expandafter\def\csname rcountfracK0e_old_chiobj_EW_5_L0-L1\endcsname{\shortstack{2/11\\\ensuremath{18.2_{-8.8}^{+14.0}\%}}}
\expandafter\def\csname rfracK0e_old_chiobj_Lambda_-5_M7\endcsname{\ensuremath{76.1_{-5.5}^{+4.8}\%}}
\expandafter\def\csname rcountfracK0e_old_chiobj_Lambda_-5_M7\endcsname{\shortstack{51/67\\\ensuremath{76.1_{-5.5}^{+4.8}\%}}}
\expandafter\def\csname rfracK0e_old_chiobj_Lambda_-5_M8\endcsname{\ensuremath{83.0_{-4.2}^{+3.5}\%}}
\expandafter\def\csname rcountfracK0e_old_chiobj_Lambda_-5_M8\endcsname{\shortstack{78/94\\\ensuremath{83.0_{-4.2}^{+3.5}\%}}}
\expandafter\def\csname rfracK0e_old_chiobj_Lambda_-5_M9\endcsname{\ensuremath{79.2_{-5.1}^{+4.3}\%}}
\expandafter\def\csname rcountfracK0e_old_chiobj_Lambda_-5_M9\endcsname{\shortstack{57/72\\\ensuremath{79.2_{-5.1}^{+4.3}\%}}}
\expandafter\def\csname rfracK0e_old_chiobj_Lambda_-5_L0-L1\endcsname{\ensuremath{37.5_{-14.7}^{+17.5}\%}}
\expandafter\def\csname rcountfracK0e_old_chiobj_Lambda_-5_L0-L1\endcsname{\shortstack{3/8\\\ensuremath{37.5_{-14.7}^{+17.5}\%}}}
\expandafter\def\csname rfracK0e_old_chiobj_Lambda_-4.7_M7\endcsname{\ensuremath{64.5_{-1.7}^{+1.7}\%}}
\expandafter\def\csname rcountfracK0e_old_chiobj_Lambda_-4.7_M7\endcsname{\shortstack{511/792\\\ensuremath{64.5_{-1.7}^{+1.7}\%}}}
\expandafter\def\csname rfracK0e_old_chiobj_Lambda_-4.7_M8\endcsname{\ensuremath{71.5_{-2.0}^{+1.9}\%}}
\expandafter\def\csname rcountfracK0e_old_chiobj_Lambda_-4.7_M8\endcsname{\shortstack{393/550\\\ensuremath{71.5_{-2.0}^{+1.9}\%}}}
\expandafter\def\csname rfracK0e_old_chiobj_Lambda_-4.7_M9\endcsname{\ensuremath{61.2_{-3.4}^{+3.3}\%}}
\expandafter\def\csname rcountfracK0e_old_chiobj_Lambda_-4.7_M9\endcsname{\shortstack{128/209\\\ensuremath{61.2_{-3.4}^{+3.3}\%}}}
\expandafter\def\csname rfracK0e_old_chiobj_Lambda_-4.7_L0-L1\endcsname{\ensuremath{11.1_{-4.7}^{+7.4}\%}}
\expandafter\def\csname rcountfracK0e_old_chiobj_Lambda_-4.7_L0-L1\endcsname{\shortstack{3/27\\\ensuremath{11.1_{-4.7}^{+7.4}\%}}}
\expandafter\def\csname rfracK0e_old_chiobj_Lambda_-4.5_M7\endcsname{\ensuremath{58.2_{-1.4}^{+1.4}\%}}
\expandafter\def\csname rcountfracK0e_old_chiobj_Lambda_-4.5_M7\endcsname{\shortstack{699/1201\\\ensuremath{58.2_{-1.4}^{+1.4}\%}}}
\expandafter\def\csname rfracK0e_old_chiobj_Lambda_-4.5_M8\endcsname{\ensuremath{58.4_{-1.6}^{+1.6}\%}}
\expandafter\def\csname rcountfracK0e_old_chiobj_Lambda_-4.5_M8\endcsname{\shortstack{575/985\\\ensuremath{58.4_{-1.6}^{+1.6}\%}}}
\expandafter\def\csname rfracK0e_old_chiobj_Lambda_-4.5_M9\endcsname{\ensuremath{33.3_{-2.4}^{+2.5}\%}}
\expandafter\def\csname rcountfracK0e_old_chiobj_Lambda_-4.5_M9\endcsname{\shortstack{120/360\\\ensuremath{33.3_{-2.4}^{+2.5}\%}}}
\expandafter\def\csname rfracK0e_old_chiobj_Lambda_-4.5_L0-L1\endcsname{\ensuremath{6.7_{-2.8}^{+4.7}\%}}
\expandafter\def\csname rcountfracK0e_old_chiobj_Lambda_-4.5_L0-L1\endcsname{\shortstack{3/45\\\ensuremath{6.7_{-2.8}^{+4.7}\%}}}
\expandafter\def\csname rfracK0e_old_chiobj_Lambda_-4.3_M7\endcsname{\ensuremath{51.9_{-1.4}^{+1.4}\%}}
\expandafter\def\csname rcountfracK0e_old_chiobj_Lambda_-4.3_M7\endcsname{\shortstack{677/1304\\\ensuremath{51.9_{-1.4}^{+1.4}\%}}}
\expandafter\def\csname rfracK0e_old_chiobj_Lambda_-4.3_M8\endcsname{\ensuremath{37.6_{-1.3}^{+1.3}\%}}
\expandafter\def\csname rcountfracK0e_old_chiobj_Lambda_-4.3_M8\endcsname{\shortstack{486/1292\\\ensuremath{37.6_{-1.3}^{+1.3}\%}}}
\expandafter\def\csname rfracK0e_old_chiobj_Lambda_-4.3_M9\endcsname{\ensuremath{15.1_{-1.5}^{+1.6}\%}}
\expandafter\def\csname rcountfracK0e_old_chiobj_Lambda_-4.3_M9\endcsname{\shortstack{83/550\\\ensuremath{15.1_{-1.5}^{+1.6}\%}}}
\expandafter\def\csname rfracK0e_old_chiobj_Lambda_-4.3_L0-L1\endcsname{\ensuremath{4.6_{-2.0}^{+3.3}\%}}
\expandafter\def\csname rcountfracK0e_old_chiobj_Lambda_-4.3_L0-L1\endcsname{\shortstack{3/65\\\ensuremath{4.6_{-2.0}^{+3.3}\%}}}
\expandafter\def\csname rfracK1s_corr_schmidt_EW_0.75_M7\endcsname{\ensuremath{94.4_{-8.0}^{+3.4}\%}}
\expandafter\def\csname rcountfracK1s_corr_schmidt_EW_0.75_M7\endcsname{\shortstack{17/18\\\ensuremath{94.4_{-8.0}^{+3.4}\%}}}
\expandafter\def\csname rfracK1s_corr_schmidt_EW_0.75_M8\endcsname{\ensuremath{100.0_{-33.1}^{+0.0}\%}}
\expandafter\def\csname rcountfracK1s_corr_schmidt_EW_0.75_M8\endcsname{\shortstack{2/2\\\ensuremath{100.0_{-33.1}^{+0.0}\%}}}
\expandafter\def\csname rfracK1s_corr_schmidt_EW_0.75_M9\endcsname{--}
\expandafter\def\csname rcountfracK1s_corr_schmidt_EW_0.75_M9\endcsname{\shortstack{0/0\\--}}
\expandafter\def\csname rfracK1s_corr_schmidt_EW_0.75_L0-L1\endcsname{--}
\expandafter\def\csname rcountfracK1s_corr_schmidt_EW_0.75_L0-L1\endcsname{\shortstack{0/0\\--}}
\expandafter\def\csname rfracK1s_corr_schmidt_EW_1_M7\endcsname{\ensuremath{76.9_{-4.0}^{+3.6}\%}}
\expandafter\def\csname rcountfracK1s_corr_schmidt_EW_1_M7\endcsname{\shortstack{93/121\\\ensuremath{76.9_{-4.0}^{+3.6}\%}}}
\expandafter\def\csname rfracK1s_corr_schmidt_EW_1_M8\endcsname{\ensuremath{100.0_{-16.5}^{+0.0}\%}}
\expandafter\def\csname rcountfracK1s_corr_schmidt_EW_1_M8\endcsname{\shortstack{5/5\\\ensuremath{100.0_{-16.5}^{+0.0}\%}}}
\expandafter\def\csname rfracK1s_corr_schmidt_EW_1_M9\endcsname{--}
\expandafter\def\csname rcountfracK1s_corr_schmidt_EW_1_M9\endcsname{\shortstack{0/0\\--}}
\expandafter\def\csname rfracK1s_corr_schmidt_EW_1_L0-L1\endcsname{--}
\expandafter\def\csname rcountfracK1s_corr_schmidt_EW_1_L0-L1\endcsname{\shortstack{0/0\\--}}
\expandafter\def\csname rfracK1s_corr_schmidt_EW_2_M7\endcsname{\ensuremath{64.8_{-1.9}^{+1.9}\%}}
\expandafter\def\csname rcountfracK1s_corr_schmidt_EW_2_M7\endcsname{\shortstack{400/617\\\ensuremath{64.8_{-1.9}^{+1.9}\%}}}
\expandafter\def\csname rfracK1s_corr_schmidt_EW_2_M8\endcsname{\ensuremath{85.5_{-4.5}^{+3.6}\%}}
\expandafter\def\csname rcountfracK1s_corr_schmidt_EW_2_M8\endcsname{\shortstack{65/76\\\ensuremath{85.5_{-4.5}^{+3.6}\%}}}
\expandafter\def\csname rfracK1s_corr_schmidt_EW_2_M9\endcsname{\ensuremath{100.0_{-5.2}^{+0.0}\%}}
\expandafter\def\csname rcountfracK1s_corr_schmidt_EW_2_M9\endcsname{\shortstack{18/18\\\ensuremath{100.0_{-5.2}^{+0.0}\%}}}
\expandafter\def\csname rfracK1s_corr_schmidt_EW_2_L0-L1\endcsname{\ensuremath{100.0_{-24.8}^{+0.0}\%}}
\expandafter\def\csname rcountfracK1s_corr_schmidt_EW_2_L0-L1\endcsname{\shortstack{3/3\\\ensuremath{100.0_{-24.8}^{+0.0}\%}}}
\expandafter\def\csname rfracK1s_corr_schmidt_EW_5_M7\endcsname{\ensuremath{43.7_{-1.5}^{+1.5}\%}}
\expandafter\def\csname rcountfracK1s_corr_schmidt_EW_5_M7\endcsname{\shortstack{492/1125\\\ensuremath{43.7_{-1.5}^{+1.5}\%}}}
\expandafter\def\csname rfracK1s_corr_schmidt_EW_5_M8\endcsname{\ensuremath{61.2_{-2.4}^{+2.4}\%}}
\expandafter\def\csname rcountfracK1s_corr_schmidt_EW_5_M8\endcsname{\shortstack{246/402\\\ensuremath{61.2_{-2.4}^{+2.4}\%}}}
\expandafter\def\csname rfracK1s_corr_schmidt_EW_5_M9\endcsname{\ensuremath{72.0_{-5.4}^{+4.8}\%}}
\expandafter\def\csname rcountfracK1s_corr_schmidt_EW_5_M9\endcsname{\shortstack{54/75\\\ensuremath{72.0_{-5.4}^{+4.8}\%}}}
\expandafter\def\csname rfracK1s_corr_schmidt_EW_5_L0-L1\endcsname{\ensuremath{41.4_{-8.7}^{+9.2}\%}}
\expandafter\def\csname rcountfracK1s_corr_schmidt_EW_5_L0-L1\endcsname{\shortstack{12/29\\\ensuremath{41.4_{-8.7}^{+9.2}\%}}}
\expandafter\def\csname rfracK1s_corr_schmidt_Lambda_-5_M7\endcsname{\ensuremath{77.8_{-4.6}^{+4.0}\%}}
\expandafter\def\csname rcountfracK1s_corr_schmidt_Lambda_-5_M7\endcsname{\shortstack{70/90\\\ensuremath{77.8_{-4.6}^{+4.0}\%}}}
\expandafter\def\csname rfracK1s_corr_schmidt_Lambda_-5_M8\endcsname{\ensuremath{85.9_{-3.8}^{+3.1}\%}}
\expandafter\def\csname rcountfracK1s_corr_schmidt_Lambda_-5_M8\endcsname{\shortstack{85/99\\\ensuremath{85.9_{-3.8}^{+3.1}\%}}}
\expandafter\def\csname rfracK1s_corr_schmidt_Lambda_-5_M9\endcsname{\ensuremath{75.5_{-6.3}^{+5.4}\%}}
\expandafter\def\csname rcountfracK1s_corr_schmidt_Lambda_-5_M9\endcsname{\shortstack{40/53\\\ensuremath{75.5_{-6.3}^{+5.4}\%}}}
\expandafter\def\csname rfracK1s_corr_schmidt_Lambda_-5_L0-L1\endcsname{\ensuremath{44.4_{-9.1}^{+9.5}\%}}
\expandafter\def\csname rcountfracK1s_corr_schmidt_Lambda_-5_L0-L1\endcsname{\shortstack{12/27\\\ensuremath{44.4_{-9.1}^{+9.5}\%}}}
\expandafter\def\csname rfracK1s_corr_schmidt_Lambda_-4.7_M7\endcsname{\ensuremath{65.2_{-2.0}^{+2.0}\%}}
\expandafter\def\csname rcountfracK1s_corr_schmidt_Lambda_-4.7_M7\endcsname{\shortstack{373/572\\\ensuremath{65.2_{-2.0}^{+2.0}\%}}}
\expandafter\def\csname rfracK1s_corr_schmidt_Lambda_-4.7_M8\endcsname{\ensuremath{68.0_{-2.5}^{+2.4}\%}}
\expandafter\def\csname rcountfracK1s_corr_schmidt_Lambda_-4.7_M8\endcsname{\shortstack{244/359\\\ensuremath{68.0_{-2.5}^{+2.4}\%}}}
\expandafter\def\csname rfracK1s_corr_schmidt_Lambda_-4.7_M9\endcsname{\ensuremath{44.3_{-4.4}^{+4.5}\%}}
\expandafter\def\csname rcountfracK1s_corr_schmidt_Lambda_-4.7_M9\endcsname{\shortstack{54/122\\\ensuremath{44.3_{-4.4}^{+4.5}\%}}}
\expandafter\def\csname rfracK1s_corr_schmidt_Lambda_-4.7_L0-L1\endcsname{\ensuremath{21.6_{-4.4}^{+5.1}\%}}
\expandafter\def\csname rcountfracK1s_corr_schmidt_Lambda_-4.7_L0-L1\endcsname{\shortstack{16/74\\\ensuremath{21.6_{-4.4}^{+5.1}\%}}}
\expandafter\def\csname rfracK1s_corr_schmidt_Lambda_-4.5_M7\endcsname{\ensuremath{61.5_{-1.6}^{+1.6}\%}}
\expandafter\def\csname rcountfracK1s_corr_schmidt_Lambda_-4.5_M7\endcsname{\shortstack{568/923\\\ensuremath{61.5_{-1.6}^{+1.6}\%}}}
\expandafter\def\csname rfracK1s_corr_schmidt_Lambda_-4.5_M8\endcsname{\ensuremath{41.4_{-2.0}^{+2.0}\%}}
\expandafter\def\csname rcountfracK1s_corr_schmidt_Lambda_-4.5_M8\endcsname{\shortstack{242/584\\\ensuremath{41.4_{-2.0}^{+2.0}\%}}}
\expandafter\def\csname rfracK1s_corr_schmidt_Lambda_-4.5_M9\endcsname{\ensuremath{25.3_{-2.8}^{+3.0}\%}}
\expandafter\def\csname rcountfracK1s_corr_schmidt_Lambda_-4.5_M9\endcsname{\shortstack{55/217\\\ensuremath{25.3_{-2.8}^{+3.0}\%}}}
\expandafter\def\csname rfracK1s_corr_schmidt_Lambda_-4.5_L0-L1\endcsname{\ensuremath{9.5_{-2.4}^{+3.1}\%}}
\expandafter\def\csname rcountfracK1s_corr_schmidt_Lambda_-4.5_L0-L1\endcsname{\shortstack{11/116\\\ensuremath{9.5_{-2.4}^{+3.1}\%}}}
\expandafter\def\csname rfracK1s_corr_schmidt_Lambda_-4.3_M7\endcsname{\ensuremath{46.0_{-1.5}^{+1.5}\%}}
\expandafter\def\csname rcountfracK1s_corr_schmidt_Lambda_-4.3_M7\endcsname{\shortstack{508/1104\\\ensuremath{46.0_{-1.5}^{+1.5}\%}}}
\expandafter\def\csname rfracK1s_corr_schmidt_Lambda_-4.3_M8\endcsname{\ensuremath{17.3_{-1.3}^{+1.3}\%}}
\expandafter\def\csname rcountfracK1s_corr_schmidt_Lambda_-4.3_M8\endcsname{\shortstack{147/849\\\ensuremath{17.3_{-1.3}^{+1.3}\%}}}
\expandafter\def\csname rfracK1s_corr_schmidt_Lambda_-4.3_M9\endcsname{\ensuremath{8.2_{-1.4}^{+1.7}\%}}
\expandafter\def\csname rcountfracK1s_corr_schmidt_Lambda_-4.3_M9\endcsname{\shortstack{25/304\\\ensuremath{8.2_{-1.4}^{+1.7}\%}}}
\expandafter\def\csname rfracK1s_corr_schmidt_Lambda_-4.3_L0-L1\endcsname{\ensuremath{3.6_{-1.2}^{+1.7}\%}}
\expandafter\def\csname rcountfracK1s_corr_schmidt_Lambda_-4.3_L0-L1\endcsname{\shortstack{6/167\\\ensuremath{3.6_{-1.2}^{+1.7}\%}}}
\expandafter\def\csname rfracK1e_corr_chiobj_EW_0.75_M6\endcsname{\ensuremath{69.1_{-4.5}^{+4.2}\%}}
\expandafter\def\csname rcountfracK1e_corr_chiobj_EW_0.75_M6\endcsname{\shortstack{76/110\\\ensuremath{69.1_{-4.5}^{+4.2}\%}}}
\expandafter\def\csname rfracK1e_corr_chiobj_EW_0.75_M7\endcsname{\ensuremath{94.4_{-8.0}^{+3.4}\%}}
\expandafter\def\csname rcountfracK1e_corr_chiobj_EW_0.75_M7\endcsname{\shortstack{17/18\\\ensuremath{94.4_{-8.0}^{+3.4}\%}}}
\expandafter\def\csname rfracK1e_corr_chiobj_EW_0.75_M8\endcsname{\ensuremath{100.0_{-33.1}^{+0.0}\%}}
\expandafter\def\csname rcountfracK1e_corr_chiobj_EW_0.75_M8\endcsname{\shortstack{2/2\\\ensuremath{100.0_{-33.1}^{+0.0}\%}}}
\expandafter\def\csname rfracK1e_corr_chiobj_EW_0.75_M9\endcsname{--}
\expandafter\def\csname rcountfracK1e_corr_chiobj_EW_0.75_M9\endcsname{\shortstack{0/0\\--}}
\expandafter\def\csname rfracK1e_corr_chiobj_EW_0.75_L0-L1\endcsname{--}
\expandafter\def\csname rcountfracK1e_corr_chiobj_EW_0.75_L0-L1\endcsname{\shortstack{0/0\\--}}
\expandafter\def\csname rfracK1e_corr_chiobj_EW_1_M6\endcsname{\ensuremath{64.9_{-2.5}^{+2.4}\%}}
\expandafter\def\csname rcountfracK1e_corr_chiobj_EW_1_M6\endcsname{\shortstack{238/367\\\ensuremath{64.9_{-2.5}^{+2.4}\%}}}
\expandafter\def\csname rfracK1e_corr_chiobj_EW_1_M7\endcsname{\ensuremath{76.9_{-4.0}^{+3.6}\%}}
\expandafter\def\csname rcountfracK1e_corr_chiobj_EW_1_M7\endcsname{\shortstack{93/121\\\ensuremath{76.9_{-4.0}^{+3.6}\%}}}
\expandafter\def\csname rfracK1e_corr_chiobj_EW_1_M8\endcsname{\ensuremath{100.0_{-16.5}^{+0.0}\%}}
\expandafter\def\csname rcountfracK1e_corr_chiobj_EW_1_M8\endcsname{\shortstack{5/5\\\ensuremath{100.0_{-16.5}^{+0.0}\%}}}
\expandafter\def\csname rfracK1e_corr_chiobj_EW_1_M9\endcsname{--}
\expandafter\def\csname rcountfracK1e_corr_chiobj_EW_1_M9\endcsname{\shortstack{0/0\\--}}
\expandafter\def\csname rfracK1e_corr_chiobj_EW_1_L0-L1\endcsname{--}
\expandafter\def\csname rcountfracK1e_corr_chiobj_EW_1_L0-L1\endcsname{\shortstack{0/0\\--}}
\expandafter\def\csname rfracK1e_corr_chiobj_EW_2_M6\endcsname{\ensuremath{58.4_{-1.6}^{+1.6}\%}}
\expandafter\def\csname rcountfracK1e_corr_chiobj_EW_2_M6\endcsname{\shortstack{541/927\\\ensuremath{58.4_{-1.6}^{+1.6}\%}}}
\expandafter\def\csname rfracK1e_corr_chiobj_EW_2_M7\endcsname{\ensuremath{64.8_{-1.9}^{+1.9}\%}}
\expandafter\def\csname rcountfracK1e_corr_chiobj_EW_2_M7\endcsname{\shortstack{400/617\\\ensuremath{64.8_{-1.9}^{+1.9}\%}}}
\expandafter\def\csname rfracK1e_corr_chiobj_EW_2_M8\endcsname{\ensuremath{85.5_{-4.5}^{+3.6}\%}}
\expandafter\def\csname rcountfracK1e_corr_chiobj_EW_2_M8\endcsname{\shortstack{65/76\\\ensuremath{85.5_{-4.5}^{+3.6}\%}}}
\expandafter\def\csname rfracK1e_corr_chiobj_EW_2_M9\endcsname{\ensuremath{100.0_{-5.2}^{+0.0}\%}}
\expandafter\def\csname rcountfracK1e_corr_chiobj_EW_2_M9\endcsname{\shortstack{18/18\\\ensuremath{100.0_{-5.2}^{+0.0}\%}}}
\expandafter\def\csname rfracK1e_corr_chiobj_EW_2_L0-L1\endcsname{\ensuremath{100.0_{-24.8}^{+0.0}\%}}
\expandafter\def\csname rcountfracK1e_corr_chiobj_EW_2_L0-L1\endcsname{\shortstack{3/3\\\ensuremath{100.0_{-24.8}^{+0.0}\%}}}
\expandafter\def\csname rfracK1e_corr_chiobj_EW_5_M6\endcsname{\ensuremath{32.3_{-1.4}^{+1.5}\%}}
\expandafter\def\csname rcountfracK1e_corr_chiobj_EW_5_M6\endcsname{\shortstack{336/1040\\\ensuremath{32.3_{-1.4}^{+1.5}\%}}}
\expandafter\def\csname rfracK1e_corr_chiobj_EW_5_M7\endcsname{\ensuremath{43.7_{-1.5}^{+1.5}\%}}
\expandafter\def\csname rcountfracK1e_corr_chiobj_EW_5_M7\endcsname{\shortstack{492/1125\\\ensuremath{43.7_{-1.5}^{+1.5}\%}}}
\expandafter\def\csname rfracK1e_corr_chiobj_EW_5_M8\endcsname{\ensuremath{61.2_{-2.4}^{+2.4}\%}}
\expandafter\def\csname rcountfracK1e_corr_chiobj_EW_5_M8\endcsname{\shortstack{246/402\\\ensuremath{61.2_{-2.4}^{+2.4}\%}}}
\expandafter\def\csname rfracK1e_corr_chiobj_EW_5_M9\endcsname{\ensuremath{72.0_{-5.4}^{+4.8}\%}}
\expandafter\def\csname rcountfracK1e_corr_chiobj_EW_5_M9\endcsname{\shortstack{54/75\\\ensuremath{72.0_{-5.4}^{+4.8}\%}}}
\expandafter\def\csname rfracK1e_corr_chiobj_EW_5_L0-L1\endcsname{\ensuremath{41.4_{-8.7}^{+9.2}\%}}
\expandafter\def\csname rcountfracK1e_corr_chiobj_EW_5_L0-L1\endcsname{\shortstack{12/29\\\ensuremath{41.4_{-8.7}^{+9.2}\%}}}
\expandafter\def\csname rfracK1e_corr_chiobj_Lambda_-5_M6\endcsname{\ensuremath{71.2_{-6.6}^{+5.8}\%}}
\expandafter\def\csname rcountfracK1e_corr_chiobj_Lambda_-5_M6\endcsname{\shortstack{37/52\\\ensuremath{71.2_{-6.6}^{+5.8}\%}}}
\expandafter\def\csname rfracK1e_corr_chiobj_Lambda_-5_M7\endcsname{\ensuremath{88.1_{-4.0}^{+3.1}\%}}
\expandafter\def\csname rcountfracK1e_corr_chiobj_Lambda_-5_M7\endcsname{\shortstack{74/84\\\ensuremath{88.1_{-4.0}^{+3.1}\%}}}
\expandafter\def\csname rfracK1e_corr_chiobj_Lambda_-5_M8\endcsname{\ensuremath{82.4_{-5.1}^{+4.1}\%}}
\expandafter\def\csname rcountfracK1e_corr_chiobj_Lambda_-5_M8\endcsname{\shortstack{56/68\\\ensuremath{82.4_{-5.1}^{+4.1}\%}}}
\expandafter\def\csname rfracK1e_corr_chiobj_Lambda_-5_M9\endcsname{\ensuremath{76.9_{-7.3}^{+6.0}\%}}
\expandafter\def\csname rcountfracK1e_corr_chiobj_Lambda_-5_M9\endcsname{\shortstack{30/39\\\ensuremath{76.9_{-7.3}^{+6.0}\%}}}
\expandafter\def\csname rfracK1e_corr_chiobj_Lambda_-5_L0-L1\endcsname{\ensuremath{52.0_{-9.8}^{+9.7}\%}}
\expandafter\def\csname rcountfracK1e_corr_chiobj_Lambda_-5_L0-L1\endcsname{\shortstack{13/25\\\ensuremath{52.0_{-9.8}^{+9.7}\%}}}
\expandafter\def\csname rfracK1e_corr_chiobj_Lambda_-4.7_M6\endcsname{\ensuremath{63.0_{-2.0}^{+1.9}\%}}
\expandafter\def\csname rcountfracK1e_corr_chiobj_Lambda_-4.7_M6\endcsname{\shortstack{382/606\\\ensuremath{63.0_{-2.0}^{+1.9}\%}}}
\expandafter\def\csname rfracK1e_corr_chiobj_Lambda_-4.7_M7\endcsname{\ensuremath{69.3_{-1.9}^{+1.8}\%}}
\expandafter\def\csname rcountfracK1e_corr_chiobj_Lambda_-4.7_M7\endcsname{\shortstack{413/596\\\ensuremath{69.3_{-1.9}^{+1.8}\%}}}
\expandafter\def\csname rfracK1e_corr_chiobj_Lambda_-4.7_M8\endcsname{\ensuremath{70.1_{-2.9}^{+2.7}\%}}
\expandafter\def\csname rcountfracK1e_corr_chiobj_Lambda_-4.7_M8\endcsname{\shortstack{185/264\\\ensuremath{70.1_{-2.9}^{+2.7}\%}}}
\expandafter\def\csname rfracK1e_corr_chiobj_Lambda_-4.7_M9\endcsname{\ensuremath{59.6_{-5.1}^{+4.9}\%}}
\expandafter\def\csname rcountfracK1e_corr_chiobj_Lambda_-4.7_M9\endcsname{\shortstack{56/94\\\ensuremath{59.6_{-5.1}^{+4.9}\%}}}
\expandafter\def\csname rfracK1e_corr_chiobj_Lambda_-4.7_L0-L1\endcsname{\ensuremath{24.3_{-4.7}^{+5.4}\%}}
\expandafter\def\csname rcountfracK1e_corr_chiobj_Lambda_-4.7_L0-L1\endcsname{\shortstack{17/70\\\ensuremath{24.3_{-4.7}^{+5.4}\%}}}
\expandafter\def\csname rfracK1e_corr_chiobj_Lambda_-4.5_M6\endcsname{\ensuremath{57.4_{-1.6}^{+1.6}\%}}
\expandafter\def\csname rcountfracK1e_corr_chiobj_Lambda_-4.5_M6\endcsname{\shortstack{529/921\\\ensuremath{57.4_{-1.6}^{+1.6}\%}}}
\expandafter\def\csname rfracK1e_corr_chiobj_Lambda_-4.5_M7\endcsname{\ensuremath{57.9_{-1.6}^{+1.6}\%}}
\expandafter\def\csname rcountfracK1e_corr_chiobj_Lambda_-4.5_M7\endcsname{\shortstack{571/987\\\ensuremath{57.9_{-1.6}^{+1.6}\%}}}
\expandafter\def\csname rfracK1e_corr_chiobj_Lambda_-4.5_M8\endcsname{\ensuremath{51.0_{-2.3}^{+2.3}\%}}
\expandafter\def\csname rcountfracK1e_corr_chiobj_Lambda_-4.5_M8\endcsname{\shortstack{248/486\\\ensuremath{51.0_{-2.3}^{+2.3}\%}}}
\expandafter\def\csname rfracK1e_corr_chiobj_Lambda_-4.5_M9\endcsname{\ensuremath{34.4_{-3.6}^{+3.8}\%}}
\expandafter\def\csname rcountfracK1e_corr_chiobj_Lambda_-4.5_M9\endcsname{\shortstack{55/160\\\ensuremath{34.4_{-3.6}^{+3.8}\%}}}
\expandafter\def\csname rfracK1e_corr_chiobj_Lambda_-4.5_L0-L1\endcsname{\ensuremath{9.2_{-2.4}^{+3.1}\%}}
\expandafter\def\csname rcountfracK1e_corr_chiobj_Lambda_-4.5_L0-L1\endcsname{\shortstack{10/109\\\ensuremath{9.2_{-2.4}^{+3.1}\%}}}
\expandafter\def\csname rfracK1e_corr_chiobj_Lambda_-4.3_M6\endcsname{\ensuremath{51.7_{-1.6}^{+1.6}\%}}
\expandafter\def\csname rcountfracK1e_corr_chiobj_Lambda_-4.3_M6\endcsname{\shortstack{519/1004\\\ensuremath{51.7_{-1.6}^{+1.6}\%}}}
\expandafter\def\csname rfracK1e_corr_chiobj_Lambda_-4.3_M7\endcsname{\ensuremath{43.3_{-1.4}^{+1.5}\%}}
\expandafter\def\csname rcountfracK1e_corr_chiobj_Lambda_-4.3_M7\endcsname{\shortstack{502/1160\\\ensuremath{43.3_{-1.4}^{+1.5}\%}}}
\expandafter\def\csname rfracK1e_corr_chiobj_Lambda_-4.3_M8\endcsname{\ensuremath{27.2_{-1.6}^{+1.7}\%}}
\expandafter\def\csname rcountfracK1e_corr_chiobj_Lambda_-4.3_M8\endcsname{\shortstack{194/713\\\ensuremath{27.2_{-1.6}^{+1.7}\%}}}
\expandafter\def\csname rfracK1e_corr_chiobj_Lambda_-4.3_M9\endcsname{\ensuremath{14.8_{-2.1}^{+2.3}\%}}
\expandafter\def\csname rcountfracK1e_corr_chiobj_Lambda_-4.3_M9\endcsname{\shortstack{38/257\\\ensuremath{14.8_{-2.1}^{+2.3}\%}}}
\expandafter\def\csname rfracK1e_corr_chiobj_Lambda_-4.3_L0-L1\endcsname{\ensuremath{3.9_{-1.3}^{+1.9}\%}}
\expandafter\def\csname rcountfracK1e_corr_chiobj_Lambda_-4.3_L0-L1\endcsname{\shortstack{6/152\\\ensuremath{3.9_{-1.3}^{+1.9}\%}}}
\expandafter\def\csname rfraccorrected_EW_0.75_M6\endcsname{\ensuremath{69.1_{-4.5}^{+4.2}\%}}
\expandafter\def\csname rcountfraccorrected_EW_0.75_M6\endcsname{\shortstack{76/110\\\ensuremath{69.1_{-4.5}^{+4.2}\%}}}
\expandafter\def\csname rfraccorrected_EW_0.75_M7\endcsname{\ensuremath{94.4_{-8.0}^{+3.4}\%}}
\expandafter\def\csname rcountfraccorrected_EW_0.75_M7\endcsname{\shortstack{17/18\\\ensuremath{94.4_{-8.0}^{+3.4}\%}}}
\expandafter\def\csname rfraccorrected_EW_0.75_M8\endcsname{\ensuremath{100.0_{-33.1}^{+0.0}\%}}
\expandafter\def\csname rcountfraccorrected_EW_0.75_M8\endcsname{\shortstack{2/2\\\ensuremath{100.0_{-33.1}^{+0.0}\%}}}
\expandafter\def\csname rfraccorrected_EW_0.75_M9\endcsname{--}
\expandafter\def\csname rcountfraccorrected_EW_0.75_M9\endcsname{\shortstack{0/0\\--}}
\expandafter\def\csname rfraccorrected_EW_0.75_L0-L1\endcsname{--}
\expandafter\def\csname rcountfraccorrected_EW_0.75_L0-L1\endcsname{\shortstack{0/0\\--}}
\expandafter\def\csname rfraccorrected_EW_1_M6\endcsname{\ensuremath{64.9_{-2.5}^{+2.4}\%}}
\expandafter\def\csname rcountfraccorrected_EW_1_M6\endcsname{\shortstack{238/367\\\ensuremath{64.9_{-2.5}^{+2.4}\%}}}
\expandafter\def\csname rfraccorrected_EW_1_M7\endcsname{\ensuremath{76.9_{-4.0}^{+3.6}\%}}
\expandafter\def\csname rcountfraccorrected_EW_1_M7\endcsname{\shortstack{93/121\\\ensuremath{76.9_{-4.0}^{+3.6}\%}}}
\expandafter\def\csname rfraccorrected_EW_1_M8\endcsname{\ensuremath{100.0_{-16.5}^{+0.0}\%}}
\expandafter\def\csname rcountfraccorrected_EW_1_M8\endcsname{\shortstack{5/5\\\ensuremath{100.0_{-16.5}^{+0.0}\%}}}
\expandafter\def\csname rfraccorrected_EW_1_M9\endcsname{--}
\expandafter\def\csname rcountfraccorrected_EW_1_M9\endcsname{\shortstack{0/0\\--}}
\expandafter\def\csname rfraccorrected_EW_1_L0-L1\endcsname{--}
\expandafter\def\csname rcountfraccorrected_EW_1_L0-L1\endcsname{\shortstack{0/0\\--}}
\expandafter\def\csname rfraccorrected_EW_2_M6\endcsname{\ensuremath{58.4_{-1.6}^{+1.6}\%}}
\expandafter\def\csname rcountfraccorrected_EW_2_M6\endcsname{\shortstack{541/927\\\ensuremath{58.4_{-1.6}^{+1.6}\%}}}
\expandafter\def\csname rfraccorrected_EW_2_M7\endcsname{\ensuremath{64.8_{-1.9}^{+1.9}\%}}
\expandafter\def\csname rcountfraccorrected_EW_2_M7\endcsname{\shortstack{400/617\\\ensuremath{64.8_{-1.9}^{+1.9}\%}}}
\expandafter\def\csname rfraccorrected_EW_2_M8\endcsname{\ensuremath{85.5_{-4.5}^{+3.6}\%}}
\expandafter\def\csname rcountfraccorrected_EW_2_M8\endcsname{\shortstack{65/76\\\ensuremath{85.5_{-4.5}^{+3.6}\%}}}
\expandafter\def\csname rfraccorrected_EW_2_M9\endcsname{\ensuremath{100.0_{-5.2}^{+0.0}\%}}
\expandafter\def\csname rcountfraccorrected_EW_2_M9\endcsname{\shortstack{18/18\\\ensuremath{100.0_{-5.2}^{+0.0}\%}}}
\expandafter\def\csname rfraccorrected_EW_2_L0-L1\endcsname{\ensuremath{100.0_{-24.8}^{+0.0}\%}}
\expandafter\def\csname rcountfraccorrected_EW_2_L0-L1\endcsname{\shortstack{3/3\\\ensuremath{100.0_{-24.8}^{+0.0}\%}}}
\expandafter\def\csname rfraccorrected_EW_5_M6\endcsname{\ensuremath{32.3_{-1.4}^{+1.5}\%}}
\expandafter\def\csname rcountfraccorrected_EW_5_M6\endcsname{\shortstack{336/1040\\\ensuremath{32.3_{-1.4}^{+1.5}\%}}}
\expandafter\def\csname rfraccorrected_EW_5_M7\endcsname{\ensuremath{43.7_{-1.5}^{+1.5}\%}}
\expandafter\def\csname rcountfraccorrected_EW_5_M7\endcsname{\shortstack{492/1125\\\ensuremath{43.7_{-1.5}^{+1.5}\%}}}
\expandafter\def\csname rfraccorrected_EW_5_M8\endcsname{\ensuremath{61.2_{-2.4}^{+2.4}\%}}
\expandafter\def\csname rcountfraccorrected_EW_5_M8\endcsname{\shortstack{246/402\\\ensuremath{61.2_{-2.4}^{+2.4}\%}}}
\expandafter\def\csname rfraccorrected_EW_5_M9\endcsname{\ensuremath{72.0_{-5.4}^{+4.8}\%}}
\expandafter\def\csname rcountfraccorrected_EW_5_M9\endcsname{\shortstack{54/75\\\ensuremath{72.0_{-5.4}^{+4.8}\%}}}
\expandafter\def\csname rfraccorrected_EW_5_L0-L1\endcsname{\ensuremath{41.4_{-8.7}^{+9.2}\%}}
\expandafter\def\csname rcountfraccorrected_EW_5_L0-L1\endcsname{\shortstack{12/29\\\ensuremath{41.4_{-8.7}^{+9.2}\%}}}
\expandafter\def\csname rfraccorrected_Lambda_-5_M6\endcsname{\ensuremath{71.2_{-6.6}^{+5.8}\%}}
\expandafter\def\csname rcountfraccorrected_Lambda_-5_M6\endcsname{\shortstack{37/52\\\ensuremath{71.2_{-6.6}^{+5.8}\%}}}
\expandafter\def\csname rfraccorrected_Lambda_-5_M7\endcsname{\ensuremath{88.1_{-4.0}^{+3.1}\%}}
\expandafter\def\csname rcountfraccorrected_Lambda_-5_M7\endcsname{\shortstack{74/84\\\ensuremath{88.1_{-4.0}^{+3.1}\%}}}
\expandafter\def\csname rfraccorrected_Lambda_-5_M8\endcsname{\ensuremath{82.4_{-5.1}^{+4.1}\%}}
\expandafter\def\csname rcountfraccorrected_Lambda_-5_M8\endcsname{\shortstack{56/68\\\ensuremath{82.4_{-5.1}^{+4.1}\%}}}
\expandafter\def\csname rfraccorrected_Lambda_-5_M9\endcsname{\ensuremath{76.9_{-7.3}^{+6.0}\%}}
\expandafter\def\csname rcountfraccorrected_Lambda_-5_M9\endcsname{\shortstack{30/39\\\ensuremath{76.9_{-7.3}^{+6.0}\%}}}
\expandafter\def\csname rfraccorrected_Lambda_-5_L0-L1\endcsname{\ensuremath{52.0_{-9.8}^{+9.7}\%}}
\expandafter\def\csname rcountfraccorrected_Lambda_-5_L0-L1\endcsname{\shortstack{13/25\\\ensuremath{52.0_{-9.8}^{+9.7}\%}}}
\expandafter\def\csname rfraccorrected_Lambda_-4.9_M6\endcsname{\ensuremath{67.8_{-3.6}^{+3.4}\%}}
\expandafter\def\csname rcountfraccorrected_Lambda_-4.9_M6\endcsname{\shortstack{118/174\\\ensuremath{67.8_{-3.6}^{+3.4}\%}}}
\expandafter\def\csname rfraccorrected_Lambda_-4.9_M7\endcsname{\ensuremath{80.3_{-2.9}^{+2.6}\%}}
\expandafter\def\csname rcountfraccorrected_Lambda_-4.9_M7\endcsname{\shortstack{167/208\\\ensuremath{80.3_{-2.9}^{+2.6}\%}}}
\expandafter\def\csname rfraccorrected_Lambda_-4.9_M8\endcsname{\ensuremath{81.3_{-4.0}^{+3.5}\%}}
\expandafter\def\csname rcountfraccorrected_Lambda_-4.9_M8\endcsname{\shortstack{87/107\\\ensuremath{81.3_{-4.0}^{+3.5}\%}}}
\expandafter\def\csname rfraccorrected_Lambda_-4.9_M9\endcsname{\ensuremath{71.2_{-6.6}^{+5.8}\%}}
\expandafter\def\csname rcountfraccorrected_Lambda_-4.9_M9\endcsname{\shortstack{37/52\\\ensuremath{71.2_{-6.6}^{+5.8}\%}}}
\expandafter\def\csname rfraccorrected_Lambda_-4.9_L0-L1\endcsname{\ensuremath{32.4_{-7.1}^{+8.0}\%}}
\expandafter\def\csname rcountfraccorrected_Lambda_-4.9_L0-L1\endcsname{\shortstack{12/37\\\ensuremath{32.4_{-7.1}^{+8.0}\%}}}
\expandafter\def\csname rfraccorrected_Lambda_-4.7_M6\endcsname{\ensuremath{63.0_{-2.0}^{+1.9}\%}}
\expandafter\def\csname rcountfraccorrected_Lambda_-4.7_M6\endcsname{\shortstack{382/606\\\ensuremath{63.0_{-2.0}^{+1.9}\%}}}
\expandafter\def\csname rfraccorrected_Lambda_-4.7_M7\endcsname{\ensuremath{69.3_{-1.9}^{+1.8}\%}}
\expandafter\def\csname rcountfraccorrected_Lambda_-4.7_M7\endcsname{\shortstack{413/596\\\ensuremath{69.3_{-1.9}^{+1.8}\%}}}
\expandafter\def\csname rfraccorrected_Lambda_-4.7_M8\endcsname{\ensuremath{70.1_{-2.9}^{+2.7}\%}}
\expandafter\def\csname rcountfraccorrected_Lambda_-4.7_M8\endcsname{\shortstack{185/264\\\ensuremath{70.1_{-2.9}^{+2.7}\%}}}
\expandafter\def\csname rfraccorrected_Lambda_-4.7_M9\endcsname{\ensuremath{59.6_{-5.1}^{+4.9}\%}}
\expandafter\def\csname rcountfraccorrected_Lambda_-4.7_M9\endcsname{\shortstack{56/94\\\ensuremath{59.6_{-5.1}^{+4.9}\%}}}
\expandafter\def\csname rfraccorrected_Lambda_-4.7_L0-L1\endcsname{\ensuremath{24.3_{-4.7}^{+5.4}\%}}
\expandafter\def\csname rcountfraccorrected_Lambda_-4.7_L0-L1\endcsname{\shortstack{17/70\\\ensuremath{24.3_{-4.7}^{+5.4}\%}}}
\expandafter\def\csname rfraccorrected_Lambda_-4.5_M6\endcsname{\ensuremath{57.4_{-1.6}^{+1.6}\%}}
\expandafter\def\csname rcountfraccorrected_Lambda_-4.5_M6\endcsname{\shortstack{529/921\\\ensuremath{57.4_{-1.6}^{+1.6}\%}}}
\expandafter\def\csname rfraccorrected_Lambda_-4.5_M7\endcsname{\ensuremath{57.9_{-1.6}^{+1.6}\%}}
\expandafter\def\csname rcountfraccorrected_Lambda_-4.5_M7\endcsname{\shortstack{571/987\\\ensuremath{57.9_{-1.6}^{+1.6}\%}}}
\expandafter\def\csname rfraccorrected_Lambda_-4.5_M8\endcsname{\ensuremath{51.0_{-2.3}^{+2.3}\%}}
\expandafter\def\csname rcountfraccorrected_Lambda_-4.5_M8\endcsname{\shortstack{248/486\\\ensuremath{51.0_{-2.3}^{+2.3}\%}}}
\expandafter\def\csname rfraccorrected_Lambda_-4.5_M9\endcsname{\ensuremath{34.4_{-3.6}^{+3.8}\%}}
\expandafter\def\csname rcountfraccorrected_Lambda_-4.5_M9\endcsname{\shortstack{55/160\\\ensuremath{34.4_{-3.6}^{+3.8}\%}}}
\expandafter\def\csname rfraccorrected_Lambda_-4.5_L0-L1\endcsname{\ensuremath{9.2_{-2.4}^{+3.1}\%}}
\expandafter\def\csname rcountfraccorrected_Lambda_-4.5_L0-L1\endcsname{\shortstack{10/109\\\ensuremath{9.2_{-2.4}^{+3.1}\%}}}
\expandafter\def\csname rfraccorrected_Lambda_-4.3_M6\endcsname{\ensuremath{51.7_{-1.6}^{+1.6}\%}}
\expandafter\def\csname rcountfraccorrected_Lambda_-4.3_M6\endcsname{\shortstack{519/1004\\\ensuremath{51.7_{-1.6}^{+1.6}\%}}}
\expandafter\def\csname rfraccorrected_Lambda_-4.3_M7\endcsname{\ensuremath{43.3_{-1.4}^{+1.5}\%}}
\expandafter\def\csname rcountfraccorrected_Lambda_-4.3_M7\endcsname{\shortstack{502/1160\\\ensuremath{43.3_{-1.4}^{+1.5}\%}}}
\expandafter\def\csname rfraccorrected_Lambda_-4.3_M8\endcsname{\ensuremath{27.2_{-1.6}^{+1.7}\%}}
\expandafter\def\csname rcountfraccorrected_Lambda_-4.3_M8\endcsname{\shortstack{194/713\\\ensuremath{27.2_{-1.6}^{+1.7}\%}}}
\expandafter\def\csname rfraccorrected_Lambda_-4.3_M9\endcsname{\ensuremath{14.8_{-2.1}^{+2.3}\%}}
\expandafter\def\csname rcountfraccorrected_Lambda_-4.3_M9\endcsname{\shortstack{38/257\\\ensuremath{14.8_{-2.1}^{+2.3}\%}}}
\expandafter\def\csname rfraccorrected_Lambda_-4.3_L0-L1\endcsname{\ensuremath{3.9_{-1.3}^{+1.9}\%}}
\expandafter\def\csname rcountfraccorrected_Lambda_-4.3_L0-L1\endcsname{\shortstack{6/152\\\ensuremath{3.9_{-1.3}^{+1.9}\%}}}
\expandafter\def\csname rfrachybrid_EW_0.75_M6\endcsname{\ensuremath{69.1_{-4.5}^{+4.2}\%}}
\expandafter\def\csname rcountfrachybrid_EW_0.75_M6\endcsname{\shortstack{76/110\\\ensuremath{69.1_{-4.5}^{+4.2}\%}}}
\expandafter\def\csname rfrachybrid_EW_0.75_M7\endcsname{\ensuremath{94.4_{-8.0}^{+3.4}\%}}
\expandafter\def\csname rcountfrachybrid_EW_0.75_M7\endcsname{\shortstack{17/18\\\ensuremath{94.4_{-8.0}^{+3.4}\%}}}
\expandafter\def\csname rfrachybrid_EW_0.75_M8\endcsname{\ensuremath{100.0_{-33.1}^{+0.0}\%}}
\expandafter\def\csname rcountfrachybrid_EW_0.75_M8\endcsname{\shortstack{2/2\\\ensuremath{100.0_{-33.1}^{+0.0}\%}}}
\expandafter\def\csname rfrachybrid_EW_0.75_M9\endcsname{--}
\expandafter\def\csname rcountfrachybrid_EW_0.75_M9\endcsname{\shortstack{0/0\\--}}
\expandafter\def\csname rfrachybrid_EW_0.75_L0-L1\endcsname{--}
\expandafter\def\csname rcountfrachybrid_EW_0.75_L0-L1\endcsname{\shortstack{0/0\\--}}
\expandafter\def\csname rfrachybrid_EW_1_M6\endcsname{\ensuremath{64.9_{-2.5}^{+2.4}\%}}
\expandafter\def\csname rcountfrachybrid_EW_1_M6\endcsname{\shortstack{238/367\\\ensuremath{64.9_{-2.5}^{+2.4}\%}}}
\expandafter\def\csname rfrachybrid_EW_1_M7\endcsname{\ensuremath{76.9_{-4.0}^{+3.6}\%}}
\expandafter\def\csname rcountfrachybrid_EW_1_M7\endcsname{\shortstack{93/121\\\ensuremath{76.9_{-4.0}^{+3.6}\%}}}
\expandafter\def\csname rfrachybrid_EW_1_M8\endcsname{\ensuremath{100.0_{-19.8}^{+0.0}\%}}
\expandafter\def\csname rcountfrachybrid_EW_1_M8\endcsname{\shortstack{4/4\\\ensuremath{100.0_{-19.8}^{+0.0}\%}}}
\expandafter\def\csname rfrachybrid_EW_1_M9\endcsname{\ensuremath{100.0_{-49.7}^{+0.0}\%}}
\expandafter\def\csname rcountfrachybrid_EW_1_M9\endcsname{\shortstack{1/1\\\ensuremath{100.0_{-49.7}^{+0.0}\%}}}
\expandafter\def\csname rfrachybrid_EW_1_L0-L1\endcsname{--}
\expandafter\def\csname rcountfrachybrid_EW_1_L0-L1\endcsname{\shortstack{0/0\\--}}
\expandafter\def\csname rfrachybrid_EW_2_M6\endcsname{\ensuremath{58.4_{-1.6}^{+1.6}\%}}
\expandafter\def\csname rcountfrachybrid_EW_2_M6\endcsname{\shortstack{541/927\\\ensuremath{58.4_{-1.6}^{+1.6}\%}}}
\expandafter\def\csname rfrachybrid_EW_2_M7\endcsname{\ensuremath{64.8_{-1.9}^{+1.9}\%}}
\expandafter\def\csname rcountfrachybrid_EW_2_M7\endcsname{\shortstack{400/617\\\ensuremath{64.8_{-1.9}^{+1.9}\%}}}
\expandafter\def\csname rfrachybrid_EW_2_M8\endcsname{\ensuremath{81.5_{-5.8}^{+4.7}\%}}
\expandafter\def\csname rcountfrachybrid_EW_2_M8\endcsname{\shortstack{44/54\\\ensuremath{81.5_{-5.8}^{+4.7}\%}}}
\expandafter\def\csname rfrachybrid_EW_2_M9\endcsname{\ensuremath{97.6_{-3.7}^{+1.5}\%}}
\expandafter\def\csname rcountfrachybrid_EW_2_M9\endcsname{\shortstack{40/41\\\ensuremath{97.6_{-3.7}^{+1.5}\%}}}
\expandafter\def\csname rfrachybrid_EW_2_L0-L1\endcsname{\ensuremath{100.0_{-33.1}^{+0.0}\%}}
\expandafter\def\csname rcountfrachybrid_EW_2_L0-L1\endcsname{\shortstack{2/2\\\ensuremath{100.0_{-33.1}^{+0.0}\%}}}
\expandafter\def\csname rfrachybrid_EW_5_M6\endcsname{\ensuremath{32.3_{-1.4}^{+1.5}\%}}
\expandafter\def\csname rcountfrachybrid_EW_5_M6\endcsname{\shortstack{336/1040\\\ensuremath{32.3_{-1.4}^{+1.5}\%}}}
\expandafter\def\csname rfrachybrid_EW_5_M7\endcsname{\ensuremath{43.7_{-1.5}^{+1.5}\%}}
\expandafter\def\csname rcountfrachybrid_EW_5_M7\endcsname{\shortstack{492/1125\\\ensuremath{43.7_{-1.5}^{+1.5}\%}}}
\expandafter\def\csname rfrachybrid_EW_5_M8\endcsname{\ensuremath{59.7_{-2.8}^{+2.8}\%}}
\expandafter\def\csname rcountfrachybrid_EW_5_M8\endcsname{\shortstack{181/303\\\ensuremath{59.7_{-2.8}^{+2.8}\%}}}
\expandafter\def\csname rfrachybrid_EW_5_M9\endcsname{\ensuremath{67.2_{-3.5}^{+3.3}\%}}
\expandafter\def\csname rcountfrachybrid_EW_5_M9\endcsname{\shortstack{129/192\\\ensuremath{67.2_{-3.5}^{+3.3}\%}}}
\expandafter\def\csname rfrachybrid_EW_5_L0-L1\endcsname{\ensuremath{18.2_{-8.8}^{+14.0}\%}}
\expandafter\def\csname rcountfrachybrid_EW_5_L0-L1\endcsname{\shortstack{2/11\\\ensuremath{18.2_{-8.8}^{+14.0}\%}}}
\expandafter\def\csname rfrachybrid_Lambda_-5_M6\endcsname{\ensuremath{71.2_{-6.6}^{+5.8}\%}}
\expandafter\def\csname rcountfrachybrid_Lambda_-5_M6\endcsname{\shortstack{37/52\\\ensuremath{71.2_{-6.6}^{+5.8}\%}}}
\expandafter\def\csname rfrachybrid_Lambda_-5_M7\endcsname{\ensuremath{88.1_{-4.0}^{+3.1}\%}}
\expandafter\def\csname rcountfrachybrid_Lambda_-5_M7\endcsname{\shortstack{74/84\\\ensuremath{88.1_{-4.0}^{+3.1}\%}}}
\expandafter\def\csname rfrachybrid_Lambda_-5_M8\endcsname{\ensuremath{81.1_{-7.2}^{+5.6}\%}}
\expandafter\def\csname rcountfrachybrid_Lambda_-5_M8\endcsname{\shortstack{30/37\\\ensuremath{81.1_{-7.2}^{+5.6}\%}}}
\expandafter\def\csname rfrachybrid_Lambda_-5_M9\endcsname{\ensuremath{79.5_{-4.7}^{+4.0}\%}}
\expandafter\def\csname rcountfrachybrid_Lambda_-5_M9\endcsname{\shortstack{66/83\\\ensuremath{79.5_{-4.7}^{+4.0}\%}}}
\expandafter\def\csname rfrachybrid_Lambda_-5_L0-L1\endcsname{\ensuremath{25.0_{-10.2}^{+14.0}\%}}
\expandafter\def\csname rcountfrachybrid_Lambda_-5_L0-L1\endcsname{\shortstack{3/12\\\ensuremath{25.0_{-10.2}^{+14.0}\%}}}
\expandafter\def\csname rfrachybrid_Lambda_-4.9_M6\endcsname{\ensuremath{67.8_{-3.6}^{+3.4}\%}}
\expandafter\def\csname rcountfrachybrid_Lambda_-4.9_M6\endcsname{\shortstack{118/174\\\ensuremath{67.8_{-3.6}^{+3.4}\%}}}
\expandafter\def\csname rfrachybrid_Lambda_-4.9_M7\endcsname{\ensuremath{80.3_{-2.9}^{+2.6}\%}}
\expandafter\def\csname rcountfrachybrid_Lambda_-4.9_M7\endcsname{\shortstack{167/208\\\ensuremath{80.3_{-2.9}^{+2.6}\%}}}
\expandafter\def\csname rfrachybrid_Lambda_-4.9_M8\endcsname{\ensuremath{77.4_{-5.7}^{+4.8}\%}}
\expandafter\def\csname rcountfrachybrid_Lambda_-4.9_M8\endcsname{\shortstack{48/62\\\ensuremath{77.4_{-5.7}^{+4.8}\%}}}
\expandafter\def\csname rfrachybrid_Lambda_-4.9_M9\endcsname{\ensuremath{74.1_{-4.2}^{+3.8}\%}}
\expandafter\def\csname rcountfrachybrid_Lambda_-4.9_M9\endcsname{\shortstack{86/116\\\ensuremath{74.1_{-4.2}^{+3.8}\%}}}
\expandafter\def\csname rfrachybrid_Lambda_-4.9_L0-L1\endcsname{\ensuremath{11.1_{-5.4}^{+9.5}\%}}
\expandafter\def\csname rcountfrachybrid_Lambda_-4.9_L0-L1\endcsname{\shortstack{2/18\\\ensuremath{11.1_{-5.4}^{+9.5}\%}}}
\expandafter\def\csname rfrachybrid_Lambda_-4.7_M6\endcsname{\ensuremath{63.0_{-2.0}^{+1.9}\%}}
\expandafter\def\csname rcountfrachybrid_Lambda_-4.7_M6\endcsname{\shortstack{382/606\\\ensuremath{63.0_{-2.0}^{+1.9}\%}}}
\expandafter\def\csname rfrachybrid_Lambda_-4.7_M7\endcsname{\ensuremath{69.3_{-1.9}^{+1.8}\%}}
\expandafter\def\csname rcountfrachybrid_Lambda_-4.7_M7\endcsname{\shortstack{413/596\\\ensuremath{69.3_{-1.9}^{+1.8}\%}}}
\expandafter\def\csname rfrachybrid_Lambda_-4.7_M8\endcsname{\ensuremath{69.4_{-3.6}^{+3.4}\%}}
\expandafter\def\csname rcountfrachybrid_Lambda_-4.7_M8\endcsname{\shortstack{120/173\\\ensuremath{69.4_{-3.6}^{+3.4}\%}}}
\expandafter\def\csname rfrachybrid_Lambda_-4.7_M9\endcsname{\ensuremath{60.5_{-3.3}^{+3.2}\%}}
\expandafter\def\csname rcountfrachybrid_Lambda_-4.7_M9\endcsname{\shortstack{135/223\\\ensuremath{60.5_{-3.3}^{+3.2}\%}}}
\expandafter\def\csname rfrachybrid_Lambda_-4.7_L0-L1\endcsname{\ensuremath{9.4_{-4.0}^{+6.4}\%}}
\expandafter\def\csname rcountfrachybrid_Lambda_-4.7_L0-L1\endcsname{\shortstack{3/32\\\ensuremath{9.4_{-4.0}^{+6.4}\%}}}
\expandafter\def\csname rfrachybrid_Lambda_-4.5_M6\endcsname{\ensuremath{57.4_{-1.6}^{+1.6}\%}}
\expandafter\def\csname rcountfrachybrid_Lambda_-4.5_M6\endcsname{\shortstack{529/921\\\ensuremath{57.4_{-1.6}^{+1.6}\%}}}
\expandafter\def\csname rfrachybrid_Lambda_-4.5_M7\endcsname{\ensuremath{57.9_{-1.6}^{+1.6}\%}}
\expandafter\def\csname rcountfrachybrid_Lambda_-4.5_M7\endcsname{\shortstack{571/987\\\ensuremath{57.9_{-1.6}^{+1.6}\%}}}
\expandafter\def\csname rfrachybrid_Lambda_-4.5_M8\endcsname{\ensuremath{56.4_{-2.7}^{+2.7}\%}}
\expandafter\def\csname rcountfrachybrid_Lambda_-4.5_M8\endcsname{\shortstack{184/326\\\ensuremath{56.4_{-2.7}^{+2.7}\%}}}
\expandafter\def\csname rfrachybrid_Lambda_-4.5_M9\endcsname{\ensuremath{33.4_{-2.4}^{+2.5}\%}}
\expandafter\def\csname rcountfrachybrid_Lambda_-4.5_M9\endcsname{\shortstack{126/377\\\ensuremath{33.4_{-2.4}^{+2.5}\%}}}
\expandafter\def\csname rfrachybrid_Lambda_-4.5_L0-L1\endcsname{\ensuremath{5.8_{-2.5}^{+4.1}\%}}
\expandafter\def\csname rcountfrachybrid_Lambda_-4.5_L0-L1\endcsname{\shortstack{3/52\\\ensuremath{5.8_{-2.5}^{+4.1}\%}}}
\expandafter\def\csname rfrachybrid_Lambda_-4.3_M6\endcsname{\ensuremath{51.7_{-1.6}^{+1.6}\%}}
\expandafter\def\csname rcountfrachybrid_Lambda_-4.3_M6\endcsname{\shortstack{519/1004\\\ensuremath{51.7_{-1.6}^{+1.6}\%}}}
\expandafter\def\csname rfrachybrid_Lambda_-4.3_M7\endcsname{\ensuremath{43.3_{-1.4}^{+1.5}\%}}
\expandafter\def\csname rcountfrachybrid_Lambda_-4.3_M7\endcsname{\shortstack{502/1160\\\ensuremath{43.3_{-1.4}^{+1.5}\%}}}
\expandafter\def\csname rfrachybrid_Lambda_-4.3_M8\endcsname{\ensuremath{31.1_{-2.1}^{+2.1}\%}}
\expandafter\def\csname rcountfrachybrid_Lambda_-4.3_M8\endcsname{\shortstack{150/482\\\ensuremath{31.1_{-2.1}^{+2.1}\%}}}
\expandafter\def\csname rfrachybrid_Lambda_-4.3_M9\endcsname{\ensuremath{15.0_{-1.4}^{+1.6}\%}}
\expandafter\def\csname rcountfrachybrid_Lambda_-4.3_M9\endcsname{\shortstack{85/567\\\ensuremath{15.0_{-1.4}^{+1.6}\%}}}
\expandafter\def\csname rfrachybrid_Lambda_-4.3_L0-L1\endcsname{\ensuremath{4.1_{-1.8}^{+3.0}\%}}
\expandafter\def\csname rcountfrachybrid_Lambda_-4.3_L0-L1\endcsname{\shortstack{3/73\\\ensuremath{4.1_{-1.8}^{+3.0}\%}}}
\expandafter\def\csname rfracpurity_retained\endcsname{\ensuremath{100.0_{-7.1}^{+0.0}\%}}
\expandafter\def\csname rcountfracpurity_retained\endcsname{\shortstack{50/50\\\ensuremath{100.0_{-7.1}^{+0.0}\%}}}
\expandafter\def\csname rfracblue_retained\endcsname{\ensuremath{4.0_{-2.0}^{+3.8}\%}}
\expandafter\def\csname rcountfracblue_retained\endcsname{\shortstack{2/50\\\ensuremath{4.0_{-2.0}^{+3.8}\%}}}
\expandafter\def\csname rfracnisp_all_shifted_corr\endcsname{\ensuremath{86.6_{-2.8}^{+2.4}\%}}
\expandafter\def\csname rfracnisp_all_shifted_old\endcsname{\ensuremath{83.7_{-3.0}^{+2.6}\%}}
\expandafter\def\csname rfracnisp_all_unshifted_corr\endcsname{\ensuremath{64.5_{-3.7}^{+3.5}\%}}
\expandafter\def\csname rfracnisp_all_unshifted_old\endcsname{\ensuremath{43.6_{-3.7}^{+3.8}\%}}
\expandafter\def\csname rfracnisp_in_sample_shifted_corr\endcsname{\ensuremath{92.4_{-2.5}^{+1.9}\%}}
\expandafter\def\csname rfracnisp_in_sample_shifted_old\endcsname{\ensuremath{89.6_{-2.8}^{+2.3}\%}}
\expandafter\def\csname rfracnisp_in_sample_unshifted_corr\endcsname{\ensuremath{66.7_{-4.0}^{+3.8}\%}}
\expandafter\def\csname rfracnisp_in_sample_unshifted_old\endcsname{\ensuremath{45.1_{-4.1}^{+4.1}\%}}
\expandafter\def\csname rfracspherex_detection\endcsname{\ensuremath{90.6_{-0.3}^{+0.3}\%}}
\expandafter\def\csname rfracspherex_clean\endcsname{\ensuremath{71.4_{-0.4}^{+0.4}\%}}
\expandafter\def\csname rfracspherex_detection_bright\endcsname{\ensuremath{97.8_{-0.2}^{+0.2}\%}}
\expandafter\def\csname rfracspherex_detection_faint\endcsname{\ensuremath{77.0_{-0.6}^{+0.6}\%}}
\expandafter\def\csname rfracnisp_all_shifted_corr\endcsname{\ensuremath{86.6_{-2.8}^{+2.4}\%}}
\expandafter\def\csname rfracnisp_all_shifted_old\endcsname{\ensuremath{83.7_{-3.0}^{+2.6}\%}}
\expandafter\def\csname rfracnisp_all_unshifted_corr\endcsname{\ensuremath{64.5_{-3.7}^{+3.5}\%}}
\expandafter\def\csname rfracnisp_all_unshifted_old\endcsname{\ensuremath{43.6_{-3.7}^{+3.8}\%}}
\expandafter\def\csname rfracnisp_in_sample_shifted_corr\endcsname{\ensuremath{92.4_{-2.5}^{+1.9}\%}}
\expandafter\def\csname rfracnisp_in_sample_shifted_old\endcsname{\ensuremath{89.6_{-2.8}^{+2.3}\%}}
\expandafter\def\csname rfracnisp_in_sample_unshifted_corr\endcsname{\ensuremath{66.7_{-4.0}^{+3.8}\%}}
\expandafter\def\csname rfracnisp_in_sample_unshifted_old\endcsname{\ensuremath{45.1_{-4.1}^{+4.1}\%}}
\expandafter\def\csname rfracspherex_detection\endcsname{\ensuremath{90.6_{-0.3}^{+0.3}\%}}
\expandafter\def\csname rfracspherex_clean\endcsname{\ensuremath{71.4_{-0.4}^{+0.4}\%}}
\expandafter\def\csname rfracspherex_detection_bright\endcsname{\ensuremath{97.8_{-0.2}^{+0.2}\%}}
\expandafter\def\csname rfracspherex_detection_faint\endcsname{\ensuremath{77.0_{-0.6}^{+0.6}\%}}

\abstract
{The fraction of late-M dwarfs with detectable H$\alpha$ rises toward the stellar--substellar boundary while their emission weakens relative to the bolometric luminosity, a contrast attributed to the fading continuum.}
{We measure how much of this rise the continuum explains and whether younger populations or longer activity lifetimes contribute.}
{We select 8939 dwarfs of type M7 or later, all but 22 of them M7--L1, from the first data release of the Dark Energy Spectroscopic Instrument (DESI), with types on the Sloan Digital Sky Survey scale. Excluding high-proper-motion targets, we compare active fractions at common thresholds in equivalent width and in $\log(\LHa)$, where $\LHa$ is the H$\alpha$-to-bolometric luminosity ratio, in spectra sensitive to each threshold. We jointly model continuum fading, emission weakening and an extra active component. Kinematics use Gaia astrometry and DESI radial velocities for dwarfs with $G<19$, free of proper-motion selection.}
{From M7 to M9 the active fraction rises by 28 points at 5\,\AA\ equivalent width. At $\log(\LHa)=-4.5$ it falls by 23.5 points. Both recur in bins of bolometric luminosity. SPHEREx and photometric bolometric fluxes agree within 0.03\,dex. Joint fits require M8 emission to be 0.12--0.15\,dex weaker than at M7. The brighter M8 dwarfs require an extra active component; the larger sample at most marginally. At $\log(\LHa)=-4.7$, inactive M7 and M8 dwarfs have larger velocity dispersions than active ones. The measured rotators, mostly M6--M7 dwarfs, have saturated H$\alpha$. \emph{Euclid} DR1 near-infrared spectra from NISP of 152 M7--L1 dwarfs, 144 of them with $S/N\ge5$, test the optical type scale, and \emph{Euclid} VIS images resolve three close binaries (0.22--0.53\arcsec) with common proper motion.}
{We confirm that the rise in equivalent-width incidence comes mostly from the fading continuum acting on weakening emission. Conditional on an age--velocity relation and uniform ages, a step model gives thin-disk activity lifetimes at $\log(\LHa)=-4.7$ of 5.9 (5.3--6.6) and 6.4 (5.2--8.0)\,Gyr at M7 and M8. At thresholds adjusted for weaker emission, the lifetimes agree, with M8 at most 3.5\,Gyr longer. Because both bins contain mostly stars, these lifetimes describe the lowest-mass stars.}


\maketitle

\section{Introduction}
\label{sec:intro}

The fraction of M dwarfs with detectable H$\alpha$ emission rises toward the stellar--substellar boundary, while their H$\alpha$ luminosity falls relative to the bolometric luminosity. H$\alpha$ is a widely measured tracer of M-dwarf magnetic activity \citep{Hawley1996, ReinersBasri2008}. Both trends were found in nearby dwarfs \citep{Gizis2000} and in spectra from the Sloan Digital Sky Survey (SDSS) and its Baryon Oscillation Spectroscopic Survey (BOSS) \citep{West2004, Schmidt2015}. The luminosity decline extends from late-M to L and T dwarfs \citep{MohantyBasri2003, ReinersBasri2010, Pineda2016}. At these temperatures the photospheres become neutral enough that magnetic fields couple weakly to the gas \citep{Mohanty2002, RodriguezBarrera2015}. The two trends are compatible because the optical continuum fades faster than the line toward later types, so a line of falling luminosity can keep or gain equivalent width; the conversions from equivalent width to luminosity were introduced for this reason \citep{Walkowicz2004, WestHawley2008, Schmidt2014}. What has not been measured at M7--M9 is how much of the rise the continuum accounts for, whether the distribution of emission strength itself changes with type, and whether the dwarfs stay active for longer.

The M7--L1 range spans the hydrogen-burning limit \citep{Kirkpatrick1999, Martin1999, Cruz2007, Kirkpatrick2021}. Its population could become younger toward later types, because brown dwarfs cool through these types while hydrogen-burning stars stay in them \citep{Burrows2001, ChabrierBaraffe2000}. Activity could last longer at lower mass: Galactic dynamics imply activity lifetimes that increase through the M sequence \citep{West2006, West2008}, and age-calibrated activity and flare samples support a connection with age \citep{Kiman2021, Medina2022}. Either would add to the continuum effect. The pseudo-continuum at H$\alpha$, the molecular-band-depressed flux against which the line is measured, fades with type, so a fixed equivalent width corresponds to a smaller $\LHa$ at later types \citep{Schmidt2014}.

Space motions separate age from these alternatives because the velocity dispersion of a disk population grows with age \citep{Nordstrom2004, Sharma2021}. Empirical lithium ages, calibrated on open clusters, are harder to apply to field dwarfs this cool: even for members cooler than 4300\,K in clusters of 60--300\,Myr, two versions of one estimator differ by 37--136\,Myr \citep{Tata2026Li}. Studies of late-M dwarfs have related activity to motion \citep{Gizis2000, Bochanski2007, Reiners2009, Kiman2019}, and L-dwarf studies extend the comparison across the boundary \citep{Schmidt2007, Faherty2009, ZapateroOsorio2007, Seifahrt2010, Burgasser2015}. Rotation is a second age-sensitive observable: slow mid-M rotators have larger dispersions \citep{Newton2016, West2015}, and H$\alpha$ and X-ray activity saturate at rapid rotation and decline at longer periods \citep{Newton2017, Wright2018}, with complementary spectroscopic constraints from CARMENES \citep{Jeffers2018}. At ultracool types, radio and optical activity can separate and follow rotation differently \citep{Berger2010, McLean2012}, and auroral emission in some objects cautions against assigning every line the same chromospheric origin \citep{Hallinan2015, Pineda2017}.

The first data release of the Dark Energy Spectroscopic Instrument (DESI DR1) contains several thousand late-M and L dwarf spectra, enough to distinguish the effects of emission strength, detection limit and age \citep{DESI_EDR2024, DESI_DR1_2025}. DESI is a 5000-fibre spectrograph on the Mayall 4\,m telescope covering 3600--9824\,\AA\ \citep{DESI2016a, DESI2022}, a range that includes H$\alpha$ and the molecular bands used for typing. We combine these spectra with Gaia DR3 astrometry \citep{GaiaPrusti2016, GaiaDR3} and rotation periods from the TESS All-Sky Rotation Survey \citep{Boyle2026}, with further periods that we measured from public ZTF and TESS light curves \citep{Bellm2019, Masci2019, Ricker2015}. SPHEREx, the NASA all-sky spectral survey, measures 0.75--5\,\um\ spectrophotometry across the sky \citep{Dore2018, Crill2020, Bock2026}. We use it to measure the near-infrared part of the bolometric flux that converts H$\alpha$ to $\LHa$ and, with colours from the Two Micron All Sky Survey (2MASS), to compare the atmospheres of fast- and slow-moving dwarfs. With Gaia parallaxes the bolometric fluxes give luminosities, which we use beside spectral type as the independent variable. Near-infrared luminosity at fixed type, from 2MASS and Gaia, picks out unresolved binaries. In its first data release (DR1), the \emph{Euclid} mission \citep{Mellier2025} provides 1.2--1.9\,\um\ slitless spectra from its Near-Infrared Spectrometer and Photometer (NISP; \citealt{Jahnke2025}) for 152 of the M7--L1 dwarfs (144 of them at $S/N\ge5$), an independent check of the type scale, and visible images from its VIS imager \citep{Cropper2025} in DR1 and in the earlier Quick Release Q1 \citep{EuclidQ1overview}, which we search for resolved companions.

Section~\ref{sec:sample} describes the sample and what its spectral types mean physically, Sect.~\ref{sec:halpha} the H$\alpha$ fractions and the model that separates continuum from emission strength, Sect.~\ref{sec:kinematics} the kinematics and lifetimes, and Sect.~\ref{sec:rotation} the rotation periods.

\section{The DESI DR1 sample and its spectral types}
\label{sec:sample}

\subsection{Selection, astrometry and targeting}
\label{subsec:selection}

In this paper $r$, $z$ and $W1_{\rm AB}$ are extinction-corrected AB magnitudes from the DESI DR1 redshift catalogue, using Legacy Imaging Surveys fluxes \citep{Dey2019} and their dust corrections \citep{Schlegel1998}. The CatWISE2020 catalogue of the Wide-field Infrared Survey Explorer (WISE) provides the Vega magnitude $W1_{\rm Vega}$ and the $W1-W2$ colour, also on the Vega system, used for the infrared comparisons.

We drew candidates from the primary entries of that catalogue. The DESI reduction and the Redrock classifier \citep{Guy2023, Redrock} assign each spectrum a class (star, galaxy or quasar). We combined two selections (Table~\ref{tab:flow}): Redrock M stars with $r-z \ge 2.0$, and red point sources of any class, most of them labelled galaxy or quasar, defined by $z < 21$, a point-source or blank Legacy Surveys morphology, and either $r-z \ge 3.0$ or a non-detected $r$ flux with $z-W1_{\rm AB} \ge 1.5$. The red point-source selection is needed because Redrock has no L-dwarf templates and labels most L dwarfs as galaxies near redshift 1.03 (Appendix~\ref{app:redrock}). Both colour limits lie below the colours of the dwarfs we select, so the spectral fit of Sect.~\ref{subsec:offset} sets the M7 boundary: 4\% of the final sample has $r-z<3.0$ (13\% at M7). In the Legacy Surveys bands, dwarfs with published M7, M8 and M9 types on the SDSS system have median $r-z$ of 3.36, 3.88 and 4.10. The mean SDSS colours of \citet{West2005}, summed from $r-i$ and $i-z$, give 3.53, 4.39 and 4.50. Our M8 and M9 dwarfs are therefore 0.4--0.5\,mag bluer than those values, while M7 dwarfs agree within the spread of \citet{West2005}, and all of these colours lie above our limits.

We matched Gaia DR3 counterparts within $2\arcsec$ after propagating Gaia positions to the DESI epoch, and use distances $1/\varpi$, with $\varpi$ the parallax, for $\varpi/\sigma_\varpi \ge 5$ \citep{BailerJones2015}. Late-M giants have the molecular bands of late-M dwarfs, so we tested each object with up to five indicators: the absolute Gaia magnitude, the reduced proper motion \citep{LepineGaidos2011, Lepine2007}, the $W1-W2$ colour \citep{Marocco2021}, and the pseudo-equivalent widths of the gravity-sensitive Na\,\textsc{i} 8183, 8195\,\AA\ doublet \citep{Schiavon1997, Lyo2004, Schlieder2012} and of Ca\,\textsc{ii} 8542\,\AA\ \citep{Mann2012}, with the thresholds of Table~\ref{tab:giants}. We removed six likely giants, none of which has a parallax that marks it as a dwarf. Of the 4287 dwarfs with $\varpi/\sigma_\varpi \ge 5$, M6 comparison bin included (Table~\ref{tab:sample}), the 4272 with a $G$ magnitude all have $M_G > 8$, which bounds the giant fraction among them below 0.09\%.

\begin{figure*}[tbp]
    \centering
    \includegraphics[width=\textwidth]{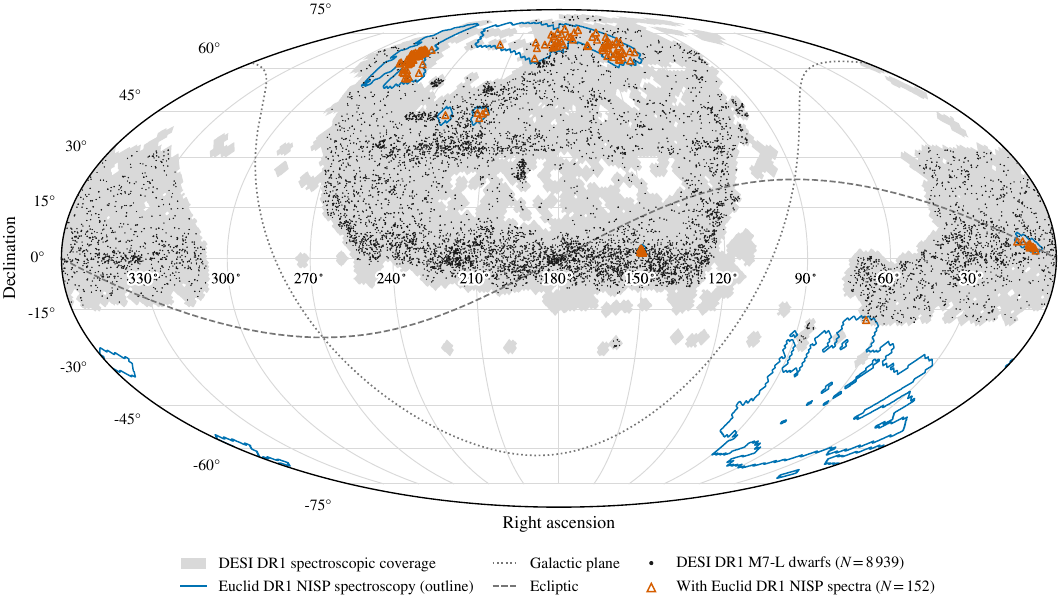}
    \caption{Equatorial Mollweide projection of the 8939 M7--L1 dwarfs (points) over the DESI DR1 spectroscopic footprint (grey; HEALPix pixels \citep{Gorski2005} of 0.84\,deg$^2$ with at least five primary spectra). The blue outline shows the \emph{Euclid} DR1 NISP spectroscopic coverage and the open vermilion triangles the 152 M7--L1 dwarfs with a \emph{Euclid} DR1 NISP spectrum (184 with the M6 comparison bin; Appendix~\ref{app:counts}); the 144 of them with $S/N\ge5$ are used in the type comparison of Sect.~\ref{subsec:validation}. The dotted and dashed curves trace the Galactic plane and the ecliptic.}
    \label{fig:skymap}
\end{figure*}

Most dwarfs were targeted by three DESI programmes, and because these select on colour, distance or motion they matter for the kinematics (Table~\ref{tab:targeting}). A colour-selected secondary programme targeted Legacy Surveys point sources whose $grz$ colours lie off the stellar locus \citep{Myers2023, Schlafly2023}. The DESI Milky Way Survey \citep[MWS;][]{Cooper2023} targeted Gaia stars within 100\,pc and a magnitude-limited stellar sample. A high-proper-motion secondary programme selected Gaia stars with $G \ge 19$ and proper motions above 100\,mas\,yr$^{-1}$. The analysis sample contains no targets of the high-proper-motion programme, whose selection on motion biases velocity dispersions upward. The 462 dwarfs also targeted by the colour-selected programme or the MWS are therefore absent, as is nearly every dwarf fainter than $G=19$ that moves faster than 100\,mas\,yr$^{-1}$ (Table~\ref{tab:targeting}). For faint dwarfs this imposes a cut in tangential velocity on the analysis sample, which biases their dispersions downward. The restored sample combines the analysis sample with these 462 dwarfs. It has no such cut but over-represents fast dwarfs: in dark time the high-proper-motion programme had a far higher fibre-assignment priority than the colour-selected targets (Table~\ref{tab:targeting}). Where the two competed for a fibre, the fast dwarf got it. Both effects vanish for dwarfs brighter than $G=19$, which the programme did not select. The dwarfs with $G<19$ (Table~\ref{tab:targeting}) therefore give kinematics free of the proper-motion selection. We use this $G<19$ sample as the reference for every kinematic result and the analysis sample for the activity fractions. We model the fractions for both (Sect.~\ref{subsec:forward}) and repeat the principal kinematic results on the analysis and restored samples (Sect.~\ref{subsec:kin_pop}). Table~\ref{tab:flow} follows the sample from selection to analysis, Table~\ref{tab:sample} lists it by type and Fig.~\ref{fig:skymap} shows its sky distribution.

\begin{table}[tbp]
\caption{Targeting programmes of the M7--L1 dwarfs.}
\label{tab:targeting}
\centering
\scriptsize
\setlength{\tabcolsep}{2pt}
\begin{tabular}{p{0.42\columnwidth}p{0.36\columnwidth}r}
\toprule
Programme (target bits) & Selection & Dwarfs \\
\midrule
Colour-selected (PSF\_OUT\_BRIGHT, PSF\_OUT\_DARK)$^a$ & $grz$ off the stellar locus & 6526 \\
Milky Way Survey (MWS) & Gaia, $d<100$\,pc or mag. limit & 1917 \\
High proper motion (FAINT\_HPM)$^a$ & $G\ge19$, $\mu>100$\,mas\,yr$^{-1}$ & 600 \\
\quad also colour-selected or MWS & & 462 \\
\quad high proper motion only & & 138 \\
None of the three & & 963 \\
\midrule
Analysis sample & without high proper motion & 8339 \\
Restored sample & analysis plus 462 & 8801 \\
$G<19$ sample & Gaia $G<19$, any programme & 1362 \\
\midrule
M7--M9 with $G\ge19$, $\mu>100$\,mas\,yr$^{-1}$ & all / analysis / restored & 550 / 36 / 417 \\
\bottomrule
\end{tabular}
\tablefoot{A dwarf can carry several bits. $d$: distance; $\mu$: proper motion. In dark time the high-proper-motion targets had a fibre-assignment priority of 1025 and the colour-selected targets 90 in the DESI targeting software (desitarget), and where two targets compete for a fibre the higher priority wins; MWS targets were observed in bright time. The $G<19$ sample includes 14 high-proper-motion targets with Gaia DR3 $G=18.95$--19.00. The last row counts the dwarfs that meet the high-proper-motion selection. Target definitions are in the DESI DR1 secondary-target documentation. \tablefoottext{a}{DESI target-bit names: PSF\_OUT\_BRIGHT and PSF\_OUT\_DARK select point sources outside the stellar locus in bright and dark time, FAINT\_HPM faint high-proper-motion stars.}}
\end{table}

\begin{table}[tbp]
\caption{Construction of the sample.}
\label{tab:flow}
\centering
\scriptsize
\setlength{\tabcolsep}{3pt}
\begin{tabular}{lr}
\toprule
Step & Spectra \\
\midrule
Primary entries of the DESI DR1 redshift catalogue & 27\,547\,223 \\
Redrock M stars with $r-z \ge 2.0$ & 53\,504 \\
Red point sources (1472 labelled galaxy or quasar) & 1573 \\
Selected (either) & 55\,077 \\
Median $S/N \ge 3$ at 8000--9000\,\AA, typed & 45\,264 \\
Best original template M7 or later, $\Delta\chi^2_{\rm type} \ge 25$ & 10\,792 \\
Likely giants removed & 10\,786 \\
Corrected type M7 or later, $\Delta\chi^2_{\rm type} \ge 25$ & 8961 \\
Quasars removed (Appendix~\ref{app:counts}) & 8939 \\
\quad without high-proper-motion targets (analysis sample) & 8339 \\
\quad H$\alpha$ measured & 4461 \\
\quad \quad with a bolometric flux & 3699 \\
\quad with space motions & 2443 \\
\bottomrule
\end{tabular}
\tablefoot{One spectrum per object (the DESI DR1 primary entry). $\Delta\chi^2_{\rm type}$: improvement of the template fit over the best earlier type (Sect.~\ref{subsec:offset}). The corrected M6 comparison bin contains 1815 dwarfs after the same steps.}
\end{table}

\begin{table*}[tbp]
\caption{The sample by spectral type on the corrected scale, with the M6 comparison bin.}
\label{tab:sample}
\centering
\small
\setlength{\tabcolsep}{4pt}
\begin{tabular}{lrrrrrrrr}
\toprule
Type & Dwarfs & $S/N\ge10$ & Redrock non-stellar & $\varpi/\sigma_\varpi \ge 5$ & Analysis & H$\alpha$ measured & Median $\ell$ [\AA] & Space motion \\
\midrule
M6$^a$ & $1\,815$ & $1\,232$ & $3$ & $1\,124$ & $1\,715$ & $1\,484$ & $1.5$ & $926$ \\
\midrule
M7 & $2\,897$ & $1\,730$ & $9$ & $1\,527$ & $2\,681$ & $2\,057$ & $3.8$ & $1\,240$ \\
M8 & $3\,505$ & $1\,335$ & $20$ & $1\,030$ & $3\,289$ & $1\,707$ & $11.4$ & $788$ \\
M9 & $1\,515$ & $465$ & $30$ & $386$ & $1\,411$ & $469$ & $13.0$ & $268$ \\
L0 & $815$ & $193$ & $233$ & $174$ & $756$ & $162$ & $15.0$ & $109$ \\
L1 & $185$ & $48$ & $185$ & $42$ & $180$ & $66$ & $17.6$ & $34$ \\
L2 & $1$ & $0$ & $1$ & $0$ & $1$ & $0$ & -- & $0$ \\
\Lbin$^b$ & $21$ & $3$ & $21$ & $4$ & $21$ & $0$ & -- & $4$ \\
\midrule
M7 and later & $8\,939$ & $3\,774$ & $499$ & $3\,163$ & $8\,339$ & $4\,461$ & $8.7$ & $2\,443$ \\
\bottomrule
\end{tabular}
\tablefoot{Dwarfs: corrected type M7 or later, $\Delta\chi^2_{\rm type} \ge 25$ relative to earlier types, likely giants and quasars removed. $S/N$ is the median per pixel over 8000--9000\,\AA. Redrock non-stellar: Redrock class galaxy or quasar in DESI DR1. Analysis: the sample without the high-proper-motion targets (Table~\ref{tab:targeting}); the last three columns count these dwarfs. Median $\ell$: median $3\sigma$ equivalent-width detection limit of the measured spectra (Sect.~\ref{subsec:halpha_method}). Space motion: Gaia astrometry and a DESI radial velocity that passes the cuts of Sect.~\ref{subsec:kin_method}.
\tablefoottext{a}{Comparison bin (Sect.~\ref{subsec:validation}).}
\tablefoottext{b}{The bin of the L6 template, which collects spectra later than L2 (Sect.~\ref{subsec:offset}).}}
\end{table*}

\subsection{An empirical type scale}
\label{subsec:offset}

We first fitted spectral types with the solar-metallicity K5--L6 templates of \citet{Kesseli2017}, as distributed with PyHammer \citep{Roulston2020}, by weighted least squares over 6000--9800\,\AA\ with a single flux scale, after convolving each DESI spectrum to the template resolution and masking the telluric O$_2$ bands. A spectrum counts as an ultracool dwarf if its median $S/N$ per pixel over 8000--9000\,\AA\ is at least 3, its best template is M7 or later, and that template improves $\chi^2$ by $\Delta\chi^2_{\rm type} \ge 25$ over the best earlier type. The library jumps from L3 to L6, and its L3 template never fits best because many of its pixels have non-positive flux, so the objects typed L6 form a bin of spectra later than L2, labelled \Lbin\ below. This bin contains 21 faint dwarfs that are redder in $z-W1$ than the L1 dwarfs.

We tested this scale on dwarfs with published types on the SDSS system from \citet{Kiman2019}, \citet{West2011} and the L dwarfs of \citet{Schmidt2015}. We call these 1353 dwarfs anchors; because Kiman et al. compile the other two samples, they are effectively on the \citet{Kiman2019} scale. We split them at random into a calibration half and a held-out half. On the held-out half the template types are late by up to 0.85 subtype at published M6--M8 and early at L0 (Fig.~\ref{fig:typescale}a). SDSS spectra of the anchors run through the same fit show the same offset, so it comes from the template library. The library's M7 and M8 templates average six and two BOSS spectra, compared with 590 at M9, so a few stars set their mean shape. Within published M7 the offset follows the pseudo-continuum colour (Spearman $\rho=0.81$ with $r-z$): the templates are redder than typical M7 dwarfs.

\subsection{Corrected types, their validation and their temperatures}
\label{subsec:validation}

\begin{figure*}[tbp]
    \centering
    \includegraphics[width=\textwidth]{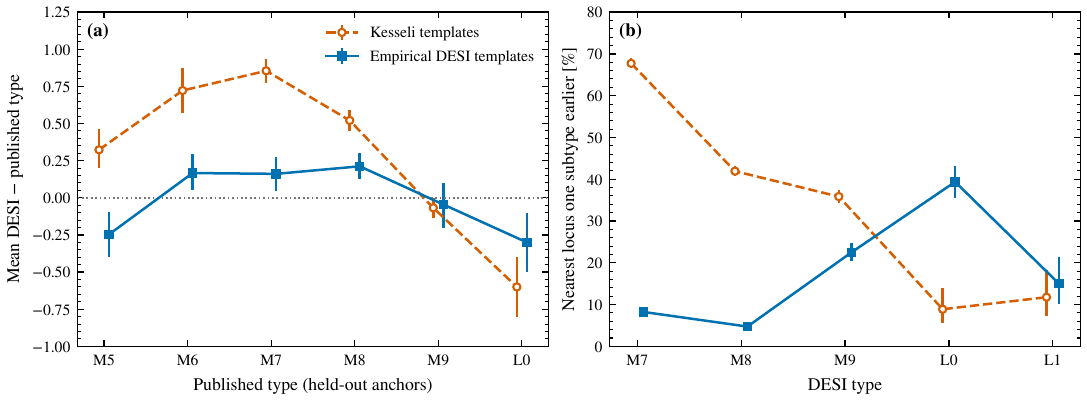}
    \caption{Validation of the spectral-type scale. (a) Mean DESI minus published type for held-out anchors, with the \citet{Kesseli2017} templates (open circles, dashed) and empirical DESI templates (filled squares, solid); bars are 95\% bootstrap intervals. (b) Fraction of dwarfs with parallaxes whose absolute $G$ magnitude is closest to the \citet{Kiman2019} mean of the subtype one earlier than assigned, with 68\% Wilson intervals. Panel (a) shows M5--L0; the six held-out L1 anchors on the SDSS system have a median offset of zero.}
    \label{fig:typescale}
\end{figure*}

Following the approach used for SDSS spectra \citep{Bochanski2007b}, we replaced the M4--L1 templates with empirical templates built from the DESI spectra of calibration anchors at $S/N \ge 10$ (per-pixel medians of at least nine spectra), and kept the \citet{Kesseli2017} templates for K5--M3, L2, L3 and L6. Fractional types come from a parabola through the $\chi^2$ minimum and the adjacent integer types. On the held-out half the corrected types have a median offset of zero at every type from M5 to L1 and mean offsets within 0.3 subtype (Fig.~\ref{fig:typescale}a), and \rfrac{type_within_one_corrected} of the held-out anchors of published type M6--L1 are within one subtype. We pay for this at M9: \rfrac{M9_exact_corrected} of held-out M9 anchors are assigned their published subtype, down from \rfrac{M9_exact_old}, because the new M9 bin scatters to both sides.

Gaia luminosities provide an independent check. On the template scale \rfrac{locus_old_M7} of the M7 dwarfs and \rfrac{locus_old_M8} of the M8 dwarfs with parallaxes sit closest to the \citet{Kiman2019} absolute-$G$ locus of the subtype one earlier; on the corrected scale these fractions drop to \rfrac{locus_corrected_M7} and \rfrac{locus_corrected_M8} (Fig.~\ref{fig:typescale}b). Six DESI dwarfs are members of Praesepe (about 0.7\,Gyr; \citealt{Brandt2015}) in both \citet{Lodieu2019} and \citet{Hunt2023}; five of them were M7 dwarfs more luminous than field M7 dwarfs on the template scale, too bright for their type at that age, and they are M6 on the corrected scale.

The near-infrared spectra test the scale beyond the optical bands used to build the templates. Of the 152 M7--L1 dwarfs with a \emph{Euclid} DR1 NISP spectrum (Fig.~\ref{fig:skymap}), 144 have $S/N\ge5$ over 1.21--1.80\,\um. We fitted these 144 spectra and the 27 of the M6 bin with the same $S/N$, 171 in all, with field-dwarf medians from the SpeX Prism Library \citep{Rayner2003, Burgasser2014}. The median infrared-minus-optical difference improves by one subtype on the corrected scale, under either of the two release calibrations described in Appendix~\ref{app:sxtest} (Table~\ref{tab:nisp_agreement}); optical and infrared types need not coincide, because gravity and clouds affect the two ranges differently \citep{Kirkpatrick2010, Cruz2018}. SPHEREx spectra confirm the order of the sequence, although the broad bands correlate equally well with the template and corrected types, leaving the choice between adjacent optical labels to the anchors. Along the optical sequence water absorption deepens and $K/J$, the ratio of the mean flux densities $f_\nu$ at 2.10--2.25 and 1.15--1.30\,\um, increases, while the 3.3\,\um\ methane band appears in the L dwarfs (Appendix~\ref{app:sxtest}).

The corrected scale is an SDSS-system scale, and it inherits the luminosity of that system. The anchors are 0.44--0.63\,mag brighter in $M_J$ than UltracoolSheet field dwarfs of the same published type \citep{UltracoolSheet, DupuyLiu2012}, and the DESI dwarfs within 40\,pc, where the magnitude limit matters little, are 0.43--0.61\,mag brighter on the corrected scale. Unresolved binaries could create such an offset without changing the temperature that goes with each label; alternatively, the SDSS spectral standards could differ from the field standards \citep{Kirkpatrick1991} so that each bin is warmer than its name. To decide between them we asked how many binaries the offset requires (Table~\ref{tab:binary}). An equal-flux pair is 0.75\,mag brighter than a single dwarf, so the median of a sample can shift by more than 0.38\,mag only if more than half of it consists of such pairs, whatever the scatter. A mixture of single dwarfs and near-equal pairs fitted to the M7--M9 dwarfs within 40\,pc needs a binary fraction above 60\% at every type if the labels correspond to field temperatures. Field very-low-mass dwarfs have a total multiplicity of 20--25\%, most of it in near-equal pairs \citep{Close2003, DucheneKraus2013}, so binaries alone cannot account for the offset. With binaries at that rate, the offset left within 40\,pc is 1.1--2.1 field subtypes, and somewhat less in the magnitude-limited anchors, where binaries are over-represented. Younger, inflated dwarfs or a selection toward luminous dwarfs would also raise the luminosity at fixed type, but we interpret most of the offset as a difference of temperature, which makes the corrected bins 1--2 field subtypes warmer than their names.

Under this interpretation, with the \citet{Pecaut2013} temperature scale and the \citet{Baraffe2015} tracks, the two bins span about 2700--3100\,K and 0.09--0.14\,M$_\odot$ at 5\,Gyr, and in a volume-limited population brown dwarfs make up at most 5\% of either bin (Table~\ref{tab:typescale}). With field temperatures the bins would contain stars of about 0.09\,M$_\odot$ with 5--9\% brown dwarfs, so in either case the lifetimes of Sect.~\ref{subsec:lifetimes} describe the lowest-mass stars.

\begin{table}[tbp]
\caption{Binary fraction needed to explain the $M_J$ offset of the corrected scale.}
\label{tab:binary}
\centering
\scriptsize
\setlength{\tabcolsep}{1.6pt}
\begin{tabular}{llrcccc}
\toprule
Sample & Type & $N$ & Median & $f_{\rm bin}$, $s=0$ & $\Delta M_{J,{\rm s}}$ at $f_{\rm pub}$ & $s$ \\
 & & & offset & (95\% low) & [mag] (95\%) & [subt.] \\
\midrule
DESI 40\,pc & M7 & 49 & $-$0.61 & 1.00 (0.95) & $-$0.62 ($-$0.71, $-$0.52) & 2.1 \\
 & M8 & 35 & $-$0.44 & 0.94 (0.67) & $-$0.37 ($-$0.45, $-$0.29) & 1.1 \\
 & M9 & 24 & $-$0.43 & 0.92 (0.62) & $-$0.36 ($-$0.44, $-$0.29) & 1.6 \\
 & M7--M9 & 108 & $-$0.52 & 1.00 (0.96) & $-$0.48 ($-$0.54, $-$0.42) & 1.7 \\
\midrule
SDSS anchors & M7 & 221 & $-$0.63 & 1.00 (0.98) & $-$0.47 ($-$0.53, $-$0.42) & 1.6 \\
 & M8 & 279 & $-$0.44 & 0.85 (0.77) & $-$0.29 ($-$0.33, $-$0.25) & 0.9 \\
 & M9 & 106 & $-$0.54 & 0.90 (0.75) & $-$0.30 ($-$0.37, $-$0.22) & 1.3 \\
 & M7--M9 & 606 & $-$0.50 & 0.99 (0.94) & $-$0.35 ($-$0.38, $-$0.32) & 1.2 \\
\bottomrule
\end{tabular}
\tablefoot{Offsets are $M_J$ minus the UltracoolSheet field value of the same type (the overluminosity $\Delta M_J$ of Sect.~\ref{subsec:binaries} is measured from the DESI sequence instead); DESI dwarfs on the corrected scale with parallax $S/N\ge10$ and $\varpi>25$\,mas, anchors on their published types. Model: single dwarfs at offset $\Delta M_{J,{\rm s}}$ with Gaussian scatter, and a fraction $f_{\rm bin}$ of near-equal pairs 0.50--0.75\,mag brighter. $f_{\rm bin}$, $s=0$: maximum-likelihood binary fraction if the labels carry field temperatures, with its 95\% lower bound. $\Delta M_{J,{\rm s}}$ at $f_{\rm pub}$: the offset of single dwarfs left when $f_{\rm bin}$ is fixed at the 22\% multiplicity of field very-low-mass dwarfs \citep{DucheneKraus2013} within 40\,pc and at 40\% in the magnitude-limited anchors, where equal pairs, 0.75\,mag brighter than single dwarfs, are over-represented by $10^{0.6\times0.75}\approx2.8$; $s$: that offset in field subtypes at that type, positive when the bin is warmer than its name (as in Table~\ref{tab:typescale}).}
\end{table}

\begin{table}[tbp]
\caption{Temperatures, masses and substellar fractions of the corrected M7 and M8 bins for a type-scale offset of $s$ field subtypes.}
\label{tab:typescale}
\centering
\scriptsize
\setlength{\tabcolsep}{3pt}
\begin{tabular}{llcccc}
\toprule
$s$ & Bin & $T_{\rm eff}$ [K] & $M$ (5\,Gyr) [M$_\odot$] & Substellar & $\tau$ [Gyr] \\
\midrule
0.0 & M7 & 2680 (2625--2745) & 0.092 & 4.8\% & 6.5 \\
0.0 & M8 & 2570 (2475--2625) & 0.087 & 8.9\% & 7.6 \\
1.0 & M7 & 2810 (2740--2930) & 0.100 & 2.3\% & 6.3 \\
1.0 & M8 & 2680 (2630--2740) & 0.092 & 4.7\% & 7.0 \\
1.6 & M7 & 2956 (2834--3070) & 0.116 & 0.7\% & 6.2 \\
1.6 & M8 & 2754 (2692--2834) & 0.096 & 3.4\% & 6.9 \\
2.1 & M7 & 3070 (2956--3130) & 0.135 & 0.0\% & 6.1 \\
2.1 & M8 & 2834 (2754--2956) & 0.102 & 2.0\% & 6.8 \\
\bottomrule
\end{tabular}
\tablefoot{$s=0$: the labels have field temperatures \citep{Pecaut2013}; $s=1.0$, 1.6 and 2.1: the offset of the corrected scale from the field sequence interpreted as a temperature offset, spanning the values of Table~\ref{tab:binary} (0.9--2.1). Masses from the \citet{Baraffe2015} tracks; substellar: mass fraction below 0.075\,M$_\odot$ in the volume-limited brown-dwarf age model; $\tau$: thin-disk step activity lifetime (Sect.~\ref{subsec:lifetimes}) with that age model at the threshold $\Lambda=-4.7$ in $\log(\LHa)$ (Sect.~\ref{subsec:halpha_method}), for dwarfs with $G<19$; with uniform ages it is 5.9\,Gyr at M7 and 6.4\,Gyr at M8.}
\end{table}

The corrected scale shifts most dwarfs one subtype earlier (Appendix~\ref{app:counts}). We keep the 1815 dwarfs now typed M6 as a comparison bin, which is incompletely selected because the sample was built from spectra that the template library typed M7 or later. Because 23 L0 anchors define the M9/L0 boundary, we also use a hybrid scale with the corrected types up to M8 and the template types from M9 on (Appendix~\ref{app:m9l0}). After we removed the six giants and 22 high-redshift and broad-absorption-line quasars whose spectra mimic L dwarfs (Appendix~\ref{app:counts}), the sample comprises 8939 dwarfs, 3774 of them at $S/N \ge 10$; 7193 of them have no earlier spectrum or published type in the catalogues we searched. We call this the M7--L1 sample: 8917 dwarfs have types M7--L1, one is L2 and 21 form the \Lbin\ bin, and none of these 22 has a measured H$\alpha$ line. All results below use the corrected types, and Appendix~\ref{app:oldscale} compares the main results on the two scales.

\subsection{Lithium absorption}
\label{subsec:lithium}

Li\,\textsc{i} at 6708\,\AA\ could test directly whether the M7 and M8 bins are mostly stars, because stars destroy their lithium early while brown dwarfs below the burning limit keep it. In the tracks of \citet{Baraffe2015}, lithium falls to 1\% of its initial abundance within 68\,Myr at 0.10\,M$_\odot$ and 123\,Myr at 0.075\,M$_\odot$. We measured the line at its vacuum wavelength of 6709.65\,\AA\ in the rest frame of each dwarf, with the radial-velocity error included in the line profile. Molecular bands set the flux around the line, so for each dwarf we scaled a median spectrum of the same corrected type, built from the other half of the sample, and replaced its core at the lithium wavelength with a cubic polynomial. Three core widths and measurements at off-line positions give the error of this continuum. Of the 8939 M7--L1 dwarfs, 954 have spectra good enough for the measurement, 907 of them in the analysis sample and 808 at M7.

\begin{figure*}[tbp]
\centering
\includegraphics[width=\textwidth]{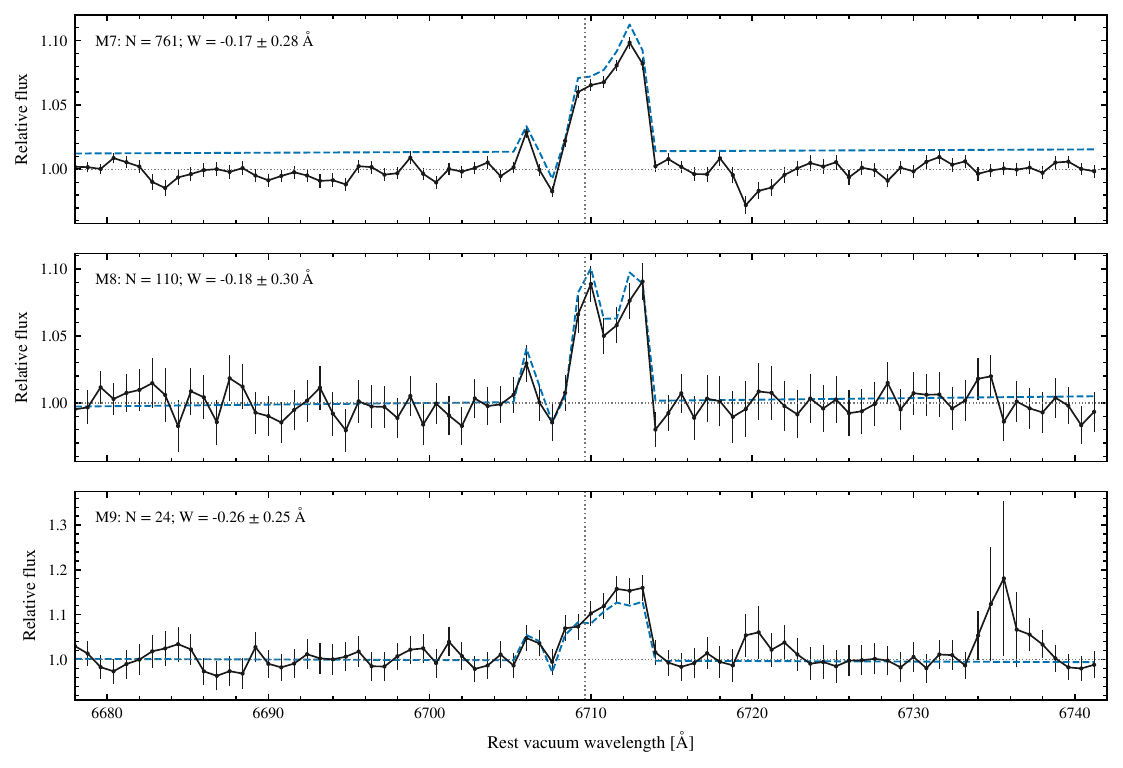}
\caption{Mean spectra around Li\,\textsc{i} of the M7, M8 and M9 dwarfs of the analysis sample with a usable measurement, each divided by its fitted baseline. Points are the mean spectrum with pixel errors, and the dashed line is the empirical spectrum of the same type before its core is replaced. The vertical dotted line marks the lithium wavelength. Each panel gives the number of dwarfs and the equivalent width.}
\label{fig:li_stacks}
\end{figure*}

These spectra are too shallow for the lines expected in late-M dwarfs. Lines injected at the lithium wavelength reach 90\% recovery in no spectrum at 0.5 or 1\,\AA. At 2\,\AA\ nine M7 spectra and one M8 spectrum reach it, and none of the ten shows a line, which limits the incidence of 2\,\AA\ lines among those nine M7 dwarfs to below 31\% (95\% confidence, allowing for recovery). Control positions away from the line give no $5\sigma$ excursion in the 941 dwarfs with a usable control. The mean spectra have equivalent widths of $-0.17\pm0.28$\,\AA\ at M7 and $-0.18\pm0.30$\,\AA\ at M8 (Fig.~\ref{fig:li_stacks}), with 95\% upper limits of 0.47 and 0.49\,\AA\ that are set by the molecular baseline. The known lithium-bearing L1 dwarf 2MASS J20575409$-$0252302 \citep{Kirkpatrick2008} shows absorption at $5.5\sigma$ when the fit allows a broader line, so DESI recovers strong lithium where it is known to exist, but a velocity offset and the dependence on the line profile prevent us from calibrating equivalent widths with this one spectrum.

A stack could not settle the question even with better baselines. Adding lithium depletion to the model of Table~\ref{tab:typescale}, 3.1\% and 4.9\% of the dwarfs retain at least half of their lithium at field M7 and M8 temperatures, fewer at the warmer temperatures, and 17--31\% when the model is weighted by bolometric luminosity, as a flux-limited sample would be. If every such dwarf had a 0.5\,\AA\ line, the stacks would change by at most 0.16\,\AA, below their limits. Retaining lithium is also a different property from being substellar: young stars keep it, and a 0.06\,M$_\odot$ brown dwarf loses it within 403\,Myr in the same tracks. The composition of the M7 and M8 bins therefore rests on the luminosities and evolutionary models of Sect.~\ref{subsec:validation}, and the DESI lithium measurements neither support nor contradict it.

\section{\texorpdfstring{H$\alpha$}{H-alpha} emission at a common detection limit}
\label{sec:halpha}

\subsection{Measurement, detection limits and thresholds}
\label{subsec:halpha_method}

We measured H$\alpha$ on the native DESI spectra in the rest frame. The equivalent width $\EW = \sum \Delta\lambda\,(F/F_{\rm c} - 1)$, with $F$ the observed and $F_{\rm c}$ the continuum flux, is positive for emission. It is summed over 6557.6--6571.6\,\AA\ (vacuum), with sidebands at 6530--6555 and 6575--6600\,\AA\ \citep{Schmidt2015}. Because the sidebands of late-M spectra contain molecular structure, the continuum is a template scaled by a linear function fitted to the sidebands, with any H$\alpha$ emission of the template replaced by a straight line. The candidate templates are the \citet{Kesseli2017} template of the original type and the solar templates within one subtype of it, and we adopt the one with the smallest sideband $\chi^2$. Measured against the empirical templates of Sect.~\ref{subsec:validation} instead, the equivalent widths are lower by a median of 0.37\,\AA, a fifth of their error. With these empirical templates, every active fraction at 5\,\AA\ and at $\log(\LHa)=-4.5$ is lower by 2.2--7.7 percentage points. The largest shift is at M9 and 5\,\AA, where it exceeds the statistical error of the fraction. The M8 and M9 differences from M7 change by 0.9--3.4 points, within their errors (Appendix~\ref{app:template}, Tables~\ref{tab:template} and \ref{tab:template_fractions}). We quote the fractions measured with the \citet{Kesseli2017} templates and treat this shift as a systematic error of their absolute level. The error $\varepsilon$ combines photon noise, a template-mismatch term correlated over about three pixels, and the spread among the candidate templates (Appendix~\ref{app:ewerr}). Measurements in five line-free windows of a random half of the spectra required these errors to be scaled up by 8--10\%. On the other half, positive $3\sigma$ excursions in these windows occur at a third of the Gaussian rate. A $3\sigma$ detection is therefore conservative.

The detection limit $\ell$ of a spectrum is the equivalent width that a line of the instrumental shape would need for a $3\sigma$ detection. The measured spectra are the 4461 spectra of the M7--L1 analysis sample with good coverage, a positive continuum and a continuum $S/N \ge 5$ at the line, as required for a measurement. We tested the detection limit $\ell$ by injecting lines into the real spectra and repeating the full measurement (Appendix~\ref{app:ewerr}). A line of strength $\ell$ is detected in close to half of the trials at every type and $S/N$, because noise scatters its measured strength almost symmetrically about the true value, below the $3\sigma$ level about half of the time. The detection rate reaches 91\% at $1.67\,\ell$. A 5\,\AA\ line is therefore recovered in 95\% of trials in spectra with $\ell\le5$\,\AA, but in 87\% at M8, where many limits fall just below 5\,\AA. Example spectra and the measurements of all dwarfs are shown in Figs.~\ref{fig:halpha_examples} and \ref{fig:halpha_individual}.

The ratio $\LHa = \chiH\,\EW$ removes the dependence of $\EW$ on the continuum \citep{Walkowicz2004, WestHawley2008}, where $\chiH$ is the ratio of the continuum flux near H$\alpha$ to the bolometric flux. Mean $\chiH$ values per type \citep{Schmidt2014} tie the conversion to the type scale, so we measured $\chiH$ for each object, as in cluster activity work \citep{Douglas2014}: the mean continuum in 6500--6550 and 6575--6625\,\AA\ divided by the bolometric flux. For the bolometric flux we integrated the DESI spectrum, scaled to Pan-STARRS1 photometry \citep{Chambers2016, Flewelling2020}, then flux-density nodes at $Y$, $J$, $H$, $K$, $W1$ and $W2$ fitted to photometry from 2MASS, the UKIRT Infrared Deep Sky Survey (UKIDSS), the VISTA Hemisphere Survey (VHS) and CatWISE \citep{Skrutskie2006, Lawrence2007, Cross2012, Marocco2021}, and a Rayleigh--Jeans tail beyond $W2$. We matched every catalogue at the position propagated with the proper motion to its own epoch, because a match at fixed position loses most dwarfs that move faster than 0.3\arcsec\,yr$^{-1}$, which are the nearest and fastest-moving.

Each comparison uses a common threshold: an equivalent width $D$ or a value $\Lambda$ of $\log(\LHa)$. It uses only spectra with a detection limit at or below that threshold ($\ell \le D$, or $\log(\chiH\ell) \le \Lambda$). A dwarf with such a spectrum is \emph{active} if its line is at least as strong and is detected at $3\sigma$ with its own error, and \emph{inactive} otherwise; inactive therefore means below the threshold, and an inactive dwarf can have a detected weaker line. Because $\chiH$ falls with type, one value of $\Lambda$ corresponds to larger equivalent widths at later types: $\Lambda=-4.7$ is about 2\,\AA\ at M7 and 6\,\AA\ at M9. Fractions have 68\% Wilson intervals \citep{Brown2001} and purity bounds 95\% Clopper--Pearson intervals \citep{ClopperPearson1934}; for model parameters the interval and its construction (bootstrap percentiles or profile likelihood) are given with each result.

All these measurements come from one epoch per dwarf, and H$\alpha$ varies \citep{Lee2010, Kruse2010}. For 268 dwarfs with a second DESI spectrum, typically a month after the first, the line strength changes by 0.07\,dex beyond the errors. Variability of this size moves 8--11\% of the dwarfs across the threshold when both spectra have limits at or below 5\,\AA\ or $\Lambda=-4.5$. The fractions therefore describe activity at a single epoch and lie a few percentage points below the fraction active in either of two epochs (Appendix~\ref{app:ewerr}).

We used photometry during the DESI exposures to assess whether flares could make quiescent dwarfs look active. We matched the individual DESI exposures to light curves from the Zwicky Transient Facility (ZTF) and the Transiting Exoplanet Survey Satellite (TESS; Appendix~\ref{app:flares}). Of the 5005 measured spectra of all M7--L1 dwarfs, high-proper-motion targets included, 326 have reliable photometry within an hour of an exposure. For 319 of them every such point agrees with the dwarf's own baseline. Six spectra have TESS coverage of every exposure, all of it quiescent. One of the other seven, the M7 dwarf Gaia DR3 1581886792134517504, was 1.26\,mag brighter in a ZTF image taken during its DESI exposure while the field stars in that image were steady. We keep it in the sample and flag it because its strong line may include a flare.

\subsection{Active fractions from M6 to L1}
\label{subsec:halpha_results}

\begin{figure*}[tbp]
    \centering
    \includegraphics[width=\textwidth]{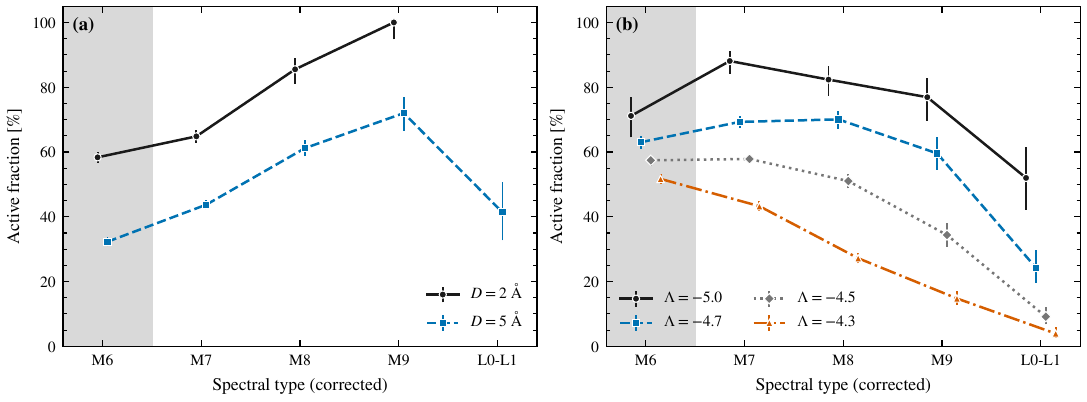}
    \caption{H$\alpha$ active fraction at a common threshold as a function of corrected spectral type; the shaded M6 bin is the comparison bin. (a) At an equivalent-width threshold $D$: the spectra used have a $3\sigma$ limit $\ell \le D$, and active means an equivalent width of at least $D$ at $\ge 3\sigma$. (b) At a threshold $\Lambda$ in $\log(\LHa)$, with $\chiH$ measured for each object. Thick bars are 68\% Wilson intervals; thin bars in (b) are 68\% ranges when $\chiH$ is redrawn from its uncertainty. Cells with fewer than ten spectra are omitted, and points are offset horizontally for visibility.}
    \label{fig:halpha}
\end{figure*}

A common threshold is needed because spectra with high limits contribute strong lines but no inactive dwarfs, and the median limit grows from 1.5\,\AA\ at M6 to 13\,\AA\ at M9 (Table~\ref{tab:sample}) as the continuum at H$\alpha$ fades and the targets get fainter. At a common equivalent width the active fraction rises steadily from M7 to M9 and falls again in the small L0--L1 bin (Fig.~\ref{fig:halpha}a, Table~\ref{tab:halpha}). At 5\,\AA\ the rise is 28 percentage points. The targeting mix changes with type, but the bright-time MWS targets alone show nearly the same rise. Figure~\ref{fig:halpha_individual}a shows the equivalent width of each dwarf against spectral type; these field values give a lower envelope for H$\alpha$ in young clusters, above the detection limit of each type.

Measured against the bolometric luminosity, the trend reverses (Fig.~\ref{fig:halpha}b): at $\Lambda=-4.5$ the active fraction falls by 23.5 percentage points from M7 to M9, and even at the lowest threshold the spectra allow, $\Lambda=-5.0$, it is flat or falling. The decline appears on the hybrid scale, with the per-type conversion of \citet{Schmidt2014} and among the bright-time MWS targets. A fixed equivalent width therefore corresponds to progressively weaker emission toward later types, while emission that is strong relative to $L_{\mathrm{bol}}$ becomes rarer.

Both trends are present in spectra whose limits are below 0.6 times the threshold. For these spectra, detection is nearly complete: a line at the threshold is detected in 91\% of injections. The M9 minus M7 differences at 5\,\AA\ and $\Lambda=-4.5$ are within 2 points of the standard ones (Table~\ref{tab:completeness}). The fractions weighted by the inverse probability of detecting each active dwarf's line are up to 4 points higher and follow the same trends. Both trends also appear in bins of bolometric luminosity, without reference to spectral type, and among the dwarfs within 60\,pc (Appendix~\ref{app:lbol}, Fig.~\ref{fig:lbol}).

\begin{table*}[tbp]
\caption{H$\alpha$ active fractions at a common threshold.}
\label{tab:halpha}
\centering\small
\setlength{\tabcolsep}{9pt}
\begin{tabular}{lccccc}
\toprule
Threshold & M6 & M7 & M8 & M9 & L0--L1 \\
\midrule

$D=1$\,\AA & \rcountfrac{corrected_EW_1_M6} & \rcountfrac{corrected_EW_1_M7} & \rcountfrac{corrected_EW_1_M8} & \rcountfrac{corrected_EW_1_M9} & \rcountfrac{corrected_EW_1_L0-L1} \\[3pt]
$D=2$\,\AA & \rcountfrac{corrected_EW_2_M6} & \rcountfrac{corrected_EW_2_M7} & \rcountfrac{corrected_EW_2_M8} & \rcountfrac{corrected_EW_2_M9} & \rcountfrac{corrected_EW_2_L0-L1} \\[3pt]
$D=5$\,\AA & \rcountfrac{corrected_EW_5_M6} & \rcountfrac{corrected_EW_5_M7} & \rcountfrac{corrected_EW_5_M8} & \rcountfrac{corrected_EW_5_M9} & \rcountfrac{corrected_EW_5_L0-L1} \\[3pt]
$\Lambda=-5.0$ & \rcountfrac{corrected_Lambda_-5_M6} & \rcountfrac{corrected_Lambda_-5_M7} & \rcountfrac{corrected_Lambda_-5_M8} & \rcountfrac{corrected_Lambda_-5_M9} & \rcountfrac{corrected_Lambda_-5_L0-L1} \\[3pt]
$\Lambda=-4.7$ & \rcountfrac{corrected_Lambda_-4.7_M6} & \rcountfrac{corrected_Lambda_-4.7_M7} & \rcountfrac{corrected_Lambda_-4.7_M8} & \rcountfrac{corrected_Lambda_-4.7_M9} & \rcountfrac{corrected_Lambda_-4.7_L0-L1} \\[3pt]
$\Lambda=-4.5$ & \rcountfrac{corrected_Lambda_-4.5_M6} & \rcountfrac{corrected_Lambda_-4.5_M7} & \rcountfrac{corrected_Lambda_-4.5_M8} & \rcountfrac{corrected_Lambda_-4.5_M9} & \rcountfrac{corrected_Lambda_-4.5_L0-L1} \\[3pt]
$\Lambda=-4.3$ & \rcountfrac{corrected_Lambda_-4.3_M6} & \rcountfrac{corrected_Lambda_-4.3_M7} & \rcountfrac{corrected_Lambda_-4.3_M8} & \rcountfrac{corrected_Lambda_-4.3_M9} & \rcountfrac{corrected_Lambda_-4.3_L0-L1} \\[3pt]
\midrule\multicolumn{6}{l}{Comparison: $\chiH$ per type from \citet{Schmidt2014}} \\
$\Lambda=-4.7$ & -- & \rcountfrac{K1s_corr_schmidt_Lambda_-4.7_M7} & \rcountfrac{K1s_corr_schmidt_Lambda_-4.7_M8} & \rcountfrac{K1s_corr_schmidt_Lambda_-4.7_M9} & \rcountfrac{K1s_corr_schmidt_Lambda_-4.7_L0-L1} \\[3pt]
$\Lambda=-4.5$ & -- & \rcountfrac{K1s_corr_schmidt_Lambda_-4.5_M7} & \rcountfrac{K1s_corr_schmidt_Lambda_-4.5_M8} & \rcountfrac{K1s_corr_schmidt_Lambda_-4.5_M9} & \rcountfrac{K1s_corr_schmidt_Lambda_-4.5_L0-L1} \\[3pt]
\bottomrule\end{tabular}
\tablefoot{Each cell gives active/total and the percentage with its 68\% Wilson interval. Only spectra whose detection limit lies at or below the threshold are included; -- marks cells without such spectra. The adopted $\Lambda$ uses $\chiH$ per object; M6 is an incompletely selected comparison bin.}
\end{table*}

\subsection{The bolometric scale}
\label{subsec:bolometric}

Because most of the bolometric flux of a late-M dwarf is emitted beyond the DESI range, the near-infrared flux sets the accuracy of $\chiH$. SPHEREx measures that part of the spectrum directly. Forward models reproduce SPHEREx spectra of M7 and M8 dwarfs, but the molecular opacities they rely on vary widely in quality across species \citep{Tata2026QORE}. We therefore integrate the measured spectrum. For 4793 typed dwarfs (M6 bin and high-proper-motion targets included; Table~\ref{tab:counts}) with clean SPHEREx spectra and a photometric bolometric flux, we replaced the photometric nodes beyond 0.98\,\um\ by the SPHEREx spectrum to 5\,\um\ and added a Rayleigh--Jeans tail (Appendix~\ref{app:sxdetail}). On our SPHEREx flux scale, which is tied to 2MASS and CatWISE, the SPHEREx values are lower by 0.02--0.03\,dex. On the 0.03\,dex brighter scale of the public library of \citet{Gagne2026}, built with their SPHEREx Photometry and Image Fitting Framework (SPIFF), the two agree within 0.006\,dex (Fig.~\ref{fig:fbol_sx}a, Table~\ref{tab:bolbudget}). The difference changes by 0.01\,dex from M7 to M9, much less than the weakening found in Sect.~\ref{subsec:forward}. We take the absolute bolometric scale to be uncertain by about 0.03\,dex, because the published bolometric fluxes of 36 dwarfs \citep{Sanghi2023} fall between our photometric and SPHEREx values (Table~\ref{tab:bolbudget}).

The choice between per-object and per-type $\chiH$ matters more, and most at M8, because the per-type values of \citet{Schmidt2014} fall by 0.38\,dex from M7 to M8 where the median of our per-object values falls by 0.21\,dex (Table~\ref{tab:chi}). With their values the M8 fraction at $\Lambda=-4.5$ is \rfrac{K1s_corr_schmidt_Lambda_-4.5_M8} instead of \rfrac{corrected_Lambda_-4.5_M8}. With this conversion, the velocity-dispersion difference between inactive and active M8 dwarfs (Sect.~\ref{subsec:kin_fixed}) is 30\% smaller (Table~\ref{tab:oldscale}). We adopt the per-object conversion: for 20 dwarfs in common it reproduces the per-object $\chiH$ of \citet{Schmidt2014} to 0.015\,dex. The difference therefore comes from their per-type means, and the SPHEREx fluxes exclude the bolometric error that would reconcile the two at M9 (Appendix~\ref{app:sxdetail}). The measured spectra without a bolometric flux, \rfrac{missing_fbol} of the total, are faint, with median detection limits of 16--23\,\AA. Assigning them type medians or SPHEREx values changes each active fraction by less than a percentage point. With either conversion and either bolometric flux, strong emission becomes rarer toward later types (Fig.~\ref{fig:fbol_sx}c).

\subsection{Continuum, weakening and an extra active population}
\label{subsec:forward}

We modelled the M8 and M9 active fractions as changes to the M7 distribution of $\log(\LHa)$: a shift in emission strength and an extra component drawn from active dwarfs. Let $S(x)$ be the probability that an M7 dwarf has $\log(\LHa)\ge x$, and $S_{\rm A}(x)$ the same probability for active M7 dwarfs, those at or above $-4.7$. The model population of a later type has a fraction $1-\phi$ drawn from $S$ and a fraction $\phi$ drawn from $S_{\rm A}$, with the weakening $\delta$ shifting both. For a target $i$ whose detection limit is at or below threshold $j$, the probability of being active is
\begin{equation}
p_{ij} = (1-\phi)\,S(x_{ij}-\delta) + \phi\,S_{\rm A}(x_{ij}-\delta),
\label{eq:joint}
\end{equation}
with $x_{ij}=\log(D\,\chi_{{\rm H}\alpha,i})$ at an equivalent width $D$ and $x_{ij}=\Lambda$ otherwise. We average $p_{ij}$ over these targets to predict the active fraction. With $\delta=\phi=0$, the M7 emission distribution is unchanged and each target's $\chiH$ converts it to equivalent width: this is the continuum term alone.

The M7 reference distribution depends on how we treat undetected lines. We built $S(x)$ from all 1695 measured M7 spectra with a $\chiH$ using the Kaplan--Meier estimator, with non-detections as upper limits \citep{Feigelson1985}. This assumes that a spectrum's detection limit is independent of its emission. Our alternative, the deep-spectrum distribution, uses the 208 M7 spectra with limits at or below $\Lambda=-4.9$ and needs no such assumption. These dwarfs are nearer than the typical M7 target, with median distance 62\,pc compared with 87\,pc for all measured M7 spectra, and are more often active. The deep-spectrum distribution lies above the observed M7 fractions and the Kaplan--Meier distribution below them. The difference between their fits measures how the inferred weakening and extra component depend on this choice. Detection limit and emission are related even within the full M7 sample: the deeper half is more often active above $\Lambda=-4.5$ by 6--8 points. Building the Kaplan--Meier distribution from either half shifts $\delta$ at M8 by 0.01\,dex and $\phi_{95}$, the one-sided 95\% upper limit on $\phi$, by up to 0.07.

We fitted $\delta$ and $\phi$, with $\phi$ between zero and one, separately for M8 and M9 to the eight differences from M7 in Table~\ref{tab:joint}. The equivalent-width and $\Lambda$ cuts are nested and have objects in common; adjacent thresholds are correlated by up to 0.75. We account for this with the objective $\chi^2=\mathbf{r}^{\mathsf T}\mathbf{C}^{-1}\mathbf{r}$, where $\mathbf{r}$ is the observed minus predicted differences and $\mathbf{C}$ their covariance. We estimated $\mathbf{C}$ from 2000 bootstraps that resample the M7, M8 and M9 dwarfs together and rebuild the M7 distribution from the same draw. Refitting each replicate with the same $\mathbf{C}$ gives the interval on $\delta$; the profile at $\Delta\chi^2=2.71$, the one-sided 95\% point at the $\phi=0$ boundary, gives the upper limit on $\phi$.

\begin{figure*}[tbp]
    \centering
    \includegraphics[width=\textwidth]{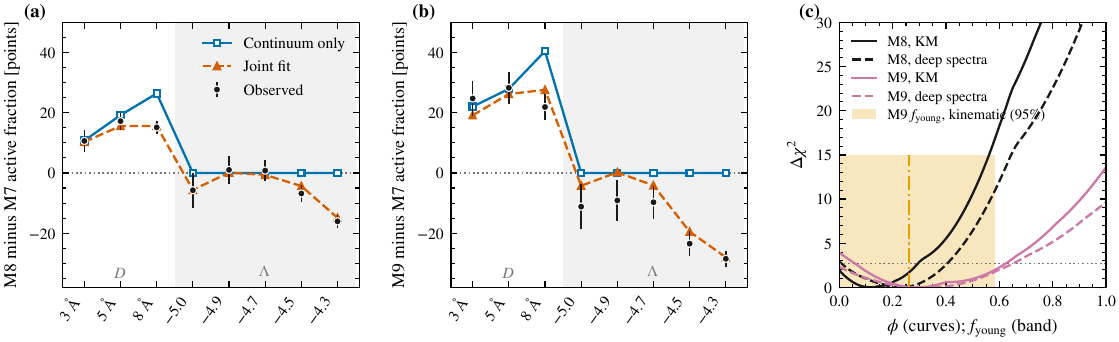}
    \caption{Joint model of the active fractions. (a, b) Active fraction at M8 and M9 minus that at M7 at three equivalent widths $D$ and five thresholds $\Lambda$ (shaded), among spectra sensitive to each threshold and with a measured $\chiH$; bars are bootstrap standard errors. Open squares: the unchanged M7 distribution of $\log(\LHa)$ converted with each target's $\chiH$ (Kaplan--Meier distribution). Triangles: the joint fit of Eq.~(\ref{eq:joint}) with the weakening $\delta$ and the weight $\phi$ of the extra active component. (c) Profile $\Delta\chi^2$ in $\phi$, with the full covariance of the eight differences, for M8 and M9 and the Kaplan--Meier (KM) and deep-spectrum M7 distributions; the dotted line marks the one-sided 95\% limit $\Delta\chi^2=2.71$, and the shaded band and dash-dotted line, drawn in the lower half of the panel, mark the 95\% range and best value of a different quantity, $f_{\rm young}$, the extra fraction of M9 dwarfs younger than 1\,Gyr from the pooled kinematics (Sect.~\ref{subsec:kin_pop}), which equals $\phi$ only under the assumption stated there.}
    \label{fig:joint}
\end{figure*}

Continuum fading accounts for the rise at 5\,\AA, but stronger lines require a change in the emission distribution. Both M7 distributions predict the observed 28-point rise from M7 to M9 at 5\,\AA\ (ratio 0.98). At 8\,\AA, however, the continuum term predicts an M9 excess of 40 points where 22 are observed; at fixed $\Lambda$ it predicts the M7 fractions where the M9 fractions are lower (Table~\ref{tab:joint}, Fig.~\ref{fig:joint}). Its fit probability is below $10^{-9}$ at both types. At M8, weakening alone fits all eight differences: $\delta=-0.12$\,dex for the Kaplan--Meier distribution and $-0.15$\,dex for the deep-spectrum distribution, with 95\% intervals excluding zero. We therefore attribute most of the M7 to M8 rise in the 5\,\AA\ fraction to the fading continuum acting on emission that itself weakens.

The extra component can represent either additional active dwarfs or a longer activity lifetime, since a longer step lifetime moves dwarfs from the inactive to the active part of the distribution in the same way. At the equivalent threshold $-4.7+\delta$, which maps onto the M7 activity cut, the component raises the active fraction by $\phi\,(1-\psi)$, where $\psi$ is the M7 active fraction at $-4.7$. At $-4.7$ itself the increase is $\phi\,[S_{\rm A}(-4.7-\delta)-S(-4.7-\delta)]$, a few percent smaller at M8 (Table~\ref{tab:joint}). For a constant star-formation rate over 10\,Gyr, the lifetime equivalent at the equivalent threshold is $\Delta\tau=\phi\,(1-\psi)\times10$\,Gyr. The fractions constrain these two interpretations together.

Evidence for this component at M8 comes from the brighter dwarfs used for the kinematics. Among dwarfs brighter than $G=19$ the same weakening is accompanied by $\phi=0.22$, with $\phi=0$ excluded at $\Delta\chi^2=5.3$. In the analysis sample, $\phi=0$ lies inside the one-sided 95\% region with the Kaplan--Meier distribution ($\Delta\chi^2=1.9$) and just outside it with the deep-spectrum distribution (2.9). The upper bounds in Table~\ref{tab:joint} allow lifetimes up to 1.5\,Gyr longer than at M7 among the brighter dwarfs and up to 1.0\,Gyr longer in the analysis sample. The brighter M8 dwarfs are more often active at 5\,\AA, while the M7 fractions agree between samples (Table~\ref{tab:twosamples}). A brighter magnitude limit favours nearby and overluminous dwarfs, which are more often active (Sect.~\ref{subsec:binaries}); these may account for the extra active component contributing to the M8 rise.

Selection also affects the type differences themselves. At 5\,\AA, the spectra with limits at or below the cut represent 42\% of the M7 analysis sample and 5\% of M9, selecting nearer and brighter dwarfs at both types. Drawing or reweighting M7 spectra to resemble the M8 and M9 spectra in detection limit, apparent magnitude, distance and overluminosity changes the M7 fraction by up to 4 points (Table~\ref{tab:selection}). This is comparable to the bootstrap errors: small compared with the differences at 5 and 8\,\AA\ and at $\Lambda=-4.5$, but large compared with the near-zero M8 difference at $-4.7$. At fixed $\Lambda$ the terms in detection limit and magnitude would enlarge the decline, and those in distance and overluminosity would reduce it by up to 3 points.

At M9 a uniform shift is a poor description of the emission distribution. The fractions require a larger weakening, $-0.27$\,dex, yet both weakening alone and the joint model give fit probabilities of 0.04 or less (Table~\ref{tab:joint}). Missed lines near the detection limit may explain part of this misfit. Restricting the fit to spectra with limits below 0.6 times the threshold makes the M9 fit acceptable; it uses about half the spectra and drops $\Lambda=-5.0$. With this selection the M8 weakening shifts by up to 0.06\,dex and still excludes zero, while its upper bound on the extra component decreases (Table~\ref{tab:joint}). The pooled kinematics constrain a separate quantity, $f_{\rm young}$, the extra fraction of M9 dwarfs younger than 1\,Gyr (Sect.~\ref{subsec:kin_pop}, Fig.~\ref{fig:joint}c). In the fractions, adding an active component improves the M9 fit because the distribution has lost more strong lines than a uniform shift removes. A weakening that grows with line strength would improve it in the same way: at M9, we interpret $\phi$ as a measure of the change in the shape of the emission distribution.

\begin{table*}[tbp]
\caption{Joint model of the M8 and M9 active fractions relative to M7.}
\label{tab:joint}
\centering
\scriptsize
\setlength{\tabcolsep}{3pt}
\begin{tabular}{llcccccccc}
\toprule
 & & \multicolumn{3}{c}{Equivalent width $D$} & \multicolumn{5}{c}{Threshold $\Lambda$ in $\log(\LHa)$} \\
Type & Row & 3\,\AA & 5\,\AA & 8\,\AA & $-5.0$ & $-4.9$ & $-4.7$ & $-4.5$ & $-4.3$ \\
\midrule
M8 & Observed & +10.6 (3.6) & +17.2 (2.8) & +15.1 (2.2) & $-$5.7 (5.9) & +1.0 (4.8) & +0.8 (3.4) & $-$6.8 (2.7) & $-$16.1 (2.2) \\
 & Continuum only & +10.7 & +19.2 & +26.4 & 0.0 & 0.0 & 0.0 & 0.0 & 0.0 \\
 & Joint fit & +10.3 & +15.6 & +15.6 & $-$5.6 & +0.1 & $-$0.6 & $-$4.3 & $-$14.9 \\
M9 & Observed & +24.7 (5.9) & +28.3 (5.4) & +21.9 (4.5) & $-$11.2 (7.5) & $-$9.1 (6.8) & $-$9.7 (5.5) & $-$23.5 (4.1) & $-$28.5 (2.7) \\
 & Continuum only & +22.0 & +27.9 & +40.4 & 0.0 & 0.0 & 0.0 & 0.0 & 0.0 \\
 & Joint fit & +19.1 & +26.2 & +27.6 & $-$4.3 & +0.2 & $-$4.2 & $-$19.5 & $-$28.0 \\
\midrule
Type & M7 distribution & \multicolumn{2}{c}{$\delta$ [dex] (95\%)} & $\phi$ & $\phi_{95}$ & $\Delta\tau_{95}$ [Gyr]$^e$ & $\chi^2_{\rm cont}$ (8) & $\chi^2_{\rm shift}$ (7), $p$ & $\chi^2_{\rm joint}$ (6), $p$ \\
\midrule
M8 & Kaplan--Meier & \multicolumn{2}{c}{$-$0.12 ($-$0.18, $-$0.09)} & 0.12 & 0.29 & 1.0 & 69 & 7.1, 0.42 & 5.2, 0.52 \\
 & Deep spectra & \multicolumn{2}{c}{$-$0.15 ($-$0.18, $-$0.08)} & 0.23 & 0.40 & 1.0 & 61 & 10.9, 0.14 & 8.0, 0.24 \\
M9 & Kaplan--Meier & \multicolumn{2}{c}{$-$0.27 ($-$0.37, $-$0.23)} & 0.29 & 0.62 & 2.2 & 150 & 17.3, 0.02 & 13.4, 0.04 \\
 & Deep spectra & \multicolumn{2}{c}{$-$0.27 ($-$0.37, $-$0.23)} & 0.29 & 0.65 & 1.7 & 109 & 21.1, 0.004 & 18.7, 0.005 \\
\midrule
\multicolumn{10}{l}{Variants at M8} \\
\multicolumn{2}{l}{\quad Restored sample, Kaplan--Meier} & \multicolumn{2}{c}{$-$0.12 ($-$0.17, $-$0.07)} & 0.11 & 0.24 & 0.9 & -- & -- & 7.0, 0.32 \\
\multicolumn{2}{l}{\quad Restored sample, deep spectra} & \multicolumn{2}{c}{$-$0.13 ($-$0.17, $-$0.08)} & 0.15 & 0.37 & 1.0 & -- & -- & 10.1, 0.12 \\
\multicolumn{2}{l}{\quad $G<19$ sample, Kaplan--Meier$^d$} & \multicolumn{2}{c}{$-$0.12 ($-$0.23, $-$0.05)} & 0.22 & 0.43 & 1.5 & -- & -- & 10.3, 0.11 \\
\multicolumn{2}{l}{\quad $\ell\le0.6$ threshold, Kaplan--Meier$^a$} & \multicolumn{2}{c}{$-$0.12 ($-$0.20, $-$0.06)} & 0.03 & 0.23 & 0.8 & -- & -- & 7.6, 0.18 \\
\multicolumn{2}{l}{\quad $\ell\le0.6$ threshold, deep spectra$^a$} & \multicolumn{2}{c}{$-$0.09 ($-$0.15, $-$0.06)} & 0.00 & 0.25 & 0.6 & -- & -- & 7.7, 0.18 \\
\multicolumn{2}{l}{\quad Kaplan--Meier, deeper half of M7$^b$} & \multicolumn{2}{c}{$-$0.13} & 0.14 & 0.32 & 1.1 & -- & -- & 6.2, 0.40 \\
\multicolumn{2}{l}{\quad Kaplan--Meier, shallower half of M7$^b$} & \multicolumn{2}{c}{$-$0.13} & 0.11 & 0.22 & 0.9 & -- & -- & 11.5, 0.07 \\
\midrule
\multicolumn{10}{l}{M9, pooled kinematics$^c$: extra fraction younger than 1\,Gyr $f_{\rm young}=0.26$, 95\% upper bound 0.59} \\
\bottomrule
\end{tabular}
\tablefoot{Upper block: active fraction at M8 or M9 minus that at M7, in percentage points, among spectra whose limit is at or below the threshold and with a measured $\chiH$, with the bootstrap standard error in parentheses; at 5\,\AA\ Table~\ref{tab:halpha}, which includes the spectra without $\chiH$, gives 17.5 points for M8. ``Continuum only'' is the unchanged M7 distribution of $\log(\LHa)$ converted with each target's $\chiH$, and ``joint fit'' adds the weakening $\delta$ and the weight $\phi$ of the extra active component (Eq.~\ref{eq:joint}; Kaplan--Meier distribution). Lower blocks: $\delta$ with its 95\% bootstrap interval; $0\le\phi\le1$; $\phi_{95}$ is the one-sided 95\% upper limit ($\Delta\chi^2=2.71$); $\Delta\tau_{95}=\phi_{95}\,(1-\psi)\times10$\,Gyr is the equivalent lengthening of the step lifetime at the equivalent threshold $-4.7+\delta$, where the extra component raises the active fraction by $\phi\,(1-\psi)$, with $\psi$ the active fraction of the M7 distribution at $\Lambda=-4.7$ (0.65 Kaplan--Meier, 0.74 deep spectra); at $\Lambda=-4.7$ itself the increase is 0.93--0.94 times as large at M8 and 0.81--0.84 at M9 (Sect.~\ref{subsec:forward}). $\chi^2$ uses the full bootstrap covariance of the differences; the degrees of freedom are in parentheses and $p$ is the fit probability. \tablefoottext{a}{A spectrum counts only if its limit is at most 0.6 times the threshold; the threshold $\Lambda=-5.0$, with fewer than ten such spectra at M7, is dropped, leaving 5 degrees of freedom.} \tablefoottext{b}{Covariance of the full Kaplan--Meier fit; the M7 spectra split at their median limit.} \tablefoottext{c}{From the pooled $\sUVW$ of M9 minus M7 in the $G<19$ sample (Sect.~\ref{subsec:kin_pop}); $f_{\rm young}$ is defined by age.} \tablefoottext{d}{Fit to M8 only; the $G<19$ sample has 237 M8 spectra with an H$\alpha$ measurement and $\chiH$.} \tablefoottext{e}{At the equivalent threshold $-4.7+\delta$.}}
\end{table*}

\begin{table}[tbp]
\caption{Selection term in the type differences of the active fraction [percentage points].}
\label{tab:selection}
\centering
\scriptsize
\setlength{\tabcolsep}{2.2pt}
\begin{tabular}{llrcccc}
\toprule
Threshold & Type & Diff. & Depth & $z$ & Distance & $z$+overlum. \\
\midrule
$D=5$\,\AA & M8 & +17.2 & +1.3 ($-$1.7, +4.8) & +0.3 & $-$0.4 & $-$0.7 ($-$3.5, +2.3) \\
 & M9 & +28.3 & +4.1 ($-$0.9, +9.0) & +0.5 & $-$2.8 & +0.5 ($-$3.5, +4.9) \\
$D=8$\,\AA & M8 & +15.1 & $-$2.3 ($-$4.0, $-$0.3) & +0.9 & +0.4 & $-$0.8 ($-$3.3, +2.0) \\
 & M9 & +21.9 & $-$1.7 ($-$4.3, +1.7) & +1.1 & $-$1.6 & $-$0.1 ($-$3.4, +3.4) \\
$\Lambda=-4.7$ & M8 & +0.8 & +2.8 ($-$0.8, +6.3) & +1.7 & +0.8 & $-$0.8 ($-$5.2, +2.7) \\
 & M9 & $-$9.7 & +0.8 ($-$1.6, +4.6) & +4.4 & $-$0.9 & $-$0.1 ($-$6.4, +5.6) \\
$\Lambda=-4.5$ & M8 & $-$6.8 & +4.1 (+1.9, +6.9) & +0.2 & $-$0.8 & $-$3.1 ($-$7.3, +0.5) \\
 & M9 & $-$23.5 & +4.2 (+1.3, +6.2) & +1.2 & $-$1.3 & $-$2.7 ($-$8.2, +2.2) \\
\bottomrule
\end{tabular}
\tablefoot{Diff.: active fraction of the type minus that of M7. The other columns give the change of the M7 fraction when the M7 spectra are restricted or reweighted to resemble the spectra of the type that reach the threshold: Depth, the deepest M7 spectra in the same proportion as for the type; $z$, reweighted in apparent $z$ magnitude; Distance, in distance (parallax $S/N\ge5$); $z$+overlum., jointly in $z$ magnitude and overluminosity (parallax $S/N\ge10$). A positive change would enlarge an M7 minus M8/M9 decline and shrink a rise. Intervals are 95\% bootstrap intervals.}
\end{table}

\section{Space motions, activity and activity lifetimes}
\label{sec:kinematics}

\subsection{Radial velocities and dispersions}
\label{subsec:kin_method}

We took radial velocities from the DESI MWS stellar pipeline \citep{Cooper2023, Koposov2024}, kept fits free of warnings with errors below 10\,\kms, and multiplied the errors by 1.34, the factor by which the formal errors underestimate the scatter between epochs for repeat targets. For 8276 of the M7--L1 dwarfs Redrock, a separate template fit to the same spectra, also gives a velocity. Its median difference from the MWS velocity runs from $-0.2$\,\kms\ at M7 to $+2.5$\,\kms\ at L0, with a scatter of 1.8\,\kms, a small fraction of the velocity dispersions below. Only four dwarfs have a Gaia radial velocity, too few for a test against an external scale. For 31 L dwarfs that lack an MWS velocity we used a template velocity calibrated on MWS velocities; they make up 1\% of the dwarfs with space motions, so the kinematics rest on the MWS velocities. Dwarfs included in the kinematic analysis require a unique Gaia DR3 match with $\varpi/\sigma_\varpi \ge 5$, renormalised unit weight error (RUWE) below 1.4 \citep{Lindegren2021} and a complete astrometric covariance matrix. Galactic velocities $(U,V,W)$ \citep{JohnsonSoderblom1987} are relative to the local standard of rest of \citet{Schonrich2010}. We computed the intrinsic dispersion of each component as the sample variance minus the mean measurement variance, with $\sUVW$ the quadrature sum of the three components, and its 95\% interval from 2000 bootstrap resamplings \citep{Efron1979} that also redraw the velocity error scale. We write $\Delta\sUVW = \sUVW^{\rm inact}-\sUVW^{\rm act}$ for the difference between inactive and active dwarfs of one type at one threshold, and compare the two groups unmatched and as one-to-one pairs matched in distance to within 10\%. The minimum detectable difference (MDD) is the difference a two-sided 5\% test would detect with 80\% probability. All kinematic results below use the dwarfs with $G<19$, free of the proper-motion selection (Sect.~\ref{subsec:selection}); Table~\ref{tab:twosamples} repeats the main results for the analysis and restored samples.

\subsection{Activity and velocity dispersion}
\label{subsec:kin_fixed}

\begin{figure*}[tbp]
    \centering
    \includegraphics[width=\textwidth]{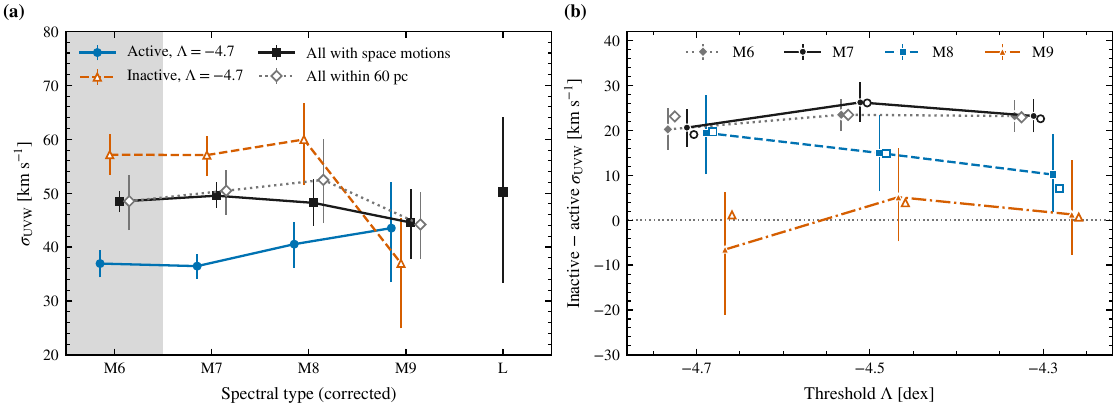}
    \caption{Velocity dispersion and H$\alpha$ activity of dwarfs with $G<19$ on the corrected types, with $\chiH$ measured for each object; the shaded M6 bin is the comparison bin. (a) Total dispersion $\sUVW$ for active (filled circles) and inactive (open triangles) dwarfs at $\Lambda = -4.7$, for all dwarfs with space motions (filled squares, including the point for all L dwarfs) and for all dwarfs within 60\,pc (open diamonds). (b) $\Delta\sUVW$ (inactive minus active) as a function of the threshold $\Lambda$, unmatched (filled symbols, lines, with 95\% bootstrap intervals) and for distance-matched pairs (open symbols; intervals in Table~\ref{tab:kin_comp}).}
    \label{fig:kinematics}
\end{figure*}

We measure larger velocity dispersions for inactive M7 and M8 dwarfs than for active dwarfs of the same type (Table~\ref{tab:kin_comp}, Fig.~\ref{fig:kinematics}; individual velocities in Fig.~\ref{fig:uvw_individual}). At $\Lambda=-4.7$, $\Delta\sUVW$ is 20.6\,\kms\ at M7 and 19.4\,\kms\ at M8, with 95\% confidence intervals that exclude zero; the M6 bin behaves like M7. Distance-matched pairs and the dwarfs within 60\,pc give similar differences (Table~\ref{tab:kin_comp}), with the selection and height comparisons in Table~\ref{tab:controls}. In bins of bolometric luminosity the inactive dwarfs are faster by 16--24\,\kms\ down to $\log(L_{\mathrm{bol}}/L_\odot)=-3.2$, and the active dwarfs move faster toward lower luminosity (Appendix~\ref{app:lbol}).

The emission cut changes the M8 comparison more than the M7 comparison. At $\Lambda=-4.3$ the M8 difference shrinks to 10.1\,\kms, whereas at M7 $\Delta\sUVW$ exceeds 20\,\kms\ at every threshold. An M8 line must be strong to count as active at this higher cut: dwarfs with moderate emission join the inactive group and reduce its dispersion difference from the active group.

\begin{table*}[tbp]
\caption{Velocity dispersions of active and inactive dwarfs with $G<19$ at a common threshold.}
\label{tab:kin_comp}
\centering
\small
\setlength{\tabcolsep}{4pt}
\begin{tabular}{llcccccc}
\toprule
Threshold & Type & $N_{\mathrm{act}}/N_{\mathrm{inact}}$ & $\sUVW^{\mathrm{act}}$ [\kms] & $\sUVW^{\mathrm{inact}}$ [\kms] & $\Delta\sUVW$ [\kms] & $\Delta\sUVW$, matched [\kms] & MDD [\kms] \\
\midrule
$\Lambda = -4.7$ & M6$^a$ & $331/199$ & $36.9\pm1.3$ & $57.1\pm1.9$ & $20.2$ ($15.6$, $25.0$) & $23.1$ ($18.2$, $27.8$) & $6.6$ \\
 & M7 & $342/172$ & $36.5\pm1.2$ & $57.1\pm1.9$ & $20.6$ ($16.3$, $24.8$) & $19.1$ ($14.3$, $24.0$) & $6.2$ \\
 & M8 & $129/52$ & $40.5\pm2.2$ & $60.0\pm3.9$ & $19.4$ ($10.3$, $27.9$) & $19.6$ ($9.2$, $30.1$) & $12.5$ \\
 & M9 & $31/19$ & $43.5\pm4.7$ & $37.0\pm5.2$ & $-6.5$ ($-21.1$, $6.3$) & $1.2$ ($-9.7$, $12.0$) & $19.7$ \\
$\Lambda = -4.5$ & M6$^a$ & $454/359$ & $35.7\pm1.1$ & $59.1\pm1.5$ & $23.5$ ($19.8$, $27.1$) & $23.4$ ($19.6$, $27.3$) & $5.2$ \\
 & M7 & $455/353$ & $35.6\pm1.0$ & $61.9\pm2.1$ & $26.2$ ($21.9$, $30.8$) & $26.1$ ($21.8$, $30.8$) & $6.4$ \\
 & M8 & $115/101$ & $40.7\pm2.6$ & $55.6\pm3.6$ & $14.9$ ($6.5$, $23.4$) & $14.9$ ($6.3$, $24.4$) & $12.2$ \\
 & M9 & $19/33$ & $38.3\pm2.8$ & $43.4\pm4.7$ & $5.1$ ($-4.7$, $16.1$) & $3.9$ ($-8.8$, $15.7$) & $15.1$ \\
$\Lambda = -4.3$ & M6$^a$ & $440/431$ & $35.3\pm1.1$ & $58.4\pm1.5$ & $23.2$ ($19.6$, $26.7$) & $22.9$ ($19.2$, $26.6$) & $5.2$ \\
 & M7 & $372/489$ & $34.7\pm1.1$ & $57.9\pm1.7$ & $23.2$ ($19.5$, $27.0$) & $22.6$ ($18.2$, $27.6$) & $5.5$ \\
 & M8 & $71/151$ & $41.2\pm3.4$ & $51.4\pm2.8$ & $10.1$ ($1.8$, $19.1$) & $7.0$ ($-2.3$, $16.3$) & $12.3$ \\
 & M9 & $11/41$ & $40.1\pm3.5$ & $41.4\pm4.1$ & $1.3$ ($-7.7$, $13.4$) & $0.7$ ($-13.4$, $13.7$) & $15.1$ \\
\midrule
$\Lambda = -4.7$, within 60\,pc & M7 & $103/48$ & $39.5\pm2.1$ & $63.6\pm3.6$ & $24.1$ ($15.3$, $31.6$) & -- & $11.8$ \\
 & M8 & $55/29$ & $43.9\pm3.4$ & $58.0\pm5.1$ & $14.1$ ($1.3$, $24.9$) & -- & $17.1$ \\
\midrule
$D = 5$\,\AA & M6$^a$ & $273/607$ & $33.9\pm1.4$ & $53.7\pm1.3$ & $19.8$ ($16.0$, $23.4$) & $20.6$ ($15.8$, $25.2$) & $5.3$ \\
 & M7 & $369/506$ & $34.6\pm1.0$ & $57.6\pm1.7$ & $23.0$ ($19.2$, $27.0$) & $22.9$ ($18.4$, $27.9$) & $5.7$ \\
 & M8 & $139/73$ & $41.2\pm2.4$ & $55.3\pm3.4$ & $14.1$ ($5.8$, $21.6$) & $13.2$ ($4.1$, $21.9$) & $11.5$ \\
\bottomrule
\end{tabular}
\tablefoot{$G<19$ sample: dwarfs with Gaia $G<19$ and DESI radial velocities, corrected types; $\Lambda$ uses $\chiH$ measured for each object. Only dwarfs whose spectral limit is at or below each threshold are included. Dispersions are given with their bootstrap standard errors; intervals are 95\% bootstrap intervals. Matched: one-to-one pairs matched in distance to within 10\%. MDD: minimum detectable difference of the unmatched comparison (two-sided 5\% test, 80\% power). Table~\ref{tab:twosamples} compares the main rows with the analysis and restored samples.
\tablefoottext{a}{Comparison bin.}}
\end{table*}

\begin{table}[tbp]
\caption{$\Delta\sUVW$ at $\Lambda=-4.7$ under selection and height controls ($G<19$).}
\label{tab:controls}
\centering
\scriptsize
\setlength{\tabcolsep}{2.5pt}
\begin{tabular}{llcc}
\toprule
Type & Sample & $N_{\rm act}/N_{\rm inact}$ & $\Delta\sUVW$ [\kms] (95\%) \\
\midrule
M7 & No cut & 342/172 & 20.6 (16.3, 24.8) \\
 & Without colour$-$selected & 324/168 & 20.5 (16.1, 24.9) \\
 & MWS only & 302/156 & 21.1 (16.4, 25.6) \\
 & Height-matched & 172/172 & 21.4 (16.0, 26.6) \\
 & Tail removed & 338/151 & 16.5 (12.8, 20.1) \\
 & Without colour$-$selected, tail removed & 320/147 & 16.4 (12.3, 20.1) \\
 & MWS only, tail removed & 299/137 & 16.7 (12.7, 20.4) \\
 & Height-matched, tail removed & 151/151 & 18.5 (13.8, 22.9) \\
\midrule
M8 & No cut & 129/52 & 19.4 (10.3, 27.9) \\
 & Without colour$-$selected & 100/40 & 21.4 (10.6, 31.0) \\
 & MWS only & 93/37 & 21.4 (9.4, 31.0) \\
 & Height-matched & 52/52 & 18.0 (7.7, 28.1) \\
 & Tail removed & 125/43 & 10.1 (3.3, 16.1) \\
 & Without colour$-$selected, tail removed & 96/32 & 10.9 (3.0, 17.8) \\
 & MWS only, tail removed & 89/29 & 9.8 (1.4, 16.7) \\
 & Height-matched, tail removed & 43/43 & 9.4 (0.8, 17.5) \\
\bottomrule
\end{tabular}
\tablefoot{Colour-selected: targets of the colour-selected programme (Table~\ref{tab:targeting}). MWS only: Milky Way Survey targets that were not also colour-selected. Height-matched: one-to-one pairs within 5\,pc or 10\% in the absolute height above the Galactic plane, computed from the distance and Galactic latitude with the Sun 20.8\,pc above the plane \citep{BennettBovy2019}. Intervals are 95\% bootstrap intervals.}
\end{table}

\begin{table}[tbp]
\caption{Velocity dispersions behind the M9 comparison.}
\label{tab:m9disp}
\centering
\scriptsize
\setlength{\tabcolsep}{3pt}
\begin{tabular}{lcr}
\toprule
Group & $\sUVW$ [\kms] (95\%) & $N$ \\
\midrule
M7, active at $\Lambda=-4.7$ & 36.5 (34.2--38.7) & 353 \\
M7, inactive at $\Lambda=-4.7$ & 56.9 (52.9--60.6) & 171 \\
M7, line of at least 2\,\AA & 35.5 (33.3--37.5) & 348 \\
M7, no line, limit at most 2\,\AA & 61.3 (57.1--65.2) & 174 \\
\quad resampled to the distances of inactive M9 & 65.4 (55.5--75.0) & 34 \\
M9, active at $\Lambda=-4.7$ & 41.0 (33.6--47.0) & 53 \\
M9, inactive at $\Lambda=-4.7$ & 37.1 (29.2--43.1) & 34 \\
\quad with a detected line & 42.1 (28.6--51.1) & 18 \\
M9 within 60\,pc, no line, too shallow for $-4.7$ & 73.1 (38.9--90.5) & 9 \\
\bottomrule
\end{tabular}
\tablefoot{Analysis sample with space motions, $\chiH$ per object. Active and inactive as in Table~\ref{tab:kin_comp}, which uses the $G<19$ sample (342 active and 172 inactive M7 dwarfs, against 353 and 171 here). This table adds 11 active and one inactive dwarf with $G\ge19$ and lacks two inactive high-proper-motion targets with Gaia DR3 $G=18.95$--19.00. Line: H$\alpha$ detected at $3\sigma$; limit: the $3\sigma$ equivalent-width limit $\ell$. Resampled: inactive M7 dwarfs drawn to the distance distribution of the inactive M9 dwarfs (median of the resampled dispersions). The last row counts M9 dwarfs within 60\,pc with no detected line whose spectra are too shallow for $\Lambda=-4.7$.}
\end{table}

\subsection{The high-velocity tail}
\label{subsec:tail}

Part of the inactive excess comes from a high-velocity tail. We define the tail as dwarfs with a space or tangential velocity of at least 100\,\kms, or a thick-disk plus halo probability of at least 0.5 in the kinematic model of \citet{Bensby2003, Bensby2014}. At $\Lambda=-4.7$ it contains 12\% of the inactive M7 dwarfs and 17\% of the inactive M8 dwarfs, compared with 1.2\% and 3.1\% of the active ones. The adopted age--velocity relation alone places 9\% of inactive M7 dwarfs and 11\% of inactive M8 dwarfs at such velocities. Most of this asymmetry is therefore the age difference between the groups, seen at its high-velocity end.

\begin{figure*}[tbp]
    \centering
    \includegraphics[width=\textwidth]{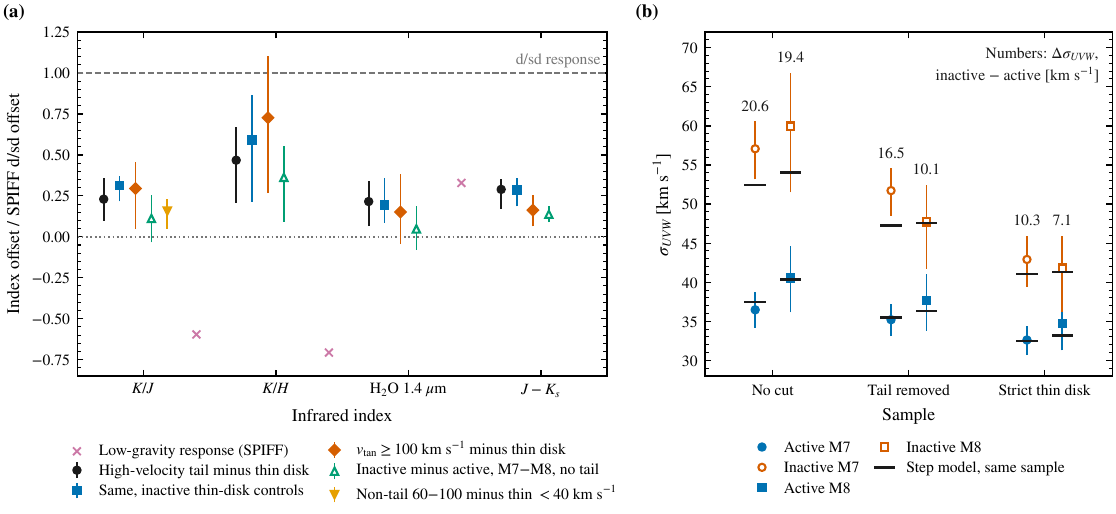}
    \caption{The high-velocity tail of the dwarfs with $G<19$. (a) Median infrared index offsets in units of the offset of SPIFF mild subdwarfs (d/sd) from field dwarfs of the same type (dashed line at 1), with 95\% bootstrap intervals: tail minus matched thin-disk dwarfs (filled circles), the same with inactive thin-disk controls (squares), a tail defined from tangential velocity alone (diamonds), inactive minus active dwarfs outside the tail (open triangles), and non-tail dwarfs at 60--100\,\kms\ minus thin-disk dwarfs below 40\,\kms\ ($K/J$ only, down-pointing triangle). Crosses show the response of SPIFF low-gravity dwarfs. (b) $\sUVW$ of active (filled) and inactive (open) M7 and M8 dwarfs with $G<19$ at $\Lambda=-4.7$ for all dwarfs, with the tail removed and in the strict thin disk, with 95\% bootstrap intervals; horizontal bars are the step model for the same sample, and the numbers give $\Delta\sUVW$.}
    \label{fig:tail}
\end{figure*}

Inside the thin disk the inactive M7 dwarfs are still older than the active ones, whereas at M8 about half of $\Delta\sUVW$ comes from the high-velocity dwarfs. Removing the tail from both groups reduces the difference at $\Lambda=-4.7$ to 16.5\,\kms\ at M7 and 10.1\,\kms\ at M8 (Table~\ref{tab:kinvariants}, Fig.~\ref{fig:tail}b). In the strict thin disk, where the thick-disk plus halo probability is below a tenth of the thin-disk probability, the M8 difference barely excludes zero. Both types have differences close to the predictions of the step model of Sect.~\ref{subsec:lifetimes} for the same cut.

The tail also differs in its atmospheres. We compared each tail dwarf with up to three dwarfs of high thin-disk probability matched in type, apparent magnitude $W1_{\rm Vega}$ ($z-3.0$ where CatWISE has no $W1$) and distance. In the combined M6--M9 sample, the tail dwarfs are bluer in 2MASS $J-K_s$ by 0.063\,mag and have lower SPHEREx $K/J$ and deeper 1.4\,\um\ water (Fig.~\ref{fig:tail}a, Table~\ref{tab:tail_tests}). This infrared result is significant at M7 alone but not at M8 alone, with 14 matched tail dwarfs. The same offsets occur with other control groups and tail definitions. In the analysis sample, independent VHS and first-epoch UKIDSS photometry gives the same $J-K$ offset. By comparison, Legacy Surveys $r-i$ and $i-z$ show no offset at a sensitivity of a few hundredths of a magnitude (Table~\ref{tab:tail_tests}). Weaker $K$-band flux relative to $J$ indicates stronger collision-induced H$_2$ absorption, which accompanies lower metallicity, while low gravity raises $K/J$. In the SPIFF library the four mild subdwarfs at M8--L2 (class d/sd, between dwarfs and subdwarfs) have $K/J$ lower by 0.069 than dwarfs of the same type, and low-gravity dwarfs higher by 0.041. The tail offset, at a median type of M6.5, is roughly a quarter of the subdwarf value.

The optical metallicity index $\zeta$ points the other way. It compares the TiO5 band at 7126--7135\,\AA\ with the CaH2 and CaH3 bands at 6814--6846 and 6960--6990\,\AA, each measured against 7042--7046\,\AA\ \citep{Lepine2007}; $\zeta$ is near 1 at solar metallicity, and a smaller TiO5 means deeper TiO. In the analysis sample the tail has higher $\zeta$, from a TiO5 band deeper by 0.013 at unchanged CaH (Table~\ref{tab:tail_tests}). The calibrations were built for earlier M dwarfs \citep{Dhital2012, Woolf2009, Mann2013, Lepine2013}, and in 76\% of our thin-disk M7 dwarfs TiO5 is below 0.25, where the band approaches saturation.

Matching in optical type also fixes the TiO strength on which $\zeta$ depends, since our types come from fits dominated by the same TiO and VO bands. If the tail dwarfs are metal-poor, that matching selects tail dwarfs cooler than their controls, because lower metallicity weakens TiO at fixed temperature. Along the thin-disk sequence the TiO5 offset corresponds to 0.4 subtype, over which $K/J$ rises by 0.009 and $J-K_s$ by 0.024. Such a temperature difference would make the tail redder, whereas it is bluer. With controls matched in absolute $G$ magnitude, which does not use the optical bands, the $\zeta$ offset vanishes and the infrared offsets remain (Table~\ref{tab:tail_tests}). We therefore base the metallicity interpretation on the infrared colours: the high-velocity dwarfs are the old, mildly metal-poor end of the disk population, from the old thin disk or a thick-disk component, as the age--metallicity relation leads one to expect. Dwarfs at 60--100\,\kms\ that fall short of the tail already show a smaller $K/J$ deficit. Earlier studies found the same link between blue $J-K_s$ and high velocity among L dwarfs \citep{Faherty2009, Schmidt2010}, and metal-poor ultracool dwarfs are known to have high velocities \citep{Zhang2017}.

\subsection{Overluminous dwarfs and unresolved binaries}
\label{subsec:binaries}

\begin{figure*}[tbp]
    \centering
    \includegraphics[width=\textwidth]{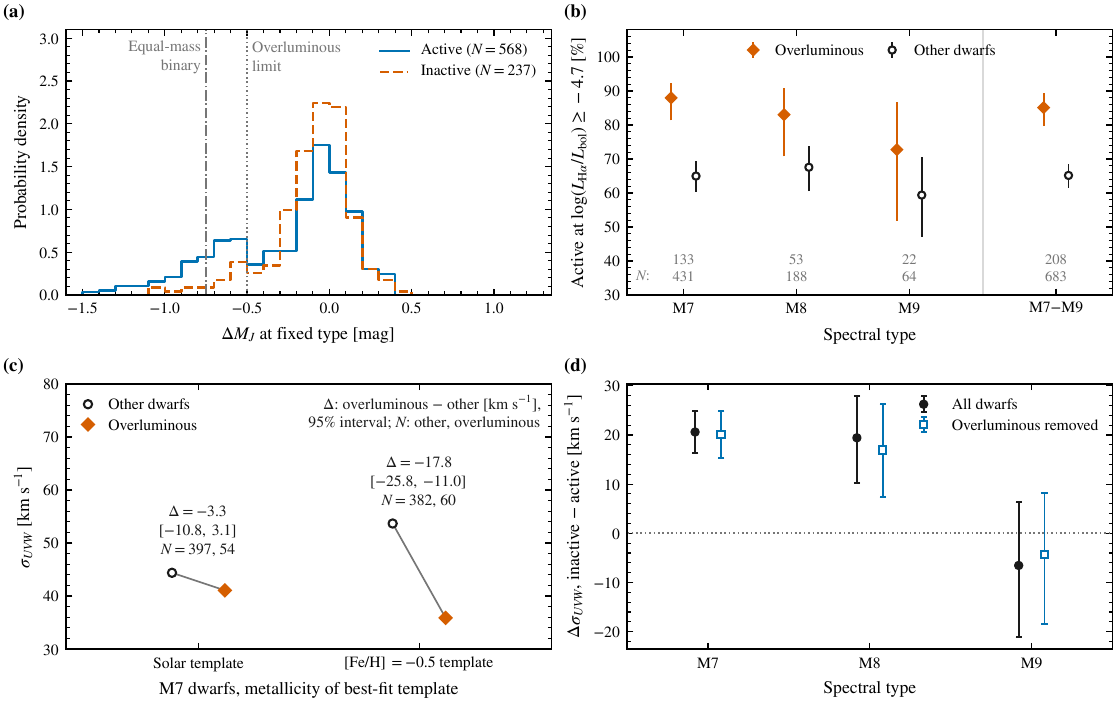}
    \caption{Overluminous dwarfs and activity. (a) Distribution of the offset $\Delta M_J$ from the $M_J$ sequence for active and inactive M7--M8 dwarfs at $\Lambda=-4.7$; the dotted line marks the overluminosity cut at $-0.5$\,mag and the dash-dotted line the offset of an equal-flux pair. (b) Active fraction at $\Lambda=-4.7$ for overluminous and other dwarfs by type and pooled, with 95\% Wilson intervals and the numbers of dwarfs; of the 954 M7--M9 dwarfs with limits at or below $-4.7$, 891 have the parallax ($S/N\ge10$) and $J$ magnitude that an $M_J$ offset needs; panels (a) and (b) use the analysis sample. (c) $\sUVW$ of overluminous and other M7 dwarfs with $G<19$, split by the metallicity of the best-fitting template, with the difference and its 95\% bootstrap interval; this panel uses all 893 such M7 dwarfs with space motions and an $M_J$ offset, whatever their activity class. (d) $\Delta\sUVW$ of dwarfs with $G<19$ at $\Lambda=-4.7$ with and without the overluminous dwarfs, with 95\% bootstrap intervals.}
    \label{fig:overlum}
\end{figure*}

Imaging and volume-limited studies establish a substantial binary population at these types \citep{Close2003, DucheneKraus2013, BardalezGagliuffi2019}. Unresolved binaries appear brighter than single dwarfs of the same type, a magnitude limit favours them, and a companion could add or dilute H$\alpha$ emission. We fitted $M_J$ as a function of corrected type for M6--L1 dwarfs with parallax $S/N\ge10$. We write $\Delta M_J$ for the offset from this DESI sequence (the offsets of Table~\ref{tab:binary} are from the field sequence) and call a dwarf overluminous when it is at least 0.5\,mag above the DESI sequence. This cut recovers most equal-flux pairs in injection tests. Overluminous dwarfs make up 15\% of the M7--M9 dwarfs with an $M_J$ offset and 23\% of those with limits at or below $\Lambda=-4.7$ (Fig.~\ref{fig:overlum}), because a common limit favours bright, nearby dwarfs. They are more often active, 85\% compared with 65\% at $\Lambda=-4.7$ (Fig.~\ref{fig:overlum}b). After stratifying by type and spectral $S/N$, the Mantel--Haenszel odds ratio is 2.9 (1.9--4.4), and a sequence in $M_G$, which avoids infrared photometry, leads to a similar value. For the sample without the overluminous dwarfs, $\Delta\sUVW$ is lower by at most 3\,\kms, whether the tail is included or omitted (Fig.~\ref{fig:overlum}d, Table~\ref{tab:kinvariants}). The inactive dwarfs on the single-star sequence are therefore also kinematically older.

The overluminous M7 dwarfs form two groups (Fig.~\ref{fig:overlum}c). In a fit over the full library of \citet{Kesseli2017}, which includes templates of non-solar metallicity, those best fitted by solar-metallicity templates have the kinematics of other M7 dwarfs, as ordinary binaries should. The 60 best fitted by the $[{\rm Fe/H}]=-0.5$ template have a dispersion of 36\,\kms, compared with 54\,\kms\ for other dwarfs with that fit. In the analysis sample their K\,\textsc{i} 7699\,\AA\ absorption (measured in 7690--7710\,\AA) is weaker by 1.05\,\AA, compared with a typical 6\,\AA. Excess luminosity, weak alkali lines and a low dispersion together point to young dwarfs of low gravity, which the template fit records as metal-poor. Among the other dwarfs, the same fit picks out the fast-moving dwarfs of the tail. The overluminous dwarfs thus appear to be mostly binaries with a young minority, both preferentially active. Resolved companions are few. Among the 173 dwarfs with usable \emph{Euclid} DR1 VIS images, fits with one and two point-spread functions resolve three close binaries whose components share a proper motion (Table~\ref{tab:vis_pairs}, Fig.~\ref{fig:vis_pairs}). The M7 dwarf TARGETID 39627853691618796 has a companion at 0.24\arcsec\ ($\Delta\mathrm{VIS}=0.52$\,mag) that HST ACS images of 2005 resolve at the same separation and position angle, within 0.01\arcsec\ and 5\degr, after both components moved 0.74\arcsec. The M6 dwarf 39633440647153130 has a companion at 0.22\arcsec\ ($\Delta\mathrm{VIS}=1.10$\,mag), 24\,au in projection at 108\,pc, and the M7 dwarf 39633509735729451 one at 0.53\arcsec\ ($\Delta\mathrm{VIS}=0.48$\,mag), 48\,au at 91\,pc; for these two the shapes of the blended images in SDSS (2000--2001) and Pan-STARRS1 (2010--2014) exclude a stationary background star. Two wider pairs, at 0.78 and 1.03\arcsec, are possible companions that the archival images cannot test. The 13 neighbours in Legacy Surveys images whose Gaia astrometry matches that of the target are candidate wide companions (Appendix~\ref{app:vis}).

\subsection{Members of young associations}
\label{subsec:banyan}

Membership in a nearby young association gives each dwarf an age that does not depend on its velocity dispersion. For the 3163 M7--L1 dwarfs with $\varpi/\sigma_\varpi\ge5$ we computed membership probabilities with BANYAN $\Sigma$ \citep{Gagne2018}, using the current association models \citep{GagneMOCA2026} and \emph{Gaia} positions, proper motions and parallaxes. We find 27 candidates (0.9\%) with a probability of at least 90\% of belonging to any association; the original BANYAN $\Sigma$ models of \citet{Gagne2018} give 13, in Argus, AB Doradus and Carina-Near. Of the 25 candidates with a DESI radial velocity, ten stay above 90\% when the velocity is included. Young association members are therefore at most about 1\% of the sample, and few enough to leave the kinematics unchanged: among the dwarfs with $G<19$ classified at $\Lambda=-4.7$, candidates are 7 of 342 active and none of 172 inactive M7 dwarfs, and one each of 129 active and 52 inactive M8 dwarfs. Removing them changes $\Delta\sUVW$ by at most 0.4\,\kms. The membership test finds only dwarfs of the modelled associations, so dwarfs of similar age that have dispersed into the field are constrained by the dispersions themselves.

\subsection{The inactive M9 dwarfs}
\label{subsec:m9}

At M9 the inactive dwarfs are no faster than the active ones: at $\Lambda=-4.7$, $\Delta\sUVW=-6.5$\,\kms, with a minimum detectable difference about equal to the M7 difference. Because the $G<19$ sample contains only 19 inactive M9 dwarfs, we examine them in the analysis sample. At M9 this threshold corresponds to about 6\,\AA, and every inactive M9 spectrum has a limit above 2\,\AA. In the analysis sample, eighteen of the 34 inactive M9 dwarfs have a detected line (median 4.4\,\AA). The 34 have the dispersion of M7 dwarfs with lines of at least 2\,\AA. By comparison, M7 dwarfs with no line at limits of 2\,\AA\ or less are much faster, including when drawn to the distances of the inactive M9 dwarfs (Table~\ref{tab:m9disp}). Most inactive M9 dwarfs are thus weak emitters below the threshold. The M9 activity groups at $\Lambda=-4.7$ cannot be compared with the M7 groups, which is why our comparison of activity with age and of the lifetimes does not include M9. Their optical gravity indicators agree: they are fainter in $M_G$ than active M9 dwarfs, the opposite of young, inflated dwarfs \citep{Cruz2009, Liu2016}, and have equal or stronger Na\,\textsc{i} and K\,\textsc{i} absorption, whose weakening signals low gravity in young brown dwarfs \citep{Gorlova2003, Allers2013}. Within 60\,pc, nine line-free M9 dwarfs whose limits lie above the $-4.7$ threshold are as fast as the line-free M7 dwarfs (Table~\ref{tab:m9disp}), which suggests that the old, line-free M9 dwarfs are concentrated among the spectra with high limits.

\subsection{Activity lifetimes at M7 and M8}
\label{subsec:lifetimes}

A longer activity lifetime at later types would make active dwarfs of later types faster when they are selected at the same threshold. At $\Lambda = -4.7$ the active M8 dwarfs do move faster than the active M7 dwarfs, but only by 4\,\kms, 1.6 times the error of the difference (Table~\ref{tab:kin_comp}). To turn dispersions into lifetimes we used a step model, in which each dwarf is active until age $\tau$ and inactive afterwards, with a constant star-formation rate over 10\,Gyr, the age--velocity relation of \citet{Aumer2009} and the asymmetric drift of \citet{Schonrich2010}. The likelihood of a dwarf's velocity given its activity class (Eq.~\ref{eq:like}) uses every kinematic dwarf with an H$\alpha$ measurement \citep{Isobe1986}. A spectrum whose limit is above $\Lambda$ contributes the probability that its line lies below its limit, estimated from the active dwarfs of its type. The high-velocity tail extends beyond a single thin disk. Our main fits therefore describe dwarfs outside this tail, with each dwarf's likelihood divided by the probability that a dwarf of its age and sky position satisfies the cut. For simulated M7 samples, the input lifetime is recovered under every cut.

At $\Lambda=-4.7$ the thin-disk lifetimes are 5.9\,Gyr at M7 (95\% interval 5.3--6.6) and 6.4\,Gyr at M8 (5.2--8.0; Table~\ref{tab:kinvariants}). Other definitions of the disk lead to 5.3--8.1\,Gyr, the longest without any tail cut at M8 (Tables~\ref{tab:kinvariants} and \ref{tab:lifetimes}). In the strict thin disk the model reproduces the dispersion of each group to within 2\,\kms, and with the looser tail cut it underpredicts the inactive M7 dispersion by 4\,\kms. The lifetime shortens as $\Lambda$ rises, because $\tau$ is the age up to which a dwarf keeps emission above $\Lambda$ and M8 emission is weaker relative to $L_{\mathrm{bol}}$ (Table~\ref{tab:lifetimes}). In the strict thin disk $\Delta\sUVW$ at M8 is barely above zero. Here the constraint on the M8 lifetime comes from the dispersion of the active dwarfs mapped through the age--velocity relation, while the inactive dwarfs alone allow almost any value. Fitted to the active M8 dwarfs alone, the lifetime is 6.0\,Gyr, close to the 5.9\,Gyr from all dwarfs (Tables~\ref{tab:kinvariants} and \ref{tab:lifetimes}).

A step is the simplest activity history. If the probability of being active instead declines over a width $w$ around $\tau$ (a logistic function of age) or exponentially (Appendix~\ref{app:like}, Table~\ref{tab:forms}), the M7 velocities favour the step, by $\Delta\ln\mathcal{L}=5.2$ for $w=1$\,Gyr and by more for wider or exponential declines. The M8 velocities hardly distinguish the histories. The mean duration of activity, which equals $\tau$ for a step, then differs between M8 and M7 at equivalent thresholds by 0.8\,Gyr for $w=1$\,Gyr and by 1.2\,Gyr for an exponential decline. A gradual decline therefore changes the comparison by less than its uncertainty.

\begin{table*}[tbp]
\caption{Velocity dispersions of active and inactive M7 and M8 dwarfs with $G<19$ at $\Lambda=-4.7$ with the high-velocity tail and the overluminous dwarfs removed.}
\label{tab:kinvariants}
\centering
\scriptsize
\setlength{\tabcolsep}{3pt}
\begin{tabular}{lllccccccc}
\toprule
Type & Sample & $N_{\rm act}/N_{\rm inact}$ & $\sigma_U$ & $\sigma_V$ & $\sigma_W$ & $\sUVW$ & $\Delta\sUVW$ (95\%) & Model $\Delta\sUVW$ & $\tau$ [Gyr] (95\%) \\
\midrule
M7 & No cut & 342/172 & 28.3/43.6 & 19.0/29.3 & 13.1/22.3 & 36.5$\pm$1.2/57.1$\pm$1.9 & 20.6 (16.3, 24.8) & 14.9 & 6.5 (5.9, 7.2) \\
 & Tail removed & 338/151 & 27.5/41.0 & 17.8/25.2 & 13.0/19.0 & 35.2$\pm$1.0/51.7$\pm$1.6 & 16.5 (12.8, 20.1) & 11.8 & 5.9 (5.3, 6.6) \\
 & Strict thin disk & 319/95 & 25.9/34.8 & 16.1/19.7 & 11.5/15.5 & 32.6$\pm$1.0/42.9$\pm$1.7 & 10.3 (6.4, 13.9) & 8.5 & 5.3 (4.5, 6.1) \\
 & Tail and overluminous removed & 258/140 & 28.1/40.5 & 17.8/25.3 & 13.6/19.2 & 36.0$\pm$1.2/51.5$\pm$1.6 & 15.5 (11.5, 19.2) & -- & 6.3 (5.6, 7.2) \\
\midrule
M8 & No cut & 129/52 & 32.8/45.2 & 19.0/32.6 & 14.5/22.2 & 40.5$\pm$2.2/60.0$\pm$3.9 & 19.4 (10.3, 27.9) & 13.7 & 8.1 (6.7, 9.7) \\
 & Tail removed & 125/43 & 30.1/34.0 & 17.6/26.7 & 14.2/20.3 & 37.6$\pm$1.8/47.7$\pm$2.7 & 10.1 (3.3, 16.1) & 11.2 & 6.4 (5.2, 8.0) \\
 & Strict thin disk & 115/34 & 28.6/31.8 & 15.2/21.6 & 12.5/16.5 & 34.7$\pm$1.7/41.8$\pm$2.5 & 7.1 (0.6, 12.1) & 8.1 & 5.9 (4.4, 8.1) \\
 & Tail and overluminous removed & 99/41 & 31.7/32.6 & 17.7/26.4 & 14.6/19.7 & 39.1$\pm$2.0/46.4$\pm$2.6 & 7.2 (0.2, 13.3) & -- & 6.8 (5.4, 8.8) \\
\bottomrule
\end{tabular}
\tablefoot{Dispersions in \kms, given as active/inactive, with bootstrap standard errors for $\sUVW$. $\Delta\sUVW$ is inactive minus active with a 95\% bootstrap interval. Tail: space or tangential velocity of at least 100\,\kms, or thick-disk plus halo probability of at least 0.5. Strict thin disk: also (thick plus halo)/thin probability below 0.1. Overluminous: at least 0.5\,mag above the $M_J$ sequence (Sect.~\ref{subsec:binaries}). Model $\Delta\sUVW$: the step model at the fitted $\tau$ under the same cut. $\tau$: censored step lifetime with a 95\% profile-likelihood interval (grid of Appendix~\ref{app:like}, rounded to 0.1\,Gyr); for the tail cuts the likelihood is truncated (Sect.~\ref{subsec:lifetimes}). The overluminous cut does not depend on velocity, so it is not truncated; because it preferentially removes young dwarfs, the last row of each type is an upper value of $\tau$.}
\end{table*}

\begin{table*}[tbp]
\caption{Step lifetimes at M7 and M8 and their difference $\Delta\tau=\tau_{\rm M8}-\tau_{\rm M7}$ [Gyr].}
\label{tab:dtau}
\centering
\scriptsize
\setlength{\tabcolsep}{4pt}
\begin{tabular}{lccccc}
\toprule
 & & \multicolumn{2}{c}{M8 at $\Lambda_{\rm M8}=-4.7$} & \multicolumn{2}{c}{M8 at $\Lambda_{\rm M8}=-4.7+\delta$} \\
Model & $\tau_{\rm M7}$ & $\tau_{\rm M8}$ & $\Delta\tau$ (95\%) & $\tau_{\rm M8}$ & $\Delta\tau$ (95\%) \\
\midrule
Uniform ages, $G<19$ (adopted) & 5.9 & 6.4 & +0.5 ($-$1.1, +2.3) & 6.8 & +0.9 ($-$0.8, +3.5) \\
Brown-dwarf ages, volume-weighted & 6.5 & 7.6 & +1.1 ($-$0.7, +3.8$^b$) & 8.0 & +1.5 ($-$0.3, +3.7$^b$) \\
Brown-dwarf ages, magnitude-weighted$^c$ & 8.2 & 9.9 & +1.8 ($-$0.2, +2.2$^b$) & 9.3 & +1.1 ($-$0.1, +2.0$^b$) \\
Uniform ages over 0--12\,Gyr & 5.8 & 6.3 & +0.6 & 6.7 & +1.0 \\
\midrule
Thick-disk component & 5.9 & 6.3 & +0.4 & 6.7 & +0.8 \\
Relation of \citet{Holmberg2009} & 6.3 & 6.8 & +0.5 & 7.2 & +0.9 \\
Strict thin disk & 5.3 & 5.9 & +0.6 & 6.2 & +0.9 \\
Single disk, no tail cut & 6.5 & 8.1 & +1.6 & 8.4 & +1.9 \\
Brown-dwarf ages, volume, $s=1.0$ & 6.3 & 7.0 & +0.7 & 7.4 & +1.1 \\
Brown-dwarf ages, volume, $s=1.6$ & 6.2 & 6.9 & +0.8 & 7.3 & +1.2 \\
Brown-dwarf ages, volume, $s=2.1$ & 6.1 & 6.8 & +0.7 & 7.2 & +1.2 \\
\midrule
Uniform ages, analysis sample$^a$ & 5.6 & 5.4 & $-$0.3 ($-$1.3, +1.8$^b$) & 5.8 & +0.2 ($-$1.0, +1.7) \\
Uniform ages, restored sample$^a$ & 6.1 & 7.2 & +1.1 ($-$0.3, +4.1$^b$) & 7.5 & +1.4 (+0.0, +3.0) \\
\bottomrule
\end{tabular}
\tablefoot{Truncated censored step-model fits with the high-velocity tail removed (Eq.~\ref{eq:like}); M7 always at $\Lambda=-4.7$; dwarfs with $G<19$ unless stated. The right-hand columns move the M8 threshold by the weakening $\delta$ of the joint model (Kaplan--Meier distribution; $-4.82$), the M8 threshold equivalent to $-4.7$ at M7; there 21 M8 dwarfs with $G<19$ are inactive, compared with 43 at $-4.7$. Intervals are 95\% percentiles of 2000 joint bootstraps that resample the M7 and M8 kinematic dwarfs, the dwarfs defining each type's activity distribution and, in the right-hand columns, $\delta$ from the joint-model bootstrap; the same replicates give the step row of Table~\ref{tab:forms}. Values at the end of the $\tau$ grid (9.95\,Gyr; 11.95\,Gyr for ages over 0--12\,Gyr; Appendix~\ref{app:like}) are lower limits. Lifetimes are rounded to 0.1\,Gyr from the 0.05\,Gyr grid of the fits, and differences were taken before rounding. The 95\% intervals of the individual lifetimes in the adopted row are 5.3--6.6\,Gyr at M7 and 5.2--8.0\,Gyr at M8 (Table~\ref{tab:kinvariants}). The middle block changes the model for both types at once; $s$ shifts the temperature of each bin by $s$ field subtypes (Table~\ref{tab:typescale}). \tablefoottext{a}{With its own $\delta$ (Table~\ref{tab:joint}).} \tablefoottext{b}{At least 2.5\% of the M8 replicates are at the end of the grid, so the upper bound is set by the end of the age range and is a lower limit.} \tablefoottext{c}{Both M8 values, 9.9 and 9.3\,Gyr, lie near the end of the grid, where the likelihood is flat (note $b$).}}
\end{table*}

A constant star-formation rate implies the same age distribution at every type. Brown dwarfs, however, pass through M7 and M8 while young, giving substellar population models age distributions that depend on type \citep{Burgasser2004, Allen2005}. We computed such distributions for the two bins from the evolutionary tracks of \citet{Baraffe2015}, the temperature scale of \citet{Pecaut2013} and the initial mass function of \citet{Chabrier2003}, weighted by volume or, as a magnitude limit does, by $L^{1.5}$. Both lengthen the lifetimes, more at M8 than at M7 (Tables~\ref{tab:lifetimes} and \ref{tab:dtau}). We disfavour the magnitude-weighted model because, in the analysis sample, it predicts pooled dispersions 9\,\kms\ below the observed one at M7. In the fit to the active dispersion the lengthening comes from the brown dwarfs: for masses of at least 0.075\,M$_\odot$, $\tau$ is within 0.2\,Gyr of the constant-rate value.

To test whether activity lasts longer at later types (Sect.~\ref{sec:intro}), we use $\Delta\tau=\tau_{\rm M8}-\tau_{\rm M7}$, with an uncertainty that includes both types. Because M8 emission is 0.12\,dex weaker at fixed incidence (Sect.~\ref{subsec:forward}), $\Lambda=-4.7$ is a stricter cut at M8. We therefore compare the M7 lifetime at $-4.7$ with the M8 lifetime at $\Lambda_{\rm M8}=-4.7+\delta=-4.82$, the equivalent threshold. One joint bootstrap resamples the M7 and M8 kinematic dwarfs together with the dwarfs that define each type's activity distribution and redraws $\delta$ from the joint-model bootstrap. At this threshold $\tau_{\rm M8}=6.8$\,Gyr and $\Delta\tau=0.9$\,Gyr ($-0.8$ to 3.5). The lifetime difference is smaller at the same threshold for both types (Table~\ref{tab:dtau}). A thick-disk component, the relation of \citet{Holmberg2009} or the strict thin disk move it by at most 0.1\,Gyr. Without a tail cut, it is longer by 1.0\,Gyr because the fast inactive M8 dwarfs are treated as old thin-disk dwarfs. Brown-dwarf cooling ages lengthen the M8 lifetime more than the M7 one. At the warmer temperatures of the type scale (Sect.~\ref{subsec:validation}) the age distributions are close to uniform and $\Delta\tau$ is 1.1--1.2\,Gyr.

\begin{table}[tbp]
\caption{Step activity lifetimes $\tau$ [Gyr] at M7 and M8 for all kinematic dwarfs with $G<19$.}
\label{tab:lifetimes}
\centering
\scriptsize
\setlength{\tabcolsep}{2pt}
\begin{tabular}{llccc}
\toprule
Type & $\Lambda$ & Censored & Unconditional$^a$ & Active $\sUVW$ \\
\midrule
M7 & $-4.7$ & 6.5 (5.9--7.2) & 6.4 & 6.0 (5.0--7.1) \\
 & $-4.5$ & 6.1 (5.4--6.8) & 5.7 & 5.6 (4.8--6.5) \\
 & $-4.3$ & 5.4 (4.7--6.1) & 4.6 & 5.2 (4.3--6.1) \\
M8 & $-4.7$ & 8.1 (6.7--9.7) & 7.4 & 8.2 (5.9--$\ge$10) \\
 & $-4.5$ & 6.9 (5.6--8.5) & 5.8 & 8.2 (5.5--$\ge$10) \\
 & $-4.3$ & 5.6 (4.5--7.0) & 3.9 & 8.6 (5.0--$\ge$10) \\
\midrule
\multicolumn{5}{l}{Censored fit at $\Lambda=-4.7$, other models (M7 / M8)} \\
\multicolumn{3}{l}{\quad With a thick-disk component$^b$} & \multicolumn{2}{l}{6.0 / 7.2} \\
\multicolumn{3}{l}{\quad Relation of \citet{Holmberg2009}} & \multicolumn{2}{l}{6.9 / 8.5} \\
\multicolumn{3}{l}{\quad Brown-dwarf ages, volume-weighted} & \multicolumn{2}{l}{7.0 / 9.0} \\
\multicolumn{3}{l}{\quad Brown-dwarf ages, magnitude-weighted} & \multicolumn{2}{l}{8.2 / 9.6} \\
\multicolumn{3}{l}{\quad M6 comparison bin} & \multicolumn{2}{l}{6.4 (5.7--7.1)} \\
\multicolumn{5}{l}{Strict thin disk, M8, $\Lambda=-4.7$, one activity group alone$^c$} \\
\multicolumn{3}{l}{\quad Active / inactive dwarfs} & \multicolumn{2}{l}{6.0 (4.5--8.1) / 9.1 (1.5--$\ge$10)} \\
\bottomrule
\end{tabular}
\tablefoot{Constant star-formation rate over 10\,Gyr and the age--velocity relation of \citet{Aumer2009} unless stated. Censored: likelihood of the velocities given the activity groups (Eq.~\ref{eq:like} without truncation), with 95\% profile-likelihood intervals. Active $\sUVW$: the step model fitted to the dispersion of the active dwarfs, with 95\% bootstrap intervals. Every fit uses the grid of Appendix~\ref{app:like}, and entries at its end (9.95\,Gyr, listed as $\ge$10) are lower limits. Lifetimes are rounded to 0.1\,Gyr from the 0.05\,Gyr grid of the fits, and differences were taken before rounding. Thin-disk lifetimes with the high-velocity tail removed are in Table~\ref{tab:kinvariants}. \tablefoottext{a}{The activity classes are fitted together with the velocities (numerator of Eq.~\ref{eq:like} alone), so the active fraction also constrains $\tau$ (Appendix~\ref{app:like}).} \tablefoottext{b}{\citet{Bensby2014}.} \tablefoottext{c}{Truncated censored likelihood of Eq.~(\ref{eq:like}) summed over one activity group only.}}
\end{table}

\subsection{Targeting, the pooled population and the L dwarfs}
\label{subsec:kin_pop}

We base the kinematics on the dwarfs brighter than $G=19$ because the larger analysis sample lacks the fast, faint dwarfs (Sect.~\ref{subsec:selection}). The differences between the samples are given in Table~\ref{tab:twosamples}. The activity fractions of the $G<19$ and analysis samples agree to within about 5 points and give the same M8 weakening, although the extra active component differs between them (Sect.~\ref{subsec:forward}). The kinematics differ at M8 alone. There the cut in tangential velocity lowers $\Delta\sUVW$ of the analysis sample by 4\,\kms. The restored sample, which over-represents fast dwarfs, agrees with the $G<19$ sample. The lifetime difference of the $G<19$ sample lies between those of the other two. The analysis sample gives a narrower interval on $\Delta\tau$, although its M8 comparison is biased downward. It has 3.4 times as many M8 kinematic dwarfs (788 compared with 230), but most of the additional dwarfs are too faint to be classified at $\Lambda=-4.7$. At this threshold the two samples have 244 and 181 classified M8 dwarfs with space motions.

Regardless of activity, the pooled dispersion of the $G<19$ sample declines from M7 to M9 by 5.0\,\kms, consistent with zero. In the analysis sample the decline is 7.1\,\kms\ and excludes zero (Table~\ref{tab:twosamples}). In the constant-rate model this decline corresponds to an extra fraction of dwarfs younger than 1\,Gyr, added to the uniform ages, of $f_{\rm young}=0.26$ at M9, at most 0.59 (95\%). The age-defined fraction $f_{\rm young}$ can be related to the activity-defined fraction $\phi$ under the step model: with $\tau\approx6$\,Gyr every dwarf younger than 1\,Gyr is active. If such dwarfs also have the $\log(\LHa)$ distribution of active M7 dwarfs, an extra young fraction is an extra active component of the same weight. Under these conditions, the kinematic bound of 0.59 lies just below the bounds of 0.62--0.65 from the M9 fractions. The 31 L dwarfs with $G<19$ and space motions have $\sUVW = 50$\,\kms, close to the 48\,\kms\ that \citet{Burgasser2015} found for nearby L dwarfs.

\begin{table*}[tbp]
\caption{Principal results on the $G<19$, analysis and restored samples.}
\label{tab:twosamples}
\centering
\scriptsize
\setlength{\tabcolsep}{4pt}
\begin{tabular}{lccc}
\toprule
Quantity & $G<19$ & Analysis & Restored \\
\midrule
Active fraction [\%], $D=5$\,\AA, M7 & 42.8 (951) & 43.7 (1125) & 43.1 (1180) \\
Active fraction [\%], $D=5$\,\AA, M8 & 66.5 (227) & 61.2 (402) & 60.0 (467) \\
Active fraction [\%], $D=5$\,\AA, M9 & 71.7 (53) & 72.0 (75) & 70.4 (81) \\
Active fraction [\%], $\Lambda=-4.7$, M7 & 68.4 (576) & 69.3 (596) & 68.9 (605) \\
Active fraction [\%], $\Lambda=-4.7$, M8 & 72.8 (195) & 70.1 (264) & 67.8 (292) \\
Active fraction [\%], $\Lambda=-4.7$, M9 & 60.7 (56) & 59.6 (94) & 56.9 (109) \\
Active fraction [\%], $\Lambda=-4.5$, M7 & 57.5 (879) & 57.9 (987) & 57.0 (1026) \\
Active fraction [\%], $\Lambda=-4.5$, M8 & 53.7 (231) & 51.0 (486) & 49.8 (574) \\
Active fraction [\%], $\Lambda=-4.5$, M9 & 35.6 (59) & 34.4 (160) & 33.3 (198) \\
$\delta$ [dex], $\phi_{95}$ at M8, Kaplan--Meier & $-$0.12, 0.43 & $-$0.12, 0.29 & $-$0.12, 0.24 \\
$\Delta\sUVW$ [\kms], $-4.7$, no cut, M7 & 20.6 (16.3, 24.8) & 20.4 (15.8, 24.5) & 20.7 (16.0, 25.4) \\
$\Delta\sUVW$ [\kms], $-4.7$, no cut, M8 & 19.4 (10.3, 27.9) & 15.5 (7.0, 22.5) & 19.4 (11.8, 26.9) \\
$\Delta\sUVW$ [\kms], $-4.7$, no cut, M9 & $-$6.5 ($-$21.1, 6.3) & $-$3.9 ($-$13.9, 5.3) & 3.0 ($-$7.9, 12.9) \\
$\Delta\sUVW$ [\kms], $-4.7$, tail removed, M7 & 16.5 (12.8, 20.1) & 16.2 (12.4, 19.9) & 15.8 (12.2, 19.3) \\
$\Delta\sUVW$ [\kms], $-4.7$, tail removed, M8 & 10.1 (3.3, 16.1) & 7.0 (1.0, 12.3) & 9.7 (4.1, 14.9) \\
$\Delta\sUVW$ [\kms], $-4.7$, tail removed, M9 & $-$3.6 ($-$15.7, 6.6) & 1.4 ($-$7.7, 9.7) & 5.6 ($-$3.5, 14.1) \\
$\Delta\sUVW$ [\kms], $-4.7$, strict thin disk, M7 & 10.3 (6.4, 13.9) & 9.7 (5.8, 13.3) & 9.7 (5.9, 13.3) \\
$\Delta\sUVW$ [\kms], $-4.7$, strict thin disk, M8 & 7.1 (0.6, 12.1) & 5.3 ($-$0.3, 10.3) & 6.3 (1.0, 11.3) \\
$\Delta\sUVW$ [\kms], $-4.7$, strict thin disk, M9 & 1.1 ($-$9.7, 10.9) & 5.1 ($-$2.5, 13.0) & 5.5 ($-$2.8, 13.5) \\
$\tau_{\rm M7}$, $\tau_{\rm M8}$ [Gyr], $\Lambda_{\rm M8}=-4.7$ & 5.9, 6.4 & 5.6, 5.4 & 6.1, 7.2 \\
$\Delta\tau$ [Gyr], $\Lambda_{\rm M8}=-4.7$ & +0.5 ($-$1.1, +2.3) & $-$0.3 ($-$1.3, +1.8$^a$) & +1.1 ($-$0.3, +4.1$^a$) \\
$\tau_{\rm M7}$, $\tau_{\rm M8}$ [Gyr], $\Lambda_{\rm M8}=-4.7+\delta$ & 5.9, 6.8 & 5.6, 5.8 & 6.1, 7.5 \\
$\Delta\tau$ [Gyr], $\Lambda_{\rm M8}=-4.7+\delta$ & +0.9 ($-$0.8, +3.5) & +0.2 ($-$1.0, +1.7) & +1.4 (+0.0, +3.0) \\
Pooled $\sUVW$, M9 minus M7 [\kms] & $-$5.0 ($-$12.4, 1.5) & $-$7.1 ($-$10.9, $-$3.5) & $-$3.0 ($-$6.5, 0.2) \\
L dwarfs, $N$ and $\sUVW$ [\kms] & 31, 50 & 147, 43 & 206, 48 \\
\bottomrule
\end{tabular}
\tablefoot{Restored: the analysis sample plus the high-proper-motion targets that the colour-selected programme or the MWS also targeted. $G<19$: every dwarf with Gaia $G<19$, including the high-proper-motion targets with Gaia DR3 $G=18.95$--19.00, 14 of them M7--L1 dwarfs (Table~\ref{tab:targeting}). Fractions with the number of spectra whose limit is at or below the threshold; $\delta$ and $\phi_{95}$ as in Table~\ref{tab:joint}; dispersion differences with 95\% bootstrap intervals; lifetimes as in Table~\ref{tab:dtau}, each sample with its own $\delta$. \tablefoottext{a}{At least 2.5\% of the M8 replicates are at the end of the grid, so the upper bound is set by the end of the age range and is a lower limit.}}
\end{table*}

\section{Rotation}
\label{sec:rotation}

\begin{figure}[tbp]
    \centering
    \includegraphics[width=\columnwidth]{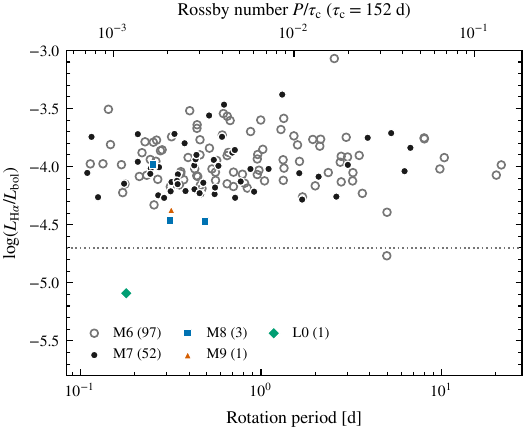}
    \caption{H$\alpha$ activity as a function of rotation period for the dwarfs whose periods belong to the target, with $\chiH$ measured for each object and corrected types (numbers in the legend count all 154; the periods beyond 7\,d belong to M6 dwarfs). The 148 dwarfs with a detected line are shown; the other six lack a usable measurement or limit. The dotted line is the threshold $\Lambda=-4.7$. The top axis gives the Rossby number for a convective turnover time of 152\,d, that of a 0.1\,M$_\odot$ dwarf \citep{Wright2018}.}
    \label{fig:rotation}
\end{figure}

Every period we measure at M7--M9 is shorter than 7\,d. With these periods we can test whether the H$\alpha$ of rapid rotators is saturated, while the spin-down of inactive dwarfs is not sampled. The periods come from the TESS All-Sky Rotation Survey \citep[TARS;][]{Boyle2026} and from public ZTF light curves \citep{Bellm2019, Masci2019} and TESS full-frame images \citep{Ricker2015} that we searched with Lomb--Scargle periodograms \citep{Lomb1976, Scargle1982, VanderPlas2018}. Appendix~\ref{app:periods} describes the search and the comparison of sources, and Fig.~\ref{fig:rotation_examples} shows example light curves. Our 154 rotators have a median period of 0.56\,d, and most of them are M6 or M7 dwarfs (Fig.~\ref{fig:rotation}). Because a photometric period needs spot modulation large enough to detect, the sample favours bright, spotted, rapidly rotating dwarfs, which are also the active ones: among 307 pairs of active and inactive M7--L1 dwarfs matched in subtype and Gaia brightness, ZTF gives periods for six active dwarfs and no inactive one (exact $p=0.031$). Flares follow the same division. Because the recovery of injected flares is incomplete, we quote rates of detected candidates. In TESS light curves the dwarfs active at 5\,\AA\ show 5.5 detected flare candidates per 1000\,d of monitoring and the inactive ones 0.58. Among 82 pairs matched in type and brightness, the active dwarfs have 18 candidates and the inactive ones three ($p=0.010$; Appendix~\ref{app:flares}).

To reach fainter and later dwarfs we fitted the TESS full-frame images with point-spread functions placed at the \emph{Gaia} positions of each dwarf and its neighbours \citep[as in][]{Han2023}, for all 1229 dwarfs with $G<18$. This adds two periods and replaces one (Table~\ref{tab:periods}). For the L0 dwarf 2MASS J12212770+0257198 it recovers a period of 4.31\,h, consistent with the period that \citet{Lambier2025} reported as possible, with the variable flux peaking within one TESS pixel of the \emph{Gaia} position (Fig.~\ref{fig:rotation_map}). Its H$\alpha$ line is detected at $5.9\sigma$ but is weak, $\log(\LHa)=-5.09$. This rapid L0 rotator is therefore inactive at every threshold in $\log(\LHa)$, as expected when the atmosphere becomes too neutral to couple to the field \citep{Mohanty2002}. The short periods we found for inactive M7--M9 dwarfs all came from neighbouring stars.
Forced photometry from the Asteroid Terrestrial-impact Last Alert System (ATLAS) extends the time coverage to 10.7\,yr for 123 dwarfs, most of them southern. It provides no additional rotator: the one significant period, 5.29\,d for an active M7 dwarf, agrees with the TARS value of 5.26\,d (Appendix~\ref{app:periods}). The same light curves contain 17 flares on 11 dwarfs (Appendix~\ref{app:flares}).

Among the 148 rotators with a detected line and a per-object $\chiH$, $\log(\LHa)$ shows no trend with period: a fit with one value per type has a slope of 0.002\,dex per dex in period ($-0.077$ to 0.080, 95\% bootstrap interval). Of the 56 M7--M9 rotators, 47 are active at $\Lambda=-4.7$, seven have limits above it and two have no usable H$\alpha$ measurement. The median $\log(\LHa)$ of the M6 and M7 rotators, about $-4.0$, is close to the saturated value of earlier M dwarfs with periods of a few days or less, $-3.8$ \citep{Newton2017}, and the active M7 rotators lie above other active M7 dwarfs, as a selection toward spotted dwarfs leads one to expect. For the convective turnover time of a 0.1\,M$_\odot$ dwarf, 152\,d \citep{Wright2018}, the M7--M9 periods correspond to Rossby numbers below 0.05 and the longest M6 periods to 0.14, the value below which the X-ray emission of fully convective stars is saturated \citep{Wright2018}. The rotators therefore lie in the saturated regime (top axis of Fig.~\ref{fig:rotation}). The two inactive rotators in Fig.~\ref{fig:rotation} are an M6 dwarf with H$\alpha$ just below the threshold and the L0 dwarf.

\section{Discussion}
\label{sec:discussion}

\subsection{Incidence, strength and age across the boundary}
\label{subsec:disc_age}

Our interpretation is that from M7 to M8 the incidence of chromospheric H$\alpha$ changes little while the emission weakens relative to $L_{\mathrm{bol}}$, and that the rise in equivalent-width incidence is mostly what the fading continuum does to that weakening distribution. At the lowest threshold the spectra allow, $\Lambda=-5.0$, the active fraction is flat or falling from M7 to M9, and the joint model requires a weakening of 0.12--0.15\,dex at M8. The model also needs an extra active component at M8 among the dwarfs brighter than $G=19$, but the evidence is inconclusive in the analysis sample (Sect.~\ref{subsec:forward}). We cannot yet say which describes the M8 population better. The $G<19$ sample is free of the proper-motion selection but small, and its brighter magnitude limit favours the overluminous dwarfs, which are more often active. The kinematic lifetimes of M7 and M8 are consistent in every sample, and each interval contains the bound from the fractions. At M9 the fractions call for a larger weakening that a uniform shift does not describe, and the pooled kinematics limit an extra young fraction to $f_{\rm young}<0.59$.

The lifetimes are measured at M7 and M8, whose populations are stellar on the evidence of Sect.~\ref{subsec:validation}. For M9 we have the fractions and the pooled dispersion, because there the threshold mixes weak emitters into the inactive group (Sect.~\ref{subsec:m9}). For the L dwarfs we have a pooled dispersion. Earlier M dwarfs stay active longer at lower mass \citep{West2008}, and we expected that trend to continue toward the hydrogen-burning limit. At equivalent thresholds, however, the M8 lifetime is consistent with the M7 one in every sample, with an upper bound 3.5\,Gyr longer in the $G<19$ sample, and the incidence at fixed $\LHa$ is flat or falling. What changes from M7 to M9 is the strength of the emission. A field that is still generated but couples less to an increasingly neutral atmosphere \citep{Mohanty2002} would produce this pattern: the same dwarfs stay active for about the same time, or a little longer, each with weaker chromospheric output, and the conversion to equivalent width hides the weakening behind a rising incidence. On the evidence here, the activity of the lowest-mass stars is set by age, as for earlier M dwarfs, while the effective temperature sets how much of it reaches the chromosphere.

The inactive M7 and M8 dwarfs are an old population, and their high-velocity end has the infrared colours of dwarfs a little poorer in metals than the thin disk. This tail explains why the kinematic ages of the inactive dwarfs exceed those covered by the age--velocity relation. For dwarfs outside the tail, a model of the strict thin disk reproduces the dispersions of both groups to within 2\,\kms. We did not expect $\Delta\sUVW$ at M8 to depend so heavily on the tail. Inside the thin disk the M7 dwarfs show an age difference close to the model prediction, but at M8 $\Delta\sUVW$ is 7.1\,\kms\ with a 95\% interval that barely excludes zero. Measuring it within the thin disk alone will therefore take a larger M8 sample. The overluminous dwarfs, mostly unresolved binaries with a young minority, are more often active, but the inactive dwarfs are older even without them. At M9 the threshold $\Lambda=-4.7$ separates weak from strong emission more than line-free from emitting dwarfs. M9 spectra with limits of 2\,\AA\ or less would show whether the line-free M9 dwarfs follow the M7 pattern.

\subsection{Dynamos and the origin of H$\alpha$}
\label{subsec:disc_dynamo}

All the dwarfs in our sample are fully convective, so the interface dynamo of solar-type stars cannot act in them. Mean-field models and simulations of fully convective interiors nevertheless generate kilogauss fields \citep{ChabrierKuker2006, Browning2008}, and the energy flux available from convection sets the field strength of rapidly rotating objects down to brown dwarfs \citep{Christensen2009, ReinersChristensen2010}. Measurements of M dwarfs show strong fields and a relation between field and rotation resembling the rotation--activity relation \citep{PhanBao2009, Morin2010, Reiners2022}. We have no field measurements for our dwarfs, but the two patterns we do see fit this picture. Rapid rotators, most of them M6 and M7 dwarfs, have saturated H$\alpha$ with no dependence on period (Sect.~\ref{sec:rotation}), and M8 dwarfs stay active about as long as M7 dwarfs while their emission weakens (Sect.~\ref{subsec:disc_age}). We read the weakening as a change in how the atmosphere responds to the field, an interpretation that field measurements of M7--M9 dwarfs could test.

The origin of the H$\alpha$ line becomes less certain at the end of our range. Besides chromospheric heating, large-scale magnetospheric currents can drive electron beams into the atmosphere and produce auroral H$\alpha$ and radio emission, as observed on the M8.5 dwarf LSR J1835+3259 \citep{Hallinan2015} and in radio pulses from late-L and T dwarfs \citep{Kao2016, Kao2018}. \citet{Pineda2017} argue that activity at the end of the main sequence changes in nature from chromospheric and coronal to auroral. Single-epoch spectra cannot separate the two mechanisms, which need time-resolved H$\alpha$ and radio data. We treat the emission at M7--M9 as chromospheric because the fractions and lifetimes behave as they do in earlier M dwarfs, and the lifetimes we measure at M7 and M8 would describe H$\alpha$ of either origin if aurorae contribute. At L0 and L1 the median sensitivity limit is $\log(\LHa)\approx-4.5$, and only 39 of 162 L0 spectra and 3 of 66 L1 spectra have a line at $3\sigma$. Those fractions measure the incidence of strong H$\alpha$ emission, whatever its source, so we draw no conclusion about an auroral contribution in L dwarfs.

\subsection{Comparison with earlier work}
\label{subsec:disc_lit}

The reversal between equivalent width and $\LHa$ \citep{Gizis2000, West2004, Schmidt2015} and the greater dynamical age of inactive M dwarfs \citep{West2006, West2008} are established results, and we confirm both. To these the DESI sample adds active fractions at common detection limits with $\chiH$ measured for each object, the share of the rise that the continuum explains (all of it at 5\,\AA, while stronger lines require emission 0.12--0.15\,dex weaker at M8), and kinematic activity lifetimes at M8 beside M7 at equivalent thresholds. Among 53 nearby M7--M9.5 dwarfs \citet{Gizis2000} found the dwarfs without strong emission to be the kinematically younger group; in the DESI sample the inactive M7 and M8 dwarfs are the older group at equivalent widths of 5\,\AA\ as well as in $\LHa$ (Table~\ref{tab:kin_comp}). At 1\,\AA\ the M7 fraction at a common threshold is \rfrac{corrected_EW_1_M7} (Table~\ref{tab:halpha}), higher than the estimate of \citet{West2004}, whose 1\,\AA\ threshold was applied to spectra of any detection limit. Later-type spectra with limits of 1\,\AA\ or less are too rare to extend the comparison with \citet{West2004} or \citet{Schmidt2015}. Our thin-disk lifetimes of 5.9--6.4\,Gyr are close to the 6--7\,Gyr at which \citet{West2006} placed the decline of activity in M7 dwarfs, and shorter than the 7--8\,Gyr that \citet{West2008} found at M5--M7; both used the vertical distribution of active stars, a different estimator with its own model dependence. Their types come from the SDSS typing that also defines our anchors \citep{West2011}. The comparison is therefore made at the same labels, and the warmer temperatures of Sect.~\ref{subsec:validation} apply to both. \citet{Kiman2021} calibrated the decline of H$\alpha$ with age on 892 M dwarfs in young associations or with white-dwarf companions. They found $\LHa$ to fall as the age to the power $-0.11$ up to 776\,Myr and more steeply afterwards, with an index of $-0.88^{+0.20}_{-0.25}$ that few calibrators older than 1\,Gyr constrain, and they had too few dwarfs later than M7 to follow the decline there. Our lifetimes sample those ages and types. A decline with their index from the saturated level of our rotators, $\log(\LHa)\approx-4.0$, would cross $-4.7$ after about 5\,Gyr, close to the 5.9 and 6.4\,Gyr of the step model. For late-M dwarfs the nearby samples of \citet{Reiners2009} and \citet{Burgasser2015} report total dispersions of 38--41\,\kms, below the 50\,\kms\ of the 317 DESI M7--M9 dwarfs with $G<19$ within 60\,pc, although our values have the measurement variance removed and the published ones include it. The excess is concentrated in the high-velocity tail, which contains a tenth of these dwarfs: with the tail removed the dispersion is 42.8\,\kms, close to the nearby samples.

\subsection{Limitations}
\label{subsec:limitations}

The lifetime comparison is limited most by the size of the sample free of the proper-motion selection. The $G<19$ sample has fewer than a third of the M8 kinematic dwarfs of the analysis sample, although three-quarters as many of those classified at $\Lambda=-4.7$ (Sect.~\ref{subsec:kin_pop}), and bounds $\Delta\tau$ at 3.5\,Gyr. The analysis sample would bound it at 1.7\,Gyr, but it lacks the fast, faint dwarfs and its M8 $\Delta\sUVW$ is 4\,\kms\ lower. A fainter sample with a modelled fibre-assignment probability would combine the size of the analysis sample with the freedom from selection of the $G<19$ sample.

The lifetimes also depend on the assumed constant star-formation rate and the \citet{Aumer2009} relation, but most changes affect both types. A thick-disk component, the relation of \citet{Holmberg2009} or another definition of the thin disk shifts each lifetime by up to 0.6\,Gyr and $\Delta\tau$ by at most 0.1\,Gyr, while fits without the tail cut give $\Delta\tau$ longer by 1.0\,Gyr (Table~\ref{tab:dtau}). Gradual declines of activity narrower than 0.5\,Gyr, which the M7 velocities allow, leave $\Delta\tau$ unchanged. Wider logistic and exponential declines, which the M7 velocities disfavour, shorten the mean M7 duration of activity by up to 1.8\,Gyr and change the M8 minus M7 difference by less than 0.4\,Gyr (Table~\ref{tab:forms}). The tail cut is partly an age cut, which the truncated likelihood corrects within the same model. The age distribution acts on the two types differently: brown-dwarf cooling ages lengthen the M8 lifetime by 1.2\,Gyr with field temperatures and by 0.4--0.6\,Gyr at the warmer temperatures implied by the type scale.

For the fractions the form of the joint model matters most. Its weakening is a uniform shift of the M7 distribution, which fits M8 but not M9, and a change of shape that removes strong lines while keeping weak ones would trade off against $\phi$. The Kaplan--Meier distribution assumes that the limit of a spectrum is independent of its emission, yet the deeper M7 spectra are more often active by 6--8 points. The half-sample distributions and the M7 fractions reweighted to the later types test this assumption, but we have not shown that they bracket the parent distribution. The bounds on $\phi$ therefore have a censoring uncertainty that we cannot quantify. The observed trends in the fractions use no such distribution. The common-limit selection adds a term of up to 4 points to the type differences, comparable to their bootstrap errors. Flares during the exposures would raise the fractions. The one flagged case has a negligible effect, although simultaneous photometry covers too few spectra to exclude flares for most of the sample. The continuum templates are the largest systematic error of the absolute fractions. At M9 and 5\,\AA\ the error is 7.7 points (Sect.~\ref{subsec:halpha_method}), more than the statistical error of that fraction. It is 10.2 points at M9 and 3\,\AA\ (Table~\ref{tab:template_fractions}). The type differences on which the joint model rests change within their errors except for M9 at $\Lambda=-5.0$.

The infrared signature of the tail is calibrated on four SPIFF d/sd dwarfs, leaving its amplitude relative to known subdwarfs uncertain by a factor of two. Because the overluminosity cut finds near-equal pairs, both activity groups still contain unequal binaries. The spectral-binary indices that find them in composite near-infrared spectra \citep{Burgasser2010, BardalezGagliuffi2014} need higher resolution than the SPHEREx channels provide. At M9, few anchors define the boundary with L0, and moving it as in the hybrid scale changes $\Delta\sUVW$ by up to 7\,\kms\ (Appendix~\ref{app:m9l0}). The uncertain bolometric zero point shifts the M9 fraction at $\Lambda=-4.5$ by 4 points, with M9 emission still weaker than M7 emission.

\section{Conclusions}
\label{sec:conclusions}

\begin{enumerate}
\item The M7 and M8 bins of the DESI sample (Table~\ref{tab:typescale}) contain mostly stars above the hydrogen-burning limit with either temperature scale considered here. Empirical templates place the spectra on the published SDSS scale, and \emph{Euclid} NISP types support the correction at M7--M8. The DESI dwarfs and the SDSS anchors are brighter in $M_J$ than field dwarfs of the same type by more than unresolved binaries can explain. We interpret most of the offset as a temperature difference that makes these bins 1--2 field subtypes warmer than their names, so the activity lifetimes describe the lowest-mass stars.
\item We confirm that the rise in equivalent-width incidence comes mostly from continuum fading, which acts on weaker emission. From M7 to M9 the active fraction rises at 5\,\AA\ and falls at $\log(\LHa)=-4.5$ (Table~\ref{tab:halpha}). The decline appears with both SPHEREx and photometric bolometric fluxes, and both trends recur in bins of bolometric luminosity and within 60\,pc (Table~\ref{tab:lbol}). Fitting all thresholds with their covariance requires M8 emission to weaken by 0.12--0.15\,dex, while a uniform weakening fails at M9. At M8 an extra active component has weight 0.22 among dwarfs brighter than $G=19$ and is at most marginal in the analysis sample (Table~\ref{tab:joint}).
\item Inactive M7 and M8 dwarfs are kinematically older than active dwarfs: at $\log(\LHa)=-4.7$ their dispersions are about 20\,\kms\ higher for dwarfs with $G<19$. Their high-velocity tail is bluer in $J-K_s$ and weaker in SPHEREx $K/J$, as expected for mildly metal-poor dwarfs, and in the strict thin disk both differences are close to the step model (Table~\ref{tab:kinvariants}). Overluminous dwarfs are mostly near-equal-flux binaries with a young minority and are more often active, and the inactive dwarfs are older without them. Candidate members of young associations make up about 1\% of the dwarfs with \emph{Gaia} parallaxes and do not change these differences.
\item Conditional on uniform ages and the age--velocity relation, the step model gives consistent M7 and M8 thin-disk activity lifetimes of 5.9 and 6.4\,Gyr at $\log(\LHa)=-4.7$ for dwarfs brighter than $G=19$. At equivalent emission thresholds an M8 lifetime equal to the M7 one, or up to 3.5\,Gyr longer, is allowed (Table~\ref{tab:dtau}). We interpret the activity of these lowest-mass stars as governed by age, while effective temperature sets how much of it reaches the chromosphere (Sect.~\ref{subsec:disc_age}).
\item Weak emission accounts for most M9 dwarfs classed as inactive at $\log(\LHa)=-4.7$, and the pooled kinematics limit an extra young M9 population to $f_{\rm young}<0.59$. The M7--M9 rotators, all with periods below 7\,d, are active or have detection limits too high for classification, and the rotators as a whole have saturated H$\alpha$ with no dependence on period.
\end{enumerate}

\begin{acknowledgements}
We thank the DESI Collaboration for making Data Release 1 publicly available. This work uses data from the \emph{Euclid} mission of the European Space Agency, from SPHEREx (NASA), from \emph{Gaia} (ESA; \url{https://www.cosmos.esa.int/gaia}), processed by the \emph{Gaia} Data Processing and Analysis Consortium, from TESS and ZTF, and from the Legacy Imaging Surveys, Pan-STARRS1, 2MASS, UKIDSS, VISTA and WISE/CatWISE2020. This research has benefitted from the UltracoolSheet, maintained by Will Best, Trent Dupuy, Michael Liu, Eugene Magnier, Aniket Sanghi, Rob Siverd, and Zhoujian Zhang. The calculations use Astropy \citep{Astropy2013, Astropy2018, Astropy2022}, NumPy \citep{Harris2020}, SciPy \citep{Virtanen2020} and Matplotlib \citep{Hunter2007}. The code behind the numbers, tables and figures will be released with the catalogue. The \emph{Euclid} DR1 spectra were used as Euclid Consortium data before the public release of \emph{Euclid} DR1, and this paper is published under the Consortium's procedures for DR1 results. %
This work has made use of data from the Asteroid Terrestrial-impact Last Alert System (ATLAS) project. ATLAS is primarily funded to search for near-Earth asteroids through NASA grants NN12AR55G, 80NSSC18K0284 and 80NSSC18K1575; byproducts of the NEO search include images and catalogues from the survey area. The ATLAS science products have been made possible through the contributions of the University of Hawaii Institute for Astronomy, the Queen's University Belfast, the Space Telescope Science Institute, the South African Astronomical Observatory, and The Millennium Institute of Astrophysics (MAS), Chile. We acknowledge the European Space Agency and a number of agencies and institutes that have supported the development of Euclid, in particular the Agenzia Spaziale Italiana, the Austrian Forschungsf\"orderungsgesellschaft funded through BMIMI, the Belgian Science Policy, the Canadian Euclid Consortium, the Deutsches Zentrum f\"ur Luft- und Raumfahrt, the DTU Space and the Niels Bohr Institute in Denmark, the French Centre National d'Etudes Spatiales, the Funda\c{c}\~{a}o para a Ci\^{e}ncia e a Tecnologia, the Hungarian Academy of Sciences, the Ministerio de Ciencia, Innovaci\'{o}n y Universidades, the National Aeronautics and Space Administration, the National Astronomical Observatory of Japan, the Netherlandse Onderzoekschool Voor Astronomie, the Norwegian Space Agency, the Research Council of Finland, the Romanian Space Agency, the Swiss Space Office (SSO) at the State Secretariat for Education, Research, and Innovation (SERI), and the United Kingdom Space Agency. A complete and detailed list is available on the Euclid web site (\url{www.euclid-ec.org/consortium/community/}). This work has made use of the Euclid Data Release 1 (DR1) from the European Space Agency (ESA) mission Euclid, available at \url{https://www.cosmos.esa.int/web/euclid/data-release-dr1}. This work has made use of the Euclid Quick Release Q1 data from the Euclid mission of the European Space Agency (ESA), 2025, \url{https://doi.org/10.57780/esa-2853f3b} \citep{EuclidQ1data}. This research makes use of ESA Datalabs (\url{datalabs.esa.int}), an initiative by ESA's Data Science and Archives Division in the Science and Operations Department, Directorate of Science \citep{Navarro2024}. This research uses services or data provided by the Astro Data Lab \citep{Fitzpatrick2014, Nikutta2020}, which is part of the Community Science and Data Center (CSDC) Program of NSF NOIRLab. NOIRLab is operated by the Association of Universities for Research in Astronomy (AURA), Inc. under a cooperative agreement with the U.S. National Science Foundation. NPB acknowledges International University, VNU-HCM for supporting this study.
E.L.M., N.S., and N.V. are supported by the European Research Council Advanced grant SUBSTELLAR, project number 101054354. NL acknowledges support from the Agencia Estatal de Investigaci\'on del Ministerio de Ciencia e Innovaci\'on (AEI-MCINN) under grants PID2022-137241NB-C41 and PID2025-176415NB-I00\@. The authors acknowledge financial support from the Agencia Estatal de Investigaci\'on (AEI) through the Severo Ochoa Centre of Excellence accreditation awarded to the Instituto de Astrof\'isica de Canarias, grant CEX2025-001609-S, funded by MICIU/AEI/10.13039/501100011033. Generative AI tools were used for code generation and manuscript editing; the authors are responsible for the scientific interpretation and the final manuscript.
\end{acknowledgements}

\bibliographystyle{aa}
\bibliography{refs_submit}

\begin{appendix}

\section{Redrock classification of ultracool dwarfs}
\label{app:redrock}

\begin{figure*}[tbp]
    \centering
    \includegraphics[width=\textwidth]{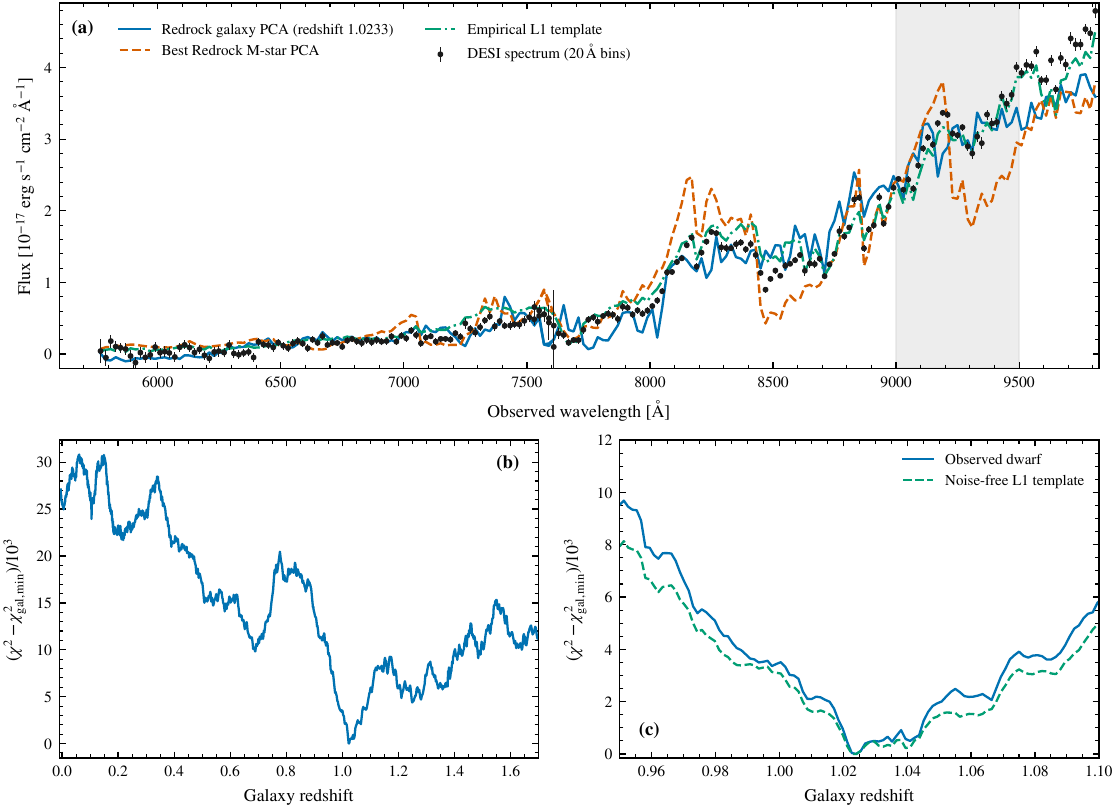}
    \caption{The Redrock galaxy alias for an L dwarf (DESI target identifier TARGETID 39628448964021073; L1 on the corrected scale, L0 on the template scale), classed as a galaxy at redshift 1.0233. (a) DESI spectrum averaged over 20\,\AA\ with the best Redrock galaxy model, the best Redrock M-star model and an empirical L1 template; the shaded band covers 9000--9500\,\AA. (b) Galaxy $\chi^2$ against trial redshift, relative to its minimum. (c) Zoom around the minimum for the observed spectrum and for a noise-free empirical L1 template passed through Redrock.}
    \label{fig:redrock}
\end{figure*}

Of the 432 dwarfs typed L on both the template and corrected scales, 353 (\rfrac{redrock_old_L}) have a Redrock galaxy or quasar label, nearly all at redshift 1.00--1.05. Among the 590 non-giant dwarfs typed M9 on the template scale and L0 on the corrected scale, 87 (\rfrac{redrock_new_L}) have such a label. Redrock compares each spectrum with principal-component bases for galaxies, quasars and stars, and its default template set has no L-dwarf basis. With the DESI DR1 template set, Redrock assigns the test spectra their catalogued classes. We inspected 30 random galaxy-labelled spectra at redshift 1.00--1.05 and found 19 late-M and L dwarfs. For them the galaxy fit beats the best stellar fit by a median $\Delta\chi^2$ of about 200, most of it from the 9000--9500\,\AA\ region (Fig.~\ref{fig:redrock}). There the M-star basis predicts a depression that L-dwarf spectra do not show, whereas a signed combination of galaxy components at redshift 1.02--1.04 follows the steep red pseudo-continuum. With that region masked, most of the dwarfs are classified as stars. Noise-free empirical L templates passed through Redrock are also classified as galaxies near redshift 1.03, and with empirical L templates included in the stellar fit, most dwarfs and every stellar control are classified as stars (Table~\ref{tab:redrock_tests}). In DESI DR1 the galaxy pile-up near redshift 1.03 contains about a thousand red point-like objects. Since about two-thirds of those we inspected are dwarfs, ultracool dwarfs plausibly make up a large part of the group. A selection restricted to stellar classes misses most L dwarfs.

\begin{table}[tbp]\caption{Redrock classification tests.}\label{tab:redrock_tests}\centering\scriptsize
\begin{tabular}{lrc}\toprule
Test & Count & Fraction (68\%)\\\midrule
Masked red interval: dwarfs restored & 14/19 & $73.7_{-11.0}^{+8.7}\%$\\
L templates added: dwarfs restored & 16/19 & $84.2_{-10.0}^{+6.6}\%$\\
L templates added: stars retained & 38/38 & $100.0_{-2.5}^{+0.0}\%$\\
\bottomrule\end{tabular}\tablefoot{Wilson intervals for the selected test spectra. The nineteen dwarfs are visually identified among thirty inspected galaxy-labelled spectra; the thirty-eight stellar controls are a separate set.}\end{table}

\section{\texorpdfstring{H$\alpha$}{H-alpha} errors and detection limits}
\label{app:ewerr}

Let $\mathbf{F}$ be the vector of pixel fluxes in the line window and the sidebands. The continuum is a template scaled by a linear function fitted to the sideband fluxes. To first order about the measured fluxes, $\EW = \mathbf{h}^{\mathsf{T}}\mathbf{F} + \mathrm{const}$, where $\mathbf{h}$ is the gradient of $\EW$ with respect to $\mathbf{F}$. The equivalent-width variance is
\begin{equation}
\varepsilon^2 = \kappa^2\left[\mathbf{h}^{\mathsf{T}}\mathbf{V}\mathbf{h} + \sum_k \omega_k\,(\EW_k - \EW)^2\right],
\label{eq:ewerr}
\end{equation}
\begin{equation}
V_{pq} = I_{pq}/\iota_p + \eta^2\,F_{{\rm c},p} F_{{\rm c},q}\,e^{-|\lambda_p-\lambda_q|/2.4\,\mbox{\scriptsize\AA}},
\end{equation}
where $\iota_p$ is the inverse variance of pixel $p$, $F_{{\rm c},p}$ the continuum and $\lambda_p$ the wavelength of pixel $p$, and $I_{pq}$ is the identity matrix. The template-mismatch amplitude $\eta$ is the scatter, 1.4826 times the median absolute deviation, of $F/F_{\rm c}-1$ in the sidebands with the median photon-noise contribution removed in quadrature, set to zero if negative; its errors are correlated over 2.4\,\AA, about three pixels. $\EW_k$ is the value measured with candidate template $k$, and $\omega_k \propto \exp(-\Delta\chi^2_k/2)$, normalised to unit sum, with $\Delta\chi^2_k$ the sideband $\chi^2$ relative to the best candidate. The factor $\kappa$, 1.08--1.10, is set from the line-free windows as described in Sect.~\ref{subsec:halpha_method}.

For the detection limit $\ell$, a line of the instrumental shape and strength $x$ is added to the spectrum. A fractional continuum error $\xi_{\rm c}$ changes $\EW$ by $-\xi_{\rm c}\,(\EW + 14\,\mbox{\AA})$, 14\,\AA\ being the width of the line window, so both $\mathbf{h}$ and $d_k = \EW_k - \EW$ are linear in $x$,
\begin{equation}
\mathbf{h} = \mathbf{h}_0 + x\,\mathbf{h}_1, \qquad d_k = d_{0k} + x\,d_{1k},
\end{equation}
where $\mathbf{h}_0$ and $d_{0k}$ are the values without the line and $\mathbf{h}_1$ and $d_{1k}$ their derivatives with respect to $x$. Equation~(\ref{eq:ewerr}) then becomes $\varepsilon^2(x) = \alpha x^2 + \beta x + \gamma$, with
\begin{align}
\alpha &= \kappa^2\Big(\mathbf{h}_1^{\mathsf{T}}\mathbf{V}\mathbf{h}_1 + \sum_k \omega_k d_{1k}^2\Big), \nonumber\\
\beta &= 2\kappa^2\Big(\mathbf{h}_0^{\mathsf{T}}\mathbf{V}\mathbf{h}_1 + \sum_k \omega_k d_{0k}d_{1k}\Big), \\
\gamma &= \kappa^2\Big(\mathbf{h}_0^{\mathsf{T}}\mathbf{V}\mathbf{h}_0 + \sum_k \omega_k d_{0k}^2\Big), \nonumber
\end{align}
$\gamma$ being the variance without a line, and
\begin{equation}
\ell = \frac{9\beta + \sqrt{81\beta^2 + 36(1-9\alpha)\gamma}}{2(1-9\alpha)}
\end{equation}
is the positive root of $x = 3\varepsilon(x)$. An observed line is judged by its own error, which can exceed $\varepsilon(\ell)$ when the line is broader than the instrumental profile or the candidate templates disagree.

\paragraph{Injection tests.} Lines of the instrumental shape were injected at 16 strengths into 6595 real spectra, 64 times each with resampled sideband residuals, and run through the full measurement. They comprise nearly all measured spectra of the M6--L sample, high-proper-motion targets included, and 41 spectra of objects outside the final sample. For 5204 of them the adopted limit is the one computed above; the other 1391 are spectra typed M9 or L on the template scale, whose adopted values use velocities for their corrected types. Because the strength detected in half of the trials is close to $\ell$ at every type and $S/N$, we apply the median detection probability as a function of $x/\ell$ to all spectra; it is 0.45 at $x=\ell$, 0.49 at $1.04\,\ell$ and 0.91 at $1.67\,\ell$. Injections of zero strength are detected in 0.03\% of trials.

\paragraph{Completeness of the fractions.} Table~\ref{tab:completeness} compares the active fractions from all spectra with detection limits at or below the threshold with two alternatives. A high-completeness estimate uses only spectra whose limit is at most 0.6 times the threshold, for which a line at the threshold is detected in at least 91\% of trials. A weighted estimate uses all spectra, with each active dwarf weighted by the inverse detection probability of a line of its measured strength in its spectrum. The weighting corrects for lines above the threshold that escape detection but not for lines that scatter across the threshold in either direction. The type differences are the same in the three versions within their errors, and at fixed $\Lambda$ the high-completeness decline is the larger one.

\begin{table}[tbp]
\caption{Active fractions [\%] with nearly complete detection.}
\label{tab:completeness}
\centering
\scriptsize
\setlength{\tabcolsep}{2.5pt}
\begin{tabular}{llcccc}
\toprule
Threshold & Version & M7 & M8 & M9 & M9$-$M7 \\
\midrule
$D=5$\,\AA & Standard & 43.9 (1099) & 61.2 (399) & 72.2 (72) & +28.3 \\
 & $\ell\le0.6$ threshold & 43.6 (910) & 59.9 (192) & 70.3 (37) & +26.6 \\
 & Weighted & 44.5 & 63.3 & 75.9 & +31.4 \\
$D=8$\,\AA & Standard & 19.4 (1244) & 34.5 (626) & 41.3 (126) & +21.9 \\
 & $\ell\le0.6$ threshold & 18.6 (1085) & 33.6 (384) & 46.3 (67) & +27.7 \\
 & Weighted & 19.6 & 36.3 & 42.6 & +23.0 \\
$\Lambda=-4.7$ & Standard & 69.3 (596) & 70.1 (264) & 59.6 (94) & $-$9.7 \\
 & $\ell\le0.6$ threshold & 75.1 (181) & 67.7 (99) & 58.8 (51) & $-$16.3 \\
 & Weighted & 70.2 & 72.1 & 62.4 & $-$7.7 \\
$\Lambda=-4.5$ & Standard & 57.9 (987) & 51.0 (486) & 34.4 (160) & $-$23.5 \\
 & $\ell\le0.6$ threshold & 61.8 (545) & 50.2 (243) & 36.5 (85) & $-$25.4 \\
 & Weighted & 58.7 & 53.5 & 36.0 & $-$22.7 \\
$\Lambda=-4.3$ & Standard & 43.3 (1160) & 27.2 (713) & 14.8 (257) & $-$28.5 \\
 & $\ell\le0.6$ threshold & 44.6 (961) & 28.4 (462) & 14.8 (155) & $-$29.8 \\
 & Weighted & 43.9 & 28.6 & 15.8 & $-$28.0 \\
\bottomrule
\end{tabular}
\tablefoot{Analysis sample, spectra with a measured $\chiH$ (as in Table~\ref{tab:joint}); numbers of spectra in parentheses. Standard: every spectrum whose limit reaches the threshold. $\ell\le0.6$ threshold: only spectra whose limit is at most 0.6 times the threshold (in equivalent width). Weighted: each active dwarf weighted by the inverse detection probability of its line.}
\end{table}

\paragraph{Repeat epochs.} Of the sample dwarfs with a measured primary spectrum, 535 have another DESI DR1 coadd with median $S/N\ge5$ at 8000--9000\,\AA, from another survey phase or programme. We retrieved the red-channel spectra of 592 of these 595 coadds from the public DESI archive and measured H$\alpha$ with the same estimator, the object's own radial velocity and template, and the resolution of each coadd. Both epochs yield a valid measurement for 295 pairs of 268 dwarfs, a median of 31\,d apart. The repeat coadds are mostly shallower than the primary ones, so few pairs have limits at the lowest thresholds in both epochs. Table~\ref{tab:repeats} lists the spread of $\log\EW$ between the epochs for the pairs with a $3\sigma$ line in both, and how often a dwarf changes class.

\begin{table}[tbp]
\caption{Activity classes in two DESI epochs.}
\label{tab:repeats}
\centering
\scriptsize
\setlength{\tabcolsep}{3pt}
\begin{tabular}{lrcccc}
\toprule
Threshold & Pairs & Active, 1st & Active, 2nd & Either & Changed class (95\%) \\
\midrule
$D=2$\,\AA & 33 & 48\% & 48\% & 48\% & 0 (0--11\%) \\
$D=5$\,\AA & 71 & 38\% & 35\% & 41\% & 6 (3--17\%) \\
$\Lambda=-4.7$ & 35 & 60\% & 57\% & 60\% & 1 (0--15\%) \\
$\Lambda=-4.5$ & 71 & 46\% & 46\% & 52\% & 8 (5--21\%) \\
\midrule
\multicolumn{6}{l}{Pairs with a $3\sigma$ line in both epochs: 71} \\
\multicolumn{6}{l}{\quad spread of $\log\EW$ between epochs: 0.13\,dex; expected from the errors: 0.11\,dex} \\
\multicolumn{6}{l}{\quad variability beyond the errors: 0.07\,dex; ratio above two: 8\%} \\
\bottomrule
\end{tabular}
\tablefoot{Pairs of a primary and a repeat DESI DR1 spectrum of the same dwarf (M6--L1), both with limits at or below the threshold and measured with the same estimator. Either: active in at least one epoch. Changed class: number of pairs active in one epoch and inactive in the other, with the exact (Clopper--Pearson) 95\% binomial interval on the fraction. Spread: 1.4826 times the median absolute deviation.}
\end{table}

\paragraph{Continuum templates.}\label{app:template} The continua of Sect.~\ref{subsec:halpha_method} come from the templates of \citet{Kesseli2017}, whose M7--M9 members we replaced for typing (Sect.~\ref{subsec:offset}). We re-measured the 4019 analysis-sample spectra whose adopted values do not use velocities for their corrected types, with the empirical templates of the corrected type and its neighbours as the candidate continua and everything else unchanged. Of these spectra, 3974 give a valid measurement with both template sets. The empirical continuum lies slightly higher under the line, giving equivalent widths lower by a median of 0.37\,\AA, about a fifth of their error. The detection limits fall by 1\% in the median and 5\% at M9 (Table~\ref{tab:template}). Table~\ref{tab:template_fractions} gives the active fractions of these spectra with both template sets at every threshold of Table~\ref{tab:joint}. Most fractions fall: by 2.2--7.7 points at 5\,\AA\ and $\Lambda=-4.5$, and by up to 10.2 points (M9 at 3\,\AA) elsewhere. A few fractions at the lowest thresholds, with the fewest spectra, rise by up to 1.8 points. Because the shift is common to the three types, the M8 minus M7 differences change by less than their bootstrap errors at every threshold, and so do the M9 minus M7 differences except at $\Lambda=-5.0$, where the change is 1.3 times the error. The catalogue gives the equivalent widths, errors and limits measured with the empirical templates for every spectrum of this test (Appendix~\ref{app:catalogue}).

\begin{table}[tbp]
\caption{H$\alpha$ measured against the empirical DESI templates instead of those of \citet{Kesseli2017}.}
\label{tab:template}
\centering
\scriptsize
\setlength{\tabcolsep}{2pt}
\begin{tabular}{lrcccccc}
\toprule
Type & $N$ & $\Delta\EW/\varepsilon$ & $\ell$ ratio & \multicolumn{2}{c}{Fraction change [points]} & \multicolumn{2}{c}{Change of diff. from M7} \\
 & & median & median & 5\,\AA & $-4.5$ & 5\,\AA & $-4.5$ \\
\midrule
M7 & 2026 & $-$0.26 & 1.00 & $-$4.3 & $-$2.2 & -- & -- \\
M8 & 1545 & $-$0.14 & 0.98 & $-$5.1 & $-$4.4 & $-$0.9 & $-$2.2 \\
M9 & 313 & $-$0.23 & 0.95 & $-$7.7 & $-$3.6 & $-$3.4 & $-$1.3 \\
\bottomrule
\end{tabular}
\tablefoot{Of the 4019 M7--L1 analysis-sample spectra whose adopted values do not use velocities for their corrected types, the 3974 with valid measurements under both template sets. The same estimator is used with the empirical template of the corrected type and its neighbours as candidate continua. $\Delta\EW/\varepsilon$: empirical minus adopted equivalent width over its error; $\ell$ ratio: empirical over adopted limit. Fraction changes: empirical minus adopted active fraction; the last two columns give the change of the M8 or M9 minus M7 difference.}
\end{table}

\begin{table*}[tbp]
\caption{Active fractions [\%] measured against the templates of \citet{Kesseli2017} (K17) and the empirical DESI templates (Emp.).}
\label{tab:template_fractions}
\centering
\scriptsize
\setlength{\tabcolsep}{3pt}
\begin{tabular}{lccrccrccrrr}
\toprule
 & \multicolumn{3}{c}{M7} & \multicolumn{3}{c}{M8} & \multicolumn{3}{c}{M9} & \multicolumn{2}{c}{Change of difference} \\
Threshold & K17 & Emp. & Change & K17 & Emp. & Change & K17 & Emp. & Change & M8$-$M7 & M9$-$M7 \\
\midrule
$D=3$\,\AA & 61.8 (920) & 57.4 (911) & $-$4.4 & 72.9 (192) & 70.9 (203) & $-$2.0 & 84.6 (39) & 74.4 (43) & $-$10.2 & +2.5 & $-$5.8 \\
$D=5$\,\AA & 44.0 (1117) & 39.7 (1116) & $-$4.3 & 60.9 (399) & 55.8 (407) & $-$5.1 & 72.2 (72) & 64.6 (79) & $-$7.7 & $-$0.9 & $-$3.4 \\
$D=8$\,\AA & 18.9 (1286) & 15.8 (1288) & $-$3.1 & 34.4 (619) & 31.3 (629) & $-$3.1 & 38.8 (116) & 34.9 (126) & $-$3.9 & 0.0 & $-$0.7 \\
$\Lambda=-5.0$ & 88.1 (84) & 88.9 (90) & +0.8 & 84.8 (66) & 86.7 (75) & +1.8 & 76.3 (38) & 67.4 (43) & $-$8.9 & +1.0 & $-$9.7 \\
$\Lambda=-4.9$ & 80.3 (208) & 74.9 (199) & $-$5.4 & 83.7 (104) & 78.8 (113) & $-$4.9 & 70.6 (51) & 70.9 (55) & +0.3 & +0.5 & +5.7 \\
$\Lambda=-4.7$ & 69.4 (594) & 66.6 (581) & $-$2.8 & 70.7 (259) & 68.5 (276) & $-$2.2 & 58.8 (85) & 56.4 (94) & $-$2.4 & +0.6 & +0.3 \\
$\Lambda=-4.5$ & 58.0 (984) & 55.8 (982) & $-$2.2 & 51.7 (468) & 47.3 (480) & $-$4.4 & 35.3 (136) & 31.7 (145) & $-$3.6 & $-$2.2 & $-$1.3 \\
$\Lambda=-4.3$ & 43.4 (1157) & 38.2 (1154) & $-$5.2 & 28.0 (678) & 24.7 (687) & $-$3.3 & 16.2 (204) & 12.8 (211) & $-$3.4 & +1.9 & +1.8 \\
\bottomrule
\end{tabular}
\tablefoot{The 3974 analysis-sample spectra of Table~\ref{tab:template} with a valid measurement against both template sets; the number of spectra whose limit is at or below the threshold is in parentheses, and $\Lambda$ uses $\chiH$ measured for each object. Because this subset leaves out the spectra measured with velocities for their corrected types and, in the rows at fixed $D$, includes spectra without $\chiH$, its K17 fractions differ from those of Table~\ref{tab:halpha} by up to 2.4 points. Changes are empirical minus K17, in percentage points; the last two columns give the change of the M8 or M9 minus M7 difference, to be compared with the bootstrap errors of Table~\ref{tab:joint}.}
\end{table*}

\section{SPHEREx spectra and bolometric fluxes}
\label{app:sxdetail}

\subsection{Forced photometry}

The SPHEREx linear variable filters give spectrophotometry at resolving power 35--130 \citep{Korngut2018}. Our SPHEREx spectra come from forced photometry on the calibrated level-2 images of the SPHEREx Quick Release 2 (QR2) from the NASA/IPAC Infrared Science Archive (IRSA), taken between April 2025 and April 2026. The extraction is independent of the SPIFF library of \citet{Gagne2026}. Positions are propagated once to a representative SPHEREx epoch with Gaia, DESI or CatWISE proper motions and held fixed across frames. The largest displacement from the frame-epoch position for 100 random targets is $0.10\arcsec$, small compared with the $6.15\arcsec$ pixels. Frames are selected from the detector footprints, with at most 25 per detector when more than 150 are available.

Each frame is fitted in an $11\times11$ pixel cutout with the nearest plane of the instrumental point-spread function, sampled at pixel centres and normalised inside the cutout (a circular Gaussian of 1.0--1.4 pixels when it is unavailable). The model contains the target, which may have negative flux, up to 80 non-negative catalogue neighbours from 2MASS \citep{Skrutskie2006}, AllWISE \citep{Wright2010, Cutri2013}, Gaia, the Legacy Surveys and CatWISE, and a background plane. The fit masks flagged pixels, and parameter errors are enlarged by $\sqrt{\chi^2_\nu}$ when $\chi^2_\nu>1.5$. Measurements are averaged in 0.05\,\um\ channels with inverse-variance weights after clipping at three median absolute deviations. Up to eight blank-sky positions at 40 or $60\arcsec$ are extracted in the same way, and their bootstrap scatter with the target's number of frames sets the error floor and the background offset of each channel. A detection requires at least six calibrated channels and a peak channel with $S/N>3$. Band ratios require $S/N\ge5$ in their windows and no quality flag (saturation, a brighter neighbour within $12\arcsec$, an unmodelled source within $6\arcsec$ contributing at least 0.2 of the target's $W2$ flux, a duplicate entry, or insufficient coverage). Fluxes are divided by photometric scale factors measured from synthetic 2MASS and CatWISE photometry of clean objects \citep{Cohen2003, Jarrett2011}, \sxScaleTwoM\ except for the detector that covers $W1$, where the factor is \sxScaleWone. Table~\ref{tab:counts} lists the numbers of detected and clean dwarfs.

\subsection{Bolometric fluxes}

\begin{figure*}[tbp]
\centering
\includegraphics[width=\textwidth]{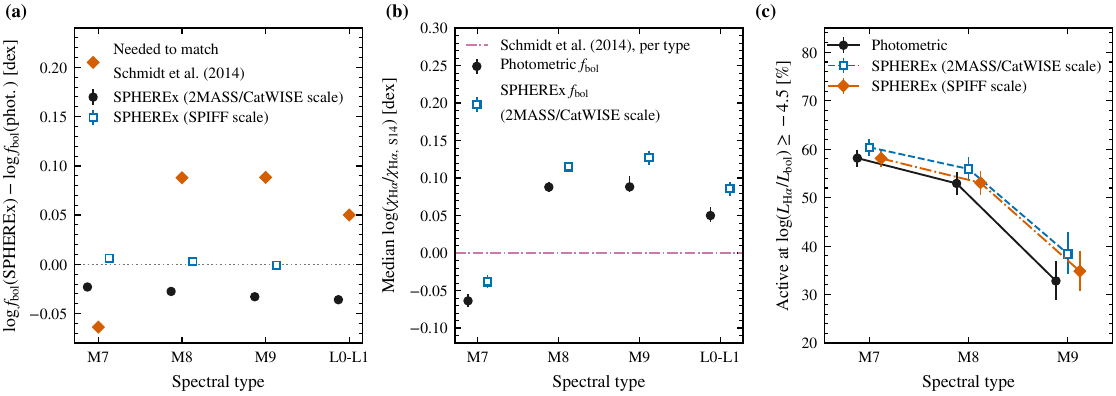}
\caption{SPHEREx bolometric fluxes. (a) Median difference between SPHEREx-based and photometric bolometric fluxes by corrected type on the 2MASS/CatWISE scale (filled circles) and the SPIFF scale (open squares), with 95\% bootstrap intervals; diamonds show the bolometric-flux error that would reconcile our $\chiH$ with the per-type values of \citet{Schmidt2014}. (b) Median ratio of our per-object $\chiH$ to the per-type values of \citet{Schmidt2014}, with photometric and SPHEREx bolometric fluxes. (c) Active fraction at $\log(\LHa)=-4.5$ on the dwarfs with both fluxes, with photometric bolometric fluxes (filled circles, solid line) and with SPHEREx fluxes on the 2MASS/CatWISE scale (open squares, dashed) and the SPIFF scale (diamonds, dash-dotted); bars are 68\% Wilson intervals.}
\label{fig:fbol_sx}
\end{figure*}

The SPHEREx bolometric flux integrates the DESI spectrum below 0.98\,\um\ and the SPHEREx spectrum on the 0.05\,\um\ grid from 0.98 to 5\,\um. Missing channels are filled from an empirical median spectrum of the same corrected type built from clean dwarfs, scaled to the observed channels; spectra with more than half the near-infrared integral filled are not used. A Rayleigh--Jeans tail is added beyond 5\,\um. Where DESI and SPHEREx overlap, their fluxes match to 0.3\% in the median. Figure~\ref{fig:fbol_sx} compares the result with the photometric bolometric flux. Table~\ref{tab:bolbudget} lists the terms of the comparison by type. In the analysis sample, with SPHEREx bolometric fluxes $\Delta\sUVW$ at M7 and M8 (Sect.~\ref{subsec:kin_fixed}) changes by up to 1.9\,\kms\ and the censored lifetimes at $-4.7$ by at most 0.5\,Gyr. In the 4.6\,\um\ CO band the L-dwarf fluxes exceed their BT-Settl fits \citep{Allard2012} by up to a factor of 1.9 and fall below their Sonora Diamondback fits \citep{Morley2024}, so an extrapolation with either model grid would introduce an atmospheric-model uncertainty that the measured spectra avoid.

\begin{table}[tbp]
\caption{Error budget of the SPHEREx-based minus photometric bolometric flux [dex].}
\label{tab:bolbudget}
\centering
\scriptsize
\setlength{\tabcolsep}{3pt}
\begin{tabular}{lccc}
\toprule
Term & M7 & M8 & M9 \\
\midrule
Median difference, 2MASS/CatWISE scale & $-$0.023 & $-$0.028 & $-$0.033 \\
Median difference, SPIFF scale & +0.006 & +0.003 & $-$0.001 \\
Zero point: SPIFF minus 2MASS/CatWISE scale & +0.029 & +0.029 & +0.030 \\
Change relative to M7 (2MASS/CatWISE scale) & +0.000 & $-$0.005 & $-$0.010 \\
Object scatter (robust) & 0.015 & 0.021 & 0.023 \\
Scatter expected from the errors & 0.020 & 0.027 & 0.029 \\
Missing-bin term (fill $<0.25$ minus all) & +0.000 & +0.000 & +0.002 \\
Median filled fraction of the 0.98--5\,\um\ flux & 0.22 & 0.21 & 0.22 \\
Shared DESI fraction of the flux & 0.15 & 0.12 & 0.11 \\
Blackbody fraction beyond 5\,\um & 0.042 & 0.046 & 0.056 \\
\midrule
Dwarfs compared & 1496 & 1382 & 548 \\
\bottomrule
\end{tabular}
\tablefoot{The two constructions share the DESI spectrum below 0.98\,\um\ and a Rayleigh--Jeans tail; their agreement tests the near-infrared integration, and the external check below tests the absolute scale. The zero point is smaller than the 0.033\,dex difference between the two SPHEREx flux scales (Appendix~\ref{app:sxdetail}) because that difference applies only to the part of the bolometric flux beyond 0.98\,\um, 85--89\% of it. The blackbody fraction uses the \citet{Pecaut2013} temperature of each type. External check, the 36 dwarfs with bolometric fluxes from \citet{Sanghi2023}: photometric minus published $+$0.031 ($+$0.027, $+$0.045), SPHEREx (2MASS/CatWISE scale) minus published $-$0.011 ($-$0.021, $-$0.003), with robust scatters of 0.025 and 0.019\,dex; 95\% bootstrap intervals.}
\end{table}

\subsection{Per-object and per-type $\chiH$}

Table~\ref{tab:chi} lists the median $\chiH$ of each type. From M6 to M9 the medians follow $\log\chiH=-5.03-0.22\,(n-7)$ for subtype M$n$ to within 0.01\,dex, and the decline flattens at L0 and L1. Our $\chiH$ lies 0.10\,dex above the per-type values of \citet{Schmidt2014} at M8 and M9 and 0.06\,dex below at M7. It is within 0.04\,dex of the per-type values that \citet{Douglas2014} measured from SDSS spectra at M6, M8 and M9, and 0.12\,dex below theirs at M7. Reconciling the two at M9 through the bolometric flux would need our fluxes to be 0.088\,dex too low, which the SPHEREx fluxes exclude on either flux scale ($p<10^{-4}$). For 20 dwarfs in common with \citet{Schmidt2014}, our continuum at H$\alpha$ matches theirs to 0.001\,dex and our $\chiH$ matches their per-object values to 0.015\,dex in the median. The difference therefore comes from their per-type means: their $\chiH$ falls by 0.38\,dex from M7 to M8, where the median of our per-object values falls by 0.21\,dex. From M8 to M9 both fall by 0.22\,dex.

\begin{table}[tbp]
\caption{Median $\log\chiH$ by corrected type, with published per-type values.}
\label{tab:chi}
\centering
\small
\setlength{\tabcolsep}{4pt}
\begin{tabular}{lccccc}
\toprule
Type & $N$ & Median & 16th to 84th percentile & S14 & D14 \\
\midrule
M6 & 1142 & $-4.80$ & $-4.86$ to $-4.76$ & -- & $-4.76$ \\
M7 & 1695 & $-5.04$ & $-5.13$ to $-4.91$ & $-4.97$ & $-4.92$ \\
M8 & 1391 & $-5.25$ & $-5.36$ to $-5.17$ & $-5.35$ & $-5.21$ \\
M9 & 407 & $-5.47$ & $-5.56$ to $-5.39$ & $-5.57$ & $-5.45$ \\
L0 & 146 & $-5.62$ & $-5.68$ to $-5.56$ & $-5.70$ & -- \\
L1 & 60 & $-5.72$ & $-5.81$ to $-5.62$ & $-5.66$ & -- \\
\bottomrule
\end{tabular}
\tablefoot{Measured spectra of the analysis sample with a bolometric flux, $\chiH$ per object. S14: mean per type of \citet{Schmidt2014}; D14: mean per type of \citet{Douglas2014}, from SDSS spectra.}
\end{table}

\subsection{Comparison with SPIFF}

The public SPIFF catalogue and binned spectra of \citet{Gagne2026}, built from QR2 data,\footnote{\url{https://zenodo.org/records/19051216}; software: \url{https://github.com/jgagneastro/SPIFF}.} contain 16\,917 spectra, including candidate and rejected objects as well as the selected library. With our positions propagated to each possible reference epoch, 557 DESI entries match 556 distinct library objects within $3\arcsec$. The matches are concentrated toward the L dwarfs (Table~\ref{tab:spiff_overlap}). The same 557 match within $2\arcsec$, while shifted control positions have no matches. The comparison tests the extraction: SPIFF fits a centroid in each exposure and integrates an oversampled point-spread function in a larger cutout, whereas our fit fixes the position and fits catalogue neighbours jointly.

\begin{figure*}[tbp]
\centering
\includegraphics[width=\textwidth]{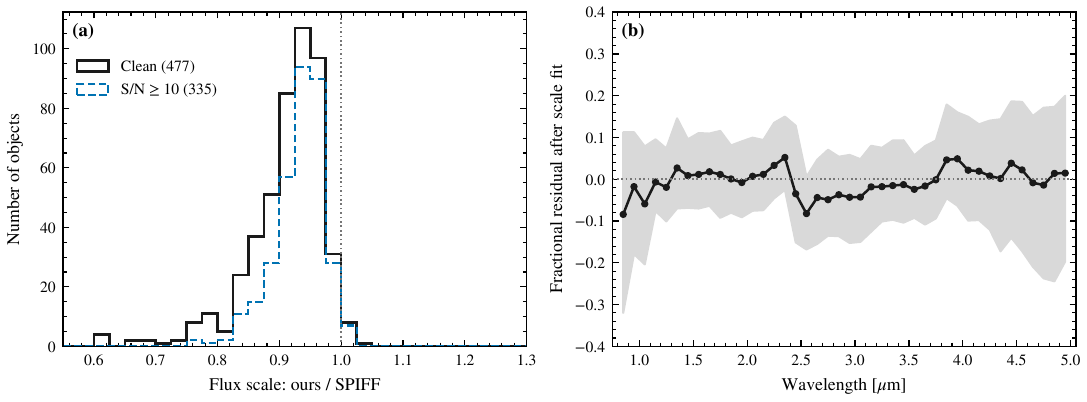}
\caption{Comparison with the public SPIFF spectra. (a) Fitted flux-scale ratio for the 477 clean common objects; the dashed histogram is the subset with median $S/N\ge10$ in both reductions. (b) Median fractional residual after fitting one scale per object, with the 16--84\% object range. The wavelength windows are common 0.1\,\um\ intervals.}
\label{fig:spiff_validation}
\end{figure*}

For each common object we averaged $f_\nu$ in disjoint 0.1\,\um\ windows between 0.8 and 5.0\,\um, required at least ten windows with SPIFF $S/N\ge5$, and fitted one multiplicative scale. Among 477 objects free of our quality flags, the median scale (ours/SPIFF) is 0.926, a 0.033\,dex difference comparable to the 8\% (0.033\,dex) absolute-calibration uncertainty discussed by \citet{Gagne2026}. After this fit the median fractional scatter of the residuals (scaled median absolute deviation) is about 0.1 (Fig.~\ref{fig:spiff_validation}). Water and $K/J$ ratios measured with identical windows match in the median, while our 3.3\,\um\ ratio is lower by 0.028 (Table~\ref{tab:spiff_indices}).

\begin{table}[tbp]\caption{Overlap with the downloadable SPIFF collection.}\label{tab:spiff_overlap}
\centering\small\begin{tabular}{lrrc}\toprule
Type & Sample & Matches & Overlap [\%]\\\midrule
M6 & 1815 & 0 & $0.0_{-0.0}^{+0.1}$ \\
M7 & 2897 & 36 & $1.2_{-0.2}^{+0.2}$ \\
M8 & 3505 & 159 & $4.5_{-0.3}^{+0.4}$ \\
M9 & 1515 & 125 & $8.3_{-0.7}^{+0.7}$ \\
L0 & 815 & 124 & $15.2_{-1.2}^{+1.3}$ \\
L1 & 185 & 95 & $51.4_{-3.7}^{+3.6}$ \\
L2 & 1 & 1 & $100.0_{-49.7}^{+0.0}$ \\
$\ge$L3 & 21 & 17 & $81.0_{-9.8}^{+7.1}$ \\
M7--L1 & 8939 & 557 & $6.2_{-0.2}^{+0.3}$ \\
\bottomrule\end{tabular}\tablefoot{68\% Wilson intervals. Matches count DESI entries; 557 entries correspond to 556 distinct SPIFF objects, including the library's candidate objects.}\end{table}
\begin{table}[tbp]\caption{Band-ratio differences relative to SPIFF.}\label{tab:spiff_indices}\centering\small
\begin{tabular}{lrc}\toprule
Ratio & Objects & Ours minus SPIFF (95\%)\\\midrule
H$_2$O 1.4\,\um & 373 & $0.007$ ($-0.002$, $0.012$) \\
$K/J$ & 401 & $0.005$ ($-0.013$, $0.023$) \\
CH$_4$ 3.3\,\um & 148 & $-0.028$ ($-0.051$, $-0.009$) \\
\bottomrule\end{tabular}\tablefoot{Median differences and 95\% object-bootstrap intervals, measured with identical band windows in the two reductions.}\end{table}

\section{Near-infrared diagnostics}
\label{app:sxtest}

\paragraph{NISP types.} The combined \emph{Euclid} NISP slitless spectra come from the spectroscopic reduction of the red grism \citep{EuclidQ1SIR}, matched within $1\arcsec$ after proper-motion propagation. The field templates are medians of SpeX spectra at each integer subtype, excluding known young objects, binaries and subdwarfs. The same 1.21--1.80\,\um\ fit applied to \emph{Euclid} Q1 dwarfs with published types \citep{Zhang2024, DominguezTagle2025, Mohandasan2026} has median offsets of zero at M7--M8 and $-1$ subtype at M9--L0; photometric Q1 searches give a complementary sample \citep{Zerjal2026}. For 21 dwarfs observed in both \emph{Euclid} Q1 and \emph{Euclid} DR1, the \emph{Euclid} DR1 type is a median of one subtype earlier (95\% interval $-1$ to zero), and object-by-object differences follow differences in continuum slope ($\rho=0.67$). We therefore report the agreement with the fitted \emph{Euclid} DR1 types and with these types shifted by one subtype (Table~\ref{tab:nisp_agreement}). On the template and corrected optical scales, respectively, the median NISP-minus-optical differences are $-2$ and $-1$ subtype with the fitted types, and $-1$ and zero with the shifted types.

\begin{table}[tbp]\caption{NISP types within one subtype of the optical type.}\label{tab:nisp_agreement}\centering\small
\begin{tabular}{lll}\toprule
NISP calibration & Original optical & Corrected optical\\\midrule
As fitted & \shortstack{75/171\\\ensuremath{43.9_{-3.7}^{+3.8}\%}} & \shortstack{111/171\\\ensuremath{64.9_{-3.7}^{+3.6}\%}}\\
Q1 shift & \shortstack{144/171\\\ensuremath{84.2_{-3.0}^{+2.6}\%}} & \shortstack{149/171\\\ensuremath{87.1_{-2.8}^{+2.3}\%}}\\
\bottomrule\end{tabular}\tablefoot{The 171 spectra of M6--L dwarfs in the final sample with NISP $S/N\ge5$. Fractions have 68\% Wilson intervals. The Q1 shift adds one subtype to the fitted \emph{Euclid} DR1 types; it is inferred from 21 common objects.}\end{table}

\begin{figure*}[tbp]
\centering
\includegraphics[width=\textwidth]{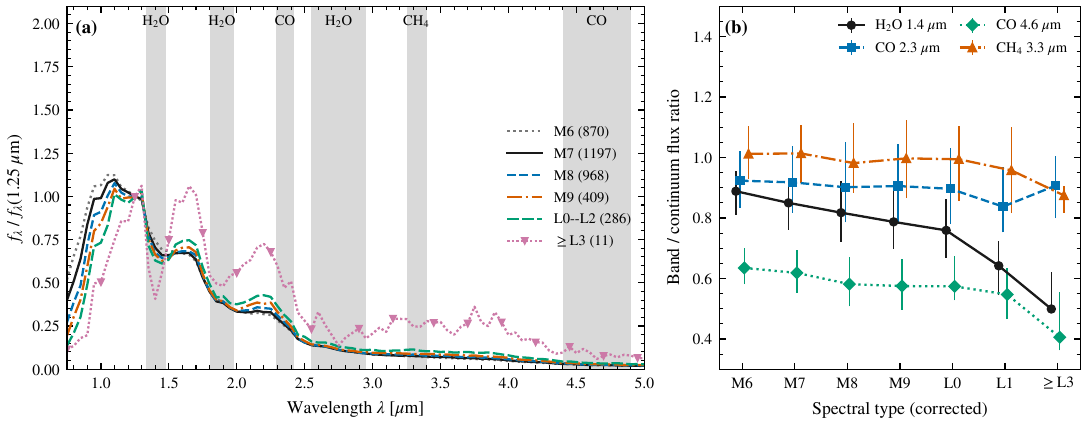}
\caption{The SPHEREx view of the optical M7--L1 sequence. (a) Median SPHEREx flux density $f_\lambda$, normalised at 1.25\,\um, for 3741 clean detections with $S/N\ge10$ at 1.20--1.30\,\um, including M6. Each plotted wavelength channel contains at least three dwarfs. (b) Band-to-continuum ratios (Table~\ref{tab:sxwin}); points are medians and bars the 16--84\% range of the dwarfs, with at least three dwarfs and $S/N\ge5$ in every contributing window.}
\label{fig:spherex}
\end{figure*}

\paragraph{The SPHEREx sequence.} Water absorption deepens and $K/J$ rises along the corrected sequence (Fig.~\ref{fig:spherex}, Table~\ref{tab:sxwin}). The 1.4\,\um\ water ratio decreases from \sxHtwoOseven\ at M7 to \sxHtwoOnine\ at M9 and \sxHtwoOLone\ at L1. Over the same types, $K/J$ rises from \sxKJseven\ to \sxKJnine\ and \sxKJLone. The correlations with corrected type, $\rho=-0.38$ and 0.49, are almost the same with template types ($-0.39$ and 0.46). Most dwarfs lie in the two adjacent M7 and M8 bins, where the object-to-object spread dominates, which explains this agreement. The 3.3\,\um\ methane ratio stays near unity from M7 through L0 and drops to 0.87 in the \Lbin\ bin, in line with the onset of the fundamental band in mid-L spectra \citep{Noll2000, Cushing2005, Sorahana2012, Stephens2009} and its appearance in SPHEREx libraries \citep{Gagne2026, Rustamkulov2026, Tu2026}. The L1 value of 0.96 (0.90--0.98) is comparable to the 0.028 offset between our methane ratio and SPIFF.

\paragraph{Activity classes at fixed type.} We compared SPHEREx indices of inactive and active dwarfs at $\Lambda=-4.7$, excluding high-proper-motion targets and flagged spectra, with minimum-cost matching within 0.5\,mag in $W1_{\rm Vega}$ and 0.4 subtype. Besides the activity classes at M6--M9, the 44 comparisons include weak-emitter and line-free M9 dwarfs and the inactive high-velocity dwarfs compared with other inactive M7--M8 dwarfs, for water, CO, methane, $K/J$, $W1-W2$ and the CO-band residual from an empirical spectrum of the same type. Median paired differences have bootstrap intervals, and probabilities come from Wilcoxon signed-rank tests, Holm-corrected over the 44 comparisons (Table~\ref{tab:nir_tests}). At M8 and M9 the median paired difference in 1.4\,\um\ water, inactive minus active, is $-0.045$ and $-0.042$, but neither comparison is significant after the Holm correction ($p=1.00$ and 0.61). Young, low-gravity M and L dwarfs differ from older field dwarfs in the near infrared, with weaker alkali lines and a peaked $H$ band \citep{Martin1996, Luhman1998, Bejar1999, LucasRoche2000, ZapateroOsorio2000, Cushing2000, Gorlova2003, McGovern2004, Slesnick2004, Slesnick2008, Lodieu2008, Scholz2012, AlvesdeOliveira2012, AlvesdeOliveira2013, Bonnefoy2014, Manjavacas2014, Muzic2014, Muzic2015, Lodieu2018}. In the SPIFF library both lower metallicity and lower gravity deepen this band, preventing water alone from distinguishing an older from a younger inactive population. Although $K/J$ moves in opposite directions for the two, its interval at M9 is too wide to decide. The NISP spectra miss the FeH, VO and K\,\textsc{i} gravity diagnostics shortward of 1.2\,\um. These spectra are available for only two active and three inactive M9 dwarfs, which is too small a sample for a matched comparison.

Within one optical type the SPHEREx water ratio follows the bolometric correction only weakly (partial Spearman $\rho=0.19$ for 1875 M7--M9 dwarfs), because non-temperature differences in band shape dominate its scatter. Water therefore gives no temperature scale at fixed optical type, and the joint model of Sect.~\ref{subsec:forward} uses the measured $\chiH$.

\begin{table}[tbp]
\caption{Infrared offsets of the high-velocity tail from matched thin-disk dwarfs ($G<19$).}
\label{tab:tail_tests}
\centering
\scriptsize
\setlength{\tabcolsep}{2.5pt}
\begin{tabular}{llrcc}
\toprule
Group & Index & $N$ & Median difference (95\%) & Holm $p$ \\
\midrule
M6--M9 & $K/J$ & 122 & $-$0.016 ($-$0.025, $-$0.007) & $7\times10^{-5}$ \\
M6--M9 & $K/H$ & 128 & $-$0.017 ($-$0.025, $-$0.008) & 0.02 \\
M6--M9 & $J-K_s$  & 120 & $-$0.063 ($-$0.077, $-$0.037) & $7\times10^{-10}$ \\
M6--M9 & H$_2$O 1.4\,\um & 120 & $-$0.041 ($-$0.064, $-$0.012) & $2\times10^{-4}$ \\
M6--M9 & CO 2.3\,\um & 110 & $-$0.003 ($-$0.014, 0.025) & 0.64 \\
M6--M9 & CO 4.6\,\um & 107 & 0.018 (0.000, 0.025) & 0.64 \\
M6--M9 & $W1-W2$ & 167 & 0.027 (0.020, 0.031) & $2\times10^{-13}$ \\
M7 & $K/J$ & 47 & $-$0.025 ($-$0.036, $-$0.012) & $2\times10^{-4}$ \\
M7 & $J-K_s$  & 47 & $-$0.090 ($-$0.112, $-$0.044) & $2\times10^{-6}$ \\
M8 & $K/J$ & 14 & $-$0.024 ($-$0.040, 0.008) & 0.47 \\
M8 & $J-K_s$  & 6 & $-$0.059 ($-$0.111, 0.001) & 0.47 \\
M7--M8$^a$ & $K/J$ & 20 & $-$0.015 ($-$0.041, 0.022) & 0.64 \\
M7--M8$^b$ & $K/J$ & 130 & $-$0.008 ($-$0.017, 0.002) & 0.09 \\
\midrule
M6--M9, no colour-selected & $J-K_s$ & 118 & $-$0.060 ($-$0.076, $-$0.037) & -- \\
M6--M9, no colour-selected & $K/J$ & 119 & $-$0.014 ($-$0.024, $-$0.007) & -- \\
M6--M9, MWS only & $J-K_s$ & 113 & $-$0.054 ($-$0.076, $-$0.037) & -- \\
M6--M9, MWS only & $K/J$ & 113 & $-$0.015 ($-$0.021, $-$0.009) & -- \\
\midrule
\multicolumn{5}{l}{Full analysis sample} \\
M6--M9 & $\zeta$ & 209 & 0.010 (0.000, 0.025) & -- \\
M6--M9 & TiO5 & 209 & $-$0.013 ($-$0.019, $-$0.002) & -- \\
M6--M9 & CaH2+CaH3 & 209 & 0.001 ($-$0.013, 0.013) & -- \\
M6--M9, $M_G$-matched$^c$ & $\zeta$ & 208 & 0.005 ($-$0.011, 0.021) & -- \\
M6--M9, $M_G$-matched$^c$ & $J-K_s$ & 158 & $-$0.053 ($-$0.075, $-$0.042) & -- \\
M6--M9, $M_G$-matched$^c$ & $K/J$ & 149 & $-$0.022 ($-$0.030, $-$0.015) & -- \\
\midrule
\multicolumn{5}{l}{Mean differences with independent photometry$^d$} \\
Pooled, 2MASS & $J-K$ & 174 & $-$0.068 ($-$0.083, $-$0.052) & -- \\
Pooled, VHS DR5 & $J-K$ & 21 & $-$0.082 ($-$0.107, $-$0.057) & -- \\
Pooled, UKIDSS, 1st-epoch $J$ & $J-K$ & 51 & $-$0.066 ($-$0.089, $-$0.042) & -- \\
Pooled, Legacy DR10 & $i-z$ & 31 & $-$0.018 ($-$0.041, +0.003) & -- \\
Pooled, Legacy DR10 & $r-i$ & 33 & $-$0.001 ($-$0.040, +0.041) & -- \\
\bottomrule
\end{tabular}
\tablefoot{Each tail dwarf is compared with up to three strict thin-disk dwarfs matched within 0.3 subtype, 0.3\,mag in $W1_{\rm Vega}$ and 0.1\,dex in distance; the difference is the tail value minus the mean of its controls. Probabilities are two-sided Wilcoxon signed-rank tests, Holm-corrected over the 13 comparisons of the first block; the rows below it are controls and carry no probability. The first two blocks use dwarfs with $G<19$; the optical-index and independent-photometry blocks use the full analysis sample. \tablefoottext{a}{Inactive tail dwarfs minus inactive thin-disk dwarfs.}\tablefoottext{b}{Inactive minus active dwarfs outside the tail.}\tablefoottext{c}{Controls matched within 0.15\,mag in $M_G$ instead of optical type.}\tablefoottext{d}{Tail dwarfs matched to thin-disk controls in exact type, $W1_{\rm Vega}$ and distance, pooled over all types including the M6 bin, each survey in its native system, from catalogues served by the Astro Data Lab \citep{Dey2019, McMahon2013, Lawrence2007}; 95\% intervals.}}
\end{table}

\begin{table}[tbp]
\caption{SPHEREx band and continuum windows.}
\label{tab:sxwin}
\centering
\scriptsize
\begin{tabular}{lll}
\toprule
Ratio & Band [\um] & Continuum [\um] \\
\midrule
H$_2$O 1.4\,\um & 1.35--1.45 & 1.15--1.30, 1.55--1.75 \\
H$_2$O 1.9\,\um & 1.85--1.95 & 1.55--1.75, 2.10--2.25 \\
CO 2.3\,\um & 2.30--2.40 & 2.10--2.25 \\
H$_2$O 2.7\,\um & 2.60--2.85 & 2.10--2.25, 3.00--3.15 \\
CH$_4$ 3.3\,\um & 3.25--3.40 & 3.00--3.15, 3.80--4.00 \\
CO 4.6\,\um & 4.45--4.85 & 3.80--4.00 \\
$K/J$ & 2.10--2.25 & 1.15--1.30 \\
$K/H$ & 2.10--2.25 & 1.55--1.75 \\
\bottomrule
\end{tabular}
\tablefoot{Each ratio is the inverse-variance mean flux density $f_\nu$ in the band divided by the mean of the continuum windows; only windows with $S/N \ge 5$ are used.}
\end{table}

\begin{table*}[tbp]\caption{Selected matched near-infrared comparisons of activity classes.}\label{tab:nir_tests}\centering\small
\begin{tabular}{llrcc}\toprule
Comparison (first minus second) & Index & Pairs & Median difference (95\%) & Holm $p$\\\midrule
M7 inactive--active & H$_2$O 1.4\,\um & 131 & $0.005$ ($-0.023$, $0.029$) & 1.00 \\
M7 inactive--active & $K/J$ & 138 & $-0.011$ ($-0.037$, $0.024$) & 1.00 \\
M8 inactive--active & H$_2$O 1.4\,\um & 49 & $-0.045$ ($-0.084$, $-0.020$) & 1.00 \\
M8 inactive--active & $K/J$ & 54 & $-0.057$ ($-0.105$, $0.055$) & 1.00 \\
M9 inactive--active & H$_2$O 1.4\,\um & 28 & $-0.042$ ($-0.078$, $-0.011$) & 0.61 \\
M9 inactive--active & $K/J$ & 31 & $0.058$ ($-0.093$, $0.140$) & 1.00 \\
M9 inactive--active & CH$_4$ 3.3\,\um & 13 & $-0.025$ ($-0.096$, $0.050$) & 1.00 \\
M9 weak emitter--active & H$_2$O 1.4\,\um & 15 & $-0.044$ ($-0.110$, $0.021$) & 1.00 \\
M9 weak emitter--active & $K/J$ & 16 & $0.069$ ($-0.144$, $0.183$) & 1.00 \\
M8 high-velocity--other inactive & H$_2$O 1.4\,\um & 8 & $-0.061$ ($-0.125$, $-0.031$) & 0.33 \\
\bottomrule\end{tabular}\tablefoot{Intervals are bootstrap intervals of the median paired difference; probabilities are Wilcoxon signed-rank tests on all pairs, Holm-corrected over 44 comparisons; no comparison is significant at the 5\% level after the correction.}\end{table*}

\FloatBarrier
\section{Resolved and comoving companions}
\label{app:vis}

VIS images at 49 sample positions were taken from the public \emph{Euclid} Q1 mosaics. The 38 targets used here, 35 of them M7--L1 dwarfs, have a catalogue match, usable unsaturated pixels at high $S/N$ and at least ten isolated, point-like comparison stars of similar magnitude. The targets show no excess elongation in second moments measured with a two-pixel Gaussian weight: three are more elongated than all of their 20 comparison stars, where 1.8 are expected (binomial $p=0.27$).

\begin{figure*}[tbp]
\centering
\includegraphics[width=\textwidth]{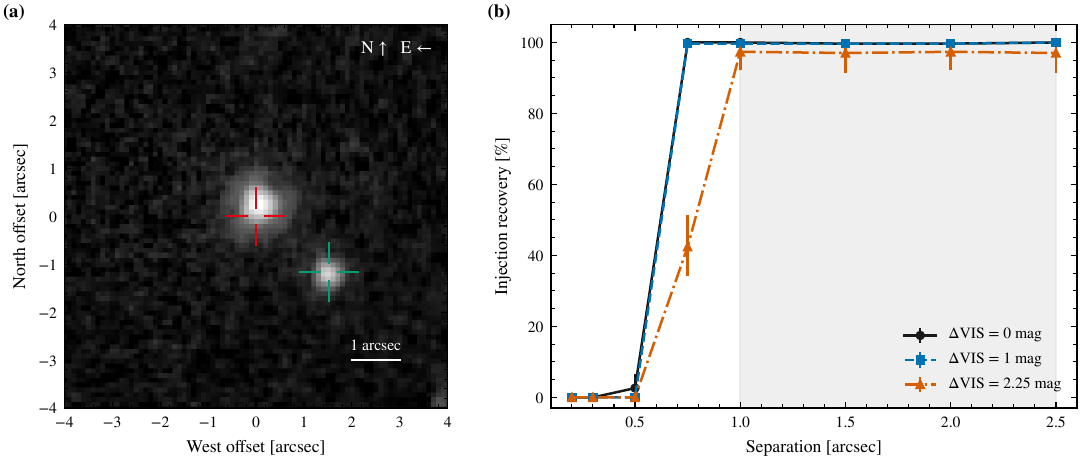}
\caption{\emph{Euclid} Q1 VIS companion search. (a) The M9 dwarf DESI TARGETID 39633456279323849 and the background star 2.05\arcsec\ from it. Crosshairs with open centres mark the propagated DESI position (red, frame centre) and the background star (green). (b) Recovery of injected point sources at three values of $\Delta\mathrm{VIS}$; intervals are 95\% target-bootstrap intervals. The companion fraction refers to 1--2.5\arcsec\ and $\Delta\mathrm{VIS}$ up to 2.25\,mag.}
\label{fig:vis}
\end{figure*}

We searched for companions as local maxima with aperture $S/N\ge8$ at 0.3--3\arcsec\ and magnitude difference $\Delta\mathrm{VIS}$ up to 2.5\,mag, rejecting extended sources. One stellar candidate is $2.05\arcsec$ from one M9 dwarf (Fig.~\ref{fig:vis}a), at $\Delta\mathrm{VIS}=2.12$\,mag. Point sources injected into each target's own image are recovered in at least 97\% of cases from 1 to 2.5\arcsec\ and up to 2.25\,mag, while recovery at $0.75\arcsec$ depends strongly on $\Delta\mathrm{VIS}$ and at $0.5\arcsec$ is close to zero. In this well-recovered domain the candidate incidence is $1/38$, or $2.6_{-1.6}^{+4.0}\%$. Stellar-looking sources in the surrounding 3--8\arcsec\ annuli predict 0.38 chance neighbours. The candidate is a background star: between Legacy Surveys images of 2017--2018 and VIS the dwarf moved 0.36\arcsec\ while the neighbour stayed in place. The neighbour is also much bluer ($r-z=1.1$ compared with 3.9).

In the \emph{Euclid} DR1 MER (merged multi-band) catalogue, 233 targets meet the sample cuts. Three have a point-like neighbour at 0.3--3\arcsec\ within 2.5\,mag in $\Delta\mathrm{VIS}$: two among the M7--L1 dwarfs and one in the M6 bin. All three lack an activity class. One is the M9 dwarf with the background star above. The neighbour of the M7 dwarf is bluer than the dwarf yet fainter in the red bands, which puts it at a different distance. The M6 pair is listed in Table~\ref{tab:vis_pairs} as a possible companion. For dwarfs without a \emph{Euclid} Q1 or \emph{Euclid} DR1 companion candidate or a SPHEREx bright-neighbour or unmodelled-source flag, $\Delta\sUVW$ at M7 in the analysis sample is 20.4\,\kms\ and the activity fractions change by at most two points (Table~\ref{tab:blend_test}).

\begin{figure*}[tbp]
\centering
\includegraphics[width=\textwidth]{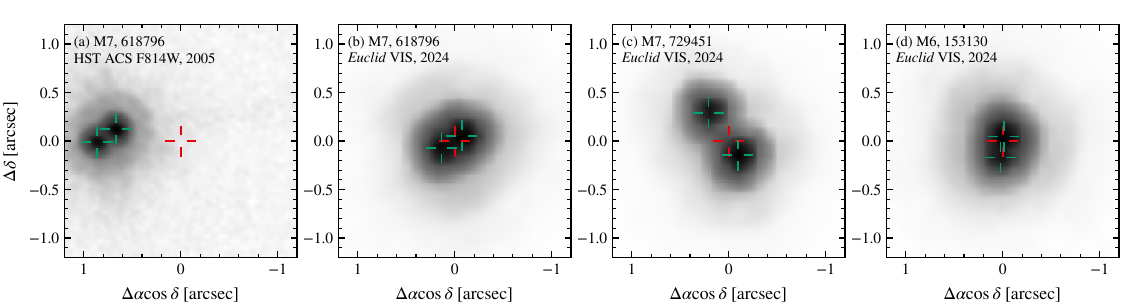}
\caption{Close pairs with common proper motion. (a, b) TARGETID 39627853691618796 in HST ACS F814W in 2005 and in \emph{Euclid} VIS in 2024. (c, d) VIS images of TARGETIDs 39633509735729451 and 39633440647153130, whose common motion is tested in SDSS and Pan-STARRS1 images (text). All panels share one grid of offsets from the \emph{Euclid} position, with north up and east left. Crosshairs with open centres mark the \emph{Euclid} position (red) and the components (green), measured in each image.}
\label{fig:vis_pairs}
\end{figure*}

\paragraph{Close pairs in the DR1 VIS images.} The DR1 MER mosaics contain VIS images of 235 targets, and 173 of them pass the cuts above. The local-maximum search finds no more candidates there than chance alignments predict, but it loses pairs closer than about 0.75\arcsec. To resolve smaller separations we fitted each target with one and with two point-spread functions built from its own comparison stars, and calibrated the improvement in fit on the same fits to those stars. The two-point-spread-function search has no completeness calibration, so we do not convert the pairs into a binary fraction. Six targets exceed the 99.5th percentile of comparison stars of similar flux, where 1.3 are expected, and they include two of the three targets with RUWE above 1.4, compared with one of the 64 with lower RUWE (Fisher $p=0.004$). Five of the six show a second source, and we tested these for common proper motion in earlier images (Table~\ref{tab:vis_pairs}). Two are possible companions, because one dwarf moves too slowly for the test and the other has no \emph{Gaia} astrometry. HST ACS images of the COSMOS field from 2005 \citep{Scoville2007, Koekemoer2007} resolve the pair TARGETID 39627853691618796 with the separation and position angle it has in VIS in 2024, within 0.01\arcsec\ and 5\degr, while both components moved 0.74\arcsec\ (Fig.~\ref{fig:vis_pairs}). For the two pairs with \emph{Gaia} proper motions, a stationary background star would have appeared at a different offset from the dwarf in SDSS \citep{York2000} images of 2000--2001 and Pan-STARRS1 images of 2010--2014, and the shapes of the blended images exclude this. In eight SDSS frames the 0.22\arcsec\ pair is elongated by about a thirtieth of what a stationary star would produce, and in Pan-STARRS1 the 0.53\arcsec\ pair is elongated along its VIS position angle, which differs from the angle a stationary star would give. Although common motion alone does not establish that a pair is bound, these three pairs therefore move together, at projected separations of 24 and 48\,au where the distance is known.

\begin{table*}[tbp]
\caption{Close pairs found in the \emph{Euclid} DR1 VIS images.}
\label{tab:vis_pairs}
\centering
\scriptsize
\setlength{\tabcolsep}{4pt}
\begin{tabular}{lcccccccl}
\toprule
TARGETID & Type & $\rho$ [\arcsec] & PA [\degr] & $\Delta\mathrm{VIS}$ & $d$ [pc] & $a_{\rm p}$ [au] & Earlier images & Common proper motion \\
\midrule
39627853691618796 & M7 & 0.24 & 120 & 0.52 & -- & -- & HST ACS, 2005 & yes, pair resolved in both epochs \\
39633440647153130 & M6 & 0.22 & 171 & 1.10 & 108 & 24 & SDSS, 2000--2001 & yes, stationary star excluded \\
39633509735729451 & M7 & 0.53 & 35 & 0.48 & 91 & 48 & Pan-STARRS1, 2010--2014 & yes, stationary star excluded \\
39633485295520779 & M6 & 1.03 & 129 & 2.12 & -- & -- & -- & possible; motion of 5\,mas\,yr$^{-1}$ too small \\
39627859718832726 & M7 & 0.78 & 146 & 4.10 & -- & -- & -- & possible; no \emph{Gaia} astrometry \\
\bottomrule
\end{tabular}
\tablefoot{$\rho$ and PA: separation and position angle of the fainter component from two-point-spread-function fits to the VIS image; $\Delta\mathrm{VIS}$: magnitude difference; $d$: \emph{Gaia} DR3 distance; $a_{\rm p}$: projected separation.}
\end{table*}

\paragraph{Comoving neighbours in the Legacy Surveys.} Through the Astro Data Lab we searched the Legacy Surveys DR10 catalogue \citep{Dey2019} for point-like neighbours at 0.5--5\arcsec\ around the targets and kept those whose Gaia DR3 proper motion and parallax agree with the target's. Thirteen pairs qualify, eight of them around M7--L1 dwarfs, at projected separations of 96--599\,au (Table~\ref{tab:comoving}). Nine have a clean primary and RUWE below 1.4 for both components. The same selection applied to fields displaced from the targets finds no comoving pair. We list them as candidate wide companions. At 0.5--1\arcsec\ the point-like neighbours, before the astrometric selection, exceed the chance expectation by $110\pm12$. Because deblending artefacts inside the seeing contribute to this excess, it is an upper bound on the number of close companions.

\begin{table}[tbp]
\caption{Candidate comoving companions in the Legacy Surveys DR10.}
\label{tab:comoving}
\centering
\scriptsize
\setlength{\tabcolsep}{3pt}
\begin{tabular}{lcrrrc}
\toprule
DESI TARGETID & Type & Sep. [\arcsec] & $\Delta z$ [mag] & $a_{\rm proj}$ [au] & Clean \\
\midrule
39627724679025085 & M7 & 4.90 & $-$0.06 & 599 & yes \\
39627866622656880 & L0 & 1.60 & $-$2.22 & 104 & yes \\
39627879008440731 & M7 & 2.29 & $-$0.17 & 196 & yes \\
39627939158953124 & M6 & 1.77 & $-$0.19 & 164 & yes \\
39628074827909931 & M8 & 1.78 & +0.16 & 108 & yes \\
39628217304222444 & M6 & 0.64 & $-$0.11 & 233 & no \\
39628263017941117 & M6 & 2.33 & +1.67 & 224 & yes \\
39628296782089643 & M7 & 2.63 & +1.81 & 401 & yes \\
39628335713617399 & M6 & 0.86 & +0.93 & 96 & no \\
39628502600780611 & M8 & 1.12 & +1.37 & 97 & no \\
39632951943629375 & M9 & 2.30 & +1.63 & 192 & yes \\
39632961238205057 & M7 & 1.04 & $-$0.91 & 113 & no \\
39633225361919126 & M6 & 3.01 & +0.09 & 353 & yes \\
\bottomrule
\end{tabular}
\tablefoot{Neighbours at 0.5--5\arcsec\ whose Gaia DR3 proper motion and parallax agree with those of the target. $\Delta z$: neighbour minus target. Clean: the target passes the image-quality selection and both components have RUWE below 1.4. Candidate physical pairs; their binding and novelty are not established.}
\end{table}

\begin{table*}[tbp]\caption{Activity and kinematics after removing identified blends and VIS candidates.}\label{tab:blend_test}\centering\small
\begin{tabular}{lcccc}\toprule
& \multicolumn{2}{c}{Active fraction [\%], 68\% Wilson}&\multicolumn{2}{c}{Inactive minus active $\sUVW$ [\kms], 95\% bootstrap}\\
Type & Before & After & Before & After\\\midrule
M6 & $63.0_{-2.0}^{+1.9}$ & $63.0_{-2.0}^{+2.0}$ & $19.8$ ($15.1$, $24.3$) & $19.8$ ($15.0$, $24.6$) \\
M7 & $69.3_{-1.9}^{+1.8}$ & $69.1_{-2.0}^{+1.9}$ & $20.4$ ($15.8$, $24.5$) & $20.4$ ($16.0$, $24.7$) \\
M8 & $70.1_{-2.9}^{+2.7}$ & $72.0_{-2.9}^{+2.7}$ & $15.5$ ($7.0$, $22.5$) & $17.7$ ($9.6$, $25.3$) \\
M9 & $59.6_{-5.1}^{+4.9}$ & $59.6_{-5.3}^{+5.0}$ & $-3.9$ ($-13.9$, $5.3$) & $-4.4$ ($-15.2$, $5.4$) \\
\bottomrule\end{tabular}\tablefoot{Every comparison uses $\Lambda=-4.7$. Before and after removal of the candidates; removal combines the bright-neighbour and unmodelled-source SPHEREx flags with the Q1 image and DR1 catalogue companion candidates.}\end{table*}

\FloatBarrier
\section{Results on the template type scale}
\label{app:oldscale}

Table~\ref{tab:oldscale} compares, for the analysis sample, the main results on the template types of \citet{Kesseli2017}, with the per-type $\chiH$ of \citet{Schmidt2014}, with those on the corrected types. The rise of the equivalent-width fraction and the fall at fixed $\LHa$ appear on both scales, and $\Delta\sUVW$ at M7 is 20--24\,\kms. At M8 $\Delta\sUVW$ has the same sign but is smaller on the corrected scale, because many template-scale M8 dwarfs are M7 on the corrected scale. At M9 the template scale gives 5.6\,\kms\ ($-0.0$ to 11.9), and the corrected scale gives $-3.9$\,\kms. Over the three thresholds and both conversions $\Delta\sUVW$ is 20--27\,\kms\ at M7 and 2--16\,\kms\ at M8.

\begin{table}[tbp]
\caption{Main results on the template and corrected type scales.}
\label{tab:oldscale}
\centering
\scriptsize
\setlength{\tabcolsep}{1pt}
\begin{tabular}{lccc}
\toprule
Quantity & Template, & Corrected, & Corrected, \\
 & $\chiH$ per type & $\chiH$ per type & $\chiH$ per object \\
\midrule
Dwarfs (M7--L1)$^a$ & 10\,786 & 8939 & 8939 \\
EW 5\,\AA\ fraction M7 & \rfrac{K0s_old_schmidt_EW_5_M7} & \rfrac{K1s_corr_schmidt_EW_5_M7} & \rfrac{K1e_corr_chiobj_EW_5_M7} \\
EW 5\,\AA\ fraction M8 & \rfrac{K0s_old_schmidt_EW_5_M8} & \rfrac{K1s_corr_schmidt_EW_5_M8} & \rfrac{K1e_corr_chiobj_EW_5_M8} \\
EW 5\,\AA\ fraction M9 & \rfrac{K0s_old_schmidt_EW_5_M9} & \rfrac{K1s_corr_schmidt_EW_5_M9} & \rfrac{K1e_corr_chiobj_EW_5_M9} \\
$\Lambda=-4.5$ fraction M7 & \rfrac{K0s_old_schmidt_Lambda_-4.5_M7} & \rfrac{K1s_corr_schmidt_Lambda_-4.5_M7} & \rfrac{K1e_corr_chiobj_Lambda_-4.5_M7} \\
$\Lambda=-4.5$ fraction M8 & \rfrac{K0s_old_schmidt_Lambda_-4.5_M8} & \rfrac{K1s_corr_schmidt_Lambda_-4.5_M8} & \rfrac{K1e_corr_chiobj_Lambda_-4.5_M8} \\
$\Lambda=-4.5$ fraction M9 & \rfrac{K0s_old_schmidt_Lambda_-4.5_M9} & \rfrac{K1s_corr_schmidt_Lambda_-4.5_M9} & \rfrac{K1e_corr_chiobj_Lambda_-4.5_M9} \\
$\Delta\sUVW$ M7 [\kms] & 23.5 (20.2, 26.6) & 23.0 (18.5, 27.2) & 20.4 (15.8, 24.5) \\
$\Delta\sUVW$ M8 [\kms] & 20.7 (16.8, 25.0) & 10.9 (5.3, 16.8) & 15.5 (7.0, 22.5) \\
$\Delta\sUVW$ M9 [\kms] & 5.6 ($-0.0$, 11.9) & $-1.7$ ($-11.7$, 8.2) & $-3.9$ ($-13.9$, 5.3) \\
$\tau$ (active $\sUVW$) M7 / M8 [Gyr] & 5.8 / 5.4 & 6.0 / 6.7 & 6.0 / 7.4 \\
Pooled $\sUVW$ M9 $-$ M7 [\kms] & $-8.4$ & $-7.1$ & $-7.1$ \\
\quad Restored sample & $-0.1$ & $-3.0$ & $-3.0$ \\
\bottomrule
\end{tabular}
\tablefoot{Dispersion differences are inactive minus active at $\Lambda=-4.7$, with 95\% bootstrap intervals; $\tau$ is the step lifetime matched to the dispersion of the active dwarfs of the analysis sample, without a tail cut (the $G<19$ values are 6.0 and 8.2\,Gyr; Table~\ref{tab:lifetimes}). It differs from the censored lifetimes of Table~\ref{tab:lifetimes}, which use the velocities of both activity groups, and from the adopted thin-disk lifetimes of Table~\ref{tab:kinvariants}, which also remove the tail and truncate the likelihood. Pooled dispersions are independent of $\chiH$.
\tablefoottext{a}{The template-scale count is after removing giants and includes the 29 quasars of Appendix~\ref{app:counts}.}}
\end{table}

\section{Sensitivity to the M9/L0 boundary}
\label{app:m9l0}

The hybrid scale isolates the boundary defined by the 23 L0 anchors: it uses the corrected type for dwarfs whose corrected type is M8 or earlier, and the template type for dwarfs typed M9 or later on the template scale. The 590 template-scale M9 dwarfs typed L0 on the corrected scale are therefore M9 on the hybrid scale. Table~\ref{tab:m9l0} compares, for the analysis sample, the results that depend on the M9 and L bins on the two scales (M6--M8 are identical). The trends in active fraction appear on both. $\Delta\sUVW$ at M9 at $\Lambda=-4.7$ is consistent with zero on both scales. One variant, the per-type conversion within 60\,pc on the hybrid scale, has a positive difference whose 95\% interval excludes zero, and on the hybrid scale the restored pooled M9 minus M7 difference also excludes zero.

\begin{table}[tbp]
\caption{Results that depend on the M9/L0 boundary.}
\label{tab:m9l0}
\centering
\scriptsize
\setlength{\tabcolsep}{2pt}
\begin{tabular}{lcc}
\toprule
Quantity & Corrected & Hybrid \\
\midrule
Dwarfs M9 / L0--L1 & 1515 / 1000 & 2105 / 405 \\
EW 5\,\AA\ fraction M9 & \rfrac{corrected_EW_5_M9} & \rfrac{hybrid_EW_5_M9} \\
EW 5\,\AA\ fraction L0--L1 & \rfrac{corrected_EW_5_L0-L1} & \rfrac{hybrid_EW_5_L0-L1} \\
$\Lambda = -4.7$ fraction M9 & \rfrac{corrected_Lambda_-4.7_M9} & \rfrac{hybrid_Lambda_-4.7_M9} \\
$\Lambda = -4.5$ fraction M9 & \rfrac{corrected_Lambda_-4.5_M9} & \rfrac{hybrid_Lambda_-4.5_M9} \\
$\Lambda = -4.3$ fraction M9 & \rfrac{corrected_Lambda_-4.3_M9} & \rfrac{hybrid_Lambda_-4.3_M9} \\
$\Delta\sUVW$ M9, $\Lambda=-4.7$ & $-3.9$ ($-13.9$, 5.3) & 1.8 ($-7.8$, 10.8) \\
\quad distance-matched & 1.5 ($-7.2$, 10.3) & 4.2 ($-4.7$, 13.3) \\
\quad per-type $\chiH$, within 60\,pc & 2.1 ($-7.4$, 12.2) & 9.1 (0.5, 17.9) \\
Pooled $\sUVW$ M9 $-$ M7 & $-7.1$ ($-10.9$, $-3.5$) & $-8.3$ ($-11.7$, $-4.9$) \\
\quad Restored sample & $-3.0$ ($-6.5$, 0.2) & $-3.2$ ($-6.5$, $-0.2$) \\
L dwarfs with space motions, $\sUVW$ & 147, 43 (36, 50) & 65, 50 (36, 63) \\
New L dwarfs & 627 & 225 \\
\bottomrule
\end{tabular}
\tablefoot{Hybrid: corrected type up to M8, template type at M9 and later. Of the L0--L1 dwarfs that leave on the hybrid scale, 590 return to M9; six corrected L1 dwarfs typed \Lbin\ on the template scale move to \Lbin, and one corrected L2 dwarf typed L1 on the template scale joins L1. Fractions use $\chiH$ per object; dispersion differences (inactive minus active) and pooled differences are in \kms\ with 95\% bootstrap intervals.}
\end{table}

\section{Activity and kinematics as a function of bolometric luminosity}
\label{app:lbol}

\begin{figure*}[tbp]
    \centering
    \includegraphics[width=\textwidth]{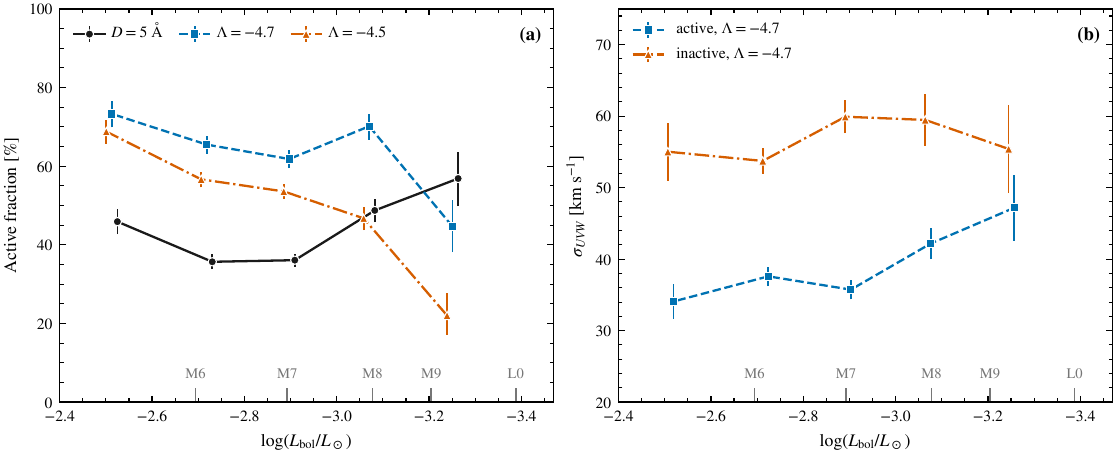}
    \caption{Activity and kinematics of the $G<19$ sample in bins of bolometric luminosity (0.2\,dex; types M6--L1 together), plotted at the median luminosity of each bin; grey ticks mark the median luminosity of each type. (a) Active fraction at an equivalent width of 5\,\AA\ and at two thresholds $\Lambda$ in $\log(\LHa)$, among spectra with limits at or below each threshold; bars are 68\% Wilson intervals. (b) Velocity dispersion of the dwarfs active and inactive at $\Lambda=-4.7$, with bootstrap standard errors.}
    \label{fig:lbol}
\end{figure*}

Spectral types are coarse labels of luminosity: at M7 and M8 the 16th to 84th percentile range of $\log L_{\mathrm{bol}}$ within one type is 0.27\,dex, more than the 0.19\,dex between the two medians. For the dwarfs of the $G<19$ sample with a Gaia parallax we therefore repeated the main comparisons with $L_{\mathrm{bol}}$ as the independent variable, in bins of 0.2\,dex that combine the types M6 to L1 (Fig.~\ref{fig:lbol}, Table~\ref{tab:lbol}). The reversal of Sect.~\ref{subsec:halpha_results} appears without reference to type. From the bin at $\log(L_{\mathrm{bol}}/L_\odot)=-2.7$ to the one at $-3.25$ the active fraction at 5\,\AA\ rises from 36\% to 57\%, while at $\Lambda=-4.5$ it falls from 57\% to 22\%. The inactive dwarfs are faster than the active ones by 16--24\,\kms\ in each of the four bins brighter than $-3.2$. In the faintest bin, which is mostly M9, the difference is consistent with zero, as in Sect.~\ref{subsec:m9}. The dispersion of the active dwarfs rises from 34 to 47\,\kms\ toward lower luminosity, while that of the inactive dwarfs stays between 54 and 60\,\kms. The active dwarfs of lower luminosity are therefore kinematically older, as expected if activity lasts longer there, although the largest step between adjacent bins is 2.5 times its error.

Luminosity also carries information within one type. With the overluminous dwarfs removed, the brighter half of the M8 dwarfs is more often active at $\Lambda=-4.5$ than the fainter half (59\% against 40\%), as is the brighter half at M7 (56\% against 49\%), while at $\Lambda=-4.7$ the halves agree. The fainter M8 half has the larger dispersion in both activity groups. Within a type, as across types, lower luminosity thus goes with weaker emission and older kinematics. Unresolved binaries below the overluminosity cut and a spread in metallicity act in the same direction, and we do not separate them from the spread in temperature within a type.

A magnitude limit favours the brighter dwarfs of each type. Within 60\,pc the rise from M7 to M9 at 5\,\AA\ is 32 points and the fall at $\Lambda=-4.5$ is 26 points, compared with 29 and 22 points for the whole $G<19$ sample (Table~\ref{tab:twosamples}), and the inactive dwarfs are faster by 24\,\kms\ at M7 and 14\,\kms\ at M8.

\begin{table*}[tbp]
\caption{Active fractions and velocity dispersions of the $G<19$ sample in bins of bolometric luminosity, within 60\,pc, and in the two luminosity halves of one type.}
\label{tab:lbol}
\centering
\small
\setlength{\tabcolsep}{4pt}
\begin{tabular}{lcccccccc}
\toprule
 & & \multicolumn{3}{c}{Active fraction [\%]} & \multicolumn{4}{c}{Kinematics at $\Lambda=-4.7$} \\
Group & Median $\log L$ & $D=5$\,\AA & $\Lambda=-4.7$ & $\Lambda=-4.5$ & $N_{\mathrm{act}}/N_{\mathrm{inact}}$ & $\sUVW^{\mathrm{act}}$ [\kms] & $\sUVW^{\mathrm{inact}}$ [\kms] & $\Delta\sUVW$ [\kms] \\
\midrule
\multicolumn{9}{l}{Bins of $\log(L_{\mathrm{bol}}/L_\odot)$, types M6--L1 together} \\
$-2.4$ to $-2.6$ & $-2.51$ & 45.9 (244) & 73.3 (176) & 68.8 (234) & 99/34 & $34.1\pm2.4$ & $55.0\pm4.0$ & 20.9 (10.6, 28.9) \\
$-2.6$ to $-2.8$ & $-2.72$ & 35.7 (636) & 65.5 (452) & 56.7 (609) & 260/149 & $37.6\pm1.4$ & $53.7\pm1.8$ & 16.1 (11.9, 20.5) \\
$-2.8$ to $-3.0$ & $-2.90$ & 36.1 (831) & 61.8 (461) & 53.6 (767) & 271/168 & $35.7\pm1.4$ & $59.9\pm2.3$ & 24.2 (18.8, 29.5) \\
$-3.0$ to $-3.2$ & $-3.07$ & 48.7 (279) & 70.2 (191) & 46.7 (276) & 127/55 & $42.2\pm2.2$ & $59.5\pm3.6$ & 17.3 (8.5, 25.4) \\
$-3.2$ to $-3.4$ & $-3.25$ & 56.9 (51) & 44.6 (56) & 22.0 (59) & 25/30 & $47.2\pm4.6$ & $55.4\pm6.2$ & 8.2 ($-$7.2, 23.5) \\
\midrule
\multicolumn{9}{l}{Within 60\,pc, by type} \\
M7 & -- & 41.3 (189) & 69.0 (168) & 59.3 (182) & 103/48 & $39.5\pm2.1$ & $63.6\pm3.6$ & 24.1 (15.3, 31.6) \\
M8 & -- & 59.2 (98) & 68.4 (95) & 43.6 (101) & 55/29 & $43.9\pm3.4$ & $58.0\pm5.1$ & 14.1 (1.3, 24.9) \\
M9 & -- & 73.3 (45) & 58.7 (46) & 33.3 (48) & -- & -- & -- & -- \\
\midrule
\multicolumn{9}{l}{Fainter and brighter half of one type} \\
M7, brighter half & -- & 42.3 (392) & 61.0 (246) & 56.1 (374) & 138/95 & $35.5\pm1.7$ & $54.4\pm2.3$ & 19.0 (13.4, 24.2) \\
M7, fainter half & -- & 34.2 (389) & 64.2 (193) & 49.0 (353) & 123/66 & $38.8\pm2.1$ & $60.8\pm3.2$ & 22.0 (14.3, 29.4) \\
M8, brighter half & -- & 64.2 (95) & 67.9 (78) & 59.4 (96) & 49/25 & $36.5\pm3.7$ & $50.7\pm5.2$ & 14.2 (0.8, 25.6) \\
M8, fainter half & -- & 59.3 (91) & 68.8 (80) & 40.0 (95) & 54/25 & $47.8\pm3.5$ & $67.4\pm6.0$ & 19.5 (5.2, 31.8) \\
\bottomrule
\end{tabular}
\tablefoot{$L_{\mathrm{bol}}$ from the photometric bolometric flux (Sect.~\ref{subsec:halpha_method}) and the Gaia parallax, for dwarfs with an uncertainty below 0.1\,dex in $\log L_{\mathrm{bol}}$; the median $\log(L_{\mathrm{bol}}/L_\odot)$ is $-2.69$ at M6, $-2.89$ at M7, $-3.08$ at M8, $-3.20$ at M9. Fractions are followed by the number of spectra whose limit is at or below the threshold. Dispersions have bootstrap standard errors and $\Delta\sUVW$ a 95\% bootstrap interval, as in Table~\ref{tab:kin_comp}; -- marks groups with fewer than ten dwarfs in a class. The halves divide each type at its median luminosity ($\log(L_{\mathrm{bol}}/L_\odot)=-2.91$ at M7 and $-3.10$ at M8) after removing the overluminous dwarfs of Sect.~\ref{subsec:binaries}.}
\end{table*}

\section{Giant indicators}
\label{app:giants}

\begin{table}[tbp]
\caption{Thresholds of the giant/dwarf indicators (Sect.~\ref{subsec:selection}).}
\label{tab:giants}
\centering
\scriptsize
\setlength{\tabcolsep}{2pt}
\begin{tabular}{llll}
\toprule
Indicator & Dwarf & Giant & Condition \\
\midrule
$M_G$ & $> 8$ & $< 4$ & $\varpi/\sigma_\varpi \ge 5$ \\
$H_z = z + 5\log\mu + 5$ & $> 10$ & $< 6$ & $\mu$ at $\ge 3\sigma$ ($\mu$ in arcsec\,yr$^{-1}$) \\
$W1-W2$ (Vega) & $> 0.15$ & $< 0.00$ & error $< 0.1$\,mag \\
Na\,\textsc{i} 8172--8210\,\AA & $> 2.89$\,\AA & $\le 2.89$\,\AA & error $< 1.5$\,\AA \\
Ca\,\textsc{ii} 8532--8552\,\AA & $\le 1.48$\,\AA & $> 1.48$\,\AA & error $< 1.5$\,\AA \\
\bottomrule
\end{tabular}
\tablefoot{An object that meets at least two giant conditions and no dwarf condition is a likely giant. Pseudo-equivalent widths are positive for absorption and measured against a linear continuum through adjacent windows. The spectral thresholds lie midway between the 95th and 5th percentiles of DESI control samples of 170 distant red giants and 250 nearby M dwarfs. $W1-W2$ is from the Legacy Surveys or CatWISE2020.}
\end{table}

Table~\ref{tab:giants} lists the thresholds of the giant test of Sect.~\ref{subsec:selection}.

\section{Sample accounting}
\label{app:counts}

\paragraph{Type changes.} Types that differ between the template and corrected scales are earlier on the corrected scale, except for template M9 dwarfs typed L0, L0 dwarfs typed L1 and one L1 dwarf typed L2 (Table~\ref{tab:counts}). Direct selection with the corrected library from all typed spectra gives the same sample of 8961 dwarfs, including the quasar contaminants.

\begin{table}[tbp]
\caption{Type changes and survey overlaps.}
\label{tab:counts}
\centering
\scriptsize
\setlength{\tabcolsep}{3pt}
\begin{tabular}{lr}
\toprule
Step & Dwarfs \\
\midrule
\multicolumn{2}{l}{Template scale to corrected scale} \\
\quad M7 $\rightarrow$ M6 (of 2452 template M7) & 1822 \\
\quad M8 $\rightarrow$ M7 & 2280 \\
\quad M9 $\rightarrow$ M8 & 1207 \\
\quad M9 $\rightarrow$ L0 (non-giant) & 591 (590) \\
\quad L0 $\rightarrow$ L1; L1 $\rightarrow$ L2 & 33; 1 \\
\quad Fail $\Delta\chi^2 \ge 25$ on the corrected scale & 3 \\
\midrule
\multicolumn{2}{l}{\emph{Euclid} DR1 NISP} \\
\quad Matched to a NISP spectroscopic source & 186 \\
\quad In the final sample (M7--L1 / M6) & 184 (152 / 32) \\
\quad Non-quasar, NISP $S/N\ge5$ & 172 \\
\quad \quad M7--L1 / M6 / outside the final sample & 144 / 27 / 1 \\
\midrule
\multicolumn{2}{l}{SPHEREx} \\
\quad Extracted (all template-typed dwarfs) & 10\,786 \\
\quad Detected & 9618 \\
\quad \quad of the 8939 M7--L1 dwarfs & 8101 \\
\quad Detected and free of quality flags & 7705 \\
\quad \quad in the M6 and M7--L1 sample, with a band ratio & 7229 \\
\quad \quad \quad $S/N\ge10$ at 1.20--1.30\,\um\ (Fig.~\ref{fig:spherex}) & 3741 \\
\quad SPHEREx bolometric flux & 5365 \\
\quad \quad with a photometric bolometric flux as well & 4793 \\
\bottomrule
\end{tabular}
\tablefoot{SPHEREx detection rates are 98\% for $W1_{\rm Vega}<15$ and 77\% for fainter targets.}
\end{table}

\paragraph{Quasars.} Template typing over 6000--9800\,\AA\ also selects quasars at redshift 5.6--6.0, whose Lyman-$\alpha$ break mimics the steep red slope of a late-M or L dwarf, a familiar degeneracy in high-redshift quasar searches \citep{Banados2016, Fan2023}. Redrock assigns a quasar solution above redshift 5.6 to 29 of the 10\,786 spectra. Of these, 26 are targets of the DESI programme for quasars near redshift 5 \citep{Yang2023, Chaussidon2023} and one is a survey-validation target \citep{DESI_SV2024}. Twenty-seven show the break at the Redrock redshift and match a published quasar of the Million Quasars catalogue \citep{Flesch2023} at the same position and redshift. The other two, targeted for their WISE variability and typed \Lbin, have no Lyman-$\alpha$ break but broad absorption troughs of Fe\,\textsc{ii} and Mg\,\textsc{ii}. One is the SDSS quasar J173049.10+585059.5 at redshift 2.052 \citep{Lyke2020}. The other has the same spectrum at redshift 2.07 by cross-correlation and no catalogue entry, which makes it a probable new low-ionisation broad-absorption-line quasar \citep{Hall2002}. All 29 lack continuum at H$\alpha$ and a significant parallax. We omit these contaminants from the dwarf sample: 22 from M7--L1 and seven from the M6 bin.

\paragraph{New dwarfs.} We sought previous classifications in SIMBAD \citep{Wenger2000}, the UltracoolSheet \citep{UltracoolSheet, Best2021}, archival optical spectra from SDSS/BOSS and the Large Sky Area Multi-Object Fiber Spectroscopic Telescope (LAMOST) \citep{York2000, Dawson2013, Cui2012}, infrared spectra from the Apache Point Observatory Galactic Evolution Experiment (APOGEE) and SpeX \citep{Majewski2017, Burgasser2014}, sixteen catalogues of ultracool dwarfs and nearby stars including \citet{West2011}, \citet{Schmidt2010, Schmidt2015}, \citet{Smart2017}, \citet{Reyle2018}, \citet{Kiman2019}, the Gaia Catalogue of Nearby Stars \citep{GCNS2021} and the Gaia DR3 ultracool-dwarf tables \citep{Sarro2023}, and the SPIFF spectra. Of the 8939 dwarfs, 5305 appear in none of these sources. Prior spectroscopy and published types are absent for 7193 dwarfs when SPIFF is included in the comparison, or 7472 without it. The 627 L dwarfs lacking an earlier spectrum include 225 typed L on both scales. Fifty of them belong to a randomly inspected subset of the original L-dwarf selection. All have the molecular bands of a late-M or L dwarf and fit their corrected template better than M8, which puts the ultracool-dwarf purity of the 627 above 92.9\% (95\% confidence). Two show excess blue flux, a possible blend or warmer companion (\rfrac{blue_retained}).

\paragraph{\emph{Euclid} and SPHEREx counts.} Table~\ref{tab:counts} lists the \emph{Euclid} and SPHEREx subsamples by survey match and available measurements. The 152 M7--L1 dwarfs with a NISP spectrum are the open triangles of Fig.~\ref{fig:skymap}, and the 171 spectra of sample dwarfs with NISP $S/N\ge5$ are compared in Table~\ref{tab:nisp_agreement}. The additional spectrum with $S/N\ge5$ in Table~\ref{tab:counts} belongs to an object outside the final sample.

\section{Catalogue description}
\label{app:catalogue}

The catalogue of the 10\,792 spectra typed on the template scale will be made available at the Centre de Donn\'ees astronomiques de Strasbourg (CDS), with flags for the six likely giants, 29 quasars, the 8939-object M7--L1 sample and the possible flare during a DESI exposure (Appendix~\ref{app:flares}). It lists identifiers and astrometry, photometry and $S/N$, template and corrected types, prior spectra and the SPIFF match, H$\alpha$ measurements and limits (and, for the 3974 spectra of Table~\ref{tab:template_fractions}, the equivalent widths, errors and limits measured against the empirical templates), $\chiH$, radial velocities and Galactic velocities, together with the SPHEREx band ratios, bolometric fluxes, overluminosity and tail flags, BANYAN $\Sigma$ association probabilities, and companion candidates used here. \emph{Euclid} DR1 measurements will be added at the public release of \emph{Euclid} DR1.

\section{Likelihood of the lifetime model}
\label{app:like}

For dwarf $i$ with velocity $\mathbf{v}_i$ and measurement covariance $\mathbf{E}_i$, the likelihood of the activity history is
\begin{equation}
\mathcal{L}_i = \frac{\sum_t p(t)\,G_i(t)\,A_i(t)}{\sum_t p(t)\,R_i(t)\,A_i(t)}, \qquad A_i(t) = a_i\,q(t) + b_i\,[1-q(t)],
\label{eq:like}
\end{equation}
with the sums over a grid of ages. Here $p(t)$ is the age distribution (a constant star-formation rate over 0--10\,Gyr, or a brown-dwarf age distribution). The function $G_i(t)=\mathcal{N}\big(\mathbf{v}_i\,|\,(0, V_{\rm a}(t), 0),\ \boldsymbol{\Sigma}(t)+\mathbf{E}_i\big)$ is the velocity distribution of dwarfs of age $t$, with the dispersions $\boldsymbol{\Sigma}(t)={\rm diag}(\sigma_U^2,\sigma_V^2,\sigma_W^2)$ of the adopted age--velocity relation and its asymmetric drift $V_{\rm a}(t)$. The function $R_i(t)$ is the probability that a dwarf of age $t$ at the position of dwarf $i$ satisfies the velocity cut (unity without a cut). The function $q(t)$ is the probability that a dwarf of age $t$ is active, $q(t)=[t<\tau]$ for the step model. The weights are $(a_i,b_i)=(1,0)$ for an active dwarf, $(0,1)$ for an inactive one, and $(\theta_i,1)$ for a spectrum whose limit is above $\Lambda$ and that has no detected line at or above $\Lambda$, with $\theta_i$ the fraction of active dwarfs of the same type, among all its measured spectra, whose $\log(\LHa)$ lies below the limit $u_i$ of that spectrum. We estimate every lifetime by maximising $\sum_i\ln\mathcal{L}_i$ (or, for the active dispersion in Table~\ref{tab:lifetimes}, matching it) on a common grid of $\tau$ from 0.20 to 9.95\,Gyr in steps of 0.05\,Gyr (to 11.95\,Gyr when ages extend to 12\,Gyr), and quote the results to 0.1\,Gyr. The 95\% profile interval is where $2\Delta\ln\mathcal{L}\le3.84$. Because the likelihood is conditional on the activity classes, $\tau$ is constrained by their velocity differences, without a constraint from the active fractions. The unconditional fit of Table~\ref{tab:lifetimes} uses the numerator of Eq.~(\ref{eq:like}) alone. It fits the classes as well, with the observed active fraction constraining $\tau$ directly. At M8 and $\Lambda=-4.7$ it gives 7.4\,Gyr instead of 8.1\,Gyr: under uniform ages, the observed active fraction of 73\% in the $G<19$ sample itself implies $\tau\approx7.3$\,Gyr. The spectra with limits at or below the threshold are the bright, nearby and binary-rich ones that are more often active (Sect.~\ref{subsec:binaries}). We therefore adopt the conditional likelihood. For the same reason $\tau$ cannot exceed the oldest age of the population, at which no age would be left for an inactive dwarf. A value at the end of the grid, 9.95\,Gyr, means that the velocities favour activity at all ages of the population.

\paragraph{Gradual activity histories.} The step is the limit of two families that let the probability of being active decline gradually: a logistic decline $q(t)=1/\{1+\exp[(t-\tau)/w]\}$ with width $w$, and an exponential decline $q(t)=\exp(-t/\tau_{\rm e})$. For a common summary we use the mean duration of activity, $T_{\rm act}=\int_0^{10\,\rm Gyr} q(t)\,{\rm d}t$, which equals $\tau$ for a step. Table~\ref{tab:forms} lists the fits.

\begin{table}[tbp]
\caption{Thin-disk lifetimes for gradual activity histories ($G<19$).}
\label{tab:forms}
\centering
\scriptsize
\setlength{\tabcolsep}{1.8pt}
\begin{tabular}{lccccc}
\toprule
History & $\Delta\ln\mathcal{L}_{\rm M7}$ & $\Delta\ln\mathcal{L}_{\rm M8}$ & $T_{\rm act}$, M7 & $T_{\rm act}$, M8 & $\Delta T_{\rm act}$ (95\%) \\
\midrule
\multicolumn{6}{l}{M8 at $\Lambda=-4.7$} \\
\quad Step & +0.0 & +0.0 & 5.9 & 6.4 & +0.5 ($-$1.1, +2.3) \\
\quad Logistic, $w=0.25$\,Gyr & $-$0.3 & +0.0 & 5.9 & 6.4 & +0.5 \\
\quad Logistic, $w=0.5$\,Gyr & $-$1.1 & +0.1 & 5.8 & 6.3 & +0.5 \\
\quad Logistic, $w=1$\,Gyr & $-$5.2 & +0.2 & 5.7 & 6.1 & +0.4 ($-$1.3, +2.4) \\
\quad Logistic, $w=2$\,Gyr & $-$19.3 & $-$0.5 & 5.4 & 5.6 & +0.2 \\
\quad Exponential & $-$28.4 & $-$1.2 & 4.1 & 4.9 & +0.8 ($-$0.9, +3.1) \\
\multicolumn{6}{l}{M8 at $\Lambda=-4.7+\delta$} \\
\quad Step & +0.0 & +0.0 & 5.9 & 6.8 & +0.9 ($-$0.8, +3.5) \\
\quad Logistic, $w=0.25$\,Gyr & $-$0.3 & +0.0 & 5.9 & 6.8 & +0.9 \\
\quad Logistic, $w=0.5$\,Gyr & $-$1.1 & +0.0 & 5.8 & 6.7 & +0.9 \\
\quad Logistic, $w=1$\,Gyr & $-$5.2 & +0.1 & 5.7 & 6.5 & +0.8 ($-$1.0, +3.0) \\
\quad Logistic, $w=2$\,Gyr & $-$19.3 & $-$0.5 & 5.4 & 6.0 & +0.5 \\
\quad Exponential & $-$28.4 & $-$1.3 & 4.1 & 5.2 & +1.2 ($-$0.7, +3.4) \\
\bottomrule
\end{tabular}
\tablefoot{Dwarfs with $G<19$, high-velocity tail removed, uniform ages, M7 at $\Lambda=-4.7$. $\Delta\ln\mathcal{L}$: maximum log-likelihood relative to the step; $T_{\rm act}=\int_0^{10\,{\rm Gyr}}q(t)\,{\rm d}t$ at the best fit [Gyr]; $\Delta T_{\rm act}$: M8 minus M7, with a 95\% interval for the step, the 1\,Gyr logistic and the exponential from the 2000 joint bootstrap replicates of Table~\ref{tab:dtau} (the same resampled dwarfs for every history).}
\end{table}

\section{Rotation periods}
\label{app:periods}

The TESS All-Sky Rotation Survey lists 137 dwarfs of the template-typed sample. We searched public ZTF light curves with Lomb--Scargle periodograms, grading a ZTF period A when two or more bands agree within 2\% or one band agrees within 5\% with a published period, and B for a single band with no contradicting band. We calibrated the significance threshold by shuffling whole nights of each light curve to preserve the correlated noise within a night, and rejected daily aliases. Aperture photometry on TESS full-frame cut-outs adds four periods, apart from the point-spread-function search described below. Table~\ref{tab:periods} lists the period sources and their contributions to the final sample. Six dwarfs have periods attributable to neighbours in the TESS pixels and are excluded. Where two sources measure the same dwarf, the periods agree or differ by a factor of two, the usual ambiguity for spots on opposite hemispheres.

\begin{table}[tbp]
\caption{Sources of the rotation periods.}
\label{tab:periods}
\centering
\scriptsize
\begin{tabular}{lr}
\toprule
Source & Dwarfs \\
\midrule
(1) TESS All-Sky Rotation Survey (TARS) & 137 \\
ZTF periods (grades A and B) & 38 \\
\quad (2) of which not in TARS or Gaia DR3 \citep{Distefano2023} & 19 \\
TESS full-frame cut-outs, aperture photometry & 4 \\
\quad (3) of which not in the rows above & 2 \\
(4) Dwarfs with a period, (1)+(2)+(3) & 158 \\
\quad with periods from two sources & 21 \\
\quad \quad matching / differing by a factor of two & 16 / 5 \\
(5) Period from a neighbour (excluded) & 6 \\
(6) Without the TESS point-spread-function search, (4)$-$(5) & 152 \\
(7) Added by that search (one L0, one M6) & 2 \\
\quad Replaced by that search (one M6)$^a$ & 1 \\
Final sample, (6)+(7) (Fig.~\ref{fig:rotation}) & 154 \\
\bottomrule
\end{tabular}
\tablefoot{\tablefoottext{a}{The TARS period of 0.246\,d is replaced by a target-centred TESS period of 0.443\,d.}}
\end{table}

The 98 dwarfs typed L on the template scale with usable ZTF light curves show no period. Injection tests limit the fraction with fast, 0.1--0.3\,mag modulation to below 5\%. These amplitudes are far above the $\sim$1\% variability common in L dwarfs \citep{Metchev2015, Radigan2014}.

Because aperture photometry in 21\,arcsec TESS pixels is dominated by neighbours below $G\approx16.5$, we fitted each sector for the 1229 dwarfs with $G<18$ with Gaussian point-spread functions fixed at the \emph{Gaia} positions of the target and its neighbours, scaled their fluxes with isolated field stars, and searched the light curves with the shuffle-calibrated threshold used for ZTF. For 1192 dwarfs at least one sector is usable. A period counts only when the variable flux peaks on the target in a map of the pixel amplitude at that period. Injected sinusoids measure the sensitivity (Table~\ref{tab:tess_inj}). A 3\% modulation is recovered in most trials up to $G=17.5$ but in a fifth at $G=17.5$--18. A 1\% modulation is lost beyond $G=17$, where a missing period therefore says little about low-amplitude rotation. The search adds two periods and replaces one (Table~\ref{tab:periods}), and 85 of the usable dwarfs are M8 or later.

\begin{table}[tbp]
\caption{Recovery of injected modulation in the TESS point-spread-function photometry.}
\label{tab:tess_inj}
\centering
\scriptsize
\begin{tabular}{lcc}
\toprule
$G$ & 1\% [\%] (95\%) & 3\% [\%] (95\%) \\
\midrule
$<16$ & 77 (59--92) & 86 (70--98) \\
16--16.5 & 35 (18--53) & 79 (60--95) \\
16.5--17 & 14 (5--24) & 68 (50--85) \\
17--17.5 & 2 (0--4) & 76 (62--88) \\
17.5--18 & 2 (0--5) & 20 (9--33) \\
\bottomrule
\end{tabular}
\tablefoot{Fraction of injected sinusoids of 1\% and 3\% semi-amplitude recovered, over all sector extractions including failed ones and before the spatial and visual checks; 95\% target-bootstrap intervals.}
\end{table}

For a longer time baseline we requested reduced-image forced photometry in the ATLAS orange ($o$) and cyan ($c$) bands \citep{Tonry2018, Smith2020, Shingles2021} for 125 dwarfs with $G<19.5$, most of them south of the equator or without usable ZTF photometry. Of these, 123 have usable light curves: 25 M6, 81 M7, 13 M8, two M9 and two L dwarfs, with $G$ from 15.4 to 19.5 and a median of 17.4. A typical light curve has 3200 $o$ points and 860 $c$ points over 10.7\,yr. With the forced positions fixed at epoch 2021.0, proper motion lowers the flux at the early and late epochs. Usable measurements have no error flags, lie away from detector edges and have normal point-spread-function shapes. We subtracted the median of each camera in each year and clipped at five times the median absolute deviation. We searched periods from 0.05 to 30\,d with Lomb--Scargle periodograms in each band. The 0.1st percentile of the false-alarm probabilities of shuffled light curves sets the threshold, $\log_{10}{\rm FAP}<-75.9$. We rejected peaks near one day, its harmonics and the lunar period, and adopted twice the peak period when a two-harmonic model at that period improves the Bayesian information criterion by more than 10.

Two dwarfs exceed the threshold. The active M7 dwarf TARGETID 39628256382553752 has a period of 5.293\,d in the $o$ band with a semi-amplitude of 3.0\%. The period agrees within 0.6\% with its TARS value of 5.26\,d. Its nearest \emph{Gaia} neighbour brighter than $G=19$ is 49\arcsec\ away. The other dwarf has a peak at 1.03\,d, which we reject as the one-day alias. Four of the 123 dwarfs have periods among our 154 rotators. Besides the 5.29\,d dwarf, two others have their highest peaks at their known periods of 0.43 and 0.46\,d, with only $\log_{10}{\rm FAP}=-29$ and $-15$. The fourth, with a period of 2.09\,d, has its highest peak elsewhere.

We injected sinusoids into the light curves of 98 of these dwarfs, with semi-amplitudes from 0.3\% to 30\% of the flux predicted from the \emph{Gaia} colour, 12 trials per dwarf and amplitude, periods drawn log-uniformly from 0.05 to 30\,d, and the flux loss from proper motion included. A trial counts as recovered when it passes the threshold and the adopted period matches the injected period or twice it within 2\%, because strong sinusoids often meet the criterion for doubling. Modulation of 10\% is recovered in 76--90\% of the trials from $G=16$ to 18.5, and 30\% in at least 78\% of the trials at every magnitude. A dwarf without an ATLAS period can still have a modulation of a few per cent. At lower amplitudes, 3\% is recovered in about half of the trials brighter than $G=17$ and in 5\% at $G=17.5$--18. A 1\% modulation is never recovered (Table~\ref{tab:atlas_inj}).

\begin{table}[tbp]
\caption{Recovery of injected modulation in the ATLAS forced photometry.}
\label{tab:atlas_inj}
\centering
\scriptsize
\begin{tabular}{lcccc}
\toprule
$G$ & $N$ & 3\% [\%] (95\%) & 10\% [\%] (95\%) & 30\% [\%] (95\%) \\
\midrule
16--16.5 & 14 & 51 (29--73) & 76 (62--85) & 84 (70--93) \\
16.5--17 & 16 & 47 (28--67) & 82 (64--95) & 78 (61--91) \\
17--17.5 & 20 & 33 (16--51) & 81 (69--92) & 86 (75--93) \\
17.5--18 & 20 & 5 (1--10) & 90 (86--94) & 90 (86--94) \\
18--18.5 & 15 & 0 & 86 (72--94) & 92 (88--94) \\
18.5--19 & 4 & 0 & 62 (25--100) & 92 (83--100) \\
19--19.5 & 8 & 0 & 5 (0--16) & 94 (89--98) \\
\bottomrule
\end{tabular}
\tablefoot{Fraction of injected sinusoids of each semi-amplitude recovered at the injected period or twice it, in $N$ dwarfs with 12 trials per amplitude, before the checks on neighbours and the visual inspection; 95\% target-bootstrap intervals. No injected 1\% modulation was recovered. The one dwarf brighter than $G=16$ is omitted.}
\end{table}

\section{Flares during the DESI exposures}
\label{app:flares}

We timed each DESI exposure of the sample dwarfs using the DESI DR1 exposure start times and durations and compared it with ZTF light curves \citep{Bellm2019, Masci2019} and with TESS light curves from the point-spread-function fits of Appendix~\ref{app:periods} and from the TESS--Gaia light curves \citep{Han2023}. For ZTF we used every measurement that passes the instrumental quality cuts, without the outlier rejection of the rotation search, which would remove flares, and estimated a baseline for each dwarf and band. A TESS sector is used only when the fit is reliable, with positive target flux and small covariance with the neighbours. A spectrum is consistent with the baseline when every matched point lies within $3\sigma$ of it, with $\sigma$ the larger of the light-curve scatter and the point error. Within 12\,h of an exposure the fraction of baseline-consistent spectra is about the same as within 1\,h, and it declines slowly at longer separations, where rotation and other variability add scatter.

\begin{figure*}[tbp]
\centering
\includegraphics[width=\textwidth]{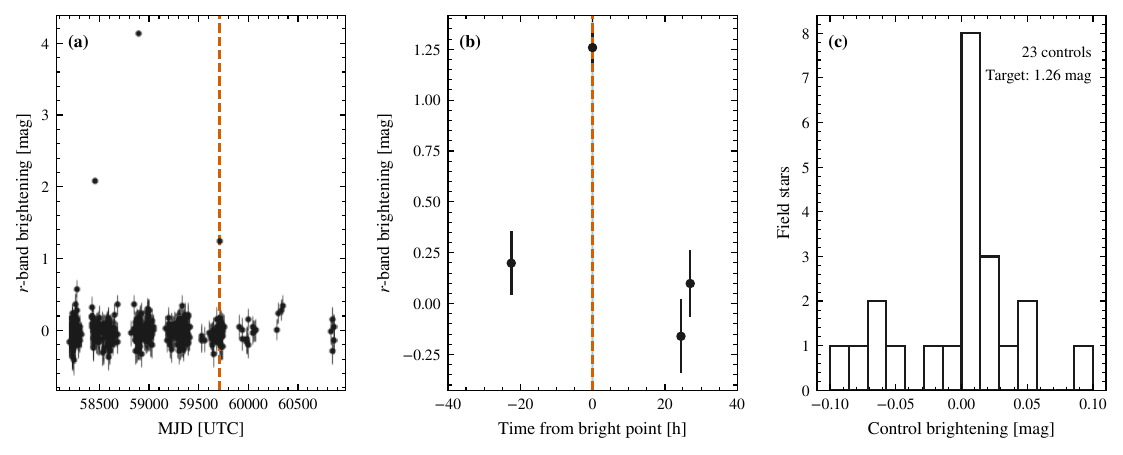}
\caption{ZTF $r$-band light curve of the M7 dwarf Gaia DR3 1581886792134517504. (a) Brightening relative to the dwarf's median magnitude over the whole survey; the dashed line marks the DESI exposure. (b) The 80\,h around the bright point; the DESI integration is narrower than the dashed line. (c) Offsets of the 23 field stars measured in the same ZTF exposure.}
\label{fig:ztf_lc}
\end{figure*}

\begin{figure}[tbp]
\centering
\includegraphics[width=\columnwidth]{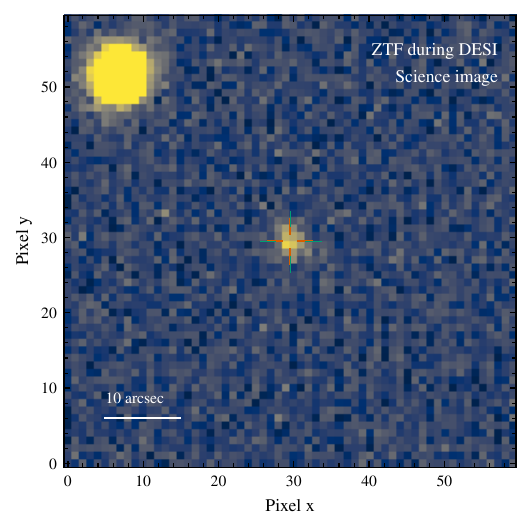}
\caption{ZTF $r$-band science image taken during the DESI exposure of Gaia DR3 1581886792134517504. Crosshairs with open centres mark the frame centre (red) and the target's position propagated with its proper motion (green).}
\label{fig:ztf_img}
\end{figure}

The M7 dwarf Gaia DR3 1581886792134517504 (DESI TARGETID 39633407788975012) has an H$\alpha$ equivalent width of $25.4\pm1.5$\,\AA. A ZTF $r$-band measurement taken during its DESI exposure is 1.26\,mag above its baseline, ten times the scatter (Fig.~\ref{fig:ztf_lc}). The 23 field stars measured in the same ZTF exposure have a median offset of 0.006\,mag and none exceeds $3\sigma$. With no reference image available for a difference image, we inspected the science image, which shows a point-like source at the target's propagated position (Fig.~\ref{fig:ztf_img}). Because the 30\,s ZTF exposure covers 3\% of the DESI integration, the duration of the event and its contribution to the coadded line are unknown. The dwarf has a resolved flare in TESS sector 14 at another epoch and two earlier ZTF points 2.1 and 4.1\,mag above its baseline (Fig.~\ref{fig:ztf_lc}a), which we did not vet in the same way; together they make a flare the likely cause. The dwarf keeps its activity classification and has a possible-flare flag in the catalogue. Its contribution to the pooled 5\,\AA\ active fraction is the difference between 804/1631 with it and 803/1630 without it.

We searched the reliable TESS sectors of the M7--L1 dwarfs for flare candidates: at least three points above $3\sigma$, consecutive apart from up to two missing cadences, a peak above $5\sigma$, baseline on both sides and a decay longer than the rise. Because the decay of a large flare can stay 2--3$\sigma$ above the baseline for hours, we followed each event until the next three points have a median below $2\sigma$, up to 0.5\,d after its last $3\sigma$ point, before requiring the baseline after it. Fits to the difference images assign each candidate to the target or a \emph{Gaia} neighbour. Each remaining candidate received two independent visual inspections and we count those that both accept, 35 candidates in 18 dwarfs (examples in Fig.~\ref{fig:flare_examples}). The broader set of candidates accepted by the first inspection and rejected by neither contains 44 in 24 dwarfs. In the pairs matched at 5\,\AA, this set gives 25 flares in active dwarfs and three in inactive dwarfs ($p=0.0005$). Table~\ref{tab:flares} gives the rates of the candidates accepted by both inspections. The efficiency of identifying each event's source from the pixels and of visual inspection is not calibrated and may differ between the groups even though their temporal recovery is the same. The rates in Table~\ref{tab:flares} are therefore lower limits to the true flare rates. The factor of about 9 between the detected rates of active and inactive dwarfs is therefore not a calibrated ratio of true flare rates. Flares injected into the real light curves, with a linear rise over $0.2\,t_{\rm d}$ and an exponential decay of time-scale $t_{\rm d}$ at random times, satisfy the temporal criteria in at most 81\% of trials up to peaks three times the quiescent flux. Their recovery rates are nearly the same in active and inactive dwarfs (Fig.~\ref{fig:flare_inj}).

\begin{figure*}[tbp]
\centering
\includegraphics[width=\textwidth]{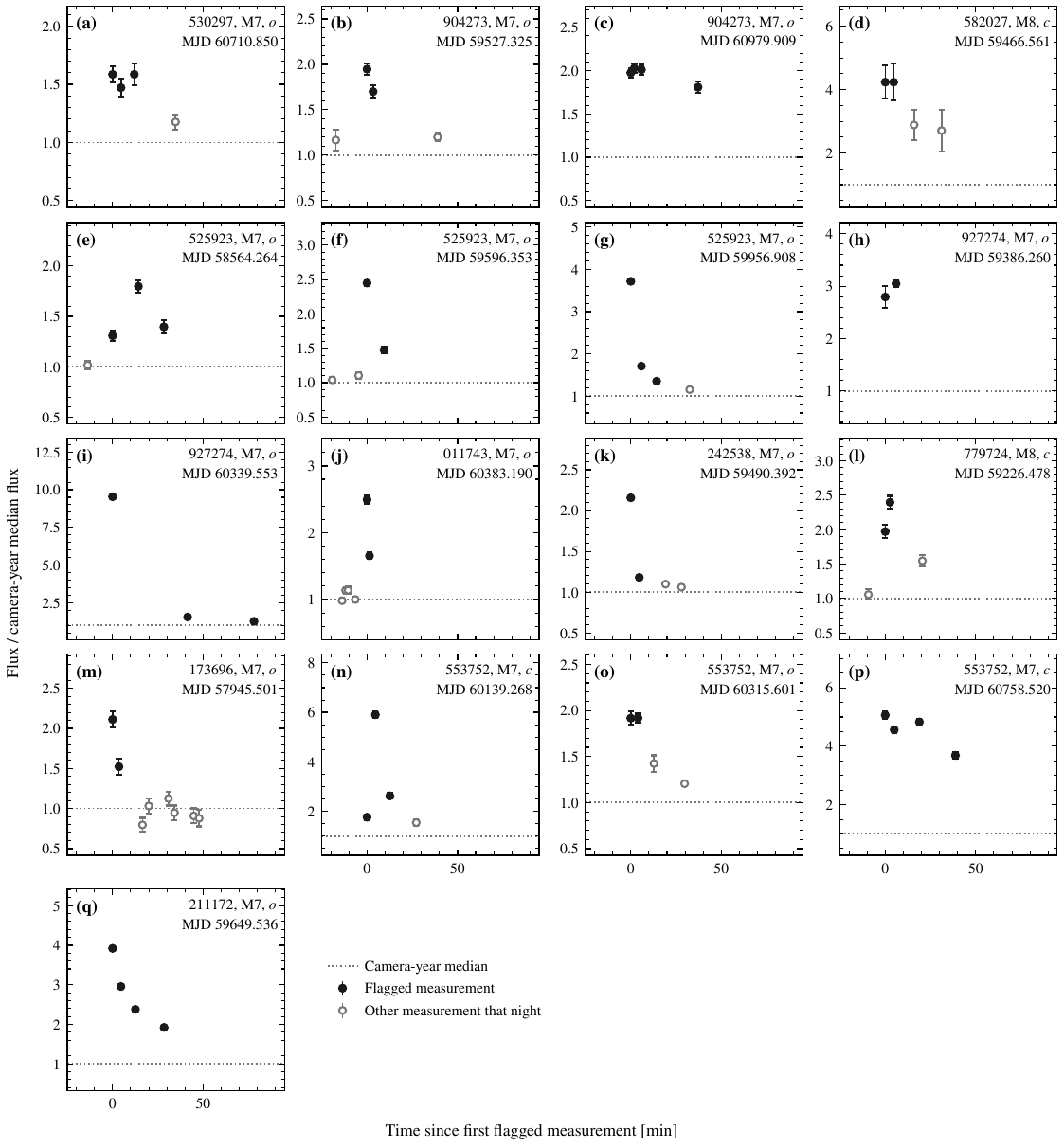}
\caption{The 17 flares of Table~\ref{tab:atlas_flares}, in the order of that table, labelled by the last six digits of the DESI TARGETID, the type and the band. Each panel shows the measurements of that night in the band of the flare, relative to the median flux of that camera in that year. Filled circles satisfy the flagging criteria of the text and open circles are the other measurements. Most nights have about four measurements in 30\,min, so few events have both a resolved rise and a decay. In panel (c) the later measurements of that night fall outside the time range shown.}
\label{fig:atlas_flares}
\end{figure*}

The ATLAS light curves of Appendix~\ref{app:periods} sample each dwarf about four times per clear night for 10.7\,yr, without the outlier clipping used for the period search. We flagged nights with at least two measurements of a dwarf in one band more than $5\sigma$ above the median of that camera in that year, with $\sigma$ the larger of the measurement error and the scatter, each detected at a signal-to-noise ratio of at least 5, and with a median reduced $\chi^2$ of the point-spread-function fit below 5, which gave 46 flagged nights. To identify instrumental brightenings that affect several dwarfs at once, we compared each flagged night with the other dwarfs measured by the same camera within 0.08\,d. We also searched for known asteroids within 60\arcsec\ with SkyBoT \citep{Berthier2006} and inspected the light curve of each flagged night. The inspection was made knowing the activity class of each dwarf, and the efficiency of these criteria was not measured. We therefore quote no ATLAS flare rate. We accept 17 flares on 11 dwarfs, with brightenings of 0.5 to 2.45\,mag (Table~\ref{tab:atlas_flares} and Fig.~\ref{fig:atlas_flares}). In the 14 of these with comparison dwarfs, one other dwarf brightened at the same time on one night only. For the three flares on TARGETID 39628256382553752 no other dwarf was measured, leaving the light-curve shape and the point-spread-function fits as the evidence. The only asteroid found near any of the 17 lies 57\arcsec\ away and moved too little to reach the dwarf. The flare at MJD 60979.9 on TARGETID 39627456142904273 is also seen by TESS, which shows a fivefold brightening within minutes and a decay over about 8\,h. ATLAS, 5\,h after the start, finds twice the quiescent flux in $o$ while the TESS band is at 1.27 times. Ten of the 11 dwarfs have H$\alpha$ equivalent widths above 5\,\AA, as do 57 of the 123 dwarfs with ATLAS light curves, and the eleventh, with 3.1\,\AA, is active at $\Lambda=-4.7$.

\begin{table}[tbp]
\caption{Flares in the ATLAS light curves.}
\label{tab:atlas_flares}
\centering
\scriptsize
\setlength{\tabcolsep}{2.5pt}
\begin{tabular}{lccccccc}
\toprule
DESI TARGETID & Type & $G$ & EW(H$\alpha$) [\AA] & Band & MJD & $\Delta m$ [mag] & $N_{\rm cmp}$ \\
\midrule
39627352002530297 & M7 & 17.64 & 3.1 & $o$ & 60710.850 & $-0.50$ & 17 (0) \\
39627456142904273 & M7 & 17.67 & 6.6 & $o$ & 59527.325 & $-0.72$ & 29 (0) \\
39627456142904273 & M7 & 17.67 & 6.6 & $o$ & 60979.909 & $-0.77$ & 11 (0) \\
39627509691582027 & M8 & 17.90 & 16.0 & $c$ & 59466.561 & $-1.57$ & 34 (0) \\
39627545372525923 & M7 & 16.44 & 17.6 & $o$ & 58564.264 & $-0.64$ & 10 (1) \\
39627545372525923 & M7 & 16.44 & 17.6 & $o$ & 59596.353 & $-0.97$ & 42 (0) \\
39627545372525923 & M7 & 16.44 & 17.6 & $o$ & 59956.908 & $-1.43$ & 12 (0) \\
39627600598927274 & M7 & 17.00 & 8.4 & $o$ & 59386.260 & $-1.21$ & 13 (0) \\
39627600598927274 & M7 & 17.00 & 8.4 & $o$ & 60339.553 & $-2.45$ & 5 (0) \\
39627673584011743 & M7 & 17.31 & 5.5 & $o$ & 60383.190 & $-0.99$ & 32 (0) \\
39627766345242538 & M7 & 16.39 & 7.5 & $o$ & 59490.392 & $-0.83$ & 27 (0) \\
39627799081779724 & M8 & 16.10 & 12.4 & $c$ & 59226.478 & $-0.95$ & 17 (0) \\
39627892753173696 & M7 & 17.51 & 22.7 & $o$ & 57945.501 & $-0.81$ & 15 (0) \\
39628256382553752 & M7 & 16.68 & 15.5 & $c$ & 60139.268 & $-1.93$ & -- \\
39628256382553752 & M7 & 16.68 & 15.5 & $o$ & 60315.601 & $-0.71$ & -- \\
39628256382553752 & M7 & 16.68 & 15.5 & $c$ & 60758.520 & $-1.76$ & -- \\
39633068155211172 & M7 & 17.38 & 9.6 & $o$ & 59649.536 & $-1.48$ & 2 (0) \\
\bottomrule
\end{tabular}
\tablefoot{MJD is the first flagged measurement. $\Delta m$ is the brightest flagged measurement relative to the median of that camera in that year. $N_{\rm cmp}$ counts the other dwarfs measured by the same camera in the same band within 0.08\,d; the number in parentheses brightened by at least 20\% and $4\sigma$, and a dash means no other dwarf was measured.}
\end{table}

\begin{table*}[tbp]
\caption{TESS flare candidates in M7--L1 dwarfs by H$\alpha$ activity.}
\label{tab:flares}
\centering
\scriptsize
\setlength{\tabcolsep}{2.5pt}
\begin{tabular}{llrrrc}
\toprule
Active if & Group & Dwarfs & Days & Flares & Rate [(1000\,d)$^{-1}$] (95\%) \\
\midrule
$\EW\ge5$\,\AA & Active & 93 & 5853 & 32 & 5.47 (3.74--7.72) \\
  & Inactive & 93 & 5148 & 3 & 0.58 (0.12--1.70) \\
$\log(\LHa)\ge-4.7$ & Active & 112 & 6607 & 33 & 4.99 (3.44--7.01) \\
  & Inactive & 46 & 2622 & 1 & 0.38 (0.01--2.13) \\
\midrule
\multicolumn{6}{l}{Pairs matched in type and Gaia $G$ (active / inactive)} \\
$\EW\ge5$\,\AA & 82 pairs & & 5068 / 4643 & 18 / 3 & difference 2.91 (0.90--5.29), $p=0.010$ \\
$\log(\LHa)\ge-4.7$ & 41 pairs & & 2477 / 2414 & 8 / 1 & difference 2.82 (0.41--5.48), $p=0.040$ \\
\bottomrule
\end{tabular}
\tablefoot{Candidates that pass the temporal, pixel-attribution and visual checks of Appendix~\ref{app:flares} in TESS sectors with reliable point-spread-function photometry; days: summed monitored time. Rates are detected-candidate rates, not corrected for the incomplete recovery of Fig.~\ref{fig:flare_inj}, with 95\% Poisson intervals. Matched pairs: same corrected type and $G$ within 0.5\,mag; the rate difference has a 95\% interval from resampling whole pairs and $p$ is a paired permutation probability.}
\end{table*}

\begin{figure*}[tbp]
\centering
\includegraphics[width=\textwidth]{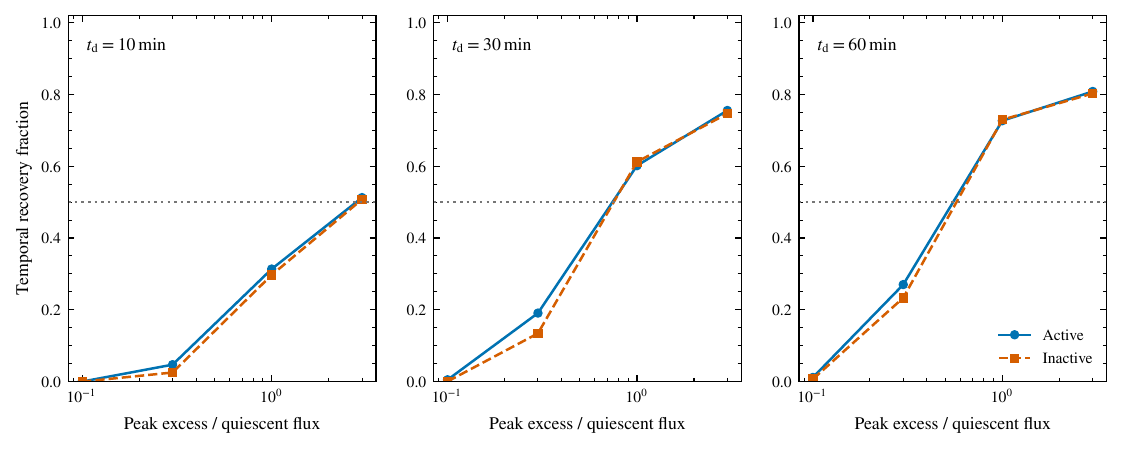}
\caption{Recovery of flares injected into the TESS light curves of M7--L1 dwarfs that are active (filled circles, solid) or inactive (squares, dashed) at 5\,\AA, after cadence averaging and re-estimation of the baseline, as a function of the peak excess over the quiescent flux, for decay times $t_{\rm d}$ of 10, 30 and 60\,min. Only the temporal criteria are applied. The dotted line marks 50\% recovery.}
\label{fig:flare_inj}
\end{figure*}

\section{Measurements of individual dwarfs}
\label{app:individual}

\begin{figure*}[tbp]
\centering
\includegraphics[width=\textwidth]{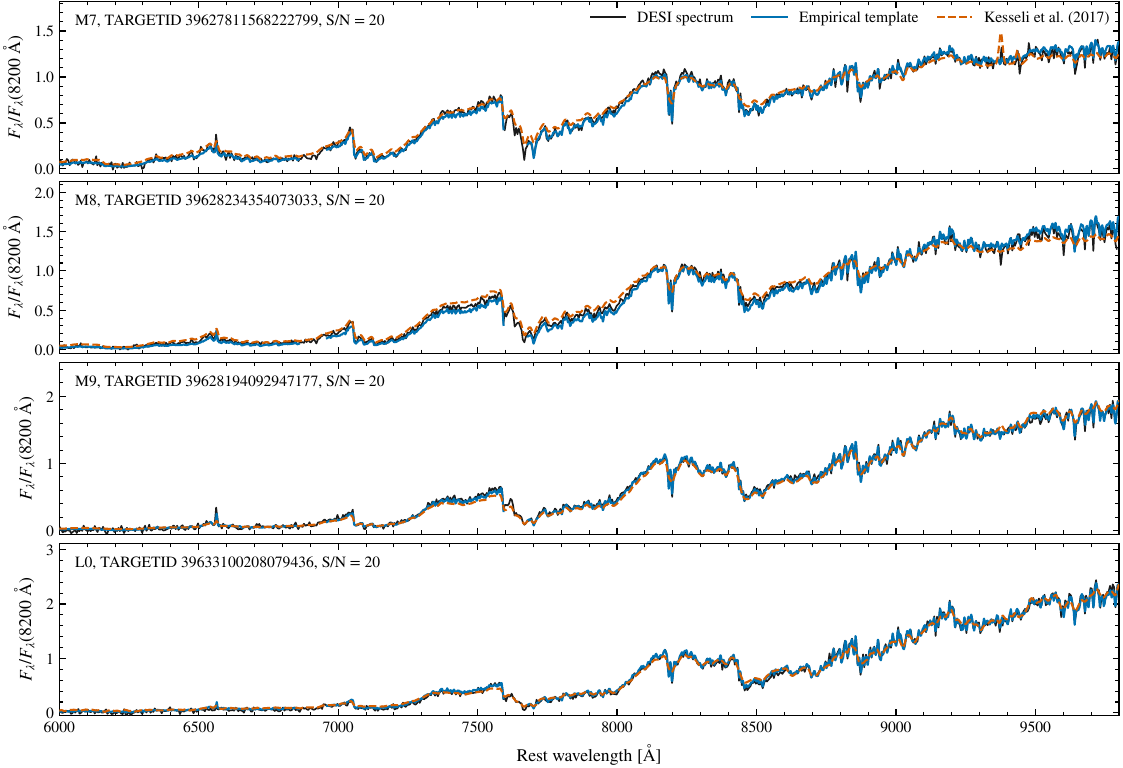}
\caption{DESI spectra (solid black) of analysis-sample dwarfs not previously known as ultracool dwarfs, averaged over 5\,\AA, at the $S/N$ closest to 20 at 8000--9000\,\AA, with the empirical template of the corrected type (blue) and the \citet{Kesseli2017} template of that type (dashed vermilion), each scaled to the spectrum.}
\label{fig:spectra_examples}
\end{figure*}

\begin{figure*}[tbp]
\centering
\includegraphics[width=\textwidth]{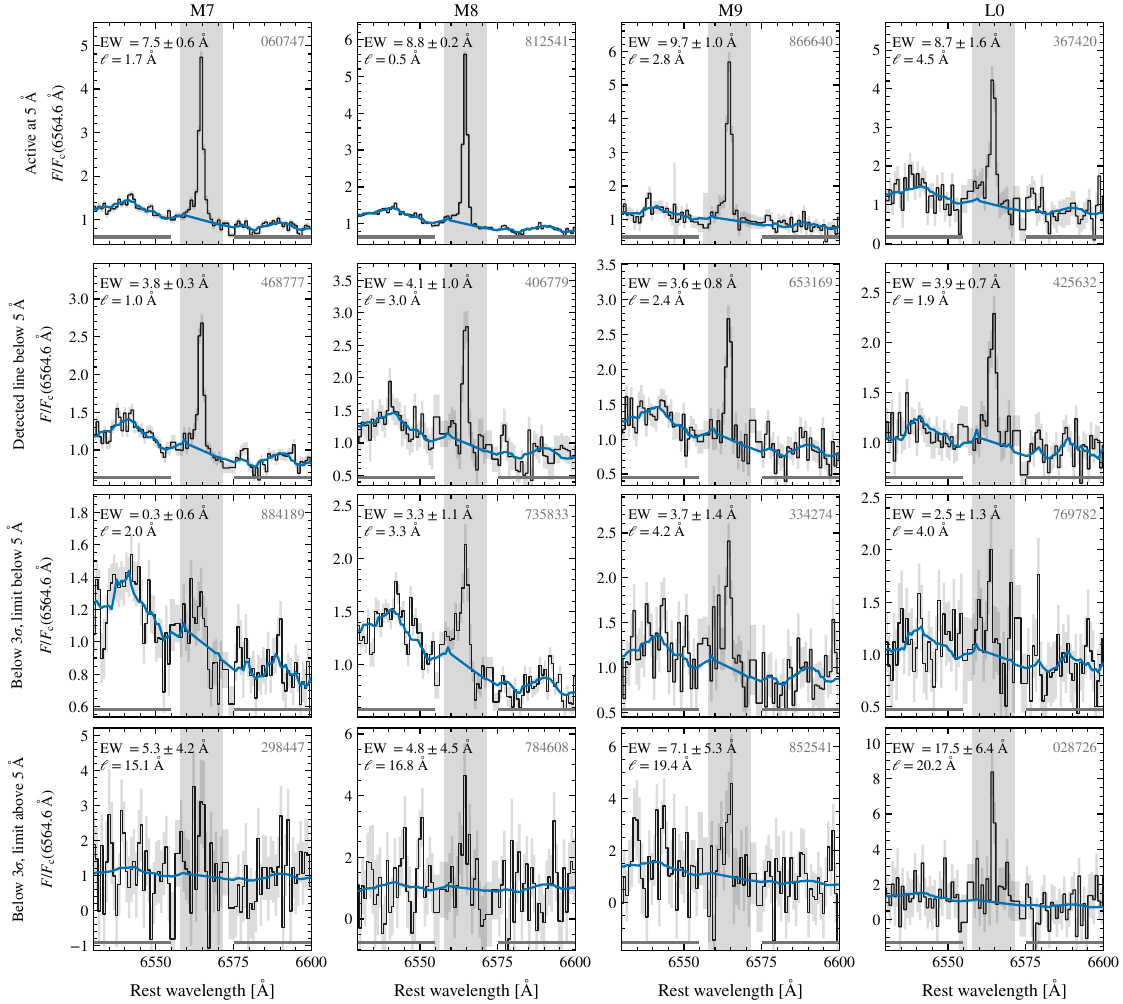}
\caption{H$\alpha$ measurements on native DESI pixels, by corrected type (columns) and outcome at $D=5$\,\AA\ (rows). Top row: active. Second row: inactive with a $3\sigma$ line weaker than 5\,\AA. Third row: inactive, line below $3\sigma$. Bottom row: line below $3\sigma$ and $\ell > 5$\,\AA, so the spectrum does not enter the 5\,\AA\ fraction. Each spectrum has an equivalent width (upper two rows) or limit (lower two rows) close to the median of its cell. Black: flux with a band for the pixel errors, divided by the continuum at 6564.6\,\AA. Blue: adopted continuum. Grey band: line window. Grey bars: sidebands. The last six digits of the TARGETID are at upper right.}
\label{fig:halpha_examples}
\end{figure*}

Figure~\ref{fig:spectra_examples} shows spectra at $S/N \approx 20$ with both template sets. Figure~\ref{fig:halpha_examples} shows the H$\alpha$ measurement at each type, one spectrum for each outcome of the classification at $D=5$\,\AA. Several spectra in its lower two rows have a peak at H$\alpha$ that does not reach $3\sigma$, which is why we define inactive by the threshold.

\begin{figure*}[tbp]
\centering
\includegraphics[width=\textwidth]{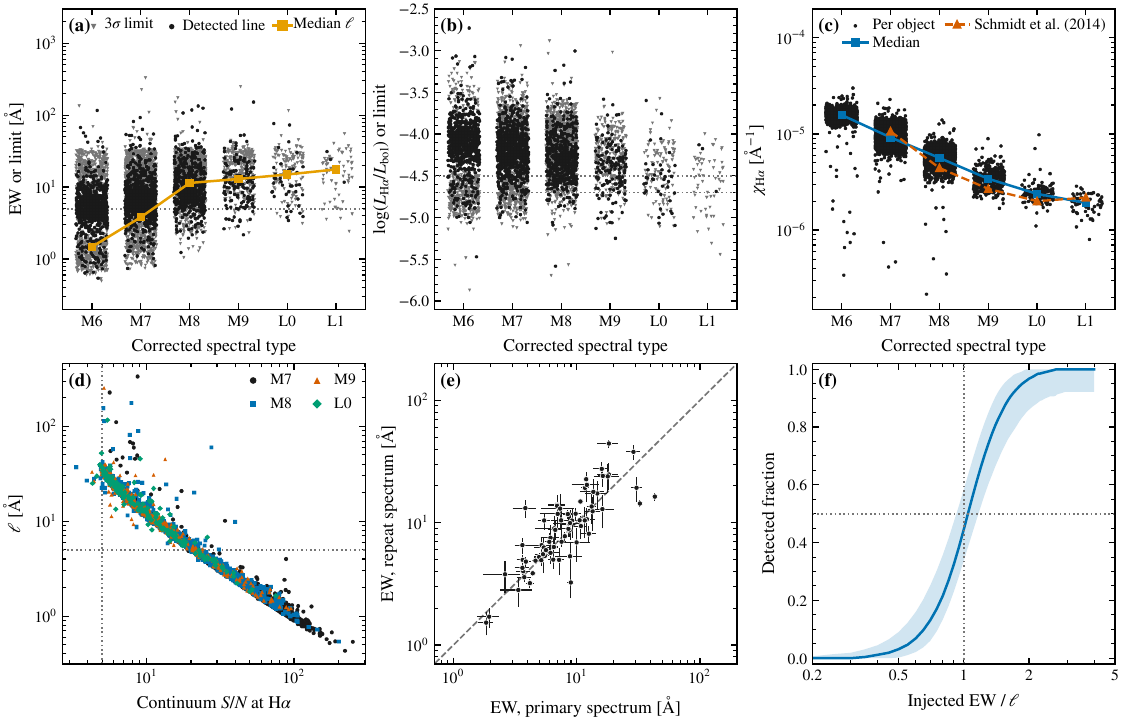}
\caption{H$\alpha$ measurements of the M6--L1 analysis sample and its tests. (a) Equivalent width of $3\sigma$ lines (circles) and limits of the other spectra (triangles), offset horizontally within each type, with the median limit of each type; the dotted line is 5\,\AA. (b) The same for $\log(\LHa)$; dotted lines are $\Lambda=-4.7$ and $-4.5$. (c) $\chiH$ for each object, the median at each type and the values of \citet{Schmidt2014}, which start at M7. (d) Limit as a function of continuum $S/N$; dotted lines mark 5\,\AA\ and the minimum $S/N$ of 5. (e) The 71 pairs of DESI spectra with a $3\sigma$ line in both epochs; the dashed line is equality. (f) Median fraction of injected lines detected as a function of strength relative to $\ell$, with the 16th--84th percentile range among spectra.}
\label{fig:halpha_individual}
\end{figure*}

\begin{figure*}[tbp]
\centering
\includegraphics[width=\textwidth]{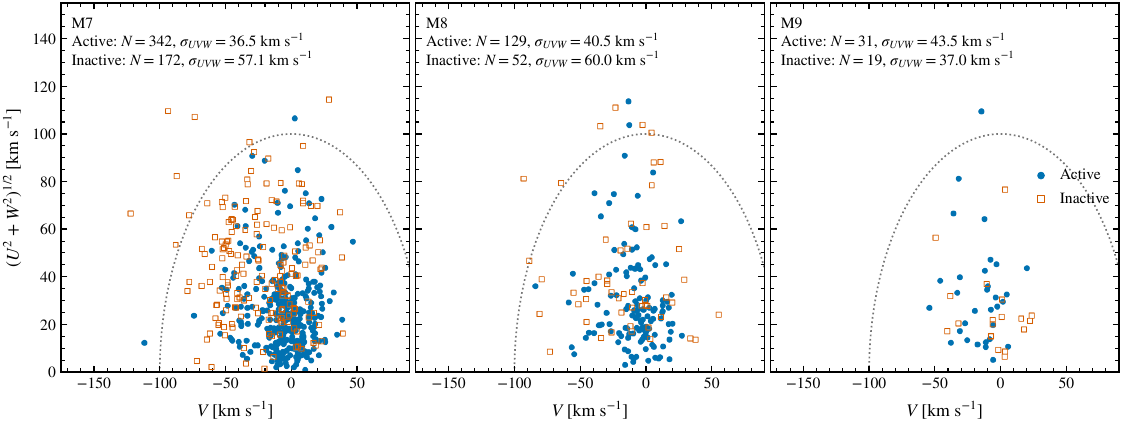}
\caption{Space motions relative to the local standard of rest for dwarfs with $G<19$, classed at $\Lambda=-4.7$: active (filled circles) and inactive (open squares). Dotted curves mark a total speed of 100\,\kms. Numbers and dispersions are those of the $G<19$ sample in the main text.}
\label{fig:uvw_individual}
\end{figure*}

The detection limit depends on the continuum $S/N$ with little dependence on type (Fig.~\ref{fig:halpha_individual}d). Its median rises toward later types (Fig.~\ref{fig:halpha_individual}a) because their spectra have lower $S/N$ at H$\alpha$. The median $\chiH$ at each type lies within 30\% of the value of \citet{Schmidt2014} (Fig.~\ref{fig:halpha_individual}c). In Fig.~\ref{fig:uvw_individual} the inactive M7 and M8 dwarfs reach larger velocities and lag in $V$, which gives their higher dispersion. The M9 groups are small and show no such difference.

\begin{figure*}[tbp]
\centering
\includegraphics[width=\textwidth]{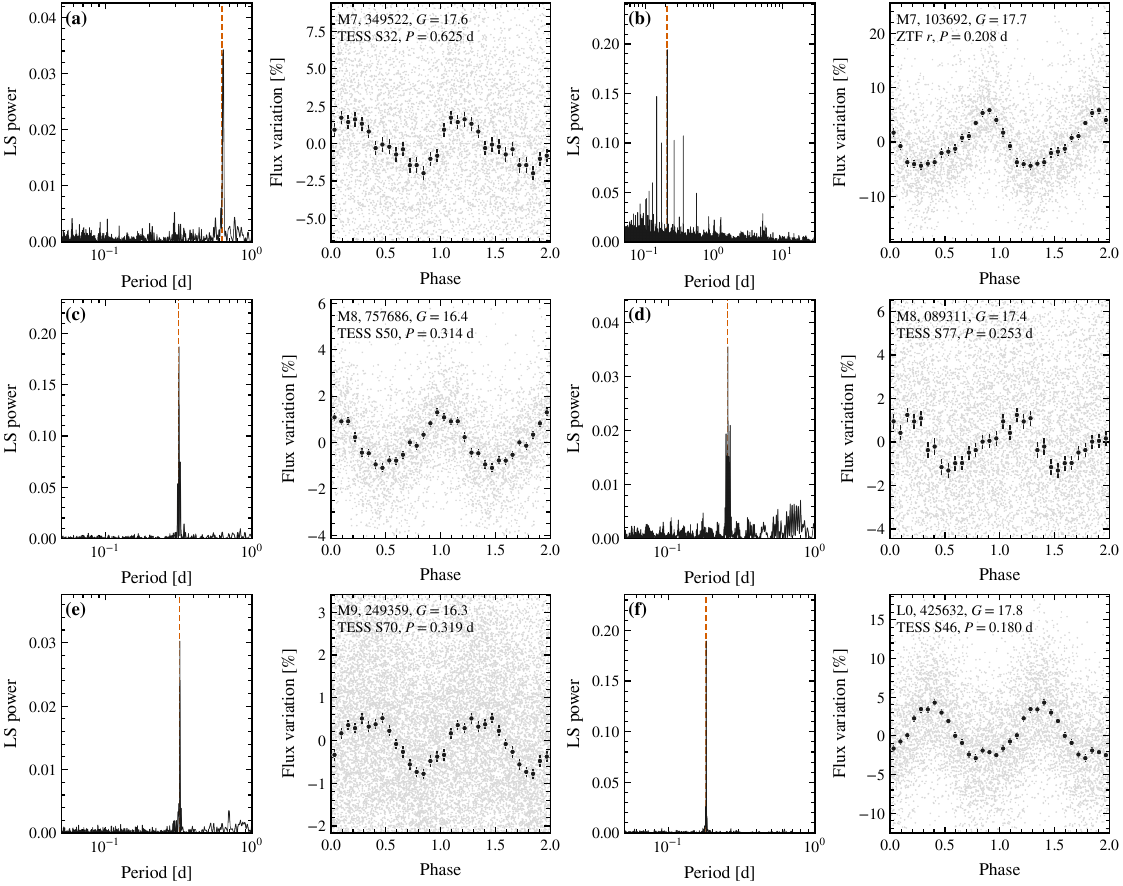}
\caption{Lomb--Scargle periodograms and folded light curves of six rotators, chosen to cover types and period sources. For each dwarf, left: periodogram, with the adopted period dashed. Right: light curve folded at that period and plotted twice (grey), with medians in 16 phase bins (black). TESS panels use the point-spread-function photometry with a one-day running median removed, which suits periods below 1\,d; a search of the same light curves without this filter covered 0.05--13\,d; their flux is relative to the target flux predicted from \emph{Gaia} photometry and the zero point of isolated field stars. The ZTF panel is relative to the median flux, and the other peaks in its periodogram are daily aliases. Panel (f) is the L0 dwarf 2MASS J12212770+0257198.}
\label{fig:rotation_examples}
\end{figure*}

\begin{figure}[tbp]
\centering
\includegraphics[width=\columnwidth]{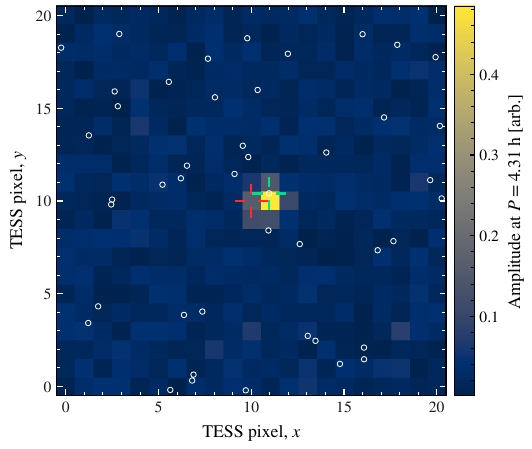}
\caption{Amplitude of each TESS pixel (21\arcsec) at the 4.31\,h period of the L0 dwarf 2MASS J12212770+0257198 in sector 46. Crosshairs with open centres mark the frame centre (red) and the position of the target (green); white rings mark \emph{Gaia} neighbours included in the fit.}
\label{fig:rotation_map}
\end{figure}

\begin{figure*}[tbp]
\centering
\includegraphics[width=\textwidth]{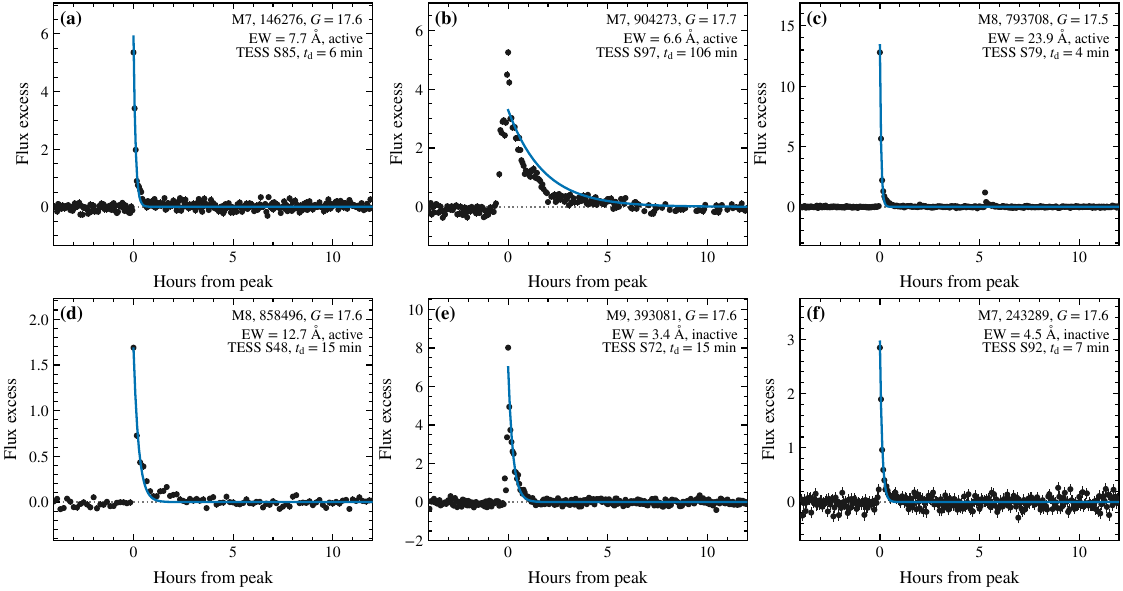}
\caption{TESS flare candidates: the strongest in two active M7 dwarfs (a, b) and two active M8 dwarfs (c, d), the only candidate on an M9 dwarf (e) and the strongest on an M7 dwarf inactive at 5\,\AA\ (f). Flux excess above the local baseline is relative to the target flux predicted from \emph{Gaia} photometry and the zero point of isolated field stars. Blue curves show the decay of the fitted flare profile, with its time-scale $t_{\rm d}$ given in each panel.}
\label{fig:flare_examples}
\end{figure*}

Figure~\ref{fig:rotation_examples} shows periods at M7--L0 from TESS and ZTF. The variable flux of the L0 dwarf peaks 0.41 TESS pixels (9\arcsec) from its \emph{Gaia} position (Fig.~\ref{fig:rotation_map}). Of the 35 TESS flare candidates, 18 are on M7 dwarfs, 16 on M8 dwarfs and one on an M9 dwarf, and one M8 dwarf, TARGETID 39633532661793708, has 13 of them (Fig.~\ref{fig:flare_examples}).

\end{appendix}

\end{document}